\documentclass[11pt, a4paper]{article}
\usepackage[T1]{fontenc}
\usepackage{a4wide, url}
\usepackage[english]{babel}
\usepackage{graphicx}
\usepackage{array}
\usepackage{natbib}
\usepackage[textfont=sc,labelfont=bf]{caption}
\usepackage{setspace}
\usepackage{multirow}
\usepackage{amsthm}
\usepackage{amsmath}
\usepackage{amsfonts}
\usepackage{amssymb}
\usepackage{bigints}
\usepackage{pgf} 
\usepackage{bm}
\usepackage{paralist}
\usepackage{hyperref}
\usepackage{pdflscape}
\usepackage{xfrac}
\usepackage[linesnumbered,ruled,vlined]{algorithm2e}
\SetKwInput{KwInput}{Input}                
\SetKwInput{KwOutput}{Output}              
\usepackage{enumitem,kantlipsum}
\usepackage{xr}
\usepackage{yfonts}
\usepackage{rotating}

\DeclareSymbolFont{YHlargesymbols}{OMX}{yhex}{m}{n}
\DeclareMathAccent{\wideparen}{\mathord}{YHlargesymbols}{"F3}

\usepackage{xcolor}
\hypersetup{
    colorlinks,
    linkcolor={red!50!black},
    citecolor={blue!50!black},
    urlcolor={blue!80!black}
}

\usepackage{comment}
\excludecomment{ext}\excludecomment{comment}

\usepackage[top=1in,bottom=1in,left=.75in,right=.75in]{geometry}

\usepackage{tabu}

\newcolumntype{L}[1]{>{\raggedright\let\newline\\\arraybackslash\hspace{0pt}}m{#1}}
\newcolumntype{C}[1]{>{\centering\let\newline\\\arraybackslash\hspace{0pt}}m{#1}}
\newcolumntype{R}[1]{>{\raggedleft\let\newline\\\arraybackslash\hspace{0pt}}m{#1}}

\DeclareMathAlphabet\mathbfcal{OMS}{cmsy}{b}{n}

\newcommand{\mbf}{\mathbf}
\newcommand{\beq}{\begin{equation}}
\newcommand{\eeq}{\end{equation}}
\newcommand{\bea}{\begin{eqnarray}}
\newcommand{\eea}{\end{eqnarray}}
\newcommand{\ba}{\begin{array}}
\newcommand{\ea}{\end{array}}
\newcommand{\bit}{\begin{itemize}}
\newcommand{\eit}{\end{itemize}}
\newcommand{\ben}{\begin{enumerate}} 
\newcommand{\een}{\end{enumerate}}
\newcommand{\bpm}{\begin{pmatrix}}
\newcommand{\epm}{\end{pmatrix}}
\newcommand{\bbm}{\begin{bmatrix}}
\newcommand{\ebm}{\end{bmatrix}}

\renewcommand{\l}{\left}
\renewcommand{\r}{\right}

\newcommand{\E}[0]{\mathbb{E}}
\newcommand{\Var}[0]{\mathbb{V}\mathrm{ar}}

\newcommand{\nn}{\nonumber}
\newcommand{\wh}{\widehat}
\newcommand{\wt}{\widetilde}

\newtheorem{ass}{Assumption}
\newtheorem{theorem}{Theorem}

\newtheorem{prop}{Proposition}
\newtheorem{rem}{Remark}
\newtheorem{lem}{Lemma}

\newtheorem*{asso}{Assumption $\mathcal O$}

\makeatletter
\newcommand{\labelO}[1]{%
  \phantomsection
  \def\@currentlabel{\ensuremath{\mathcal O}}%
  \label{#1}%
}
\makeatother

\graphicspath{{Figures/Submission/}}

\title{\textsc{\large 	
Principal Component Analysis \\ \vskip .3cm
for High-Dimensional Approximate Factor Models in Time Series:\\ 
Assumptions, Asymptotic Theory, and Identification  }}
\date{ }
\author{}

\begin{document}
\maketitle

\begin{center}\vspace{-1.5cm}
 Matteo Barigozzi$^\dag$ \\[.1cm]

\small This version: \today
\end{center}

\begin{abstract}



We consider estimation of large approximate factor models in high-dimensional panels of stationary time series using Principal Component Analysis (PCA). We review the key results establishing the necessary and sufficient conditions for consistency and asymptotic normality of the estimators, which hold when both the cross-sectional dimension $n$ and the sample size $T$ tend to infinity. Special emphasis is placed on identification. First, we show that the common and idiosyncratic components are identified only in the limit  $n\to\infty$. Second, we discuss the restrictions required to uniquely determine factors and loadings and examine their consequences for statistical inference. 

\vspace{0.5cm}


\end{abstract}

\renewcommand{\thefootnote}{$\dag$} 
\thispagestyle{empty}

\footnotetext{Department of Economics - Universit\`a di Bologna. Email: matteo.barigozzi@unibo.it 
\\
I thank Haeran Cho, Philipp Gersing, Daniele Massacci, and Esther Ruiz for helpful comments and discussions.} 

\renewcommand{\thefootnote}{\arabic{footnote}}

\section{Introduction}

Consider a $T$-dimensional realization of $n$ zero-mean time series: $\{x_{it},\,i=1,\ldots, n,\, t=1,\ldots, T\}$.  We say that $x_{it}$ follows an $r$-factor model if
\begin{align}
x_{it}&=\bm\lambda_i^\prime \mbf F_t +  e_{it}, \quad i=1,\ldots, n, \quad t=1,\ldots, T,\label{eq:SDFM1R_base}
\end{align}
where $\bm\lambda_i:=(\lambda_{i1}\cdots\lambda_{ir})^\prime$ and $\mbf F_t:=(F_{1t}\cdots F_{rt})^\prime$ are the $r$-dimensional vectors of loadings for series $i$ and of factors at time $t$, respectively, and $r\ll \min(n,T)$. We call $ e_{it}$ the idiosyncratic component and 
${C}_{it}:=\bm \lambda_i^\prime \mbf F_t$ the common component. 

While for small fixed $n$ it can be reasonably assumed that there is no idiosyncratic cross-sectional correlation and, thus, we say that the model is \textit{exact}, in a high-dimensional setting the idiosyncratic components are likely to be weakly cross-sectionally correlated, since, even if the common factors capture the main covariances, some weaker local covariances are likely to remain unexplained. In this case we say that the  factor model is \textit{approximate}. 

In an approximate factor model the common and idiosyncratic components can be identified, provided that the $r$ eigenvalues of the common component covariance matrix diverge as $n\to\infty$, while the $n$ idiosyncratic eigenvalues are bounded for all $n\in\mathbb N$ \citep{chamberlainrothschild83}. Hence, identification of the model, and then also of the number of factors $r$, is possible only when letting $n\to\infty$, a feature sometimes known as the \textit{blessing of dimensionality}.  However, it is well known that, unless further $r^2$ restrictions are imposed, the loadings and factors are identified only up to post- and pre-multiplication by an invertible $r\times r$ matrix and its inverse.

An approximate factor model is then characterized by an eigengap in the population covariance matrix that widens as $n \to \infty$. Conversely, if the population covariance exhibits a widening eigengap as $n \to \infty$, the data can be said to follow an approximate factor model---see, e.g., \citet[Theorem 4]{chamberlainrothschild83}, or \citet[Theorem 2]{gersing2023weak}, or \citet[Proposition 1]{BH25}, for various proofs. This is precisely the setting in which Principal Component Analysis (PCA) is most effective for dimensionality reduction, as it projects the data onto a subspace spanned by directions of maximum variation, by construction.
It is therefore not surprising that PCA is the most natural and common way to estimate factor models. Unlike classical Maximum Likelihood methods, PCA is a non-parametric approach that relies solely on moment conditions rather than distributional assumptions. Furthermore, it can be applied to weakly stationary time series without any modification from its standard formulation for independent data.

Specifically, PCA can be formulated as the minimization of a quadratic loss yielding estimated loadings proportional to the $r$ normalized eigenvectors corresponding to the $r$ largest eigenvalues of the $n\times n$ sample covariance matrix, whose $(i,j)$th entry is given by $T^{-1}\sum_{t=1}^T x_{it} x_{jt}$, $i,j=1,\ldots, n$. The estimated factors are then obtained by linear projection of the data onto these estimated loadings. 

However, because the factors and the loadings are in general not uniquely identified, to guarantee the existence of a unique solution, the minimizers of the PCA objective must satisfy an additional set of $r^2$ constraints. There are two standard ways of imposing these constraints, 
each leading to estimated loadings that are equivalent up to a rescaling of the corresponding eigenvectors.
 \begin{compactenum}
 \item [A1.] If we impose the estimated loadings to be orthogonal and the estimated factors to be orthonormal, then the eigenvectors are rescaled by the square-root of the corresponding eigenvalues \citep[see, e.g.,][]{mardia1979multivariate,FGLR09,DGRfilter}.
 \item [A2.] If we impose the estimated loadings to be orthonormal and the estimated factors to be orthogonal, then the eigenvectors are rescaled by $\sqrt n$ \citep[see, e.g.,][]{stockwatson02JASA,stockwatson02JBES}.
 \end{compactenum}
 
Definition A1 is the classical and most popular one and it is our main focus. 
Its theoretical properties, namely consistency and asymptotic normality, were established by \citet{bai03} when formulating the estimators using an equivalent definition with the estimated factors written as $\sqrt T$ times the $r$ normalized eigenvectors corresponding to the $r$ largest eigenvalues of the $T\times T$ matrix having entries given by $n^{-1}\sum_{i=1}^n x_{it} x_{is}$, $t,s=1,\ldots, T$ (see also \citealp{FLM13}). The estimated loadings are then obtained by linear projection of the data onto the estimated factors. We shall denote this definition as B1 (while a fourth definition, B2, is obtained analogously from A2).
 
Some important issues arise from the above overview. First, while we need $n\to\infty$ to identify the model, the sample size $T$ plays no role for identification of the model. However, we still need $T\to\infty$ for estimation. It follows that, on the one hand, we need to consider double asymptotics $n,T\to\infty$ to ensure consistency, but, on the other hand,  $n$ and $T$ do not have a symmetric role and should not be treated on equal grounds.  This implies that carrying out asymptotic analysis using the classical definition A1 (or A2) rather than the alternative B1 (or B2) could have benefits in terms of interpretation and for deriving the asymptotic properties of the estimators. In fact, there are cases in which considering PCA based on the definition of approach A1 (or A2) is clearly the most natural way forward for developing the theory. This is the case: (a) in the presence of non-stationarities in the data, due, e.g., to change-points \citep{BCF18}, exogenous \citep{massacci2017} or endogenous regime shifts \citep{urga2024estimation,BM22}, unit roots \citep{bai04,baing04,OW19,BLL2,barigozzi2022testing},  time-varying loadings  \citep{motta2011locally,bates_plagborgmoller_stock_watson_2013,pelger_xiong_2022,mikkelsen_hillebrand_urga_2018}; (b) when dealing with PCA based on the spectral density matrix  \citep{FHLR00,FHLR05,FHLZ17}; (c) when considering factor models for matrix and tensor data, which are estimated via PCA on every mode (\citealp{YuHe2022,chang2023modelling,chen2023statistical,chen2024rank,barigozzi2026tensor}, \citealp{barigozzi2025tail});
(d) when dealing with factors which are pervasive along the time dimension so that the idiosyncratic components are white noise (\citealp{LamYao2012,ZRY,chen2020constrained}; \citealp{chen2022factor}); (e) when estimating exact factor models with $n$ being fixed so that only the loadings, but not the factors, can be estimated consistently via PCA \citep{anderson1963asymptotic}; (f) when considering data and/or factors with non-zero mean \citep{lettau2020estimating}.

A second issue relates to identification of the loadings and the factors. Since, in general, these are not identified, it is not clear what the target of the PC  estimators is when it comes to defining consistency. This means that no inference can be carried out, unless one can provide a clear theoretical study of the impact and the meaning of any imposed identifying assumption on the true loadings and factors. This topic is not much explored in the  literature of high-dimensional factor models for time series with few exceptions as, e.g., \citet{baing13} and \citet{BW15}.

\paragraph{Contributions and outline.} 
In this paper we provide a full treatment of PCA for factor models, by reviewing, discussing, completing, and extending existing asymptotic results, and, thus, also clarifying the above issues. The main contributions are as follows.\medskip

\noindent
\textit{PCA formulations.} Section~\ref{sec:ABCD} shows that the two standard PC estimators---based respectively on the $n\times n$ and $T\times T$ sample covariance matrices---are formally equivalent. This elementary equivalence is often left implicit; making it explicit clarifies how PCA follows directly from the corresponding eigenvalue--eigenvector decompositions, without resorting to iterative procedures. Appendix~\ref{sec:otherPCA} gives the analogous derivation for the alternative definition A2.\medskip

\noindent
\textit{Assumptions.} Section~\ref{sec:ass} presents a streamlined set of assumptions for deriving consistency and asymptotic normality in a time-series setting, and explains the role of each condition. Appendix~\ref{sec:hannan} reviews primitive conditions ensuring consistent estimation of the covariance matrix.\medskip

\noindent
\textit{Asymptotic theory.} In Sections~\ref{sec:cons1}-\ref{sec:cent} we first derive consistency assuming only orthogonality between common
and idiosyncratic components, and using direct arguments based on Weyl's inequality and the Davis--Kahan theorem. Then, we show that we can obtain sharper convergence rates by adding more assumptions involving higher-order dependences. Third, we provide several alternative representations for the linear transformation matrix linking true
and estimated loadings/factors. Last, under the extended assumptions, we prove asymptotic normality and show that, under full identification of loadings and factors, the leading term of the asymptotic expansion coincides with the Ordinary Least Squares (OLS) expansion. We also cover in detail the case of non-zero mean data.
Appendices~\ref{sec:cmppca} and~\ref{sec:cmppcaCD} extend these conclusions to the equivalent PCA formulations B1, A2, and B2.\medskip

\noindent
\textit{Identification.} We consider two distinct notions of identification. The first concerns the decomposition of the observed process into common and idiosyncratic components. Section~\ref{sec:blessing} shows that, under the maintained assumptions, this decomposition is identified only in the limit $n\to\infty$. This is the blessing of dimensionality result. The second concerns the loadings and factors themselves, which are identified only up to an invertible transformation unless additional restrictions are imposed. Section~\ref{sec:II} studies the restrictions commonly used in exploratory factor analysis (see also Appendix \ref{app:calH}). Finally, in Section \ref{sec:ottavo} we  show that, although, in principle, these restrictions should enable valid inference, they either slow convergence---thereby affecting asymptotic normality---or are overly stringent, as they effectively would hold only if the factors were deterministic or if we interpreted all results as conditional on a realization of the factors.\medskip

\noindent
\textit{Proofs.} Appendix~\ref{app:mainproof} contains the proofs of the main results, i.e., the proofs of 
Propositions \ref{prop:L}, \ref{prop:HHAT}, \ref{th:CLTcent}, \ref{corol:K00}, and \ref{corol:H00}, which are entirely or partially new; the remaining proofs are collected in Appendices~\ref{sec:otherproofs} and~\ref{sec:lemma}.

\paragraph{Notation.}
Throughout, $\Vert\cdot\Vert$ denotes the Euclidean norm for vectors and the spectral norm for matrices, while $\Vert\cdot\Vert_F$ denotes the Frobenius norm. We write $\text{tr}(\cdot)$ for the trace, $\mbf I_q$ for the $q\times q$ identity matrix, and $\mbf 0_q$ for a zero vector of dimension $q$. For a matrix $\bm A$, $[\bm  A]_{ij}$ denotes its $(i,j)$th entry. Eigenvalues are sorted in decreasing order, and $\mu_j(\bm  A)$ denotes the $j$th largest eigenvalue. If $\bm  A$ is positive definite, $(\bm  A)^{1/2}$ denotes its symmetric square root. Finally, $O_{\mathrm P}$, $o_{\mathrm P}$, and $\mathrm P\text{-}\lim$ denote boundedness and convergence in probability; $\to_d$ denotes convergence in distribution; $O_{\mathrm{m.s.}}$ and  $\mathrm {m.s.}\text{-}\lim$ denote boundedness and convergence in mean square.

\section{Estimation of factor models via Principal Component Analysis}\label{sec:ABCD}

In this section we present PCA estimation of a factor model. We assume that we observe $n$ zero-mean time series over $T$ periods following the factor model, as in \eqref{eq:SDFM1R_base}, which in vector notation reads as
\[
\mbf x_t = \bm\Lambda\mbf F_t +\bm  e_t,\quad t=1,\ldots, T, 
\]
where $\mbf x_t:=(x_{1t}\cdots x_{nt})^\prime$ and $\bm e_{t}:=( e_{1t}\cdots e_{nt})^\prime$ are $n$-dimensional vectors of observables and idiosyncratic components at time $t$, respectively, and  $\bm\Lambda:=(\bm\lambda_1\cdots\bm\lambda_n)^\prime$ is the $n\times r$ matrix of factor loadings. 
We call $\bm{C}_t:=\bm\Lambda\mbf F_t$ the vector of common components.
Furthermore, let $\bm X:=(\mbf x_1\cdots\mbf x_T)^\prime$ and $\bm  E:=(\bm e_1\cdots\bm  e_T)^\prime$ be $T\times n$ matrices of observables and idiosyncratic components, respectively, and $\bm F:=(\mbf F_1\cdots\mbf F_T)^\prime$ be the $T\times r$ matrix of factors. Equivalently, we can write \eqref{eq:SDFM1R_base} in vector notation as:
\[
\bm x_i = \bm F\bm\lambda_i +\bm e_i,\quad i=1,\ldots, n, 
\]
where $\bm x_i:=(x_{i1}\cdots x_{iT})^\prime$ and $\bm e_i:=(e_{i1}\cdots e_{iT})^\prime$ are the $T$-dimensional vectors of the $i$th observable time series and its idiosyncratic component.

The idea of PCs as the $r$ directions of best fit in an $n$-dimensional space is originally due to \citet{pearson1901} and the use of PCA for factor analysis dates back to \citet{hotelling1933analysis}. The PC estimators of $\bm\Lambda$ and $\bm F$ are defined as:
\begin{align}
\l(\wh{\bm\Lambda},\wh{\bm F}\r)&=\arg\min_{{\bm L},{\bm {G}}} \frac 1{nT}\sum_{i=1}^n\sum_{t=1}^T (x_{it}-
{\bm \ell}_i^\prime {\mbf G}_t)^2\label{eq:minimizza}\\
&=\arg\min_{{\bm L},{\bm {G}}} \frac 1T\sum_{t=1}^T \l\Vert\mbf x_t -\bm L\mbf G_t \r\Vert^2
=\arg\min_{{\bm L},{\bm {G}}} \frac 1n\sum_{i=1}^n \l\Vert\bm x_i -\bm G\bm \ell_i \r\Vert^2
=\arg\min_{{\bm L},{\bm {G}}} \frac 1{nT}\l\Vert \bm X-
 {\bm G}\bm L^\prime\r\Vert ^2_F\nn\\
&=
\arg\min_{{\bm L},{\bm {G}}} \frac 1{nT} \text{tr}\l\{
\l(\bm X-
{\bm G}\,{\bm L}^\prime \r)
\l(\bm X-
{\bm G}\,{\bm L}^\prime \r)^\prime
\r\}=
\arg\min_{{\bm L},{\bm {G}}} \frac 1{nT} \text{tr}\l\{
\l(\bm X-
{\bm G}\,{\bm L}^\prime \r)^\prime
\l(\bm X-
{\bm G}\,{\bm L}^\prime \r)
\r\},\nn
\end{align}
where $\bm L:=(\bm\ell_1\cdots\bm\ell_n)^\prime\in\mathbb R^{n\times r}$ and $\bm G:=(\mbf G_1\cdots\mbf G_T)^\prime\in\mathbb R^{T\times r}$ are generic loadings and factor matrices.

Clearly, for a given estimator of the loadings, the estimator of the factors is obtained by least squares and vice versa. This motivates the classical way to derive the solution of \eqref{eq:minimizza} by using an alternating least squares approach \citep[and references therein]{bai2020simpler}. However, there is a simpler way to derive the solution. Indeed, we can restate \eqref{eq:minimizza} in two equivalent ways which depend either only on the loadings or only on the factors. First, by substituting ${\bm G}=\bm X{\bm L}({\bm L}^\prime{\bm L})^{-1}$ in \eqref{eq:minimizza}, we obtain the equivalent optimization problem:
\begin{align}
\min_{{\bm L}}\frac 1{nT} \text{tr}\l\{
\bm X\l(\mbf I_n-{\bm L}({\bm L}^\prime{\bm L})^{-1}{\bm L}^\prime \r)\bm X^\prime
\r\}
=\max_{{\bm L}}\frac 1{n} \text{tr}
\l\{
({\bm L}^\prime{\bm L})^{-1/2}
{\bm L}^\prime
\frac{\bm X^\prime\bm X}{T}
{\bm L}
({\bm L}^\prime{\bm L})^{-1/2}
\r\},\label{eq:maxforni}
\end{align}
which gives an estimator of the loadings, while the factors can be estimated in a second step by linear projection onto that estimator.
Second, by substituting ${\bm L}=\bm X^\prime {\bm G}({\bm G}^\prime{\bm G})^{-1}$ in \eqref{eq:minimizza}, we obtain yet another equivalent formulation of the optimization problem in  \eqref{eq:minimizza} or \eqref{eq:maxforni}:
\begin{align}
\min_{{\bm G}}\frac 1{nT} \text{tr}\l\{
\bm X^\prime\l(\mbf I_T-{{\bm G}({\bm G}^\prime{\bm G})^{-1}{\bm G}^\prime} \r)\bm X
\r\}
=\max_{{\bm G}}\frac 1{T} \text{tr}
\l\{
({\bm G}^\prime{\bm G})^{-1/2}{{\bm G}^\prime}
\frac{\bm X\bm X^\prime}{n}
{{\bm G}}({\bm G}^\prime{\bm G})^{-1/2}
\r\},\label{eq:maxbai}
\end{align}
which gives an estimator of the factors, while the loadings can be estimated in a second step by linear projection onto that estimator.

From \eqref{eq:maxforni} and \eqref{eq:maxbai} we see that the PC estimators are related to eigenvectors, but the above solutions are not unique unless we fix the scale of those eigenvectors.
Indeed, the original problem in \eqref{eq:minimizza} does not have a unique solution either. Indeed, given any solution $(\wh{\bm\Lambda}, \wh{\bm F})$, we can find infinite other equivalent solutions given by $(\wh{\bm\Lambda}{\bm K},\wh{\bm F}({\bm K}^\prime)^{-1})$, where ${\bm K}$ is any $r\times r$ invertible matrix. Therefore, to ensure uniqueness of the PCA solution, we must solve \eqref{eq:minimizza} subject to $r^2$ identifying constraints. The two most common choices are: (a) $n^{-1}\bm L'\bm L$ being diagonal with distinct entries sorted in descending order and $T^{-1}\bm G'\bm G= \mbf I_r$, or (b) $n^{-1}\bm L'\bm L=\mbf I_r$ and $T^{-1}\bm G'\bm G$ being diagonal with distinct entries sorted in descending order. These restrictions are convenient because they imply normalization of the eigenvectors (see below) and are also consistent with the typical identifying assumptions made on the true loadings and factors in exploratory factor analysis (see Section 
\ref{sec:II}). Here we focus on the first identification conditions (a), and derive the PC estimators using two equivalent approaches, A1 and B1.  Alternative estimators are obtained when imposing the identification conditions (b). These are presented in approaches A2 and B2---which are equivalent---in Appendix \ref{sec:otherPCA}.

\paragraph{Approach A1 solving \eqref{eq:maxforni}.}  Denote the $n\times n$  sample covariance matrix obtained by averaging over time as (note that  $\E[\mbf x_t]=\mbf 0_n$ by assumption)
\beq\label{eq:nncov}
\wh{\bm\Gamma}^x:=\frac 1T\sum_{t=1}^T \mbf x_t\mbf x_t^\prime = \frac{\bm X^\prime\bm X}{T},
\eeq
having its $r$ largest eigenvalues in the $r\times r$ diagonal matrix $\wh{\mbf M}^x$ (sorted in descending order) and the corresponding normalized eigenvectors as the columns of  the $n\times r$ matrix $\wh{\mbf V}^x$. 
Then, a solution of  \eqref{eq:maxforni} is found as follows.

If we impose $n^{-1} {{\bm L}^\prime {\bm L}}$ to be diagonal, then by construction each column of ${\bm L}
({\bm L}^\prime{\bm L})^{-1/2}$ is normalized and, by definition of eigenvectors and eigenvalues, the value of the objective function in \eqref{eq:maxforni} must be the sum of the $r$ largest eigenvalues of $\wh{\bm\Gamma}^x$ divided by $n$, i.e., it must give $n^{-1}\text{tr}({\wh{\mbf M}^x})$. So, our estimator $\wh{\bm\Lambda}$ must be such that $\wh{\bm\Lambda}
(\wh{\bm\Lambda}^\prime\wh{\bm\Lambda})^{-1/2}$ is the matrix of the corresponding normalized eigenvectors, that is:
\begin{align}
&(\wh{\bm\Lambda}^\prime\wh{\bm\Lambda})^{-1/2}
\wh{\bm\Lambda}^\prime
\frac{\bm X^\prime\bm X}{nT}
\wh{\bm\Lambda}
(\wh{\bm\Lambda}^\prime\wh{\bm\Lambda})^{-1/2}=\frac{\wh{\mbf M}^x}{n}=\wh{\mbf V}^{x\prime}
\frac{\bm X^\prime\bm X}{nT}
\wh{\mbf V}^x.\label{eq:evecXX2}
\end{align}
Therefore, from \eqref{eq:evecXX2}  we must have $\wh{\bm\Lambda} = \wh{\mbf V}^x(\wh{\bm\Lambda}^\prime\wh{\bm\Lambda})^{1/2}$. So, by linear projection, the estimator of the factors is obtained as $\wh{\bm F}:= \bm X\wh{\bm\Lambda}(\wh{\bm\Lambda}^\prime\wh{\bm\Lambda})^{-1}=\bm X \wh{\mbf V}^x(\wh{\bm\Lambda}^\prime\wh{\bm\Lambda})^{-1/2}$. Moreover, by imposing also the identifying constraint $T^{-1}\wh{\bm F}^\prime\wh{\bm F}=\mbf I_r$, it follows that we must have
$\wh{\bm\Lambda}^\prime\wh{\bm\Lambda} = \wh{\mbf M}^x$, which is diagonal as expected.

Summing up, it follows that the PC estimators are given by:
\begin{align}
&\wh{\bm\Lambda}:=\wh{\mbf V}^x(\wh{\mbf M}^x)^{1/2},\label{eq:estL}\\
&\wh{\bm F}:=\bm X\wh{\bm\Lambda}(\wh{\bm\Lambda}^\prime\wh{\bm\Lambda})^{-1}=\bm X\wh{\mbf V}^x(\wh{\mbf M}^x)^{-1/2}. 
\label{eq:estF}
\end{align}
The factors defined in this way are the classical normalized PCs of $\bm X$. Their definition dates back to \citet{hotelling1933analysis} (see also \citealp[Chapter 4]{lawleymaxwell71}, \citealp[Chapter 9.3]{mardia1979multivariate}, \citealp[Chapter 7.2]{jolliffe2002principal}). In high-dimensional factor analysis this approach is considered, for example, by \citet{FGLR09} and \citet{DGRfilter}, who prove consistency although not with the sharpest possible rate. 

\paragraph{Approach B1 solving \eqref{eq:maxbai}.}  Now, consider the $T\times T$ sample covariance matrix obtained by averaging over the cross-section (note that $\E[\bm x_i]=\mbf 0_T$ by assumption)
\beq\label{eq:TTcov}
\wt{\bm\Gamma}^x:=\frac 1n\sum_{i=1}^n \bm x_i\bm x_i^\prime = \frac{\bm X\bm X^\prime}{n},
\eeq
having its $r$ largest eigenvalues collected in the $r\times r$ diagonal matrix $\wt{\mbf M}^x$ (sorted in descending order) and the corresponding normalized eigenvectors as columns of  the $T\times r$ matrix $\wt{\mbf V}^x$. 
Then, another solution of  \eqref{eq:maxbai}, equivalent to A1,  is found as follows.

If we impose $T^{-1}{{\bm G}^\prime{\bm G}}=\mbf I_r$, then each column of $T^{-1/2}{ {\bm G}}$ is normalized and, by definition of eigenvectors and eigenvalues, the value of the objective function in the above maximization must be the sum of the $r$ largest eigenvalues of $\wt{\bm\Gamma}^x$ divided by $T$,  i.e., it must give $T^{-1}\text{tr}(\wt{\mbf M}^x)$. 
So, our estimator $\wt{\bm F}$ must be such that $T^{-1/2}{\wt{\bm F}}$ is the matrix of the corresponding normalized eigenvectors,
that is:
\begin{align}
&\frac{\wt{\bm F}^\prime}{\sqrt T}
\frac{\bm X\bm X^\prime}{nT}
\frac{\wt{\bm F}}{\sqrt T}
=\frac{\wt{\mbf M}^x}{T}=\wt{\mbf V}^{x\prime}
\frac{\bm X\bm X^\prime}{nT}
\wt{\mbf V}^x.\nn
\end{align}
Therefore, the PC estimators of the factors and the loadings (obtained by linear projection) are
\begin{align}
&\wt{\bm F}:=\wt{\mbf V}^x\sqrt T,\label{eq:estFbai}\\
&\wt{\bm \Lambda}:= \bm X^\prime\wt{\bm F}(\wt{\bm F}^\prime\wt{\bm F} )^{-1} = \frac{\bm X^\prime \wt{\mbf V}^x}{\sqrt T} 
,\label{eq:estLbai}
\end{align}
And, as expected, we get $n^{-1}{\wt{\bm \Lambda}^\prime\wt{\bm \Lambda}}=T^{-1}{\wt{\mbf M}^x}$, which  is diagonal. These are the estimators defined by \citet{bai03}, who proves consistency and asymptotic normality.

\paragraph{Equivalence of A1 and B1.} As long as the number of factors $r$ is such that $r<\min(n,T)$, the $r$ largest eigenvalues of
$\bm X^\prime \bm X$ and $\bm X \bm X^\prime$  coincide, i.e., 
\beq\label{eq:evaleq}
\frac{\wh{\mbf M}^x}n=\frac{\wt{\mbf M}^x}T.
\eeq
and the objective function in \eqref{eq:maxforni} and \eqref{eq:maxbai} has the same value in its minimum. Since approaches A1 and B1 are based on the same identification conditions, the solutions \eqref{eq:estL}-\eqref{eq:estF} and \eqref{eq:estFbai}-\eqref{eq:estLbai} must then coincide, i.e.,  it must be that
\beq\label{eq:stessi}
\wh{\bm F}=\wt{\bm F} \; \text{ and }\; \wh{\bm \Lambda}=\wt{\bm \Lambda}.
\eeq

\paragraph{Centering and standardizing.}
If the observed time series do not have zero mean, then, the derivation above is unchanged provided that we replace each element of $\bm X$ with its centered version $x_{it}- \bar{x}_i$, $i=1,\ldots, n$, $t=1,\ldots T$, with $
\bar {x}_i:=T^{-1}\sum_{t=1}^T  x_{it}$. However, while centering has no effect on the asymptotic properties of the estimated loadings, it does have important effects on the asymptotic properties  of the estimated factors (see Section \ref{sec:cent}).

Moreover, to avoid scale effects due to data having different units of measure, each element of $\bm X$ should always be standardized. Therefore, in practice, the above derivation of PC estimators should be applied to $\frac{x_{it}-\bar x_i}{s_i}$, $i=1,\ldots, n$, $t=1,\ldots T$, with $s_i^2:=T^{-1}\sum_{t=1}^T (x_{it}-\bar x_i)^2$.  Notice that this last operation makes the idiosyncratic components heteroskedastic even if those of the non-standardized data were homoskedastic.

Finally, note that both centering and standardizing must always be carried out along the time dimension even when working with a $T\times T$ covariance. This is because the cross-sectional units are unlikely to be identically distributed.

\paragraph{Alternative derivation of PC estimators.}
Consider the Singular Value Decomposition (SVD) isolating the $r$ largest singular values 
\beq
\frac{\bm X}{\sqrt{nT}} = \bm U \bm D\bm V^\prime + \bm Z, \label{eq:SVD}
\eeq
where $ \bm U$ and $ \bm V$ are matrices of dimensions $T\times r$ and $n\times r$, respectively, and have orthonormal columns, $\bm D$ is $r\times r$ diagonal and contains the $r$ largest singular values of $(nT)^{-1/2}\bm X$, and $\bm Z$ is $T\times n$ such that $\bm Z^\prime\bm U=\mbf 0_{n\times r}$ and $\bm Z\bm V=\mbf 0_{T\times r}$. By solving \eqref{eq:minimizza} using an alternating least squares approach jointly with \eqref{eq:SVD} and by imposing orthogonal loadings and orthonormal factors, \citet{bai2020simpler} show that the estimator of the loadings is $\check{\bm\Lambda}:= \sqrt n\, \bm V \bm D$, while the estimator of the factors is $\check{\bm F}:=\sqrt T\, \bm U$. 

The estimators $\check{\bm\Lambda}$ and $\check{\bm F}$ are the PC estimators written in terms of the SVD representation \eqref{eq:SVD}.
To see this, first, we recall that $\bm U\bm D\bm V^\prime$ is the finite-sample best rank-$r$ approximation of $(nT)^{-1/2}\bm X$,
meaning that it is a solution of $\min_{\bm A:\mathrm{rk}(\bm A)\le r}\Vert (nT)^{-1/2}\bm X-\bm A\Vert_F$ \citep{EY36}.
Moreover, 
by the orthogonality conditions defining \eqref{eq:SVD},
\beq\label{eq:smeo}
\frac{\wh{\bm\Gamma}^x}n=\frac{\bm X^\prime \bm X}{nT} = \bm V \bm D^2 \bm V^\prime + \bm Z^\prime\bm Z\;\text{ and }\;
\frac{\wt{\bm\Gamma}^x}T=\frac{\bm X \bm X^\prime}{nT} = \bm U \bm D^2 \bm U^\prime + \bm Z\bm Z^\prime,
\eeq
Thus, from \eqref{eq:smeo} and by using also \eqref{eq:evaleq}, it is straightforward to see that
\beq\label{eq:giovine}
\bm D^2 =\frac{\wh{\mbf M}^x}n=  \frac{\wt{\mbf M}^x}T,\quad  \bm U=\wt{\mbf V}^x, \quad \bm V=\wh{\mbf V}^x. 
\eeq
Therefore, from \eqref{eq:giovine} we get $\check{\bm\Lambda}=\wh{ \mbf V}^{x}(\wh{\mbf M}^{x})^{1/2}=\wh{\bm\Lambda}$  as in approach A1, and 
$\check{\bm F}=\sqrt T\, \wt{\mbf V}^x=\wt{\bm F}$  as in approach B1, and by linear projection
\begin{align}
\wh {\bm F} &=  \bm X\wh{\bm\Lambda}(\wh{\bm\Lambda}^\prime\wh{\bm\Lambda})^{-1}=
\bm X  \wh{ \mbf V}^{x}(\wh{\mbf M}^{x})^{-1/2}=
\sqrt {nT}\, \bm U\bm D\bm V^\prime  \bm V  ( \sqrt n  \bm D)^{-1} = \sqrt T \bm U = \wt{\bm F},\nn\\
\wt{\bm\Lambda} &=\bm X^\prime \wt{\bm F}(\wt{\bm F}^\prime\wt{\bm F})^{-1} =\frac{\bm X^\prime \wt{\mbf V}^{x} }{\sqrt T}=\frac{\sqrt{nT}\,\bm V\bm D\bm U^\prime \bm U}{\sqrt T}
=\sqrt n \bm V\bm D= \wh{\bm\Lambda}.\nn
\end{align}
As already stated in \eqref{eq:stessi}, approaches A1 and B1 coincide. 

\begin{rem}  \label{rem:SVD}
By means of the SVD representation in \eqref{eq:SVD} and its connection with PCA we can immediately derive a first proof of consistency of the PC-estimated rescaled common component.
Denote by $\sigma_j(\bm A)$ the $j$th largest singular value of a generic matrix $\bm A$. 
Then, from our assumptions of weakly correlated idiosyncratic components and
pervasive factors, we can prove that
$\mathrm P\text{-}\lim_{n,T\to\infty}(nT)^{-1/2}\sigma_{1}(\bm E)=0$,
$\mathrm P\text{-}\lim_{n,T\to\infty}(nT)^{-1/2}\sigma_{r}(\bm F\bm\Lambda^{\prime})>0$, and $\sigma_{r+1}(\bm F\bm\Lambda^{\prime})=0$ for all $n,T\in\mathbb N$.\footnote{
Under Assumption \ref{ass:common}, 
$
(nT)^{-1}\sigma_r^2({\bm F\bm\Lambda^\prime}) 
= (nT)^{-1} \mu_r(
{\bm F}{\bm\Lambda^\prime\bm\Lambda \bm F^\prime})
= (nT)^{-1} \mu_r(
{\bm F^\prime\bm F}{\bm\Lambda^\prime\bm\Lambda })
\to_{\mathrm P}
\mu_r(\bm\Gamma^F\bm\Sigma_\Lambda)>0
$, as $n,T\to\infty$, and the number of factors is $r$ for any fixed $n$ and $T$ large enough, so $\text{rk}(\bm F\bm\Lambda^\prime)= r$.
Under Assumption \ref{ass:idio}, 
$
(nT)^{-1}\sigma_1^2(\bm E) = (nT)^{-1}\mu_1(\bm E^\prime\bm E) =O_{\mathrm P}(\max(n^{-1},T^{-1/2}))
$ (see Lemma \ref{lem:LLN} for details).}
Moreover, from the above derivation $(nT)^{-1/2}\wh{\bm F}\wh{\bm\Lambda}^\prime=\bm U\bm D\bm V^\prime$.
So, $(nT)^{-1/2}\bm X$ is asymptotically equivalent to the rescaled true rank $r$ common component $(nT)^{-1/2}\bm F\bm\Lambda^\prime$, and the PC-estimated common component gives the finite sample best rank-$r$ approximation of $(nT)^{-1/2}\bm X$.
By recalling also that $\Vert \bm Z\Vert = (nT)^{-1/2}\sigma_{r+1}(\bm X)$, by definition of SVD in \eqref{eq:SVD},
it follows that, as $n,T\to\infty$,
\begin{align}
\left\Vert
\frac{\widehat{\bm F}\wh{\bm\Lambda}^\prime-\bm F\bm\Lambda^{\prime}}{\sqrt{nT}}\right\Vert
&\le 
\l\Vert\frac{\widehat{\bm F}\wh{\bm\Lambda}^\prime-\bm X}{\sqrt{nT}}\right\Vert+
\l\Vert\frac{\bm X-\bm F\bm\Lambda^{\prime}}{\sqrt{nT}}\right\Vert
= \l\Vert \bm U\bm D\bm V^\prime-\frac{\bm X}{\sqrt{nT}} \r\Vert
+\l\Vert
\frac{\bm E}{\sqrt{nT}}\r\Vert
= \l\Vert \bm Z \r\Vert
+\l\Vert
\frac{\bm E}{\sqrt{nT}}\r\Vert\nn\\
&
=\frac{\sigma_{r+1}(\bm X)}{\sqrt{nT}}+
\sqrt{\frac{\mu_1(\bm E^\prime\bm E)}{nT}}
=\frac{\sigma_{r+1}(\bm X)}{\sqrt{nT}}+
\frac{\sigma_1(\bm E)}{\sqrt{nT}}
=
\frac{\sigma_{r+1}(\bm F\bm\Lambda^\prime+\bm E)}{\sqrt{nT}}
+ \frac{\sigma_1(\bm E)}{\sqrt{nT}}\nn\\
&\le  \frac{2 \sigma_1(\bm E)}{\sqrt{nT}}+\frac{\sigma_{r+1}(\bm F\bm\Lambda^\prime)}{\sqrt{nT}}=o_{\mathrm P}(1),\nn
\end{align}
because of Weyl's inequality for singular values \citep[Theorem 6]{MK04}. This simple proof of consistency, 
however, does not provide consistency of the loading and factor spaces, nor the sharp convergence rates needed for asymptotic inference. These require a more refined  analysis, which is developed in the following sections.
\end{rem}

\section{Assumptions}\label{sec:ass}
We adapt to the high-dimensional setting the approach taken in most works in time series analysis. First, we define the model for an infinite dimensional process $\{x_{it},\, i\in\mathbb N,\, t\in\mathbb Z\}$. Second, the properties and the assumptions of the model are defined for the $n$-dimensional sub-process $\{x_{it},\, i=1,\ldots, n,\, t\in\mathbb Z\}$ either holding for all $n\in\mathbb N$ or in the limit $n\to\infty$. Third, the properties of the sample moments are assumed (or derived from primitive conditions) for a given $nT$-dimensional realization  $\{x_{it},\, i=1,\ldots, n,\, t=1,\ldots, T\}$ and hold in the limit $n,T\to\infty$.

An infinite dimensional zero-mean process follows an $r$-factor model if its elements satisfy 
\begin{align}
x_{it}&=\bm\lambda_i^\prime \mbf F_t +  e_{it}, \quad i\in\mathbb N, \quad t\in\mathbb Z.\label{eq:SDFM1R}
\end{align}
This is the equivalent of \eqref{eq:SDFM1R_base}, but when we have infinite time series, so now $i\in\mathbb N$. 

We start by characterizing the common component by means of the following assumption.

\begin{ass}[\textsc{common component}]\label{ass:common} $\,$
\begin{compactenum}[(a)]
	\item  $\lim_{n\to\infty}\Vert n^{-1}\bm\Lambda^\prime\bm\Lambda-\bm\Sigma_{\Lambda}\Vert=0$, where $\bm\Sigma_{\Lambda}$ is $r\times r$ positive definite, and, for all $i\in\mathbb N$, 
	$\Vert\bm\lambda_i\Vert\le M_\Lambda$ for some finite positive real $M_\Lambda$ independent of $i$.  
\item For all $t\in\mathbb Z$, $\E[\mbf F_{t}]=\mbf 0_r$ and $\bm\Gamma^F:=\E[\mbf F_t\mbf F_t^\prime]$ is $r\times r$ positive definite and $\Vert\bm\Gamma^F\Vert\le M_F$ for some finite positive real $M_F$ independent of $t$.
	\item 
	\begin{inparaenum} 
	\item [(i)] For all $t\in\mathbb Z$, $\E[\Vert \mbf F_t\Vert^4]\le K_F$ for some finite positive real $K_F$ independent of $t$; \\
	\item	[(ii)] $\mathrm P\text{-}\lim_{T\to\infty}\l\Vert T^{-1}\bm F^\prime\bm F -\bm\Gamma^F\r\Vert=0$.
	\end{inparaenum}
	\item There exists an integer $N$ such that for all $n> N$, $r$ is a finite positive integer, independent of $n$. 
\end{compactenum}
\end{ass}

Part (a) is also assumed in \citet[Assumption B]{bai03}. It implies that the loadings matrix has asymptotically maximum column rank $r$. Moreover, for any given $n\in\mathbb N$, all the factors have a finite contribution to each series (upper bound on $\Vert\bm\lambda_i\Vert$) and are cross-sectionally pervasive (full-rank of $\bm\Sigma_\Lambda$). 

Part (b) is imposing zero mean and weak stationarity of the factors. The former requirement is generalized to non-zero mean in Section \ref{sec:cent}, while the latter is assumed only for simplicity. In fact, what is really needed for deriving our results is just part (c) which is assumed also in \citet[Assumption A]{bai03}. In part (c-i) we assume finite 4th order moments of the factors, and, in particular, we are bounding
$\E[\Vert \mbf F_t\Vert^4]=\sum_{j=1}^r\sum_{k=1}^r \E[F_{jt}^2F_{kt}^2]$. 
Part (c-ii) is a high-level one and it requires introducing the sample size $T$. It is therefore useful to discuss which are the primitive conditions under which it can hold. Clearly, part (c-i) is a necessary condition for part (c-ii) to hold. Given part (b), part (c-ii) simply says that the sample covariance matrix of the factors is a consistent estimator of its population counterpart $\bm\Gamma^F$. In  Appendix \ref{sec:hannan} we give sufficient conditions for such a consistency result to hold, possibly even in mean-square. We show that this is the case if the process $\{\mbf F_t\}$ is either linear, or has summable 4th order cumulants, or has bounded physical dependence \citep{wu05}, or is strongly mixing. There it is also shown that the consistency rate is $\sqrt T$, as it should be expected, and that we can also allow for conditionally heteroskedastic factors.

In principle, we do not have to require weak stationarity in part (b) for part (c-ii) to hold, and we can even have $\bm\Gamma^F$ that depends on $t$ as long as $M_F$ in part (b) is independent of $t$ (hence the reason for stating it explicitly). This is, for example the case when we are in the presence of structural breaks/change-points (see, e.g., \citealp{BCF18} and \citealp{duan2022quasi}) or regime shifts (\citealp{massacci2017,BM22}).

Notice that, on the one hand, we consider deterministic, i.e., non-random, loadings as in classical factor models.\footnote{As in \citet{bai03}, the generalization to the case of random loadings is possible provided that: we assume convergence in probability in part (a), for all $i\in\mathbb N$, $\E[\Vert \bm\lambda_i\Vert^4]\le K_\Lambda$, for some finite positive real $K_\Lambda$ independent of $i$, and $\{\bm\lambda_i,\, i\in\mathbb N\}$ is an independent sequence.} But, on the other hand, we explicitly treat the factors as random variables since they are time series. This has important implications when it comes to understanding the implications of various identifying assumptions (see Section \ref{sec:II}).

Part (d) implies the existence of a finite number $r$ of factors independent of $n$.  This implicitly means that we must have $r\le N$ with $N$ being the minimum number of common components we need to be able to identify $r$ \citep[Assumption 3]{LDA2}. In fact, due to the presence of an idiosyncratic component, $r$ can be identified only as $n\to\infty$ (see Section \ref{sec:blessing}).  Therefore, in practice, to estimate the model we must always work with $n$ series such that $n>r$. Moreover, since PCA is based on an $n\times n$ sample covariance matrix computed summing over $T$ observations, or, equivalently, on a $T\times T$ matrix computed summing over $n$ variables, we must also have $T>r$. It is then common to directly assume $r<\min(n,T)$.

To characterize the idiosyncratic component, we make the following assumption.

\begin{ass}[\textsc{idiosyncratic component}]\label{ass:idio}
  $\,$
\begin{compactenum}[(a)]
	\item For all $i\in\mathbb N$ and all $t\in\mathbb Z$, $\E[ e_{it}]= 0$.
	\item For all $i,j\in\mathbb N$, all $t\in\mathbb Z$, and all $k\in\mathbb Z$, $\gamma^ e_{ij,k}:=\E[ e_{it} e_{j,t-k}]$ such that
	$\vert \gamma^ e_{ij,k}\vert\le \rho^{\vert k\vert} M_{ij}$, where $\rho$ and $M_{ij}$ are finite positive reals independent of $t$ such that $0\le \rho <1$, $M_{ii}\le C_ e$, $\sum_{j=1,i\ne j}^n M_{ij}\le M_ e$, and $\sum_{i=1, j\ne i}^n M_{ij}\le M_ e$ for some finite positive reals $C_ e$ and $M_{ e}$ independent of  $i$, $j$, and $n$. 
\item 
\begin{inparaenum}
\item [(i)]  For all $i=1,\ldots, n$, all $t=1,\ldots, T$, and all $n,T\in\mathbb N$, $\E[ e_{it}^4]\le Q_ e$ for some finite positive real $\phantom {(i)}\, Q_ e$ independent of $i$ and $t$;\\
\item [(ii)] for all $j=1,\ldots, n$, all $s=1,\ldots, T$, and all $n,T\in\mathbb N$,
\[
\E\l[\l\vert\frac 1{\sqrt{nT}} \sum_{i=1}^n\sum_{t=1}^T\l\{ e_{is} e_{jt}-\E[ e_{is} e_{jt}]\r\} \r\vert^2\r]
\le K_ e
\]
for some finite positive real $K_ e$ independent of $j$, $s$, $n$, and $T$.
\end{inparaenum}
 \end{compactenum}
 \end{ass}

By part (a), we assume that the idiosyncratic components  have zero mean and are cross-sectionally heteroskedastic, which, as we noticed already, is generally the case given that the data are standardized before estimation. It follows that for any $n\in\mathbb N$ the idiosyncratic covariance matrix is defined as $\bm\Gamma^ e:=\E[\bm e_t\bm e_t^\prime]$ with entries $\gamma_{ij,0}^ e$, $i,j=1,\ldots,n$. 

Part (b) has a twofold purpose. First, it limits the degree of serial correlation of the idiosyncratic components, implying weak stationarity of idiosyncratic components. Indeed, they have summable autocovariances (see Lemma \ref{lem:Gxi}(iii)):
\[
\sup_{n,T\in\mathbb N}\max_{i=1,\ldots, n}\frac 1{T}\sum_{t,s=1}^T \vert\E_{}[ e_{it} e_{is}]\vert \le \frac{C_ e(1+\rho)}{1-\rho}.
\]
The same condition follows directly from \citet[Assumption C.4]{bai03} although it is not stated explicitly.
Second, part (b) implicitly bounds the L1 norm  $\Vert \bm\Gamma^ e\Vert_1:=\max_{i=1,\ldots, n}\sum_{j=1}^n \vert \E[e_{it}e_{jt}]\vert$ as in \citet{FLM13} and \citet{FHLZ17}, and, as a consequence, limits the degree of cross-sectional correlation between idiosyncratic components, which is usually assumed in approximate factor models. Moreover, it implies the moment conditions (see Lemmas \ref{lem:Gxi}(i), \ref{lem:Gxi}(ii), \ref{lem:GxiBAI}(i), \ref{lem:GxiBAI}(ii)):
\begin{align}
&\sup_{n,T\in\mathbb N} \frac 1{T}\sum_{t,s=1}^T \l \vert\frac 1n \sum_{i=1}^n \E_{}[ e_{it} e_{is}]\r \vert \le\frac{C_ e (1+\rho)}{1-\rho}, \label{eq:baiC2}\\
&\sup_{n,T\in\mathbb N}\max_{t=1,\ldots, T}\frac 1{n}\sum_{i,j=1}^n \vert \E_{}[ e_{it} e_{jt}] \vert \le C_ e+M_{ e},
\label{eq:baiC3}\\
&\sup_{n,T\in\mathbb N}\frac 1{nT}\sum_{i,j=1}^n\sum_{t,s=1}^T \vert\E_{}[ e_{it} e_{js}]\vert \le \frac{(C_ e+M_ e)(1+\rho)}{1-\rho},\label{eq:bai}\\
&\sup_{n,T\in\mathbb N} \max_{t=1,\ldots, T}\sum_{s=1}^T  \l \vert\frac 1n \sum_{i=1}^n \E_{}[ e_{it} e_{is}]\r\vert \le\frac{C_ e (1+\rho)}{1-\rho}, \label{eq:baiE1}\\
&\sup_{n,T\in\mathbb N} \max_{\substack{i=1,\ldots, n\\t=1,\ldots, T}} \sum_{j=1}^n \vert \E[ e_{it}  e_{jt}]  \vert\le C_ e+M_ e, \label{eq:baiE2}
\end{align}
which coincide with the conditions required by \citet[Assumptions C.4, C.3, C.2, E.1, and E.2, respectively]{bai03}. 
Furthermore, part (b) implies directly that
\begin{align}
&\sup_{n,T\in\mathbb N} \max_{s=1,\ldots,T} \l\vert \frac 1n \sum_{i=1}^n \E[ e_{is}^2]\r\vert \le C_ e,\label{eq:baiC2bis}\\
&\sup_{T\in\mathbb N}
\max_{t=1,\ldots, T} \vert \E_{}[ e_{it} e_{jt}] \vert \le M_{ij},\label{eq:baiC3bis}
\end{align}
which are the
first conditions in \citet[Assumptions C.2 and C.3, respectively]{bai03}. 

Part (c-i) assumes finite and summable 4th order moments of the idiosyncratic components, it is weaker than what is assumed by \citet[Assumption C.1]{bai03} where finite 8th order moments are required. Part (c-ii) gives summability conditions over the cross-section and time dimensions for the 4th order cumulants of $\{ e_{it}\}$. Indeed, it can be equivalently written as:
\begin{align}
\sup_{n,T\in\mathbb N}\max_{\substack{j=1,\ldots, n\\s=1,\ldots, T}}
\E&\l[\l\vert\frac 1{\sqrt{nT}} \sum_{i=1}^n\sum_{t=1}^T\l\{ e_{is} e_{jt}-\E[ e_{is} e_{jt}]\r\} \r\vert^2\r]\nn\\
&=\sup_{n,T\in\mathbb N}\max_{\substack{j=1,\ldots, n\\s=1,\ldots, T}}\frac 1{nT}\sum_{i,k=1}^n\sum_{t,u=1}^T \l\{\E\l[ e_{is} e_{jt} e_{ks} e_{ju}\r]-\E[ e_{is} e_{jt}]\E[ e_{ks} e_{ju}]\r\}\le K_ e.\label{eq:bordellobai}
\end{align} 
If we set $j=i$ in \eqref{eq:bordellobai} we get a condition similar to that assumed by \citet[Assumption C.5]{bai03}, where, however, $T=1$, and the sums of 8th cross-cumulants are bounded.

Part (c-ii) has two more important consequences. First, it implies that (see Lemma \ref{lem:GxiBAI}(vi)):
\begin{align}
\sup_{n,T\in\mathbb N}\max_{j=1,\ldots, n}\E\l[\l\Vert\frac 1{\sqrt {nT}}\sum_{s=1}^T\sum_{k=1}^n \bm\lambda_k
\{
e_{ks}e_{js}-\E[e_{ks}e_{js}]
\}
\r\Vert^2
\r]\le M_{\Lambda}^2K_e.\label{eq:frascati24}
\end{align}
This result is crucial to prove asymptotic normality of the estimated loadings. In particular, without \eqref{eq:frascati24} consistency of the loadings  would hold but with a slower rate, $\min(\sqrt n,\sqrt T)$ which is slower than  $\min(n,\sqrt T)$ derived in Proposition \ref{prop:L2}. Such a slower rate is what prevents the proof of asymptotic normality of the estimated loadings from going through. 

Second, part (c-ii)  implies also that (setting $n=1$ therein):
\begin{align}
\sup_{n,T\in\mathbb N}
\max_{\substack{i,j=1,\ldots,n\\s=1,\ldots, T}}
&\E\l[\l\vert\frac 1{\sqrt {T}} \sum_{t=1}^T\l\{ e_{is} e_{jt}-\E[ e_{is} e_{jt}]\r\} \r\vert^2\r]\nn\\
&=\sup_{n,T\in\mathbb N}
\max_{\substack{i,j=1,\ldots,n\\s=1,\ldots, T}}\frac 1{T}\sum_{t=1}^T\sum_{u=1}^T\l\{\E[ e_{is} e_{jt} e_{is} e_{ju}]-\E[ e_{is} e_{jt}]\E[ e_{is} e_{ju}]
\r\}\le  {K_ e}, \label{eq:bordellobai2}
\end{align}
which means that the sample (auto)covariances between $\{ e_{it}\}$ and $\{ e_{jt}\}$ are $\sqrt T$-consistent estimators of their population counterparts. In particular, by setting $s=t$ in \eqref{eq:bordellobai2}, we see that we can consistently estimate the $(i,j)$th entry of the idiosyncratic covariance matrix $\bm\Gamma^ e$ with its usual sample counterpart.
 Like for the factors, this assumption is a high-level one and requires introducing the sample size $T$. In Appendix \ref{sec:hannan} we provide a series of possible primitive conditions on the process $\{ e_{it}\}$ which guarantee part (c-ii) to hold.

Last, we notice that we assumed serially homoskedastic idiosyncratic components just for simplicity. Indeed, if we allowed for serial heteroskedasticity so that we had $\gamma_{ij,t,t-k}^ e:=\E[ e_{it} e_{j,t-k}]$, then all our proofs would still hold provided that in part (b) we required $\sup_{t\in\mathbb Z} \vert \gamma_{ij,t,t-k}^ e\vert\le \rho^{|k|} M_{ij}$, and then we replaced  $\gamma_{ij,k}^ e$ with $\gamma_{ij,t,t-k}^ e$ throughout. As for the factors, conditional heteroskedasticity can also be allowed for (see Appendix \ref{sec:hannan} for details).

We then introduce the following identifying assumption.

\begin{ass}[\textsc{distinct eigenvalues}]
\label{ass:eval}
The $r$ eigenvalues of  $\bm\Sigma_\Lambda\bm\Gamma^F$ are distinct.
\end{ass}

The same assumption is in \citet[Assumption G]{bai03}. Since the $r$ eigenvalues of  $\bm\Sigma_\Lambda\bm\Gamma^F$ are given by $\lim_{n\to\infty} n^{-1}\mu_j^{C}$, $j=1,\ldots,r$, Assumption \ref{ass:eval} can be equivalently stated by requiring $\overline C_j>\underline C_{j+1}$, $j=1,\ldots, r-1$, in \eqref{eq:lindiv}.  This assumption is needed so that the eigenvectors of the sample covariance matrix, $\wh{\bm\Gamma}^x$, which, by definition, coincide (up to a scale) with the PC estimator of the loadings,  identify, asymptotically as $n,T\to\infty$, the same directions (up to a sign) as the eigenvectors of the common component covariance, $\bm\Lambda\bm\Gamma^F\bm\Lambda^\prime$, which, in turn, span the same space as the columns of $\bm\Lambda$.

We then turn to the issue of dependence between common and idiosyncratic components. Given the nature of PCA, it is natural to assume contemporaneously orthogonal, i.e., uncorrelated, common and idiosyncratic components:
\begin{asso}[\textsc{orthogonality}]
\labelO{eq:ortho} For all $i,j\in\mathbb N$ and all $t\in\mathbb Z$,
$\E[{C}_{it} e_{jt}]=0.$
\end{asso}
This assumption is enough if we just want to prove consistency of the PC estimators, provided we
make primitive assumptions on the processes $\{\mbf F_t\}$ and $\{ e_{it}\}$ as those in Appendix \ref{sec:hannan} (see the proofs of Proposition \ref{prop:L} and Lemma \ref{lem:covarianze}(i-b)). In this way, the sample covariance between common and idiosyncratic components is a $\sqrt T$-consistent estimator of the moment condition in Assumption~\ref{eq:ortho}. Alternatively, this latter condition can be directly assumed (see \eqref{eq:oslo2} below and its discussion).

However, to prove asymptotic normality of the PC estimators, we need, besides orthogonality,  additional conditions relating to 4th-order dependencies.  To this end, we assume the following. 

\begin{ass}[\textsc{independence}]
\label{ass:ind} The processes
$\{ e_{it},\, i\in\mathbb N,\, t\in\mathbb Z\}$ and $\{\mbf F_{t}, t\in\mathbb Z\}$ are mutually independent. 
\end{ass}

This assumption is stronger than what is usually assumed. It obviously implies contemporaneous orthogonality as stated in Assumption~\ref{eq:ortho}, but it deserves a more detailed discussion. Specifically, it has three main implications.

First,  from Assumptions \ref{ass:idio}(c-ii) and \ref{ass:ind} it follows that (see Lemma \ref{lem:GxiBAI}(iv)):
\begin{align}
\sup_{n,T\in\mathbb N}\max_{t=1,\ldots, T}
\E\l[\l\Vert\frac 1{\sqrt {n T} }\sum_{s=1}^T\sum_{k=1}^n \mbf F_{s} \{e_{ks}e_{kt}-\E[e_{ks}e_{kt}] \}
\r\Vert^2\r] \le r M_F K_e,\label{eq:frascati24b}
\end{align}
which is equivalent to what is assumed by \citet[Assumption F.1]{bai03}. This inequality is the analogue,
 for the factors, of  condition \eqref{eq:frascati24}, which we derived for the loadings from Assumption \ref{ass:idio}(c-ii). 
It is a crucial requirement as it is necessary to prove asymptotic normality of the estimated factors.  In particular, without \eqref{eq:frascati24b} consistency of the factors would hold, but with a rate, $\min(\sqrt n,\sqrt T)$ slower than the rate  $\min(\sqrt n, T)$ derived in Proposition \ref{prop:F2}. Such a slower rate is what prevents the proof of asymptotic normality of the estimated factors from going through. 
While proving \eqref{eq:frascati24} is easy since the loadings are deterministic, to prove \eqref{eq:frascati24b} it must be that $\E[ e_{ks} e_{kt} e_{iu} e_{it}F_{js}F_{ju}]=\E[ e_{ks} e_{kt} e_{iu} e_{it}]\,\E[F_{js}F_{ju}]$ for all $i,k=1,\ldots, n$, all $t,s,u=1,\ldots, T$, and all $j=1,\ldots, r$, this is why we need independence in Assumption \ref{ass:ind} if we want to prove  \eqref{eq:frascati24b},  rather than just assuming it. 

Second, Assumption \ref{ass:ind} implies that 
 (see \eqref{eq:oslo3} in the proof of Lemma \ref{lem:GxiBAI}(v))
\begin{align}\label{eq:oslo2}
&\sup_{n,T\in\mathbb N}\max_{i=1,\ldots, n}\E\l[\l\Vert\frac 1{\sqrt {T}}\sum_{t=1}^T\mbf F_t e_{it}\r\Vert^2\r] \le \frac{r M_F(C_ e+M_ e)(1+\rho)}{1-\rho},
\end{align}
which, in turn, implies
the same condition as in \citet[Assumption D]{bai03}, indeed,
\beq\label{eq:BaiD}
\!\!\!\sup_{n,T\in\mathbb N}\E\l[\frac 1n\sum_{i=1}^n \l\Vert\frac 1{\sqrt {T}}\sum_{t=1}^T\mbf F_t e_{it} \r\Vert^2\r] \le 
\sup_{n,T\in\mathbb N} \max_{i=1,\ldots, n} \E\l[\l\Vert\frac 1{\sqrt {T}}\sum_{t=1}^T\mbf F_t e_{it} \r\Vert^2\r] \le\frac{r M_F(C_ e+M_ e)(1+\rho)}{1-\rho}.
\eeq
Notice that  \eqref{eq:BaiD} allows for weak dependence between the factors and the idiosyncratic components,  but just at the level of 4th-order moments, while it still implies contemporaneous orthogonality. Indeed, as $T\to\infty$, by Chebychev's inequality, \eqref{eq:BaiD}  implies that $\Vert T^{-1}\sum_{t=1}^T\mbf F_t  e_{it}\Vert =O_{\mathrm P}(T^{-1/2})$. Moreover, under the assumptions on serial dependence of the factors and the idiosyncratic components allowing for Assumptions \ref{ass:common}(c-ii) and \ref{ass:idio}(c-ii) to hold, we have the Law of Large Numbers 
$\Vert T^{-1}\sum_{t=1}^T\mbf F_t  e_{it}-\E[\mbf F_t  e_{it}]\Vert=O_{\mathrm P}(T^{-1/2})$ (this is implied also by Assumption \ref{ass:CLT}(a) below). Hence, by uniqueness of the limit, it follows that we must have $\E[\mbf F_t  e_{it}] = \mbf 0_r$ for all $i\in\mathbb N$ and $t\in\mathbb Z$, and, thus, Assumption~\ref{eq:ortho} still holds.

Crucially, the moment condition in \eqref{eq:oslo2} is needed to prove not only asymptotic normality, but also consistency of the PC estimators (see Lemma \ref{lem:covarianze}(i-a)) so either it is assumed, or it is derived from Assumption \ref{ass:ind}, or it could be ensured by making further primitive assumptions on the processes $\{\mbf F_t\}$ and $\{ e_{it}\}$, as those discussed in Appendix \ref{sec:hannan}.

Third,  from Assumptions \ref{ass:common}(a), \ref{ass:idio}(b), and \ref{ass:ind} it follows that (see Lemma \ref{lem:GxiBAI}(v)):
\begin{align}
\sup_{n,T\in\mathbb N}\E\l[\l\Vert\frac 1{\sqrt{nT}}\sum_{t=1}^T\sum_{j=1}^n\mbf F_t{\bm\lambda}_j^\prime  e_{jt} \r\Vert_F^2\r] 
\le 
\frac{r^2 M_\Lambda^2 M_F(C_ e+M_ e)(1+\rho)}{1-\rho},
\label{eq:BaiF2}
\end{align}
which is the same condition as in \citet[Assumption F.2]{bai03}. Once again  condition \eqref{eq:BaiF2} is crucial for proving asymptotic normality, thus, it must be either assumed or derived from Assumption \ref{ass:ind}.

The trade-off is then clear. Either we make Assumption \ref{ass:ind}, which, although strong, is intuitive and it implies Assumption~\ref{eq:ortho} as well as the required moment inequalities in \eqref{eq:frascati24b}, \eqref{eq:BaiD}, and \eqref{eq:BaiF2}. Or we directly assume the less intuitive \eqref{eq:frascati24b}, \eqref{eq:BaiD}, and \eqref{eq:BaiF2} as in  \citet[Assumptions F.1, D, and F.2, respectively]{bai03}, which imply also Assumption~\ref{eq:ortho}. Here we choose the former approach in order to keep working with more intuitive assumptions.
Notice that our choice coincides with \citet[Assumption D]{baing04}, where, in the presence of unit roots, they assume independence between the first differences of the factors and the idiosyncratic components, for simplicity.

In order to prove asymptotic normality of the PC estimators it is also common to introduce the following assumptions.

\begin{ass}[\textsc{Central limit theorems}]
\label{ass:CLT}
$\,$
\begin{compactenum}[(a)]
\item For all $i\in\mathbb N$, as $T\to\infty$,
	$
	\frac 1{\sqrt T}\sum_{t=1}^T
	 \mbf F_t  e_{it}
	\to_d
	\mathcal N\l(\mbf 0_r, \bm\Phi_i
	\r),
	$
	with $\bm\Phi_i:=\lim_{T\to\infty}\frac 1 T\sum_{t,s=1}^T
	\E_{}[\mbf F_t\mbf F_s^\prime e_{it} e_{is}]$.
\item For all $t\in\mathbb Z$, as $n\to\infty$,
	$
	\frac 1{\sqrt n}\sum_{i=1}^n
	 \bm\lambda_i   e_{it}
	\to_d
	\mathcal N\l(\mbf 0_r, \bm\Gamma_t
	\r),
	$
	with $\bm\Gamma_t:=\lim_{n\to\infty}\frac 1 n\sum_{i,j=1}^n 
	\bm\lambda_i\bm\lambda_j^\prime\E_{}[ e_{it} e_{jt}]$.
\end{compactenum}
\end{ass}

Part (a) is also assumed by \citet[Assumption F.4]{bai03}. It can be derived from more primitive assumptions. For example, we could assume strong mixing factors and idiosyncratic components and such that, for all $t\in\mathbb Z$ and all $i\in\mathbb N$, we strengthen Assumptions \ref{ass:common}(c-i) and \ref{ass:idio}(c-i) to: $\E[\Vert\mbf F_{t}\Vert^{4+\epsilon}]\le K_F$ and  $\E[| e_{it}|^{4+\epsilon}]\le Q_ e$, for some $\epsilon>0$.  
Then, $\{\mbf F_t e_{it}\}$ is also strong mixing because of \citet[Theorem 5.1.a]{bradley05} and with finite 4th order moments because of Assumption \ref{ass:ind} and part (a) would follow from \citet[Theorem 1.4]{ibra62}. For alternative primitive conditions see also Appendix \ref{sec:hannan}. Notice also that, under Assumption \ref{ass:ind}, in $\bm\Phi_i$ in part (a) we actually have 
	$\E_{}[\mbf F_t\mbf F_s^\prime e_{it} e_{is}]=\E_{}[\mbf F_t\mbf F_s^\prime]\,\E[ e_{it} e_{is}]$

 Part (b) is also assumed by \citet[Assumption F.3]{bai03}. Clearly, it holds if we assumed cross-sectionally uncorrelated idiosyncratic components, i.e., if $\bm\Gamma^ e$ were diagonal as in an exact factor model. In general, to derive part (b) from primitive assumptions we could either introduce an ordering of the $n$ cross-sectional items and a related notion of spatial dependence and derive it from the properties of stationary mixing random fields \citep{bolthausen1982}, or of cross-sectional martingale difference sequences \citep{KP13}. Or we could apply results on exchangeable sequences, which are instead independent of the ordering, and are in turn obtained by virtue of the Hewitt-Savage-de Finetti theorem (\citealp[Theorem 4]{austern2022limit}). In any case, we would also need  to strengthen Assumption \ref{ass:idio}(c-i)  by asking that for all $t\in\mathbb Z$ and all $i\in\mathbb N$, $\E[| e_{it}|^{4+\epsilon}]\le Q_ e$, for some $\epsilon>0$.
%

A last assumption is needed for proving asymptotic normality. 
\begin{ass}[\textsc{Rates}]
\label{ass:rates}
As $n,T\to\infty$,
$\sqrt T/n\to0$ and $\sqrt n/T\to0$.
\end{ass}

This is also assumed in \citet[Theorems 1 and 2]{bai03}. It is mild as it allows $n/T\to \gamma$ for any $\gamma\in(0,\infty)$, as well as moderately unbalanced regimes with $n/T\to0$ or $n/T\to\infty$, as $n,T\to\infty$.

\begin{table}[t!]
\caption{Assumptions for PCA}\label{tab:ass}
\centering
\scriptsize{
\begin{tabular}{l | l}
\hline
\hline
\citet{bai03} & This paper\\
\hline\\[-5pt]
Assumption A & Assumptions \ref{ass:common}(c-i) and \ref{ass:common}(c-ii).\\[3pt]
Assumption B&Assumption \ref{ass:common}(a).\\[3pt]
Assumption C.1 & Assumption \ref{ass:idio}(c-i), but only 4th order moments need to be finite.\\[3pt]
Assumption C.2 & Condition \eqref{eq:baiC2}, implied by Assumption \ref{ass:idio}(b), and \eqref{eq:baiC2bis}, implied by Assumption \ref{ass:idio}(b) (see Lemma \ref{lem:GxiBAI}(i)).\\[3pt]
Assumption C.3 & Condition \eqref{eq:baiC3}, implied by Assumption \ref{ass:idio}(b), and \eqref{eq:baiC3bis}, implied by Assumption \ref{ass:idio}(b) (see Lemma \ref{lem:Gxi}(ii)).\\[3pt]
Assumption C.4 & Condition \eqref{eq:bai}, implied by Assumption \ref{ass:idio}(b) (see Lemma \ref{lem:Gxi}(i)).\\[3pt]
Assumption C.5 & Assumption \ref{ass:idio}(c-ii), but only 4th order cumulants need to be summable.\\[3pt]
Assumption D & Condition \eqref{eq:BaiD}, implied by Assumption \ref{ass:ind} (see Lemma  \ref{lem:LLN}(i)).\\[3pt]
Assumption E.1 & Condition \eqref{eq:baiE1}, implied by Assumption \ref{ass:idio}(b) (see Lemma \ref{lem:GxiBAI}(ii)).\\[3pt]
Assumption E.2& Condition \eqref{eq:baiE2}, implied by Assumption \ref{ass:idio}(b) (see Lemma \ref{lem:GxiBAI}(iii)).\\[3pt]
Assumption F.1& Condition \eqref{eq:frascati24b}, implied by Assumptions \ref{ass:common}(b), \ref{ass:idio}(c-ii), and \ref{ass:ind} (see Lemma \ref{lem:GxiBAI}(iv)).\\[3pt]
Assumption F.2& Condition \eqref{eq:BaiF2}, implied by Assumptions \ref{ass:common}(a), \ref{ass:idio}(b), and \ref{ass:ind} (see Lemma \ref{lem:GxiBAI}(v)).\\[3pt]
Assumption F.3& Assumption \ref{ass:CLT}(b).\\[3pt]
Assumption F.4& Assumption \ref{ass:CLT}(a).\\[3pt]
Assumption G& Assumption \ref{ass:eval}.\\[3pt]
N.A. & Assumptions \ref{ass:common}(b), \ref{ass:common}(d), \ref{ass:idio}(a), and \ref{ass:rates}.\\
\hline
\hline
\end{tabular}
}
\end{table}

In Table \ref{tab:ass} we summarize the relation between our Assumptions \ref{ass:common}-\ref{ass:rates} and those made by \citet{bai03}. Assumptions similar to ours  are also found in other works on PCA for factor models with some important differences, though. First, \citet{stockwatson02JASA} assume, like here, only summable 4th-order moments, but do not impose any other cross-moment conditions since they just prove consistency but not asymptotic normality. Second, \citet{FLM13} impose stronger assumptions by requiring exponentially decaying tails of the distributions of factors and idiosyncratic components. This is done in order to derive uniform consistency for loadings and factors.

\section{Identification of the approximate factor model}\label{sec:blessing}

In this section we show that, under Assumptions \ref{ass:common} and \ref{ass:idio} together with Assumption~\ref{eq:ortho}, the common and idiosyncratic components are asymptotically identified as $n\to\infty$. 

First, for any  $n\in\mathbb N$, let  $\bm{C}_t:=\bm\Lambda\mbf F_t$ denote the $n$-dimensional vector of common components with covariance matrix $\bm\Gamma^{C}:=\E[\bm{C}_t\bm{C}_t^\prime]=\bm\Lambda\bm\Gamma^F \bm\Lambda^\prime$. Under Assumptions \ref{ass:common}(a) and \ref{ass:common}(b), $\bm\Gamma^{C}$  has rank $r$ and in Lemma \ref{lem:Gxi}(iv) we prove that its $j$th largest eigenvalue $\mu_j^{C}$, $j=1,\ldots, r$, is such that 
\beq\label{eq:lindiv}
\underline C_j\!\le \lim_{n\to\infty} \frac{\mu_{j}^{C}}n\le\! \overline C_j,
\eeq
for some finite positive reals $\underline C_j$ and $\overline C_j$ independent of $n$. 
Thus, we restrict attention to pervasive, or strong, factors. Rates of eigenvalue divergence slower than $n$ would require imposing some ordering or structure on the cross-sectional units \citep{BH25}. In the absence of such an ordering, the natural rate in Assumption \ref{ass:common}(a) is $\sqrt n$, which can be interpreted as a form of cross-sectional stationarity.

Second, in Lemma \ref{lem:Gxi}(v) we prove that, under Assumptions \ref{ass:idio}(a) and \ref{ass:idio}(b), the largest eigenvalue of the idiosyncratic covariance matrix $\bm\Gamma^e:=\E[\bm e_t\bm e_t^\prime]$ is such that 
\beq\label{eq:evalidio}
 \sup_{n\in\mathbb N}\mu_{1}^ e  \le C_ e+M_ e,
 \eeq
 where $C_ e$ and $M_ e$ are defined in Assumption \ref{ass:idio}(b). This is the essence of an approximate factor model 
 as opposed to an exact factor model, which imposes the restrictive assumption that $\bm\Gamma^e$ is diagonal. Condition \eqref{eq:evalidio} was originally assumed by \citet{chamberlainrothschild83}. 
 
 Third, let $\bm\Gamma^x:=\E[\mbf x_t\mbf x_t^\prime]$, and denote by $\mu_j^x$, $j=1,\ldots,n$, its $j$th largest eigenvalue. By Assumption~\ref{eq:ortho}, $\bm\Gamma^x=\bm\Gamma^{C}+\bm\Gamma^ e$. Hence, Weyl's inequality \citep[Theorem 1]{MK04}, together with \eqref{eq:lindiv} and \eqref{eq:evalidio}, implies that (see Lemma \ref{lem:Gxi}(vi)):
\begin{align}
&\underline C_r\le \lim_{n\to\infty} \frac{ \mu_{r}^x}n\le \overline C_r\;\text{ and }\;\sup_{n\in\mathbb N}\mu_{r+1}^x \le C_ e+M_ e.\label{eq:evalidioX}
\end{align}
Note that \eqref{eq:evalidioX} is equivalent to saying that under \eqref{eq:lindiv} and \eqref{eq:evalidio} there exists an eigengap in the spectrum of $\bm\Gamma^x$ that widens with $n$. Furthermore, the converse is also true, i.e., if \eqref{eq:evalidioX} holds, then \eqref{eq:lindiv} and \eqref{eq:evalidio} hold and $\{x_{it},\, i\in\mathbb N,\, t\in\mathbb Z\}$ admits an approximate factor representation (\citealp[see][Theorem 4, for the original proof]{chamberlainrothschild83}, and \citealp[Theorem 2]{gersing2023weak}, or \citealp[Proposition 1]{BH25} for more recent and complete proofs). This means that the approximate factor model is, in fact, not merely a statistical model; it is a general representation of an infinite-dimensional stationary vector time series.

The above representation result implies that the presence of $r$ factors can be detected by detecting an eigengap in the spectrum of $\bm\Gamma^x$ that widens with $n$. Most methods for determining the number of factors $r$ rely precisely on this feature \citep[see, e.g.,][]{baing02,ABC10,ahnhorenstein13}.  

Consequently, once $r$ has been determined, the common component can be disentangled from the idiosyncratic component, and the model can be regarded as identified. To see this, suppose first that we observe an infinite-dimensional cross-section and that $\bm\Gamma^x$ is known. In that case, the common component can be recovered by projecting the data onto its first $r$ population PCs.

In practice, however, only a finite number $n$ of time series is observed. Thus, even if $\bm\Gamma^x$ were known, the PCs would provide only an approximation to the true common component. This approximation, which depends only on $n$, reflects the signal-to-noise ratio between the pervasive common component and the bounded idiosyncratic component. Furthermore, since $\bm\Gamma^x$ is in fact unknown, one must use sample PCs, which adds an estimation error depending on $T$.

Summing up, identification (and estimation) of an approximate factor model are possible only if we let $n\to\infty$, meaning that the model is fully identified only if we observed an infinite dimensional cross-section. Thus, although these are large-panel models, they enjoy a blessing of dimensionality rather than suffering from its curse.

\section{Consistent estimation of the loading and factor spaces - part 1}\label{sec:cons1}
Our first asymptotic result is derived using only the properties of the sample covariance and its eigenvalues and eigenvectors (see Appendix \ref{prop:Lproof} for a proof).
\begin{prop}\label{prop:L}
Under Assumptions \ref{ass:common} through \ref{ass:eval}, Assumption~\ref{eq:ortho}, and either of the following:
\begin{inparaenum}
\item [(A)] Assumptions \ref{ass:Wold} or \ref{ass:Hannan} or \ref{ass:Wu} in Appendix \ref{sec:hannan} on serial dependence of the factor process $\{\mbf F_t,\, t\in\mathbb Z\}$;
\item [(B)] the moment condition in \eqref{eq:oslo2} and Assumption \ref{ass:common}(c-ii) holding with the rate $\sqrt T$;
\end{inparaenum}
it follows that, as $n,T\to\infty$,\footnote{We say parts (b) and (d) hold uniformly in $i$ or $t$ when the constants involved in the $O_{\mathrm P}$ bound do not depend on $i$ and $t$. This does not mean uniform consistency, which would imply statements like 
$\max_{i=1,\ldots, n}\Vert\wh{\bm \lambda}_i^\prime-\bm\lambda_i^\prime\bm{\mathcal H}\Vert=o_{\mathrm P}(1)$ nor
$\max_{t=1,\ldots,T}\Vert\wh{\mbf F}_t-\bm{\mathcal H}^{-1}\mbf F_t\Vert=o_{\mathrm P}(1)$.} 
\begin{compactenum}[(a)]
\item $\min(n,\sqrt T)\l\Vert\frac{\wh{\bm\Lambda}-\bm\Lambda\bm{\mathcal H}}{\sqrt n}\r\Vert= O_{\mathrm P}(1)$;
\item $ {\min(\sqrt n,\sqrt T)} \l\Vert\wh{\bm \lambda}_i^\prime-\bm\lambda_i^\prime\bm{\mathcal H}\r\Vert= O_{\mathrm P}(1)$, uniformly in $i$;
\item $\min(\sqrt n,\sqrt T)\l\Vert\frac {\wh{\bm F}-\bm F(\bm{\mathcal H}^{-1})^\prime}{\sqrt T}\r\Vert= O_{\mathrm P}(1)$;
\item ${\min(\sqrt n,\sqrt T)} \l\Vert\wh{\mbf F}_t-\bm{\mathcal H}^{-1}\mbf F_t\r\Vert= O_{\mathrm P}(1)$, uniformly in $t$;
\end{compactenum}
where $\bm{\mathcal H} :=(\bm\Gamma^F)^{1/2} \mbf K{\mbf J}$ is an $r\times r$ finite and invertible matrix, with $\mbf K$ having as columns the normalized eigenvectors of $(\bm\Gamma^F)^{1/2}(n^{-1}\bm \Lambda^\prime\bm \Lambda) (\bm\Gamma^F)^{1/2}$, and $\mbf J$ being an $r\times r$ diagonal matrix with entries $\pm 1$ which depend on $T$.
\end{prop}
The bound derived in part (a) is tighter than the one derived by \citet[Proposition P]{FGLR09} and \citet[Proposition 1.ii]{bai2020simpler} where the rate is $\min(\sqrt n,\sqrt T)$.
Parts (b), (c), and (d) follow directly from part (a). In particular, part (c) coincides with the result by 
\citet[Proposition 1.i]{bai2020simpler}, while parts (b) and (d) coincide with the results by \citet[Proposition 2.i and 2.ii]{DGRfilter}. The latter adopt a  technique similar to ours for their proof with the main difference that they assume independence between factors and idiosyncratic components while we just require orthogonality.

The  proof makes use of the Cauchy-Schwarz inequality. Therefore, with the exception of part (a), the derived rates are not the sharpest possible. Hence, although this result improves over the consistency proof given in Remark \ref{rem:SVD}, it still does not provide the sharpest rates which allow us to derive asymptotic normality and make inference. These can be derived at the cost of imposing additional moment conditions (see Section \ref{sec:cons2}). Nevertheless, Proposition \ref{prop:L}  is a useful and important result, because,
besides providing an intuitive and quick proof of consistency of PC, which is often enough for many purposes, it is also needed for proving successive results.

Part (a) is based essentially on a result on the $n\times n$ sample covariance matrix $\wh{\bm\Gamma}^x$ which in turn is determined by the serial dependence of the considered stochastic process, as explained in Appendix \ref{sec:hannan}.
Its proof is based on four main steps. 

First, we prove that the rescaled sample covariance matrix $n^{-1}\wh{\bm\Gamma}^x$ is a $\sqrt T$-consistent estimator of the rescaled population covariance matrix $n^{-1}\bm\Gamma^x$ for all $n\in\mathbb N$. This result holds under a minimal and mild set of assumptions, namely Assumptions \ref{ass:common} and \ref{ass:eval} jointly with Assumption~\ref{eq:ortho} plus either the conditions (A) (see Lemma \ref{lem:covarianze}(i-b)) or the conditions (B) (see Lemma \ref{lem:covarianze}(i-a)).
We stress that under conditions (A) or (B) Assumption \ref{ass:ind} of independence between factors and idiosyncratic components is not needed. 

Second, we prove that the rescaled sample covariance $n^{-1}\wh{\bm\Gamma}^x$ actually converges to the rescaled common component covariance $n^{-1}\bm\Gamma^{C}$ with rate $\min(n,\sqrt T)$ (see Lemma \ref{lem:covarianze}(ii)). This result follows from Assumption \ref{ass:idio} which implies that the idiosyncratic covariance matrix $\bm\Gamma^ e$ has bounded eigenvalues for all $n\in\mathbb N$ (see \eqref{eq:evalidio}). It is clear that, in fact, Assumption \ref{ass:idio}(b) is not strictly needed since it is enough to make the weaker assumption of bounded idiosyncratic eigenvalues.  

Third, we prove that the matrices of  sorted eigenvalues, $\wh{\mbf M}^x$, and corresponding normalized eigenvectors, $\wh{\mbf V}^x$, of $\wh{\bm\Gamma}^x$ converge to the matrices of sorted eigenvalues,  ${\mbf M^{C}}$, and corresponding normalized eigenvectors, ${\mbf V^{C}}$, of $\bm\Gamma^{C}$. Specifically, from Weyl's inequality we get $n^{-1}\Vert \wh{\mbf M}^x-\mbf M^{C}\Vert=O_{\mathrm P}(\max(n^{-1},T^{-1/2}))$  (see Lemma \ref{lem:covarianze}(iii) and \citealp[Theorem 1]{MK04}), while from Davis-Kahan theorem we get $\Vert \wh{\mbf V}^x-\mbf V^{C}{\mbf J}\Vert =O_{\mathrm P}(\max(n^{-1},T^{-1/2}))$, where $\mbf J$ is an $r\times r$ diagonal matrix with entries $\pm 1$ depending only on $T$, and accounting for the fact that a sample and a population eigenvector identify asymptotically the same one-dimensional subspaces, but might point in opposite directions (see Lemma \ref{lem:covarianze}(iv) and \citealp[Corollary 1]{yu15}). Notice that for this result to hold we must require $n\to\infty$, so that we can apply Davis-Kahan theorem using the $r$ eigenvalues of $\lim_{n\to\infty}n^{-1}\bm\Lambda\bm\Gamma^F\bm\Lambda^\prime$, which are the same as the eigenvalues of $\bm\Sigma_\Lambda\bm\Gamma^F$, and  are distinct by Assumption \ref{ass:eval}. If we kept $n$ fixed, we would consistently estimate the eigenvectors $\mbf V^{C}$ only up to a rotation \citep[Theorem 2]{yu15}, unless we also assumed distinct eigenvalues of $\bm\Gamma^C=\bm\Lambda\bm\Gamma^F\bm\Lambda^\prime$ for all $n\in\mathbb N$ (see Section \ref{sec:II}).

Fourth, since $\bm\Gamma^{C} =\bm\Lambda\bm\Gamma^F\bm\Lambda^\prime = \mbf V^{C}\mbf M^{C}\mbf V^{{C}\prime}$, the columns of the true loadings matrix $\bm\Lambda$ must span the same space as the normalized eigenvectors of the common component covariance matrix. In other words, it must be that $\bm\Lambda(\bm\Gamma^F)^{1/2}\mbf  K=\mbf V^{C}(\mbf M^{C})^{1/2}$ for some finite invertible $r\times r$ matrix $\mbf K$ with columns given by the normalized eigenvectors of $(\bm\Gamma^F)^{1/2} (n^{-1}\bm\Lambda^\prime\bm\Lambda) (\bm\Gamma^F)^{1/2}$  (see Lemma \ref{lem:KO1}(iii)), i.e., $\mbf K$ is a rotation. 

Summing up, given that the PC estimator of the loadings is $\wh{\bm\Lambda}=\wh{\mbf V}^x(\wh {\mbf M}^x)^{1/2}$ (see \eqref{eq:estL}), part (a) follows, with  
$\bm{\mathcal H}:=(\bm\Gamma^F)^{1/2} \mbf K{\mbf J}$. Thus, the true loadings are recovered up to: (i) a scale given by $(\bm\Gamma^F)^{1/2}$; (ii) a rotation $\mbf K$, and (iii) a diagonal matrix of signs ${\mbf J}$. Note that both $(\bm\Gamma^F)^{1/2}$ and $\mbf K$ do not depend on the sample size $T$, but depend only on population quantities as well as $n$.  The only dependence on $T$ is in the sign matrix ${\mbf J}$, but such dependence can be easily fixed. 

The following result gives a series of more intuitive asymptotically equivalent expressions for $\bm{\mathcal H}$ (see Appendix \ref{corol:ovvioproof} for a proof).
\begin{prop}\label{corol:ovvio}
Under Assumptions \ref{ass:common} through \ref{ass:eval}, Assumption~\ref{eq:ortho}, and either of the following:
\begin{inparaenum}
\item [(A)] Assumptions \ref{ass:Wold} or \ref{ass:Hannan} or \ref{ass:Wu} in Appendix \ref{sec:hannan} on serial dependence of the factor process $\{\mbf F_t,\, t\in\mathbb Z\}$;
\item [(B)] the moment condition in \eqref{eq:oslo2} and Assumption \ref{ass:common}(c-ii) holding with the rate $\sqrt T$;
\end{inparaenum}
it follows that, as $n,T\to\infty$,
\begin{compactenum}
\item [(a)] $\min(n,\sqrt T)\l\Vert\bm{\mathcal H} - (\bm\Lambda^\prime\bm\Lambda)^{-1}\bm\Lambda^\prime\wh{\bm\Lambda}  \r\Vert = O_{\mathrm P}(1)$;
\item [(b)] $\min(\sqrt n,\sqrt T)\l\Vert\bm{\mathcal H}^{-1} - \wh{\bm F}^\prime \bm F (\bm F^\prime\bm F)^{-1} \r\Vert = O_{\mathrm P}(1)$;
\item [(c)] $\min(n,\sqrt T)\l\Vert\bm{\mathcal H}^{-1} - (\wh{\bm\Lambda}^\prime\wh{\bm\Lambda})^{-1}\wh{\bm\Lambda}^\prime{\bm\Lambda}  \r\Vert = O_{\mathrm P}(1)$, where we recall that $\wh{\bm\Lambda}^\prime\wh{\bm\Lambda}=\wh{\mbf M}^x$;
\item [(d)] $\min(\sqrt n,\sqrt T)\l\Vert\bm{\mathcal H} - {\bm F}^\prime \wh{\bm F} (\wh{\bm F}^\prime\wh{\bm F})^{-1} \r\Vert = O_{\mathrm P}(1)$, where we recall that $\wh{\bm F}^\prime \wh{\bm F} = T\, \mbf I_r$.
\end{compactenum}
\end{prop}

The proof is an immediate consequence of Proposition \ref{prop:L}(a) and \ref{prop:L}(c). Since the assumptions made are the same as those made for Proposition \ref{prop:L}, these results are directly applicable to  the  consistency  results proved therein. 

Proposition \ref{corol:ovvio} shows that asymptotically the matrix $\bm{\mathcal H}$ is obtained by linear projection of the estimated loadings onto the true loadings (part (a)) or of the true factors onto the estimated ones (part (d)). Similarly, asymptotically $\bm{\mathcal H}^{-1}$ is obtained by linear projection of the estimated factors onto the true factors (part (b)) or by linear projection of the true loadings onto the estimated ones (part (c)). The derived rates are not the sharpest possible for two reasons.  First, the rates in part (a) and (b) are obtained by using the Cauchy-Schwarz bound $n^{-1}\Vert\bm\Lambda^\prime(\wh{\bm\Lambda}-\bm\Lambda\bm{\mathcal H})\Vert\le n^{-1/2}\Vert \bm\Lambda\Vert\, n^{-1/2}\Vert\wh{\bm\Lambda}-\bm\Lambda\bm{\mathcal H} \Vert$, which, therefore, inherits the bound obtained for the loadings in Proposition \ref{prop:L}(a).  Second, the rates in parts (b) and (d) are slower than those in parts (a) and (c) since the rates in Proposition \ref{prop:L}(c) are not the sharpest possible. As shown, in Proposition \ref{prop:LLFF} below, it is possible to define a different transformation matrix, which is asymptotically equivalent to  $\bm{\mathcal H}$, but  which  converges to the projection matrices in Proposition \ref{corol:ovvio} at a faster rate. 

Finally, consider the following spectral decomposition: \beq\label{eq:U0V0U0}
(\bm\Gamma^F)^{1/2}\bm\Sigma_\Lambda(\bm\Gamma^F)^{1/2} =: \bm\Upsilon_0\bm V_0\bm\Upsilon_0^\prime,
\eeq
where $\bm\Upsilon_0$ is the $r\times r$ matrix having as columns the normalized eigenvectors of $(\bm\Gamma^F)^{1/2}\bm\Sigma_\Lambda(\bm\Gamma^F)^{1/2}$, and $\bm V_0$ is the $r\times r$ matrix of corresponding eigenvalues sorted in descending order.
Then, we prove a result, which is the analogue of the result in \citet[Proposition 1]{bai03}, and is needed for the next section. It gives the limit of the space spanned by the estimated loadings (see Appendix \ref{app:KKKc1} for a proof). 

\begin{prop}\label{prop:KKK}
Under Assumptions \ref{ass:common} through \ref{ass:eval}, Assumption~\ref{eq:ortho}, and either of the following:
\begin{inparaenum}
\item [(A)] Assumptions \ref{ass:Wold} or \ref{ass:Hannan} or \ref{ass:Wu} in Appendix \ref{sec:hannan} on serial dependence of the factor process $\{\mbf F_t,\, t\in\mathbb Z\}$;
\item [(B)] the moment condition in \eqref{eq:oslo2};
\end{inparaenum}
it follows that, as $n,T\to\infty$,
$$
\l\Vert \frac{\wh{\bm\Lambda}^\prime\bm\Lambda}{n}-\bm V_0\bm{\mathcal J}_0\bm\Upsilon_0^\prime  (\bm\Gamma^F)^{-1/2}\r\Vert = o_{\mathrm P}(1),
$$
where the columns of $\bm\Upsilon_0$ are the normalized eigenvectors of $(\bm\Gamma^F)^{1/2}\bm\Sigma_\Lambda(\bm\Gamma^F)^{1/2}$, $\bm V_0$ is the $r\times r$ diagonal matrix containing the corresponding eigenvalues, and $\bm{\mathcal J}_0$ is an $r\times r$ diagonal matrix with entries $\pm 1$ independent of $n$ and $T$.
If Assumption \ref{ass:common}(a)  holds with rate $\sqrt n$ and Assumption \ref{ass:common}(c-ii)  holds with rate $\sqrt T$, then the above holds with rate $\min(\sqrt n,\sqrt T)$. Furthermore, $\bm V_0\bm{\mathcal J}_0\bm\Upsilon_0^\prime  (\bm\Gamma^F)^{-1/2}$ is finite and invertible.
\end{prop}

\section{Consistent estimation of the loading and factor spaces - part 2}\label{sec:cons2}

By definition of $\wh{\bm\Lambda}$ in \eqref{eq:estL}  $\wh{\mbf V}^x=\wh{\bm\Lambda}(\wh{\mbf M}^x)^{-1/2}$.
 Thus, from \eqref{eq:evecXX2}, we obtain
\beq\label{eq:start}
\frac{\bm X^\prime\bm X}{nT}\wh{\bm\Lambda}=\wh{\bm\Lambda}\frac{\wh{\mbf M}^x}{n}.
\eeq
Then, substituting $\bm X^\prime\bm X=(\bm\Lambda\bm F^\prime+\bm E^\prime)^\prime(\bm F\bm\Lambda^\prime+\bm E)$ into \eqref{eq:start}, 
and letting 
\beq\label{eq:acca}
\wh{\mbf H}:=
\frac{\bm F^\prime\bm F}{T}
\frac{\bm\Lambda^\prime\wh{\bm\Lambda}}{n}
\l(\frac{\wh{\mbf M}^x}{n}\r)^{-1},
\eeq
we obtain (notice that, as $n,T\to\infty$, $\wh{\mbf H}$  is well defined because of Lemma \ref{lem:HO1}(i))
\begin{align}
\wh{\bm\Lambda}-\bm\Lambda\wh{\mbf H}=&\, \l(\frac{\bm\Lambda\bm F^\prime\bm  E\wh{\bm\Lambda}}{nT}
+\frac{\bm E^\prime\bm F\bm\Lambda^\prime\wh{\bm\Lambda}}{nT}
+\frac{\bm E^\prime\bm  E\wh{\bm\Lambda}}{nT}\r)\l(\frac{\wh{\mbf M}^x}{n}\r)^{-1}
.\label{eq:start4}
\end{align}
This expression conveys the same message as Davis-Kahan theorem. Since the distance between the space spanned by the (rescaled) sample eigenvectors and the true ones is equal to the sample covariance matrix divided by the sample eigenvalues, by bounding the sample covariance matrix we obtain a bound for the estimation error of the eigenvectors, provided the sample eigenvalues are all positive at least asymptotically, as ensured by Lemma \ref{lem:MO1}(iv).

By taking the $i$th row of  \eqref{eq:start4}, we get
\begin{align}
\wh{\bm\lambda}_i^\prime-{\bm\lambda}_i^\prime\wh{\mbf H}=&\,
\l(
\underbrace{\frac 1{nT}{\bm\lambda}_i^\prime\sum_{t=1}^T\sum_{j=1}^n\mbf F_t e_{jt}\wh{\bm\lambda}_j^\prime}_{\text{(1.a)}}
+\underbrace{\frac 1{nT} \sum_{t=1}^T  e_{it}\mbf F_t^\prime\sum_{j=1}^n\bm\lambda_j\wh{\bm\lambda}_j^\prime}_{\text{(1.b)}}
+\underbrace{\frac 1{nT} \sum_{t=1}^T\sum_{j=1}^n e_{it} e_{jt} \wh{\bm\lambda}_j^\prime}_{\text{(1.c)}}
\r) \l(\frac{\wh{\mbf M}^x}{n}\r)^{-1}.\label{eq:sviluppoLambda}
\end{align}

The following bounds hold for the terms in \eqref{eq:sviluppoLambda} (see Appendix \ref{prop:loadproof} for a proof).

\begin{lem}\label{prop:load}
Under Assumptions \ref{ass:common} through \ref{ass:ind}, as $n,T\to\infty$,
\begin{inparaenum}[(a)]
\item $\sqrt {nT}\l\Vert \text{\upshape (1.a)}\r\Vert = O_{\mathrm P}(1)$;
\item $\sqrt T \l\Vert \text{\upshape (1.b)}\r\Vert = O_{\mathrm P}(1)$;
\item $\min(n,\sqrt{nT})\l\Vert \text{\upshape (1.c)}\r\Vert = O_{\mathrm P}(1)$;
\end{inparaenum}
uniformly in $i$.
\end{lem}

Consistency follows immediately (see Appendix \ref{prop:L2proof} for a proof).

\begin{prop}\label{prop:L2}
Under Assumptions \ref{ass:common} through \ref{ass:ind}, as $n,T\to\infty$,
\begin{compactenum}[(a)]
\item $\min(n,\sqrt {T})\l\Vert \wh{\bm\lambda}_i-\wh{\mbf H}^\prime{\bm\lambda}_i\r\Vert = O_{\mathrm P} (1)$,
uniformly in $i$;
\item $\min(n,\sqrt {T}) \l\Vert \frac {\wh{\bm\Lambda}-{\bm\Lambda}\wh{\mbf H}}{\sqrt n}\r\Vert=O_{\mathrm P}(1)$.
\end{compactenum}
\end{prop}

Part (a) refines the result in Proposition \ref{prop:L}(b). Part (b) has the same rate as in Proposition \ref{prop:L}(a). As shown in Proposition \ref{prop:LLFF}(a) below, the  matrix $\wh{\mbf H}$ is asymptotically equivalent to the projection matrix of the estimated loadings onto the true ones.  

Turning to the estimated factors, from \eqref{eq:estL} and \eqref{eq:estF}, we have
\begin{align}\label{eq:FFF}
\wh{\bm F}&=\bm X\wh{\bm\Lambda}(\wh{\bm\Lambda}^\prime \wh{\bm\Lambda})^{-1}=\bm F\bm\Lambda^\prime \wh{\bm\Lambda}(\wh{\bm\Lambda}^\prime \wh{\bm\Lambda})^{-1}+\bm E\wh{\bm\Lambda}(\wh{\bm\Lambda}^\prime \wh{\bm\Lambda})^{-1}= \frac{\bm F\bm\Lambda^\prime \wh{\bm\Lambda}}n\l(\frac {\wh{\mbf M}^x}n\r)^{-1} + 
\frac{\bm E\wh{\bm\Lambda}}{n}\l(\frac {\wh{\mbf M}^x}n\r)^{-1}.
\end{align}
Then, substituting on the right-hand-side of  \eqref{eq:FFF} 
$\bm{\Lambda}$ with
$(\bm\Lambda-\wh{\bm\Lambda}\wh{\mbf H}^{-1})+\wh{\bm\Lambda}\wh{\mbf H}^{-1},$ in the first term and 
$\wh{\bm\Lambda}$ with
$
\wh{\bm\Lambda}-\bm\Lambda\wh{\mbf H}+\bm\Lambda\wh{\mbf H}
$
in the second term, we obtain (notice that, as $n,T\to\infty$, $\wh{\mbf H}^{-1}$  is well defined because of Lemma \ref{lem:HO1}(ii))
\begin{align}
\wh{\bm F}-\bm F (\wh{\mbf H}^{-1})^\prime&=\l(\frac{\bm F(\bm\Lambda-\wh{\bm\Lambda}\wh{\mbf H}^{-1})^\prime\wh{\bm\Lambda}}{n}+
\frac{\bm  E(\wh{\bm\Lambda}-\bm\Lambda \wh{\mbf H})}{n}+
\frac{\bm E\bm\Lambda\wh{\mbf H}}{n}\r)\l(\frac {\wh{\mbf M}^x}n\r)^{-1}.\label{eq:FFF2}
\end{align}
By taking the $t$th row of \eqref{eq:FFF2}, we get
\begin{align}
\wh{\mbf F}_t^\prime -\mbf F_t^\prime (\wh{\mbf H}^{-1})^\prime&=\l(
\underbrace{\mbf F_t^\prime\frac{(\bm\Lambda-\wh{\bm\Lambda}\wh{\mbf H}^{-1})^\prime\wh{\bm\Lambda}}{n}}_{\text{ (2.a)}}+
\underbrace{\bm e_t^\prime \frac{(\wh{\bm\Lambda}-\bm\Lambda \wh{\mbf H})}{n}}_{\text{(2.b)}}+
\underbrace{\bm e_t^\prime \frac{\bm\Lambda\wh{\mbf H}}{n}}_{\text{(2.c)}}
\r)\l(\frac {\wh{\mbf M}^x}n\r)^{-1}.\label{eq:sviluppoFactor}
\end{align}

The following bounds hold for the terms in \eqref{eq:sviluppoFactor} (see Appendix \ref{prop:factorproof} for a proof). 

\begin{lem}\label{prop:factor}
Under Assumptions \ref{ass:common} through \ref{ass:ind}, as $n,T\to\infty$,
\begin{inparaenum}[(a)]
\item $\min(n,\sqrt{nT},T)\Vert \text{2.a}\Vert=O_{\mathrm {P}}(1)$;\linebreak
\item $\min(n,\sqrt{nT},T)\Vert \text{2.b}\Vert=O_{\mathrm {P}}(1)$;
\item $\sqrt n\Vert \text{2.c}\Vert=O_{\mathrm {P}}(1)$;
\end{inparaenum}
uniformly in $t$.
\end{lem}

Consistency follows immediately  (see Appendix \ref{prop:F2proof} for a proof). 

\begin{prop}\label{prop:F2}
Under Assumptions \ref{ass:common} through \ref{ass:ind}, as $n,T\to\infty$
\begin{compactenum}[(a)]
\item $\min(\sqrt n,{T})\l\Vert \wh{\mbf F}_t-\wh{\mbf H}^{-1}{\mbf F}_t \r\Vert = O_{\mathrm P} (1)$,
uniformly in $t$;
\item $\min(\sqrt n, {T}) \l\Vert \frac{\wh{\bm F}-{\bm F}(\wh{\mbf H}^{-1})^\prime}{\sqrt T}\r\Vert=O_{\mathrm P}(1)$.
\end{compactenum}
\end{prop}

Parts (a) and (b) refine the results in Proposition \ref{prop:L}(c)-\ref{prop:L}(d). As shown in Proposition \ref{prop:LLFF}(b) below, the  matrix $\wh{\mbf H}^{-1}$ is asymptotically equivalent to the projection matrix of the estimated factors onto the true ones.

Finally, we derive a result analogous to the result in \citet[Theorem 1]{baing02} and \citet[Lemma A.1]{bai03} (see Appendix \ref{prop:sumFproof} for a proof).

\begin{prop}\label{prop:sumF}
Under Assumptions \ref{ass:common} through \ref{ass:ind}, as $n,T\to\infty$,
$$
\min(n,T)\l\{\frac 1T\sum_{t=1}^T \l\Vert \wh{\mbf F}_t-\wh{\mbf H}^{-1}{\mbf F}_t  \r\Vert^2\r\}=O_{\mathrm P}(1).
$$
\end{prop}

 This result is often needed when using the estimated quantities for further analysis, e.g., in factor augmented regressions \citep{baing06}. It is also needed to prove consistency for the selection of the number of factors based on Information Criteria \citep{baing02,ABC10}.

\section{The role of \texorpdfstring{$\wh{\mbf H}$}{H-hat} and its relation with \texorpdfstring{$\pmb{\mathcal H}$}{cal H}}\label{sec:Hhat}

To understand the meaning of $\wh{\mbf H}$ in Propositions \ref{prop:L2} and \ref{prop:F2}, we can derive a series of more intuitive asymptotically equivalent expressions (see Appendix \ref{prop:LLFFproof} for a proof).

\begin{prop}\label{prop:LLFF}
Under Assumptions \ref{ass:common} through \ref{ass:ind}, as $n,T\to\infty$,
\begin{compactenum}[(a)]
\item $\min(n,\sqrt{nT}, {T}) \l\Vert  \wh{\mbf H} - (\bm\Lambda^\prime\bm\Lambda)^{-1}
{\bm\Lambda}^\prime\wh{\bm\Lambda}
\r\Vert=O_{\mathrm P}(1)$;
\item $\min( n,\sqrt{nT}, {T}) \l\Vert\wh{\mbf H}^{-1}-
\wh{\bm F}^\prime{\bm F}(\bm F^\prime\bm F)^{-1}
\r\Vert=O_{\mathrm P}(1)$;
\item $
\min(n,\sqrt{nT}, {T})\l\Vert  \wh{\mbf H}^{-1} - (\wh{\bm\Lambda}^\prime\wh{\bm\Lambda})^{-1}
\wh{\bm\Lambda}^\prime{\bm\Lambda}
\r\Vert=O_{\mathrm P}(1)$, where we recall that $\wh{\bm\Lambda}^\prime\wh{\bm\Lambda}=\wh{\mbf M}^x$;
\item $\min(n,\sqrt{nT}, {T}) \l\Vert \wh{\mbf H} -  {\bm F}^\prime\wh{\bm F}(\wh{\bm F}^\prime\wh{\bm F})^{-1}\r\Vert= O_{\mathrm P}(1)$, where we recall that $\wh{\bm F}^\prime \wh{\bm F} = T\, \mbf I_r$;
\item $\min(n,\sqrt{nT}, {T})\l\Vert  \wh{\mbf H}^{-1} -({\bm\Lambda}^\prime\wh{\bm\Lambda})^{-1} \bm\Lambda^\prime\bm\Lambda
\r\Vert=O_{\mathrm P}(1)$;
\item  $\min( n,\sqrt{nT}, {T}) \l\Vert\wh{\mbf H}-
\bm F^\prime\bm F (\wh{\bm F}^\prime{\bm F})^{-1}
\r\Vert=O_{\mathrm P}(1)$;
\item $\min( n,\sqrt{nT}, {T}) \l\Vert  \frac{(\wh{\mbf H}^{-1})^\prime\wh{\bm\Lambda}^\prime\wh{\bm\Lambda}\wh{\mbf H}^{-1}}n-\frac{\bm\Lambda^\prime\bm\Lambda}n\r\Vert=O_{\mathrm P}(1)$, where we recall that $\wh{\bm\Lambda}^\prime\wh{\bm\Lambda}=\wh{\mbf M}^x$;
\item $\min( n,\sqrt{nT}, {T}) \l\Vert \frac{\wh{\mbf H} \wh{\bm F}^\prime\wh{\bm F}\wh{\mbf H}^\prime}T-  \frac{{\bm F}^\prime{\bm F}}T\r\Vert=O_{\mathrm P}(1)$, where we recall that $\wh{\bm F}^\prime \wh{\bm F} = T\, \mbf I_r$.
\end{compactenum}
\end{prop}
This result, which is similar to what proved by \citet[Lemma 3.i]{bai2020simpler}, shows that asymptotically the matrix $\wh{\mbf H}$ is obtained by linear projection of the estimated loadings onto the true loadings (part (a)) or of the true factors onto the estimated ones (part (d)). Similarly, asymptotically $\wh{\mbf H}^{-1}$ is obtained by linear projection of the estimated factors onto the true factors (part (b)) or by linear projection of the true loadings onto the estimated ones (part (c)). Parts (e)-(h) are less intuitive, but still useful for proving further results. 

Next, we derive another asymptotically equivalent expression for $\wh{\mbf H}$ (see Appendix \ref{prop:HHATproof} for a proof).

\begin{prop}\label{prop:HHAT} 
Under Assumptions \ref{ass:common} through \ref{ass:ind}, as $n,T\to\infty$,
\begin{compactenum}
\item [(a)] $\min(n,\sqrt{T})\l\Vert \wh{\mbf H} - \l(\frac{\bm F^\prime\bm F}{T}\r)^{1/2}\wh{\mbf Q}\r\Vert = O_{\mathrm P}(1)$;
\item [(b)] $\min(n,\sqrt{T})\l\Vert\wh{\mbf H}^{-1} - \wh{\mbf Q}^\prime \l(\frac{\bm F^\prime\bm F}T\r)^{-1/2}\r\Vert =O_{\mathrm P}(1)$;
\end{compactenum}
where the columns of $\wh{\mbf Q}$ are the normalized eigenvectors of $(T^{-1}\bm F^\prime\bm F)^{1/2}(n^{-1}\bm \Lambda^\prime\bm \Lambda) (T^{-1}\bm F^\prime\bm F)^{1/2}$.
\end{prop}

According to this result and Propositions \ref{prop:L2} and \ref{prop:F2},  the loadings and the factors can be consistently estimated  up to: (i) a scale $(T^{-1}{\bm F^\prime\bm F})^{1/2}$ and (ii) a rotation $\wh{\mbf Q}$. The proof of this result rests on four intermediate  results, which we summarize here because they are relevant for many derivations in this paper. First and second, from Proposition \ref{prop:LLFF}, it follows that (see also \eqref{eq:warwick3} and \eqref{eq:warwick4} in the proof of Proposition \ref{prop:HHAT})
\begin{align}
&\l\Vert\wh{\mbf H} - \frac{\bm F^\prime\bm F}{T} (\wh{\mbf H}^{-1})^{\prime}\r\Vert =O_{\mathrm P}\l(\max\l(\frac 1n,\frac{1}{\sqrt{nT}},\frac 1T\r)\r),\label{eq:starop1}\\
&\l\Vert\frac{\wh{\mbf M}^x}{n} - \wh{\mbf H}^{-1}\frac{\bm F^\prime\bm F}{T} \frac{\bm\Lambda^\prime\bm\Lambda}{n} \wh{\mbf H}\r\Vert=O_{\mathrm P}\l(\max\l(\frac 1n,\frac{1}{\sqrt{nT}},\frac 1T\r)\r).\label{eq:starop2}
\end{align}
Third, by substituting \eqref{eq:starop1} into \eqref{eq:starop2} (see also \eqref{eq:madre} in the proof of Proposition \ref{prop:HHAT}) 
\begin{align}
\l\Vert
\frac{\wh{\mbf M}^x}{n} - \wh{\mbf H}^{-1} \frac{\bm F^\prime\bm F}{T} \frac{\bm\Lambda^\prime\bm\Lambda}{n}\frac{\bm F^\prime\bm F}{T} (\wh{\mbf H}^{-1})^{\prime}
\r\Vert=O_{\mathrm P}\l(\max\l(\frac 1n,\frac{1}{\sqrt{nT}},\frac 1T\r)\r).\label{eq:starop3}
\end{align}
Last, the $r$ non-zero eigenvalues of 
$n^{-1} \bm\Lambda(T^{-1}\bm F^\prime\bm F) \bm\Lambda^\prime$, collected into the diagonal $r\times r$ matrix $n^{-1}\wh{\mbf M}^C$ coincide with the eigenvalues of $(T^{-1}\bm F^\prime\bm F)^{1/2}(n^{-1}\bm \Lambda^\prime\bm \Lambda) (T^{-1}\bm F^\prime\bm F)^{1/2}$, and are such that (see also \eqref{eq:mistake} in the proof of Proposition \ref{prop:HHAT})
\beq\label{eq:starop3bis}
\l\Vert \frac{\wh{\mbf M}^C}n-\frac{\wh{\mbf M}^x}n\r\Vert = O_{\mathrm P}\l(\max\l(\frac 1n,\frac{1}{\sqrt{T}}\r)\r).
\eeq
Notice that \eqref{eq:starop3bis} has a slow rate $\sqrt T$ due to the fact that to prove it we need to bound the sample covariance of the idiosyncratic component (see also \eqref{eq:mistake3}  in the proof of Proposition \ref{prop:HHAT}). Then, part (b) follows directly from \eqref{eq:starop3} and \eqref{eq:starop3bis}. 

We then prove a result which is the counterpart of Proposition \ref{prop:KKK} and it gives the limit of the space spanned by the estimated factors. It follows essentially from the definition of $\wh{\mbf H}$ in \eqref{eq:acca} and Propositions \ref{prop:KKK} and \ref{prop:LLFF}(f) (see Appendix \ref{prop:KKKbisproof} for a proof). 
\begin{prop}\label{prop:KKKbis}
Under Assumptions \ref{ass:common} through \ref{ass:ind}, as $n,T\to\infty$,
$$
\l\Vert \frac{\wh{\bm F}^\prime\bm F}{T}-\bm{\mathcal J}_0\bm\Upsilon_0^\prime  (\bm\Gamma^F)^{1/2}\r\Vert = o_{\mathrm P}(1),
$$
where the columns of $\bm\Upsilon_0$ are the normalized eigenvectors of $(\bm\Gamma^F)^{1/2}\bm\Sigma_\Lambda(\bm\Gamma^F)^{1/2}$, and $\bm{\mathcal J}_0$ is an $r\times r$ diagonal matrix with entries $\pm 1$ independent of $n$ and $T$. If Assumption \ref{ass:common}(a)  holds with rate $\sqrt n$ and Assumption \ref{ass:common}(c-ii) holds with rate $\sqrt T$, then the above holds with rate $\min(\sqrt n,\sqrt T)$. Furthermore, $\bm{\mathcal J}_0\bm\Upsilon_0^\prime  (\bm\Gamma^F)^{1/2}$ is finite and invertible.
\end{prop}

Thanks to this result we can derive a useful limit for $\wh{\mbf H}$, which depends only on population quantities (see Appendix \ref{cor:sempliceproof} for a proof). 

\begin{prop}\label{cor:semplice}
Under Assumptions \ref{ass:common} through \ref{ass:ind}, as $n,T\to\infty$,
\begin{inparaenum}
\item [(a)] $\l\Vert\wh{\mbf H}-  (\bm\Gamma^F)^{1/2}\bm\Upsilon_0 \bm{\mathcal J}_0   \r\Vert =o_{\mathrm P}(1)$;
\item [(b)] $\l\Vert\wh{\mbf H}^{-1} -  \bm{\mathcal J}_0\bm\Upsilon_0^\prime  (\bm\Gamma^F)^{-1/2} \r\Vert =o_{\mathrm P}(1)$;
\end{inparaenum}
where the columns of $\bm\Upsilon_0$ are the normalized eigenvectors of $(\bm\Gamma^F)^{1/2}\bm\Sigma_\Lambda(\bm\Gamma^F)^{1/2}$, and $\bm{\mathcal J}_0$ is an $r\times r$ diagonal matrix with entries $\pm 1$ independent of $n$ and $T$. If Assumption \ref{ass:common}(a)  holds with rate $\sqrt n$ and Assumption \ref{ass:common}(c-ii) holds with rate $\sqrt T$, then the above holds with rate $\min(\sqrt n,\sqrt T)$. 
\end{prop}

Part (a) is analogous to the result in \citet[Lemma 3.ii]{bai2020simpler}. According to this result and Propositions \ref{prop:L2} and \ref{prop:F2},  the loadings and the factors are consistently estimated  up to: (i) a scale $(\bm\Gamma^F)^{1/2}$, (ii) a rotation, $\bm{\Upsilon}_0$, and (iii) a diagonal matrix of signs $\bm{\mathcal  J}_0$.  

Proposition \ref{cor:semplice} can be proved either by directly combining Propositions \ref{prop:LLFF}(d) and \ref{prop:KKKbis}, or from Proposition \ref{prop:HHAT}, by noticing that Assumptions \ref{ass:common}(a) and \ref{ass:common}(c-ii) imply $\Vert(T^{-1}\bm F^\prime\bm F)^{1/2}(n^{-1}\bm \Lambda^\prime\bm \Lambda) (T^{-1}\bm F^\prime\bm F)^{1/2}-(\bm\Gamma^F)^{1/2} \bm\Sigma_\Lambda(\bm\Gamma^F)^{1/2}\Vert =o_{\mathrm P}(1)$.
Since the limiting matrix has distinct eigenvalues by Assumption \ref{ass:eval},
and since signs are collected in $\bm{\mathcal J}_0$, the Davis-Kahan theorem
\citep[Corollary 1]{yu15} implies that the corresponding normalized eigenvectors satisfy
$\Vert\wh{\mbf Q}-\bm\Upsilon_0\bm{\mathcal J}_0\Vert=o_{\mathrm P}(1)$.

Propositions \ref{prop:HHAT} and \ref{cor:semplice} convey the same message. The difference is in the way we approximate $\wh{\mbf H}$. In Proposition \ref{prop:HHAT} the rate is $\min(n,\sqrt{T})$, and the limiting quantity is random and depends on $n$ and $T$. In Proposition \ref{cor:semplice} we obtain a deterministic limiting quantity independent of $n$ and $T$, but the rate is slower and based on the rates at which Assumptions \ref{ass:common}(a) and \ref{ass:common}(c-ii) are satisfied, which are, obviously, $\sqrt n$ and $\sqrt T$, respectively. 

Finally, we can compare $\bm{\mathcal H}$ and $\wh{\mbf H}$ by means of the results in Propositions \ref{corol:ovvio} and \ref{prop:LLFF}. 
The main difference is that while $\bm{\mathcal H}$ depends only on population quantities (but for an irrelevant sign indeterminacy), $\wh{\mbf H}$ depends both on population and estimated quantities. 
Such difference has a series of important consequences: (i) $\wh{\mbf H}$ is finite and invertible only asymptotically as $n,T\to\infty$ (see Lemma \ref{lem:HO1}), while $\bm{\mathcal H}$ is always finite and invertible (see Lemma \ref{lem:HO1bis}); (ii)  $\wh{\mbf H}$ can be interpreted only asymptotically (see Propositions \ref{prop:LLFF}, \ref{prop:HHAT}, and \ref{cor:semplice}), while $\bm{\mathcal H}$ has a well defined expression for all $n$ and $T$ as given in Proposition \ref{prop:L}; (iii) any identification restriction imposed on the loadings and/or the factors will constrain $\wh{\mbf H}$ only asymptotically, while we can derive exact expressions for $\bm{\mathcal H}$ (see Section \ref{sec:II} and Appendix \ref{app:calH}, respectively).
Still, $\bm{\mathcal H}$ and $\wh{\mbf H}$ are asymptotically equivalent. This can be seen in two ways. 

First, by comparing  Proposition \ref{corol:ovvio}(a) and Proposition \ref{prop:LLFF}(a),
it follows that
\begin{align}
\l\Vert \bm{\mathcal H}-\wh{\mbf H}\r\Vert &\le 
\l\Vert\bm{\mathcal H} - (\bm\Lambda^\prime\bm\Lambda)^{-1}\bm\Lambda^\prime\wh{\bm\Lambda}  \r\Vert 
+ 
\l\Vert  \l(\bm\Lambda^\prime\bm\Lambda\r)^{-1}
{\bm\Lambda}^\prime\wh{\bm\Lambda}- \wh{\mbf H}
\r\Vert 
\nn\\
&= O_{\mathrm P}\l(\max\l(\frac 1n,\frac 1{\sqrt T}\r)\r)+O_{\mathrm P}\l(\max\l(\frac 1n,\frac 1{\sqrt{n T}},\frac 1T\r)\r) = O_{\mathrm P}\l(\max\l(\frac 1n,\frac 1{\sqrt T}\r)\r).\label{eq:accheeq}
\end{align}
It is important to stress that $\bm{\mathcal H}$ approximates the projection matrix of the estimated loadings onto the true loadings at a slower rate than $\wh{\mbf H}$. Indeed, while to prove Proposition \ref{prop:LLFF}(a) we are able to use a tighter bound than the one obtained for the loadings in Proposition \ref{prop:L2}(b), by bounding $ n^{-1}\Vert\bm\Lambda^\prime(\wh{\bm\Lambda}-\bm\Lambda\wh{\mbf H})\Vert$, we already noticed that to prove Proposition \ref{corol:ovvio}(a), we can only use a looser bound, which therefore inherits the weaker bound obtained for the loadings in Proposition \ref{prop:L}(a). 

Second, since the columns of $\mbf K$ in the definition of $\bm{\mathcal H}$ (see Proposition \ref{prop:L}) are the normalized eigenvectors of $(\bm\Gamma^F)^{1/2} (n^{-1}\bm\Lambda^\prime\bm\Lambda) (\bm\Gamma^F)^{1/2}$,
then, by Assumptions \ref{ass:common}(a) and \ref{ass:eval}, and the
Davis-Kahan theorem \citep[Corollary 1]{yu15},
it must hold that  $\lim_{n\to\infty}\Vert \mbf K-\bm\Upsilon_0\bm{\mathcal J}^*\Vert =0$ for some $r\times r$ diagonal matrix $\bm{\mathcal J}^*$ with entries $\pm 1$ and independent of $n$, and letting $\bm{\mathcal J}_0$ be such that $\Vert\bm{\mathcal J}^*\mbf J -\bm{\mathcal J}_0\Vert=o_{\mathrm P}(1)$ as $n,T\to\infty$, we get $\Vert \mbf K\mbf J-\bm\Upsilon_0\bm{\mathcal J}_0\Vert=o_{\mathrm P}(1)$ (see also \eqref{eq:ups} in the proof of Proposition \ref{prop:KKK}).
As a consequence of this result, of the definition of $\bm{\mathcal H}$ in Proposition \ref{prop:L}, and of Proposition \ref{cor:semplice}(a), it immediately follows that
\begin{align}
\l\Vert \bm{\mathcal H}-\wh{\mbf H}\r\Vert &\le \l\Vert  \bm{\mathcal H}-(\bm\Gamma^F)^{1/2}\bm\Upsilon_0\bm{\mathcal J}_0\r\Vert +\l\Vert (\bm\Gamma^F)^{1/2}\bm\Upsilon_0\bm{\mathcal J}_0-\wh{\mbf H}\r\Vert = o_{\mathrm P}(1),\label{eq:accheeq2}
\end{align}
thus, providing a proof of the  asymptotic equivalence of $\bm{\mathcal H}$ and $\wh{\mbf H}$ alternative to the proof in \eqref{eq:accheeq}. This proof, however, relies on a looser bound than the one used in \eqref{eq:accheeq}, indeed, it would deliver the slower rate $\min(\sqrt n,\sqrt T)$ inherited from Proposition \ref{cor:semplice}.

\section{Asymptotic normality}\label{sec:AN}
Asymptotic normality of the PC estimator of the loadings follows from Lemma \ref{prop:load}.

\begin{theorem}\label{th:CLTL}
Under Assumptions \ref{ass:common} through \ref{ass:rates}, as $n,T\to\infty$, for any given $i=1,\ldots,n$,
$$
\sqrt T\l(\wh{\bm\lambda}_i-\wh{\mbf H}^\prime{\bm\lambda}_i\r) 
 \to_d\mathcal N\l(\mbf 0_r,  
\bm\Upsilon_0^\prime(\bm\Gamma^F)^{1/2}\bm\Theta_i^{\text{\tiny \upshape OLS}} (\bm\Gamma^F)^{1/2} \bm\Upsilon_0
\r),
$$
with $\bm\Upsilon_0$ defined in  \eqref{eq:U0V0U0} and $\bm\Theta_i^{\text{\tiny \upshape OLS}}:=(\bm\Gamma^F)^{-1}
\bm\Phi_i
(\bm\Gamma^F)^{-1}$ with $\bm\Phi_i$ defined in Assumption \ref{ass:CLT}(a).
\end{theorem}

\paragraph{Proof.} From  the transpose of \eqref{eq:sviluppoLambda} for any $i=1,\ldots, n$, we get 
\begin{align}
\sqrt T\l(\wh{\bm\lambda}_i-\wh{\mbf H}^\prime{\bm\lambda}_i \r)
&=\sqrt T \l(\frac{\wh{\mbf M}^x}{n}\r)^{-1} \bm{\mathcal H}^\prime \l\{\text{(1.a)+(1.b)+(1.c)}\r\}+ \sqrt T \l(\frac{\wh{\mbf M}^x}{n}\r)^{-1} \l\{\text{(1.d)+(1.e)+(1.f)}\r\}\nn\\
&=\sqrt T\l(\frac{\wh{\mbf M}^x}{n}\r)^{-1} \bm{\mathcal H}^\prime\text{(1.b)} + O_{\mathrm {P}}\l(\max\l(\frac {\sqrt T} n, \frac 1{\sqrt{n}},\frac 1{\sqrt T}\r)\r) \nn\\
&= \l(\frac{\wh{\mbf M}^x}{n}\r)^{-1}\frac{ \bm{\mathcal H}^\prime \bm\Lambda^\prime\bm\Lambda}{n} \l(\frac 1{\sqrt T}\sum_{t=1}^T \mbf F_t e_{it}\r)
+ O_{\mathrm {P}}\l(\max\l(\frac {\sqrt T} n, \frac 1{\sqrt{n}},\frac 1{\sqrt T}\r)\r)\nn\\
&=  \l(\frac{\wh{\mbf M}^x}{n}\r)^{-1}\frac{ \wh{ \bm\Lambda}^\prime\bm\Lambda}{n} \l(\frac 1{\sqrt T}\sum_{t=1}^T \mbf F_t e_{it}\r)
+ O_{\mathrm {P}}\l(\max\l(\frac {\sqrt T} n, \frac 1{\sqrt{n}},\frac 1{\sqrt T}\r)\r)\nn\\
&=\wh{\mbf H}^\prime \l(\frac{\bm F^\prime\bm F}{T}\r)^{-1} \l(\frac 1{\sqrt T}\sum_{t=1}^T \mbf F_t e_{it}\r)
+ O_{\mathrm {P}}\l(\max\l(\frac {\sqrt T} n, \frac 1{\sqrt{n}},\frac 1{\sqrt T}\r)\r),\label{eq:finaleL}
\end{align}
because of Proposition \ref{corol:ovvio}(a) and Lemma \ref{prop:load}, the definition of $\wh{\mbf H}$ in \eqref{eq:acca}, and since
$\Vert n ({\wh{\mbf M}^x})^{-1}\Vert=O_{\mathrm P}(1)$
and
$\Vert\bm{\mathcal H}\Vert=O(1)$ by Lemmas \ref{lem:MO1}(iv) and \ref{lem:HO1bis}(i), respectively. 

Define  
\[
\wh{\bm\lambda}_i^{\text{\tiny OLS}}  := \l(\frac{\bm F^\prime\bm F}{T}\r)^{-1} \l(\frac 1{ T}\sum_{t=1}^T \mbf F_t x_{it}\r),
\]
which is the unfeasible OLS estimator of $\bm\lambda_i$ when regressing $x_{it}$ onto $\mbf F_t$. 
Then, since by Assumption \ref{ass:rates} $\sqrt T/n\to 0$ as $n,T\to\infty$, from \eqref{eq:finaleL}, using  Proposition \ref{cor:semplice}(a),
\begin{align}\label{eq:CLTL}
\sqrt T\l(\wh{\bm\lambda}_i-\wh{\mbf H}^\prime{\bm\lambda}_i\r) 
&= \l\{ \underset{n,T\to\infty}{\text{P-lim}} \wh{\mbf H}^\prime \r\} \l(\frac{\bm F^\prime\bm F}{T}\r)^{-1} \l(\frac 1{\sqrt T}\sum_{t=1}^T \mbf F_t e_{it}\r)
+ o_{\mathrm {P}}(1)\nn\\
&=
\bm{\mathcal J}_0\bm\Upsilon_0^\prime  (\bm\Gamma^F)^{1/2} \l(\frac{\bm F^\prime\bm F}{T}\r)^{-1} \l(\frac 1{\sqrt T}\sum_{t=1}^T \mbf F_t e_{it}\r)+o_{\mathrm P}(1)\nn\\
&=
\bm{\mathcal J}_0\bm\Upsilon_0^\prime  (\bm\Gamma^F)^{1/2}  \sqrt T\l(\wh{\bm\lambda}_i^{\text{\tiny OLS}}-{\bm\lambda}_i\r)  +o_{\mathrm P}(1).
\end{align}
Moreover, as $T\to\infty$, 
\beq\label{eq:thetaOLS}
\sqrt T\l(\wh{\bm\lambda}_i^{\text{\tiny OLS}}-{\bm\lambda}_i\r) =(\bm\Gamma^F)^{-1} \l(\frac 1{\sqrt T}\sum_{t=1}^T \mbf F_t e_{it}\r)+o_{\mathrm P}(1)
 \to_d\mathcal N\l(\mbf 0_r,  (\bm\Gamma^F)^{-1}
\bm\Phi_i
(\bm\Gamma^F)^{-1}
\r),
\eeq
by Assumptions \ref{ass:common}(b), \ref{ass:common}(c-ii), and \ref{ass:CLT}(a), and Slutsky's theorem.
The proof follows by substituting \eqref{eq:thetaOLS} into  \eqref{eq:CLTL}, using again Slutsky's theorem, and since $\bm{\mathcal J}_0$ being a diagonal sign matrix plays no role in the asymptotic covariance. $\Box$

\paragraph{} Similarly, for the PC estimator of the factors asymptotic normality follows from Lemma \ref{prop:factor}.

\begin{theorem}\label{th:CLTF}
Under Assumptions \ref{ass:common} through \ref{ass:rates}, as $n,T\to\infty$, for any given $t=1,\ldots,T$,
$$
\sqrt n\l(\wh{\mbf F}_t-\wh{\mbf H}^{-1}{\mbf F}_t\r) \to_d\mathcal N\l(\mbf 0_r, \bm\Upsilon_0^\prime(\bm\Gamma^F)^{-1/2}\bm\Pi_t^{\text{\tiny \upshape OLS}}(\bm\Gamma^F)^{-1/2}\bm\Upsilon_0\r),
$$
with $\bm\Upsilon_0$ defined in  \eqref{eq:U0V0U0}
and $\bm\Pi_t^{\text{\tiny \upshape OLS}}:=(\bm\Sigma_\Lambda)^{-1}\bm\Gamma_t(\bm\Sigma_\Lambda)^{-1}$ with $\bm\Gamma_t$ defined in Assumption \ref{ass:CLT}(b).
\end{theorem}

\paragraph{Proof.} From  the transpose of \eqref{eq:sviluppoFactor}, for any $t=1,\ldots, T$, we get
\begin{align}
\sqrt n\l(\wh{\mbf F}_t-\wh{\mbf H}^{-1}{\mbf F}_t\r )
&=\sqrt n \l(\frac{\wh{\mbf M}^x}{n}\r)^{-1} \l\{\text{(2.a)+(2.b)+(2.c)}\r\}\nn\\
&=\sqrt n \l(\frac{\wh{\mbf M}^x}{n}\r)^{-1}\text{(2.c)}+ O_{\mathrm {P}}\l(\max\l(\frac 1{\sqrt n},\frac {\sqrt n} T, \frac 1{\sqrt{T}}\r)\r)\nn\\
&= \l(\frac{\wh{\mbf M}^x}{n}\r)^{-1} \wh{\mbf H}^\prime \l(\frac 1{\sqrt n}\sum_{i=1}^n \bm \lambda_i e_{it}\r)
+ O_{\mathrm {P}}\l(\max\l(\frac 1{\sqrt n},\frac {\sqrt n} T, \frac 1{\sqrt{T}}\r)\r)\nn\\
&=\l(\frac{\wh{\bm\Lambda}^\prime\wh{\bm\Lambda}}{n}\r)^{-1} \wh{\mbf H}^\prime \l(\frac 1{\sqrt n}\sum_{i=1}^n \bm\lambda_i e_{it}\r)+ O_{\mathrm {P}}\l(\max\l(\frac 1{\sqrt n},\frac {\sqrt n} T, \frac 1{\sqrt{T}}\r)\r)\nn\\
&= \l(\frac{\wh{\mbf H}^\prime{\bm\Lambda}^\prime{\bm\Lambda}\wh{\mbf H}}{n}\r)^{-1} \wh{\mbf H}^\prime \l(\frac 1{\sqrt n}\sum_{i=1}^n \bm\lambda_i e_{it}\r)+ O_{\mathrm {P}}\l(\max\l(\frac 1{\sqrt n},\frac {\sqrt n} T, \frac 1{\sqrt{T}}\r)\r)\nn\\
& = \wh{\mbf H}^{-1} \l(\frac{\bm\Lambda^\prime\bm\Lambda}{n}\r)^{-1}\l(\frac 1{\sqrt n}\sum_{i=1}^n \bm\lambda_i e_{it}\r)+ O_{\mathrm {P}}\l(\max\l(\frac 1{\sqrt n},\frac {\sqrt n} T, \frac 1{\sqrt{T}}\r)\r),
\label{eq:finaleF}
\end{align}
because of Proposition \ref{prop:L2}(b) and Lemma \ref{prop:factor},
the definition of $\wh{\bm\Lambda}$ in \eqref{eq:estL}, and since $\Vert n({\wh{\mbf M}^x})^{-1}\Vert=O_{\mathrm P}(1)$ by Lemma \ref{lem:MO1}(iv). 
Define,
\[
\wh{\mbf F}_t^{\text{\tiny OLS}}:=\l(\frac{\bm\Lambda^\prime\bm\Lambda}{n}\r)^{-1}\l(\frac 1{ n}\sum_{i=1}^n \bm\lambda_i x_{it}\r),
\]
which is the unfeasible OLS estimator of $\mbf F_t$ when regressing $x_{it}$ onto $\bm\lambda_i$.  Then, since by Assumption \ref{ass:rates} $\sqrt n/T\to 0$ as $n,T\to\infty$, from \eqref{eq:finaleF}, using  Proposition \ref{cor:semplice}(b),
\begin{align}\label{eq:CLTF}
\sqrt n\l(\wh{\mbf F}_t-\wh{\mbf H}^{-1}{\mbf F}_t\r) 
& = \l\{ \underset{n,T\to\infty}{\text{P-lim}} \wh{\mbf H}^{-1} \r\}
 \l(\frac{\bm\Lambda^\prime\bm\Lambda}{n}\r)^{-1}\l(\frac 1{\sqrt n}\sum_{i=1}^n \bm\lambda_i e_{it}\r)+o_{\mathrm P}(1)\nn\\
 &=  \bm{\mathcal J}_0\bm\Upsilon_0^\prime  (\bm\Gamma^F)^{-1/2}  \l(\frac{\bm\Lambda^\prime\bm\Lambda}{n}\r)^{-1}\l(\frac 1{\sqrt n}\sum_{i=1}^n \bm\lambda_i e_{it}\r)+o_{\mathrm P}(1)\nn\\
 &= \bm{\mathcal J}_0\bm\Upsilon_0^\prime  (\bm\Gamma^F)^{-1/2} \sqrt n\l(\wh{\mbf F}_t^{\text{\tiny OLS}}-\mbf F_t\r)+ o_{\mathrm P}(1).
\end{align}
Moreover, as $n\to\infty$, 
\beq\label{eq:gammaOLS}
\sqrt n\l(\wh{\mbf F}_t^{\text{\tiny OLS}}-\mbf F_t\r) = (\bm\Sigma_\Lambda)^{-1} \l(\frac 1{\sqrt n}\sum_{i=1}^n \bm\lambda_i e_{it}\r)
+o(1)
\to_d\mathcal N\l(\mbf 0_r,  (\bm\Sigma_\Lambda)^{-1}
\bm\Gamma_t
(\bm\Sigma_\Lambda)^{-1}
\r),
\eeq
by Assumptions \ref{ass:common}(a) and \ref{ass:CLT}(b), and Slutsky's theorem. 
The proof follows by substituting \eqref{eq:gammaOLS} into  \eqref{eq:CLTF}, using again Slutsky's theorem, and since $\bm{\mathcal J}_0$ being a diagonal sign matrix plays no role in the asymptotic covariance. $\Box$ 

\paragraph{}
Theorems \ref{th:CLTL} and \ref{th:CLTF} show explicitly the relationship between the PC estimators and the unfeasible OLS estimators. 
It is crucial to stress that, in order to prove these results without needing to impose too restrictive bounds on the rates of divergence of $n$ and $T$, we have to make use of the rates derived in Propositions \ref{prop:L2} and \ref{prop:F2}, which are sharper than those derived in Proposition \ref{prop:L}. Those results, however, require making stronger moment assumptions (all implied by Assumption \ref{ass:ind}) and the use of the matrix $\wh{\mbf H}$ rather than $\bm{\mathcal H}$. 

Due to the presence of the unknown random matrix $\wh{\mbf H}$, which depends on both the unknown true factors and loadings and their estimators, Theorems \ref{th:CLTL} and \ref{th:CLTF} cannot be used to make inference on the loadings or the factors. This is the case even when $r=1$, so that $\wh{\mbf H}$ is a scalar, because factors and loadings are still consistently estimated only up to an unknown scale. In particular, the presence of $\wh{\mbf H}$ has two effects. First, the locations of the asymptotic distribution of $\wh{\bm\lambda}_i$ and $\wh{\mbf F}_t$ are random.  
Second,  the asymptotic covariance matrix cannot be estimated consistently, since any such estimator would require consistent estimators of $\bm\Gamma^F$ and/or $\bm\Sigma_\Lambda$, but these cannot be estimated consistently. Indeed, the candidate estimators  $T^{-1}\wh{\bm F}^\prime \wh{\bm F}$ and $n^{-1}\wh{\bm \Lambda}^\prime \wh{\bm \Lambda}$ have probability limits which still depend on $\wh{\mbf H}$ (see Proposition \ref{prop:LLFF}(g) and \ref{prop:LLFF}(h)). 

The same relation between the PC estimators and their unfeasible OLS counterparts can be derived using the alternative approaches B1, A2, and B2, employing linear transformations which are asymptotically equivalent to $\wh{\mbf H}$. The CLTs under those approaches are 
derived in Appendices \ref{sec:cmppca} and \ref{sec:cmppcaCD}, and are summarized in Table \ref{tab:H}.


\begin{sidewaystable}[htbp]
\caption{Asymptotic normality}\label{tab:H}
\centering
\scriptsize{
\begin{tabular}{l | l  | c | c | c }
\hline
\hline
&&&&\\
\eqref{eq:CLTL} $\sqrt T(\wh{\bm\lambda}_i-\wh{\mbf H}^\prime{\bm\lambda}_i)\to_d\mathcal N\l(\mbf 0_r,\mbf H_{0,A}^\prime \bm\Theta_i^{\text{OLS}}\mbf H_{0,A}\r) $&&&&\\
\eqref{eq:ANLbis} $\sqrt T(\wh{\bm\lambda}_i-\wh{\mbf H}^\prime{\bm\lambda}_i)\to_d\mathcal N\l(\mbf 0_r,\mbf H_{1,A}^\prime \bm\Theta_i^{\text{OLS}}\mbf H_{1,A}\r) $&$\wh{\mbf H}$ 	&  $\l(\frac{\bm F^\prime\bm F}{T}\r)^{1/2}\wh{\mbf Q}$	& $\mbf H_{0,A}:=(\bm \Gamma^F)^{1/2} \bm\Upsilon_0 \bm{\mathcal J}_0$ 	& $\mbf H_{1,A}:=(\bm\Sigma_\Lambda)^{-1/2}\bm\Upsilon_1  \bm{\mathcal J}_1(\bm V_0)^{1/2}$\\[3pt]
\eqref{eq:CLTF} $\sqrt n(\wh{\mbf F}_t-\wh{\mbf H}^{-1}{\mbf F}_t)\to_d\mathcal N\l(\mbf 0_r,\mbf H_{0,A}^{-1} \bm\Pi_t^{\text{OLS}}(\mbf H_{0,A}^{-1})^\prime\r)$&	rate			& $\min(n,\sqrt{T})$		& $\min(\sqrt n,\sqrt T)$								& $\min(\sqrt n,\sqrt T)$\\[3pt]
\eqref{eq:ANFbis} $\sqrt n(\wh{\mbf F}_t-\wh{\mbf H}^{-1}{\mbf F}_t)\to_d\mathcal N\l(\mbf 0_r,\mbf H_{1,A}^{-1} \bm\Pi_t^{\text{OLS}}(\mbf H_{1,A}^{-1})^\prime\r)$&				&Proposition \ref{prop:HHAT}	& Proposition \ref{cor:semplice} 						& \eqref{eq:ennesimaespansione}\\[3pt]
\hline
&&&&\\[-3pt]
\eqref{eq:ANLBAI1} $\sqrt T(\wt{\bm\lambda}_i-\wt{\mbf H}^{-1}{\bm\lambda}_i)\to_d\mathcal N\l(\mbf 0_r,\mbf H_{0,B}^{-1}\bm\Theta_i^{\text{OLS}}(\mbf H_{0,B}^{-1})^\prime\r)$&&&&\\
\eqref{eq:ANLBAI} $\sqrt T(\wt{\bm\lambda}_i-\wt{\mbf H}^{-1}{\bm\lambda}_i)\to_d\mathcal N\l(\mbf 0_r,\mbf H_{1,B}^{-1}\bm\Theta_i^{\text{OLS}}(\mbf H_{1,B}^{-1})^\prime\r)$&$\wt{\mbf H}$	&$ \l(\frac{\bm F^\prime\bm F}{T}\r)^{-1/2} \wh{\mbf Q}$& $\mbf H_{0,B}:=(\bm\Gamma^F)^{-1/2} \bm\Upsilon_0\bm{\mathcal J}_0$	& $\mbf H_{1,B}:=(\bm\Sigma_\Lambda)^{1/2}\bm\Upsilon_1 \bm{\mathcal J}_1 (\bm V_0)^{-1/2}$\\[3pt]
\eqref{eq:ANFBAI1} $\sqrt n(\wt{\mbf F}_t-\wt{\mbf H}^{\prime}{\mbf F}_t)\to_d\mathcal N\l(\mbf 0_r,\mbf H_{0,B}^\prime\bm\Pi_t^{\text{OLS}} \mbf H_{0,B}\r)$&	rate			& $\min(n,\sqrt{T})$		& $\min(\sqrt n,\sqrt T)$								& $\min(\sqrt n,\sqrt T)$\\[3pt]
\eqref{eq:ANFBAI} $\sqrt n(\wt{\mbf F}_t-\wt{\mbf H}^{\prime}{\mbf F}_t)\to_d\mathcal N\l(\mbf 0_r,\mbf H_{1,B}^\prime\bm\Pi_t^{\text{OLS}} \mbf H_{1,B}\r)$&&\eqref{eq:sameQ}&\eqref{eq:Htildelim2}& \eqref{eq:Htildelim}\\[3pt]
\hline
&&&&\\[-3pt]
\eqref{eq:ANLSW1} $\sqrt T(\wideparen{\bm\lambda}_i-\wideparen{\mbf H}^{\prime}{\bm\lambda}_i)\to_d\mathcal N\l(\mbf 0_r,\mbf H_{0,C}^{\prime}\bm\Theta_i^{\text{OLS}}\mbf H_{0,C}\r)$&&&&\\
\eqref{eq:ANLSW2} $\sqrt T(\wideparen{\bm\lambda}_i-\wideparen{\mbf H}^{\prime}{\bm\lambda}_i)\to_d\mathcal N\l(\mbf 0_r,\mbf H_{1,C}^{\prime}\bm\Theta_i^{\text{OLS}}\mbf H_{1,C}\r)$&$\wideparen{\mbf H}$	&$ \l(\frac{\bm\Lambda^\prime\bm\Lambda}{n}\r)^{-1/2} \wideparen {\mbf Q}$& $\mbf H_{0,C}:=(\bm\Gamma^F)^{1/2} \bm\Upsilon_0\bm{\mathcal J}_0(\bm V_0)^{-1/2}$ & $\mbf H_{1,C}:=(\bm\Sigma_\Lambda)^{-1/2} \bm\Upsilon_1\bm{\mathcal J}_1$\\[3pt]
\eqref{eq:ANFSW1} $\sqrt n(\wideparen{\mbf F}_t-\wideparen{\mbf H}^{-1}{\mbf F}_t ) \to_d\mathcal N\l(\mbf 0_r, \mbf H_{0,C}^{-1}\bm\Pi_t^{\text{\tiny \upshape OLS}}(\mbf H_{0,C}^{-1})^\prime\r)$&	rate			& $\min(n,\sqrt{T})$		& $\min(\sqrt n,\sqrt T)$								& $\min(\sqrt n,\sqrt T)$\\[3pt]
\eqref{eq:ANFSW2} $\sqrt n(\wideparen{\mbf F}_t-\wideparen{\mbf H}^{-1}{\mbf F}_t ) \to_d\mathcal N\l(\mbf 0_r, \mbf H_{1,C}^{-1}\bm\Pi_t^{\text{\tiny \upshape OLS}}(\mbf H_{1,C}^{-1})^\prime\r)$&&\eqref{eq:sameQsw}&\eqref{eq:semplicesw2}& \eqref{eq:semplicesw}\\[3pt]
\hline
&&&&\\[-3pt]
\eqref{eq:ANLD1} $\sqrt T(\bar{\bm\lambda}_i-\bar{\mbf H}^{-1}{\bm\lambda}_i)\to_d\mathcal N\l(\mbf 0_r,\mbf H_{0,D}^{-1}\bm\Theta_i^{\text{OLS}}(\mbf H_{0,D}^{-1})^\prime\r)$
&&&&\\
\eqref{eq:ANLD2} $\sqrt T(\bar{\bm\lambda}_i-\bar{\mbf H}^{-1}{\bm\lambda}_i)\to_d\mathcal N\l(\mbf 0_r,\mbf H_{1,D}^{-1}\bm\Theta_i^{\text{OLS}}(\mbf H_{1,D}^{-1})^\prime\r)$&$\bar{\mbf H}$	&$ \l(\frac{\bm\Lambda^\prime\bm\Lambda}{n}\r)^{1/2} \wideparen {\mbf Q}$& $\mbf H_{0,D}:=(\bm\Gamma^F)^{-1/2} \bm\Upsilon_0\bm{\mathcal J}_0(\bm V_0)^{1/2}$ & $\mbf H_{1,D}:=(\bm\Sigma_\Lambda)^{1/2} \bm\Upsilon_1\bm{\mathcal J}_1$\\[3pt]
\eqref{eq:ANFD1} $\sqrt n(\bar{\mbf F}_t-\bar{\mbf H}^{\prime}{\mbf F}_t ) \to_d\mathcal N\l(\mbf 0_r, \mbf H_{0,D}^{\prime}\bm\Pi_t^{\text{\tiny \upshape OLS}}\mbf H_{0,D}\r)$&	rate			& $\min(n,\sqrt{T})$		& $\min(\sqrt n,\sqrt T)$								& $\min(\sqrt n,\sqrt T)$\\[3pt]
\eqref{eq:ANFD2} $\sqrt n(\bar{\mbf F}_t-\bar{\mbf H}^{\prime}{\mbf F}_t ) \to_d\mathcal N\l(\mbf 0_r, \mbf H_{1,D}^{\prime}\bm\Pi_t^{\text{\tiny \upshape OLS}}\mbf H_{1,D}\r)$
&&\eqref{eq:sameQswD}&\eqref{eq:sempliceD2}& \eqref{eq:sempliceD}\\
&&&&\\
\hline
\hline
\end{tabular}

\begin{tabular}{p{.8\textwidth}}
$\wh{\mbf Q}$ are normalized eigenvectors of $(T^{-1}\bm F^\prime\bm F)^{1/2}(n^{-1}\bm \Lambda^\prime\bm \Lambda) (T^{-1}\bm F^\prime\bm F)^{1/2}$; 
$\wideparen {\mbf Q}$ are the normalized eigenvectors of $(n^{-1}\bm \Lambda^\prime\bm \Lambda)^{1/2} (T^{-1}\bm F^\prime\bm F) (n^{-1}\bm \Lambda^\prime\bm \Lambda)^{1/2}$; \\
$\bm\Upsilon_0$ are normalized eigenvectors of $(\bm\Gamma^F)^{1/2}\bm\Sigma_\Lambda(\bm\Gamma^F)^{1/2}$; 
$\bm\Upsilon_1$ are normalized eigenvectors of $(\bm\Sigma_\Lambda)^{1/2}\bm\Gamma^F(\bm\Sigma_\Lambda)^{1/2}$;\\
$\bm V_0=\lim_{n\to\infty} n^{-1}\mbf M^{C}$ are eigenvalues of $(\bm\Gamma^F)^{1/2}\bm\Sigma_\Lambda(\bm\Gamma^F)^{1/2}$ and $(\bm\Sigma_\Lambda)^{1/2}\bm\Gamma^F(\bm\Sigma_\Lambda)^{1/2}$;\\
$\bm{\mathcal J}_0$ and $\bm{\mathcal J}_1$ are diagonal with entries $\pm 1$.
\end{tabular}
}
\end{sidewaystable}

\section{The case of non-zero mean data}\label{sec:cent}
We extend the previous results to the case in which the data have non-zero mean. There are three cases that can generate non-zero mean data and are worth investigating. These are collected into the following assumption.

\begin{ass}\label{ass:center} The factor model \eqref{eq:SDFM1R} reads as
\begin{compactenum}
\item [(a)]  $x_{it}=\alpha_i+\bm\lambda_i^\prime \mbf F_t + e_{it}$, $i\in\mathbb N$, $t\in\mathbb Z$, 
where $\bm\lambda_i$, $\mbf F_t$, and $e_{it}$ still satisfy Assumptions \ref{ass:common}-\ref{ass:CLT}, and  $\alpha_i$ is a scalar with $\vert \alpha_i\vert\le M_\alpha$ for some finite positive real $M_\alpha$ independent of $i$, so that $\E[x_{it}]=\alpha_i$;
\item [(b)]  $x_{it}=\bm\lambda_i^\prime\bm\mu +\bm\lambda_i^\prime \mbf F_t+ e_{it}$, $i\in\mathbb N$, $t\in\mathbb Z$, 
where $\bm\lambda_i$, $\mbf F_t$, and $e_{it}$ still satisfy Assumptions \ref{ass:common}-\ref{ass:CLT}, and $\bm\mu$ is an $r$-dimensional vector such that $\Vert \bm\mu\Vert\le M_\mu$ for some finite positive real $M_\mu$, so that $\E[x_{it}]=\bm\lambda_i^\prime\bm \mu$;
\item [(c)]  $x_{it}=\alpha_i+\bm\lambda_i^\prime\bm\mu +\bm\lambda_i^\prime \mbf F_t+ e_{it}$, $i\in\mathbb N$, $t\in\mathbb Z$, 
where $\bm\lambda_i$, $\mbf F_t$, and $e_{it}$ still satisfy Assumptions \ref{ass:common}-\ref{ass:CLT}, $\alpha_i$ satisfies case (a) and $\bm\mu$ satisfies case (b), so that $\E[x_{it}]=\alpha_i+\bm\lambda_i^\prime\bm \mu$.
\end{compactenum}
\end{ass}

Part (a) is relevant for macroeconometrics, where $x_{it}$ is differenced data, e.g., growth rate of GDP, and $\alpha_i$ then the slope of the linear trend of the raw data, i.e., levels of GDP  \citep{baing04,BLL2}. Parts (b) and (c) are relevant in financial econometrics. If $x_{it}$ denotes an excess return and $\bm\mu$ is interpreted as the vector of risk premia, then $\bm\lambda_i^\prime\bm\mu$ is the risk premium implied by the factor exposures $\bm\lambda_i$, while $\alpha_i$ is the pricing error of asset $i$  \citep{chamberlainrothschild83,lettau2020estimating}.

In all cases, the PC estimators of the loadings and the factors are obtained by now using centered data, i.e., on 
$\bm X-\bar{\bm X}$, where $\bar{\bm X}$ is a $T\times n$ matrix with all rows equal to $\bar{\mbf x}^\prime:=T^{-1}\sum_{t=1}^T \mbf x_t^\prime$.
Namely,  the estimator of the loadings is given by \eqref{eq:estL}, but when using eigenvectors and eigenvalues of the sample covariance matrix $T^{-1}(\bm X-\bar {\bm X})^\prime (\bm X-\bar {\bm X})$, while the estimator of the factors \eqref{eq:estF} is now given by $\wh{\bm F}:=(\bm X-\bar{\bm X})\wh{\bm \Lambda}(\wh{\bm \Lambda}^\prime \wh{\bm \Lambda})^{-1}$, with rows $\wh{\mbf F}_t^\prime:=(\mbf x_t-\bar{\mbf x})^\prime\wh{\bm \Lambda}(\wh{\bm \Lambda}^\prime \wh{\bm \Lambda})^{-1}$, $t=1,\ldots, T$. Furthermore, an estimator of $\bm \mu$ is obtained as $\wh{\bm \mu}:= (\wh{\bm\Lambda}^\prime \wh{\bm\Lambda})^{-1}\wh{\bm\Lambda}^\prime \bar{\mbf x}$.
The asymptotic distribution of these new PC estimators is given in the following result (see Appendix \ref{app:CLTcentproof} for a proof).

\begin{prop}\label{th:CLTcent}
Let 
$$
\wh{\mbf H}_c:= \frac{(\bm F-\bar{\bm F})^\prime(\bm F-\bar{\bm F})}{T}
\frac{\bm\Lambda^\prime\wh{\bm\Lambda}}{n}
\l(\frac{\wh{\mbf M}^x}{n}\r)^{-1},
$$
where $\bar{\bm F}$ is a $T\times r$ matrix with all rows equal to $\bar{\mbf F}^\prime:=T^{-1}\sum_{t=1}^T \mbf F_t^\prime$.

\begin{compactenum}
\item[(a)] 
Under Assumptions \ref{ass:common} through \ref{ass:rates}, and \ref{ass:center}(a) or \ref{ass:center}(b) or \ref{ass:center}(c),  for any given $i=1,\ldots,n$,   as $n,T\to\infty$,
$$
\sqrt T\l(\wh{\bm\lambda}_i-\wh{\mbf H}_c^\prime{\bm\lambda}_i\r) 
 \to_d\mathcal N\l(\mbf 0_r,  
\bm\Upsilon_0^\prime(\bm\Gamma^F)^{1/2}\bm\Theta_i^{\text{\tiny \upshape OLS}} (\bm\Gamma^F)^{1/2} \bm\Upsilon_0
\r),
$$
where $\bm\Theta_i^{\text{\tiny \upshape OLS}}$ is defined in Theorem \ref{th:CLTL}.
\item [(b)]  Under Assumptions \ref{ass:common} through \ref{ass:CLT}, \ref{ass:Wold} or \ref{ass:Wu} in Appendix \ref{sec:hannan}, 
and \ref{ass:center}(a) or \ref{ass:center}(b) or \ref{ass:center}(c),
for any given $t=1,\ldots, T$, as $n,T\to\infty$,
$$
\sqrt{\frac{nT}{n+T}}
\l(\wh{\mbf F}_t - \wh{\mbf H}_c^{-1}\mbf F_t \r)\to_d \mathcal N\l(\mbf 0_r, \bm\Upsilon_0^\prime(\bm\Gamma^F)^{-1/2} \l\{\frac 1{1+\gamma}\l(\bm\Pi_t^{\text{\tiny \upshape OLS}}+\gamma\, \bm\Omega_0  \r)\r\} (\bm\Gamma^F)^{-1/2}\bm\Upsilon_0 \r),
$$
where $\gamma:=\lim_{n,T\to\infty} n/T$, $\bm\Pi_t^{\text{\tiny \upshape OLS}}$ is defined in Theorem \ref{th:CLTF},
and
$\bm\Omega_0:=\sum_{h=-\infty}^\infty \E[\mbf F_t\mbf F_{t-h}^\prime]$.\smallskip

\item [(c-i)] Under Assumptions \ref{ass:common} through \ref{ass:CLT}, \ref{ass:Wold} or \ref{ass:Wu} in Appendix \ref{sec:hannan}, 
and \ref{ass:center}(b), as $n,T\to\infty$, 
$$
\sqrt T\l(\wh{\bm \mu}-\wh{\mbf H}^{-1}_c\bm\mu\r)
\to_d \mathcal N\l(\mbf 0_r,\bm\Upsilon_0^\prime  (\bm\Gamma^F)^{-1/2}
\bm\Omega_0
(\bm\Gamma^F)^{-1/2}
\bm\Upsilon_0\r).
$$

\item [(c-ii)] Under Assumptions \ref{ass:common} through \ref{ass:rates},
and \ref{ass:center}(b), for any given $t=1,\ldots, T$, as $n,T\to\infty$, 
$$
\sqrt n\l(\wh{\mbf F}_t+\wh{\bm \mu}- \wh{\mbf H}^{-1}_c \mbf F_t - \wh{\mbf H}^{-1}_c\bm \mu \r)\to_d \mathcal N\l(\mbf 0_r, \bm\Upsilon_0^\prime(\bm\Gamma^F)^{-1/2}\bm\Pi_t^{\text{\tiny \upshape OLS}}(\bm\Gamma^F)^{-1/2}\bm\Upsilon_0\r),
$$
where $\bm\Pi_t^{\text{\tiny \upshape OLS}}$ is defined in Theorem \ref{th:CLTF}.
\item [(d)] 
Under Assumptions \ref{ass:common} through \ref{ass:ind}, and \ref{ass:center}(c), as $n,T\to\infty$, 
$$
\min( n,\sqrt T)\l\Vert\wh{\bm \mu}-\wh{\mbf H}^{-1}_c\bm\mu - \wh{\mbf H}^{-1}_c\l(\frac{\bm\Lambda^\prime\bm\Lambda}n\r)^{-1}
\l(\frac 1n\sum_{i=1}^n  \bm\lambda_i  \alpha_i\r) 
\r\Vert = O_{\mathrm P}(1).
$$
\end{compactenum}
\end{prop}

The main result underlying the proof is the following. Under Assumption \ref{ass:center}(a) we consistently estimate the elements of $\bm\alpha:=(\alpha_1\cdots  \alpha_n)^\prime$ by means of $\bar {\mbf x}$ (the proof is omitted since it is standard). Under Assumption \ref{ass:center}(b) $\bar {\mbf x}$ can be used, jointly with the estimated loadings, to estimate $\bm\mu$ by linear projection. However, under Assumption \ref{ass:center}(c), $\bm\alpha$ and $\bm\mu$ cannot be separately identified, so no consistent estimator of those quantities can be derived via PCA.

The proof of parts (a) and (b) follows the same steps used for Lemmas \ref{prop:load} and \ref{prop:factor}, but when using centered data so that $(\mbf x_t-\bar{\mbf x})=\bm\Lambda(\mbf F_t-\bar{\mbf F}) +(\bm e_t-\bar{\bm e})$, which holds for all cases in Assumption \ref{ass:center}.
In part (a), the asymptotic distribution of the PC estimator of the loadings is the same as the one given in Theorem \ref{th:CLTL} (see also \citealp[Theorem 1.2]{lettau2020estimating}). This is because we are modifying the sample covariance matrix by using a consistent estimator of the mean.

The distribution of the PC estimator of the factors in part (b), however, requires more restrictions to hold since, while we center data along the time dimension,  we estimate the factors by means of a cross-sectional regression.  This implies that now the asymptotic expansion used in the proof of Theorem \ref{th:CLTF} contains the extra term $-\wh{\mbf H}_c^{-1}\bar{\mbf F} $, which is $O_{\mathrm P}(T^{-1/2})$. So, unless $n/T\to 0$, i.e., $\gamma=0$, the new asymptotic distribution has to account for this term (see also \citealp[Theorem 5B]{Mao2024StatisticalAnalysis},  for the case of deterministic factors). However, if $\gamma>0$, the rate in the CLT determining the asymptotic distribution depends on both $n$ and $T$ and the asymptotic covariance is going to be different from that in the zero-mean case.  In particular, to compute the asymptotic covariance we must ensure summability of the autocovariances of $\{\mbf F_t\}$, i.e., that $\Vert\bm\Omega_0\Vert = O(1)$. This can be either directly assumed or derived, for example, from Assumptions \ref{ass:Wold} or \ref{ass:Wu} in Appendix \ref{sec:hannan}, where $\{\mbf F_t\}$ is assumed to be a linear process with summable coefficients. 

For part (c-i) note that for given loadings, we have $\bar{\mbf x}=\bm\Lambda \bar{\mbf F}+\bm\Lambda \bm\mu +\bar {\bm e}$. So $\bm\mu$ can be estimated by linear projection of $\bar{\mbf x}$ onto $\bm\Lambda$ since $\Vert \bar{\mbf F} \Vert= O_{\mathrm P}(T^{-1/2})$. The presence of $\wh{\mbf H}_c$ is because we use estimated rather than true  loadings.

Part (c-ii) follows directly by combining part (b) and (c-i). It shows that the PC estimator of the non-zero mean factors $\mbf G_t:=\mbf F_t+\bm\mu$  has the same asymptotic distribution as the PC estimator of the factors as given in Theorem \ref{th:CLTF} (see also \citealp[Theorem 1.3]{lettau2020estimating}). Indeed, under Assumption \ref{ass:center}(b), we can write $x_{it}=\bm\lambda_i^\prime \mbf G_t + e_{it}$, and the non-zero mean factors  $\mbf G_t$ can then be estimated as 
$\wh{\mbf G}_t:=\wh{\mbf F}_t+\wh{\bm\mu} = (\wh{\bm\Lambda}^\prime \wh{\bm\Lambda})^{-1}\wh{\bm\Lambda}^\prime \mbf x_t$.  Notice that in this case, differently from the case of zero-mean factors in part (b), the requirement $\sqrt n/T\to 0$, as $n,T\to\infty$, in Assumption \ref{ass:rates} is sufficient.

Under part (d) we see that, if Assumption \ref{ass:center}(c) holds, then $\wh{\bm\mu}$ is an asymptotically biased estimator of $\bm\mu$. In particular, in this case we have $\bar{\mbf x}=\bm\Lambda \bar{\mbf F}+\bm\alpha+\bm\Lambda \bm\mu +\bar {\bm e}$ so by noticing again that $\Vert \bar{\mbf F} \Vert= O_{\mathrm P}(T^{-1/2})$ we see that by projecting $\bar{\mbf x}$ onto $\bm\Lambda$ we do not recover $\bm\mu$ but we recover $\bm\mu+(\bm\Lambda^\prime\bm\Lambda)^{-1}\bm\Lambda^\prime\bm\alpha$, the bias being given by the coefficient of the projection of $\bm\alpha$ onto the loadings. In general, the bias does not vanish asymptotically, since  $\Vert n^{-1}\sum_{i=1}^n \bm\lambda_i\alpha_i\Vert= O(1)$ for all $n\in\mathbb N$, because of Assumptions \ref{ass:common}(a) and \ref{ass:center}(c). In this case PCA cannot help because, while this identification issue is related to the mean, estimation is run on demeaned data. 

\section{Identification of loadings and factors}\label{sec:II}

In this section we study the implications of standard identifying restrictions on the true loadings and factors. These restrictions are statistical: they select one representative among observationally equivalent representations, and therefore belong to exploratory rather than confirmatory factor analysis. The following are the identifying conditions typically found in the literature.\footnote{
Other identifying conditions, not considered here, are either imposed by constraining the dynamic evolution of the conditional mean or the conditional covariance of the factors (see, e.g., \citealp{BW15}, and \citealp{SentanaFiorentini2001}, respectively), or
are related to confirmatory factor analysis \citep[see, e.g.,][conditions PC2 and PC3]{baing13}.}
\begin{compactenum}
\item [(A)] $n^{-1}\bm\Lambda^\prime\bm\Lambda$ diagonal for all $n\in\mathbb N$ (\citealp[condition (c) page 122]{AR56}; \citealp[PC1]{baing13}).
\item [(B)] $n^{-1}\bm\Lambda^\prime(\text{diag}(\bm\Gamma^ e))^{-1}\bm\Lambda$ diagonal for all $n\in\mathbb N$ (\citealp[condition (d) page 122]{AR56}; \citealp[Chapter 2]{lawleymaxwell71};
\citealp[Chapter 9.2]{mardia1979multivariate}; \citealp[IC3]{baili16}).
\item  [(C)] $\bm\Gamma^F=\mathbf I_r$  (\citealp[Equation (2)]{hotelling1933analysis}; 
\citealp[Section 4]{AR56};
\citealp[Chapter 2]{lawleymaxwell71};
\citealp[Chapter 9.2]{mardia1979multivariate}; \citealp[Cases 1 and 2]{RT82}; \citealp[Chapter 7.1]{jolliffe2002principal}; \citealp[Chapter 14.2]{anderson2003introduction}).
\item  [(D)] $T^{-1}\bm F^\prime\bm F=\mathbf I_r$   for all $T\in\mathbb N$ (\citealp[Chapter 14.2]{anderson2003introduction}; \citealp[IC3]{baili16}; \citealp[Assumption 1]{onatski2012asymptotics}; \citealp[PC1]{baing13}; 
\citealp[Assumption 1]{Freyaldenhoven2022}).
\end{compactenum}
Restrictions (A)--(B) concern the loadings, while (C)--(D) concern the factors. In maximum likelihood estimation of exact factor models, (B) is often used instead of (A) because loadings and idiosyncratic variances are estimated jointly. In PCA, however, the idiosyncratic covariance is not directly estimated, so we focus on (A). Since classical treatments usually keep $n$ fixed, (A) is imposed for all $n$; in high-dimensional settings it is also natural to consider its limiting version, $\bm\Sigma_\Lambda$ diagonal.

The choice between (C) and (D) depends on whether factors are treated as random or deterministic. In maximum likelihood estimation this distinction is largely immaterial for the asymptotic normality of the loadings (\citealp[Chapter 14.2]{anderson2003introduction}; \citealp{AFP87}; \citealp{baili16}). In PCA it is more consequential: if factors are random, (D) is generally not credible, whereas (C) remains natural. We nevertheless also consider (D) below to make its implications explicit.

We now study the implications of these restrictions for the true loadings and factors. A key distinction is between restrictions imposed for all fixed $n>N$, with $N$ defined in Assumption \ref{ass:common}(d), and restrictions imposed only in the limit $n\to\infty$. The following assumption is used throughout this section.

\begin{ass}\label{ass:sign_eval}
For all $n>N$, with $N$ defined in Assumption \ref{ass:common}(d),
\begin{compactenum}[(a)]
\item $[\mbf K^\prime \bm \Upsilon_0]_{jj}\ge 0$ and 
$[ \bm \Upsilon_0^\prime(\bm\Gamma^F)^{1/2}(\bm\Sigma_\Lambda)^{1/2} \bm \Upsilon_1]_{jj}\ge 0$
 for all $j=1,\ldots, r$, where $\mbf K$, $\bm\Upsilon_0$, and $\bm \Upsilon_1$ are the $r\times r$ matrices having as columns the normalized eigenvectors of $(\bm\Gamma^F)^{1/2}(n^{-1}{\bm\Lambda^\prime\bm\Lambda})(\bm\Gamma^F)^{1/2}$, 
 $(\bm\Gamma^F)^{1/2}\bm\Sigma_\Lambda(\bm\Gamma^F)^{1/2}$, 
 and
 $(\bm\Sigma_\Lambda)^{1/2}\bm\Gamma^F(\bm\Sigma_\Lambda)^{1/2}$,
 respectively;
\item $\mu_j^{C} > \mu_{j+1}^{C}$ for all $j=1,\ldots, r-1$, and $n^{-1}\mu_r^C\ge C_\Lambda$, for some finite positive real $C_\Lambda$ independent of $n$.
\end{compactenum}
\end{ass}

Part (a) assumes that the angle between the columns of $\mbf K$ and $\bm\Upsilon_0$ and of $(\bm\Gamma^F)^{1/2}\bm\Upsilon_0$ and $(\bm\Sigma_\Lambda)^{1/2} \bm\Upsilon_1$ is always less than 90$^{\circ}$, thus fixing the sign of the eigenvectors or their rescaled version. This can always be assumed without loss of generality. In particular,  part (a),  together with Assumptions \ref{ass:common}(a) and Davis-Kahan theorem \citep[Corollary 1]{yu15},  implies that the columns of $\mbf K$ and $\bm\Upsilon_0$ coincide asymptotically as $n\to\infty$ (see also the proof of Proposition \ref{prop:KKK}). It also implies that $\bm{\mathcal J}_0$, defined in Proposition \ref{prop:KKK}, and $\bm{\mathcal J}_1$, defined in \eqref{eq:KKKBai} in Appendix \ref{sec:cmppca}, coincide.

Part (b) is instead crucial when we impose constraints on $\bm\Lambda$ holding for all $n\in\mathbb N$. It is immediate to see that this assumption implies the weaker asymptotic requirement of distinct eigenvalues of $\bm\Sigma_\Lambda\bm\Gamma^F$ in Assumption \ref{ass:eval}. Moreover, it imposes positive definiteness of $n^{-1}\bm\Lambda^\prime \bm\Lambda$ for all $n\in\mathbb N$, thus strengthening Assumption \ref{ass:common}(a).

The next results show the implications of various identifying constraints for the true loadings and factors (see Appendix \ref{corol:K00proof} for a proof). 
\begin{prop}\label{corol:K00}
Under Assumptions \ref{ass:common} and \ref{ass:sign_eval},\footnote{
Hereafter, when we write ``for all $n\in\mathbb N$'', we always mean that $n>N$ with $N$ defined in Assumption \ref{ass:common}(d).
}
\begin{compactenum}

\item [(I.a)] if $\bm\Gamma^F=\mbf I_r$ and 
$\bm\Lambda$  is unrestricted, 
then:
\begin{compactenum}
\item [(i.1)]  $\bm\lambda_i^\prime=\mbf v_i^{C\prime}(\mbf M^{C})^{1/2}\mbf K^\prime$  for all $i=1,\ldots, n$ and all $n\in\mathbb N$; 
\item [(ii.1)] $\mbf  F_t=\mbf K(\mbf M^{C})^{-1/2} \mbf V^{{C}\prime}\bm{C}_t$ for all $t\in\mathbb Z$ and all $n\in\mathbb N$;
\item [(i.2)]  $\lim_{n\to\infty}\Vert \bm\lambda_i^\prime-\bm p_{i0}^\prime (\bm V_0)^{1/2} \bm\Upsilon_0^\prime \Vert=0$  for all $i\in\mathbb N$;
\item [(ii.2)] $\text{\upshape m.s.-}\lim_{n\to\infty}\Vert \mbf F_t-\bm\Upsilon_0(\bm V_0)^{-1/2}  \bm W_{\infty}^\prime\bm C_{t,\infty}\Vert=0$ for all $t\in\mathbb Z$;
\end{compactenum}

\item [(II.a)] if $\bm\Gamma^F=\mbf I_r$ and  
$n^{-1}\bm\Lambda^\prime\bm \Lambda$  is diagonal for all $n\in\mathbb N$, 
then:
\begin{compactenum}
\item [(i.1)] $\bm\lambda_i^\prime=\mbf v_i^{C\prime}(\mbf M^{C})^{1/2}\bm S$  for all $i=1,\ldots, n$ and all  $n\in\mathbb N$; 
\item [(ii.1)] $\mbf  F_t=\bm S(\mbf M^{C})^{-1/2}\mbf V^{{C}\prime}\bm{C}_t$ for all $t\in\mathbb Z$ and all $n\in\mathbb N$; 
\item [(i.2)] $\lim_{n\to\infty}\Vert \bm\lambda_i^\prime-\bm p_{i0}^\prime (\bm V_0)^{1/2} \bm {\mathcal S}_0 \Vert=0$ for  all  $i\in\mathbb N$; 
\item [(ii.2)] $\text{\upshape m.s.-}\lim_{n\to\infty}\Vert\mbf  F_t-\bm{\mathcal S}_0(\bm V_0)^{-1/2}  \bm W_{\infty}^{\prime}\bm{C}_{t,\infty}\Vert=0$  for all $t\in\mathbb Z$;
\end{compactenum}

\item [(III.a)] if $\bm\Gamma^F=\mbf I_r$ and 
$\bm\Sigma_\Lambda$  is diagonal, 
then:
\begin{compactenum}
\item [(i)] $\lim_{n\to\infty}\Vert \bm\lambda_i^\prime-\bm p_{i0}^{\prime}(\bm V_0)^{1/2}\bm{\mathcal S}_0\Vert=0$ for all $i\in\mathbb N$;
\item [(ii)] $\text{\upshape m.s.-}\lim_{n\to\infty}\Vert\mbf  F_t-\bm{\mathcal S}_0(\bm V_0)^{-1/2}  \bm W_{\infty}^{\prime}\bm{C}_{t,\infty}\Vert=0$  for all $t\in\mathbb Z$;
\end{compactenum}
 
\item [(IV.a)] if $T^{-1} \bm F^\prime\bm F=\mbf I_r$ for all $T\in\mathbb N$ and 
$\bm\Lambda$  is unrestricted,  
then:
\begin{compactenum}
\item [(i.1)] $\bm\lambda_i^\prime=\wh{\mbf v}_i^{C\prime}(\wh{\mbf M}^{C})^{1/2}\wh{\mbf Q}^\prime$ for all $i=1,\ldots, n$ and all $n,T\in\mathbb N$;
\item [(ii.1)] $\mbf  F_t=\wh{\mbf Q}(\wh{\mbf M}^{C})^{-1/2} \wh{\mbf V}^{{C}\prime}\bm{C}_t$  for all $t=1,\ldots, T$ and all $n,T\in\mathbb N$;
\item [(i.2)] $\lim_{T\to\infty}\Vert \bm\lambda_i^\prime-{\mbf v}_i^{C\prime}({\mbf M}^{C})^{1/2}{\mbf K}^\prime\Vert =0$ for all $i=1,\ldots, n$ and all $n\in\mathbb N$;
\item [(ii.2)] $\text{\upshape m.s.-}\lim_{T\to\infty}\Vert\mbf  F_t-{\mbf K}({\mbf M}^{C})^{-1/2} {\mbf V}^{{C}\prime}\bm{C}_t\Vert=0$  for all $t\in\mathbb N$ and all $n\in\mathbb N$;
\item [(i.3)]  $\lim_{n,T\to\infty}\Vert \bm\lambda_i^\prime-\bm p_{i0}^\prime (\bm V_0)^{1/2} \bm\Upsilon_0^\prime \Vert=0$  for all $i\in\mathbb N$;
\item [(ii.3)] $\text{\upshape m.s.-}\lim_{n,T\to\infty}\Vert \mbf F_t- \bm\Upsilon_0(\bm V_0)^{-1/2}  \bm W_{\infty}^\prime\bm C_{t,\infty}\Vert=0$ for all $t\in\mathbb N$;
\end{compactenum}

\item [(V.a)] if $T^{-1} \bm F^\prime\bm F=\mbf I_r$ and 
$n^{-1}\bm\Lambda^\prime\bm \Lambda$  is diagonal for all $n,T\in\mathbb N$, 
then:
\begin{compactenum}
\item [(i.1)] $\bm\lambda_i^\prime=\wh{\mbf v}_i^{C\prime}(\wh{\mbf M}^{C})^{1/2}\wh{\bm S}$ for all $i=1,\ldots, n$ and all $n,T\in\mathbb N$;
\item [(ii.1)] $\mbf  F_t=\wh{\bm S}(\wh{\mbf M}^{C})^{-1/2} \wh{\mbf V}^{{C}\prime}\bm{C}_t$  for all $t=1,\ldots, T$ and all $n,T\in\mathbb N$;
\item [(i.2)] $\lim_{T\to\infty}\Vert \bm\lambda_i^\prime-{\mbf v}_i^{C\prime}({\mbf M}^{C})^{1/2}{\bm S}^\prime\Vert =0$ for all $i=1,\ldots, n$ and all $n\in\mathbb N$;
\item [(ii.2)] $\text{\upshape m.s.-}\lim_{T\to\infty}\Vert\mbf  F_t-{\bm S}({\mbf M}^{C})^{-1/2} {\mbf V}^{{C}\prime}\bm{C}_t\Vert=0$  for all $t\in\mathbb N$ and all $n\in\mathbb N$;
\item [(i.3)]  $\lim_{n,T\to\infty}\Vert \bm\lambda_i^\prime-\bm p_{i0}^\prime (\bm V_0)^{1/2} \bm{\mathcal S}_0 \Vert=0$  for all $i\in\mathbb N$;
\item [(ii.3)] $\text{\upshape m.s.-}\lim_{n,T\to\infty}\Vert \mbf F_t- \bm{\mathcal S}_0(\bm V_0)^{-1/2}  \bm W_{\infty}^\prime\bm C_{t,\infty}\Vert=0$ for all $t\in\mathbb N$;
\end{compactenum}

\item [(VI.a)] if $T^{-1} \bm F^\prime\bm F=\mbf I_r$ for all $T\in\mathbb N$ and 
$\bm\Sigma_\Lambda$  is diagonal, 
then:
\begin{compactenum}
\item [(i)] $\lim_{n,T\to\infty}\Vert \bm\lambda_i^\prime-\bm p_{i0}^{\prime}(\bm V_0)^{1/2}\bm{\mathcal S}_0\Vert=0$ for all $i\in\mathbb N$;
\item [(ii)] $\text{\upshape m.s.-}\lim_{n,T\to\infty}\Vert\mbf  F_t-\bm{\mathcal S}_0(\bm V_0)^{-1/2}  \bm W_{\infty}^\prime\bm C_{t,\infty}\Vert=0$  for all $t\in\mathbb N$;
\end{compactenum}

\item [(I.b)] if $\bm\Sigma_\Lambda=\mbf I_r$ and 
$\bm F$ is unrestricted, 
then:
\begin{compactenum}
\item [(i)] $\lim_{n\to\infty} \l\Vert\bm\lambda_i^\prime - \bm p_{i0}^{\prime}   \bm\Upsilon_1^\prime \r \Vert=0$ for all $i\in\mathbb N$;
\item [(ii)] $\text{\upshape m.s.-}\lim_{n\to\infty}\Vert\mbf  F_t-\bm \Upsilon_1 \bm W_{\infty}^\prime\bm C_{t,\infty}\Vert =0$ for all $t\in\mathbb Z$;
\end{compactenum}

\item[(II.b)]  if $\bm\Sigma_\Lambda=\mbf I_r$ and 
$T^{-1} \bm F^\prime\bm F$ is diagonal for all $T\in\mathbb N$,  
then:
\begin{compactenum}
\item [(i)] $\text{\upshape P-}\lim_{n,T\to\infty}\Vert\bm\lambda_i^\prime- \bm p_{i0}^\prime \bm{\mathcal S}_0\Vert=0$  for all $i\in\mathbb N$;
\item [(ii)] $\text{\upshape P-}\lim_{n,T\to\infty}\Vert\mbf  F_t-\bm {\mathcal S}_0 \bm W_{\infty}^\prime\bm C_{t,\infty}\Vert =0$ for all $t\in\mathbb N$;
\end{compactenum}

\item[(III.b)]  if  $\bm\Sigma_\Lambda=\mbf I_r$ and 
$\bm\Gamma^F$ is diagonal, 
then:
\begin{compactenum}
\item [(i)] $\lim_{n\to\infty}\Vert\bm\lambda_i^\prime- \bm p_{i0}^\prime \bm{\mathcal S}_0\Vert=0$  for all $i\in\mathbb N$;
\item [(ii)] $\text{\upshape m.s.-}\lim_{n\to\infty}\Vert\mbf  F_t-\bm {\mathcal S}_0 \bm W_{\infty}^\prime\bm C_{t,\infty}\Vert =0$ for all $t\in\mathbb Z$;
\end{compactenum}

\item [(IV.b)] if  $n^{-1}\bm\Lambda^\prime\bm \Lambda=\mbf I_r$  for all $n\in\mathbb N$ and 
$\bm F$ is unrestricted, 
then:
\begin{compactenum}
\item [(i.1)] $\bm\lambda_i^\prime=  n \mbf v_i^{C\prime} (\mbf M^{C})^{-1/2}\mbf K^\prime(\bm\Gamma^F)^{1/2}$ for all $i=1,\ldots, n$ and all $n\in\mathbb N$;
\item [(ii.1)] $\mbf F_t=  (\bm\Gamma^F)^{1/2}\mbf K(\mbf M^{C})^{-1/2}\mbf V^{C\prime}\bm C_t$  for all $t\in\mathbb Z$ and all $n\in\mathbb N$;
\item [(i.2)] $\lim_{n\to\infty} \l\Vert\bm\lambda_i^\prime - \bm p_{i0}^{\prime}   \bm\Upsilon_1^\prime \r \Vert=0$ for all $i\in\mathbb N$;
\item [(ii.2)] $\text{\upshape m.s.-}\lim_{n\to\infty}\Vert\mbf  F_t-\bm \Upsilon_1 \bm W_{\infty}^\prime\bm C_{t,\infty}\Vert =0$ for all $t\in\mathbb Z$;
\end{compactenum}

\item [(V.b)] if  $n^{-1}\bm\Lambda^\prime\bm \Lambda=\mbf I_r$ and 
$T^{-1} \bm F^\prime\bm F$ is diagonal for all $n,T\in\mathbb N$,  
then:
\begin{compactenum}
\item [(i.1)] $\bm\lambda_i^\prime =\sqrt n \wh{\mbf v}_i^{C\prime} \wh{\bm S}$ for all $i=1,\ldots, n$ and all $n,T\in\mathbb N$;
\item [(ii.1)] $\mbf F_t=n^{-1/2}\wh{\bm S} \wh{\mbf V}^{C\prime}\bm C_t$  for all $t=1,\ldots,T$ and all $n,T\in\mathbb N$; 
\item [(i.2)] $\text{\upshape P-}\lim_{T\to\infty}\Vert \bm\lambda_i^\prime -\sqrt n {\mbf v}_i^{C\prime} {\bm S}\Vert=0$ for all $i=1,\ldots, n$ and all $n\in\mathbb N$;
\item [(ii.2)] $\text{\upshape P-}\lim_{T\to\infty}\Vert\mbf F_t-n^{-1/2}{\bm S} {\mbf V}^{C\prime}\bm C_t\Vert=0$ for all $t\in\mathbb N$ and all $n\in\mathbb N$;
\item [(i.3)] $\text{\upshape P-}\lim_{n,T\to\infty}\Vert \bm\lambda_i^\prime-\bm p_{i0}^\prime \bm{\mathcal S}_0\Vert=0$ for all $i\in\mathbb N$;
\item [(ii.3)] $\text{\upshape P-}\lim_{n,T\to\infty}\Vert \mbf F_t-\bm {\mathcal S}_0 \bm W_{\infty}^\prime\bm C_{t,\infty}\Vert=0$ for all $t\in\mathbb N$;
\end{compactenum}

\item[(VI.b)]  if  $n^{-1}\bm\Lambda^\prime\bm \Lambda=\mbf I_r$  for all $n\in\mathbb N$ and 
$\bm\Gamma^F$ is diagonal, 
then:
\begin{compactenum}
\item [(i.1)] $\bm\lambda_i^\prime = \sqrt n  \mbf v_i^{C\prime}\bm S$ for all $i=1,\ldots, n$ and all $n\in\mathbb N$;
\item [(ii.1)] $\mbf F_t=n^{-1/2}\bm S \mbf V^{C\prime}\bm C_t$  for all $t\in\mathbb Z$ and all $n\in\mathbb N$;
\item [(i.2)] $\lim_{n\to\infty}\Vert \bm\lambda_i^\prime-\bm p_{i0}^\prime \bm{\mathcal S}_0\Vert=0$ for all $i\in\mathbb N$;
\item [(ii.2)] $\text{\upshape m.s.-}\lim_{n\to\infty}\Vert \mbf F_t-\bm {\mathcal S}_0 \bm W_{\infty}^\prime\bm C_{t,\infty}\Vert=0$ for all $t\in\mathbb Z$;
\end{compactenum}
\end{compactenum}
and if we replace Assumption \ref{ass:common}(c-ii) with Assumption  \ref{ass:Wold}, or \ref{ass:Hannan}, or \ref{ass:Wu} in Appendix \ref{sec:hannan},
the limits in (II.b) and (V.b) hold also in mean-square. We used the following notation: 
the columns of $\mbf V^C$ and $\wh{\mbf V}^C$ are the normalized eigenvectors of $\bm\Lambda\bm\Gamma^F\bm\Lambda^\prime$ (with eigenvalues $\mbf M^C$) and of 
$\bm\Lambda(T^{-1}\bm F^\prime\bm F)\bm\Lambda^\prime$ (with eigenvalues $\wh{\mbf M}^C$), respectively,
the columns of ${\mbf K}$ and $\wh{\mbf Q}$ are the normalized eigenvectors of $(\bm\Gamma^F)^{1/2}(n^{-1}\bm \Lambda^\prime\bm \Lambda) (\bm\Gamma^F)^{1/2}$ (with eigenvalues $n^{-1}\mbf M^C$)
and of $(T^{-1}\bm F^\prime\bm F)^{1/2}(n^{-1}\bm \Lambda^\prime\bm \Lambda) (T^{-1}\bm F^\prime\bm F)^{1/2}$ (with eigenvalues $n^{-1}\wh{\mbf M}^C$), respectively,
the columns of $\bm\Upsilon_0$ and $\bm\Upsilon_1$ are the normalized eigenvectors of  $(\bm\Gamma^F)^{1/2}\bm \Sigma_\Lambda (\bm\Gamma^F)^{1/2}$ (with eigenvalues $\bm V_0$) and of
$(\bm \Sigma_\Lambda)^{1/2}\bm\Gamma^F(\bm \Sigma_\Lambda)^{1/2}$ (with eigenvalues $\bm V_0$), respectively,
$\bm S$ is an $r\times r$ diagonal matrix with entries $\pm 1$ depending only on $n$, 
$\wh{\bm S}$ is an $r\times r$ diagonal matrix with entries $\pm 1$ depending on $n$ and $T$, 
$\bm{\mathcal S}_0$ is an $r\times r$ diagonal matrix with entries $\pm 1$ independent of $n$ and $T$; $\bm p_{i0}:=\lim_{n\to\infty}\sqrt n \mbf v_i^C$, $i\in\mathbb N$, 
$\bm W_{\infty}:=\lim_{n\to\infty}n^{-1/2}\bm P_{n,\infty}$, with 
$\bm P_{n,\infty}:=(\bm p_{10}\cdots \bm p_{n0})^\prime$, and
$\bm{C}_{t,\infty}:=\text{\upshape m.s.-}\lim_{n\to\infty} n^{-1/2}\bm C_t$, $t\in\mathbb Z$,  such that, as $n\to\infty$, the following hold:
$\Vert \bm p_{i0}\Vert = O(1)$, 
$n^{-1/2}\Vert\bm P_{n,\infty}\Vert = O(1)$, $\Vert\bm W_{\infty}\Vert=O(1)$, and $\Vert \bm C_{t,\infty}\Vert = O_{\mathrm{m.s.}}(1)$.  

\end{prop}

First note that, unless we constrain $T^{-1}\bm F^\prime\bm F$, the sample size $T$ plays no role, since in population the factors are a stochastic process so $t\in\mathbb Z$ and, then, we consider only cases with either $n$ fixed or $n\to\infty$ (parts (I.a), (II.a), (III.a), (I.b), (III.b), (IV.b), (VI.b)). If instead we constrain also $T^{-1}\bm F^\prime\bm F$, then we either consider the case of $n$ and $T$ fixed or $n,T\to\infty$ (parts (IV.a), (V.a), (VI.a), (II.b), (V.b)). Moreover, in three cases we can derive limits for fixed $n$ and $T\to\infty$  (parts (IV.a), (V.a), (V.b)).
Overall it is clear that $n$ and $T$ have no symmetric roles. This is apparent by looking in detail at the implications of the various conditions considered, which are also summarized in Table \ref{tab:ident}. The following are the main implications of Proposition \ref{corol:K00}.

\begin{compactenum}
\item[(i)] From parts (I.a), (IV.a), (I.b), and (IV.b) it is obvious that it is not enough to impose orthonormal factors or loadings to achieve identification. This is because the imposed conditions are only $r(r+1)/2$ so identification can be achieved only up to a rotation accounting for the remaining $r(r-1)/2$ degrees of freedom. Note that part (I.b) is the weakest since identification requires also  $n\to\infty$.

\item[(ii)]  Parts (II.a) and (VI.b) allow one to identify the loadings and the factors up to a sign for any given fixed $n$ and regardless of $T$. In particular, the factors are identified as the population PCs of the common component. These are the classical identification conditions for time series. 

\item[(iii)]  Parts (V.a) and (V.b) allow one to identify the loadings and the factors up to a sign for any given fixed $n$ and $T$. Provided we let $n,T\to\infty$, these have the same implications as parts (II.a) and (VI.b), respectively. But, given that they impose stronger conditions on the factors, they also imply that, for any fixed $n$ and $T$, the true loadings are random while the true factors are the sample PCs of the common component. Such restrictions are, in general, not credible in time series. Indeed, if $\bm F$ is stochastic, then $\mathrm P(T^{-1}\bm F^\prime\bm F=\mbf I_r)=0$, so for the restrictions to hold the factors have to be deterministic, implying that the common components, and so the loadings, are deterministic too. For these reasons parts (V.a) and (V.b), should never be considered in a time series context, unless we interpret all asymptotic results as conditional on a realization $\mbf F_1,\ldots, \mbf F_T$ of the factors.

\item[(iv)]  Parts (VI.a) and (II.b) not only require deterministic factors to hold, but do not even allow for identification unless we also let $n,T\to\infty$, so also these constraints should never be considered in a time series context, unless we interpret all asymptotic results as conditional on a realization $\mbf F_1,\ldots, \mbf F_T$ of the factors.

\item [(v)] Finally, parts (III.a) and (III.b) require $n\to\infty$ for achieving identification and do not depend on $T$. As shown below, in this case the factors are identified as the PCs of the infinite dimensional process of common components. They are the most realistic conditions which should be employed consistently with the fact that the approximate factor model  is identified only when $n\to\infty$ (see Section \ref{sec:blessing}).
\end{compactenum}

Following \citet{gersing_id}, who first derived it, part (III.a) can be reformulated as follows. First, let $\bm\Lambda_{n,\infty}$ be the $n\times r$ matrix with rows $\bm\lambda_{i,\infty}^\prime:=\lim_{n\to\infty} \mbf v_i^{C\prime}(\mbf M^C)^{1/2}$.  
Then, by definition of $\bm p_{i0}$ and Lemma \ref{lem:Vzero}(i),
\beq\label{eq:phil1}
\bm\lambda_{i,\infty}^\prime=\bm p_{i0}^\prime (\bm V_0)^{1/2},
\eeq
and part (III.a.i) is  equivalent to $\lim_{n\to\infty}\Vert \bm\lambda_i^\prime -\bm\lambda_{i,\infty}^\prime  \bm{\mathcal S}_0 \Vert=0.$ Note that the rows of $\bm\Lambda_{n,\infty}$ do not depend on $n$, so
 the sequence $\{\bm\Lambda_{n,\infty},\, n\in\mathbb N\}$ is nested as $n$ grows.

Second, by definition of $\bm W_\infty$ and $\bm p_{i0}$, 
\begin{align}
\bm W_{\infty}^\prime \bm W_{\infty}=\lim_{n\to\infty} \frac 1n \bm P_{n,\infty}^\prime \bm P_{n,\infty}=\lim_{n\to\infty} \frac 1n\sum_{i=1}^n \bm p_{i0}\bm p_{i0}^\prime 
=\lim_{n\to\infty} \sum_{i=1}^n \mbf v_{i}^C\mbf v_{i}^{C\prime} =\lim_{n\to\infty}  \mbf V^{C\prime}\mbf V^C= \mbf I_r.\label{eq:Pinfnrom}
\end{align}
 Now let $\mbf F_{t,\infty}:=\text{m.s.-}\lim_{n\to\infty}(\bm\Lambda_{n,\infty}^\prime \bm\Lambda_{n,\infty})^{-1}\bm\Lambda_{n,\infty}^\prime \bm C_t$. Then, since $\bm\Lambda_{n,\infty} = \bm P_{n,\infty} (\bm V_0)^{1/2}$ for all $n\in\mathbb N$, from \eqref{eq:Pinfnrom}  it follows that  $n^{-1}\bm\Lambda_{n,\infty}^\prime \bm\Lambda_{n,\infty}=\bm V_0$ for all $n\in\mathbb N$. So
 \beq\label{eq:phil2}
\mbf F_{t,\infty}=(\bm V_0)^{-1/2} \text{m.s.-}\lim_{n\to\infty}\frac{\bm P_{n,\infty}^\prime\bm C_t}{n}=(\bm V_0)^{-1/2}\bm W_{\infty}^\prime\bm C_{t,\infty}.
 \eeq
 So part (III.a.ii) is equivalent to $\text{m.s.-}\lim_{n\to\infty}\Vert \mbf F_t-\bm{\mathcal S}_0\mbf F_{t,\infty}\Vert=0$.  Note that $\mbf F_{t,\infty}$ does not depend on $n$ and it is obtained as a weighted average of infinitely many common components. Specifically, the elements of $\mbf F_{t,\infty}$ are the normalized PCs of the infinite dimensional vector $\bm C_{t,\infty}$ which is the mean-squared limit of the $n$-dimensional vector $n^{-1/2}\mbf C_t$. Indeed, by definition of $\bm W_{\infty}$ and $\bm C_{t,\infty}$, and by Lemma \ref{lem:Vzero}(i), 
\begin{align}
\bm W_{\infty}^\prime \E[\bm C_{t,\infty}\bm C_{t,\infty}^\prime]\bm W_{\infty}&=
\lim_{n\to\infty} \l\{\frac{ \bm P_{n,\infty}^\prime}{\sqrt n} \frac{\bm\Gamma^C}{n}\frac{\bm P_{n,\infty}}{\sqrt n} \r\}\nn\\
&=\lim_{n\to\infty} \l\{ \frac 1{n^2} \sum_{i,j=1}^n \bm p_{i0} [\bm\Gamma^C]_{ij} \bm p_{j0}^{\prime}  \r\}=\lim_{n\to\infty} \l\{ \frac 1n \sum_{i,j=1}^n \mbf v_{i}^C [\bm\Gamma^C]_{ij} \mbf v_{j}^{C\prime}  \r\}\nn\\
&=\lim_{n\to\infty} \l\{\mbf V^{C\prime} \frac{\bm\Gamma^C}{n} \mbf V^C\r\}=\lim_{n\to\infty} \l\{ \frac{\mbf M^C}{n}\r\}
= \bm V_0,\label{eq:Pinfevec}
\end{align}
which implies $\E[\mbf F_{t,\infty}\mbf F_{t,\infty}^\prime]=\mbf I_r$.
Moreover, because of \eqref{eq:Pinfnrom} and \eqref{eq:Pinfevec}, the columns of $\bm W_{\infty}$ are the normalized eigenvectors of the infinite dimensional matrix $\E[\bm C_{t,\infty}\bm C_{t,\infty}^\prime]=\lim_{n\to\infty} n^{-1} \bm\Gamma^C$  which is the covariance matrix of the infinite dimensional process $\bm C_{t,\infty}$. An analogous reasoning could be done for part (III.b), but in that case the factors would be identified with the non-normalized PCs of $\bm C_{t,\infty}$, having covariance matrix $\bm V_0$.

Finally, since the columns of $\mbf V^C$ and $\bm\Lambda$ span the same space, then 
$\bm\Lambda=\mbf V^{C}\mbf V^{C\prime}\bm\Lambda$, so that 
$\bm\lambda_i^\prime=\mbf v_i^{C\prime}\mbf V^{C\prime}\bm\Lambda$ and $C_{it} = \bm\lambda_i^\prime \mbf F_t=   \mbf v_i^{C\prime}\mbf V^{C\prime} \bm\Lambda \mbf F_t= \mbf v_i^{C\prime}\mbf V^{C\prime}\bm C_t$ for all $i=1,\ldots, n$ and all $n\in\mathbb N$. Therefore, since $C_{it}$ does not depend on $n$ and, by definition, $\bm P_{n,\infty}=\sqrt n\mbf V^{C}$ for all $n\in\mathbb N$, from \eqref{eq:phil1} and \eqref{eq:phil2}, we obtain
\begin{align}
C_{it} = 
\text{m.s.-}\lim_{n\to\infty} C_{it}&= \text{m.s.-}\lim_{n\to\infty}  \mbf v_i^{C\prime} {\mbf V^{C\prime}\bm C_t }= \text{m.s.-}\lim_{n\to\infty} \sqrt n \mbf v_i^{C\prime} \frac{\mbf V^{C\prime}\bm C_t }{\sqrt n}= \text{m.s.-}\lim_{n\to\infty} \sqrt n \mbf v_i^{C\prime}\frac{\bm P_{n,\infty}^{\prime}\bm C_t }{n}\nn\\
& =\bm p_{i0}^\prime  \bm W_{\infty}^\prime\bm C_{t,\infty}=\bm p_{i0}^\prime (\bm V_0)^{1/2} (\bm V_0)^{-1/2} \bm W_{\infty}^\prime\bm C_{t,\infty} = \bm\lambda_{i,\infty}^\prime\mbf F_{t,\infty}.\nn
\end{align}
We conclude that, in general, for all $i\in\mathbb N$,  there exists an invertible $r\times r$ matrix $\bm{\mathcal H}_{\infty}$, independent of $n$ and $T$, such that
\beq\label{eq:phil3}
C_{it}=\bm\lambda_i^\prime \mbf F_t=\bm\lambda_i^\prime\bm{\mathcal H}_{\infty}\bm{\mathcal H}_{\infty}^{-1} \mbf F_t=  \bm\lambda_{i,\infty}^\prime\mbf F_{t,\infty}.
\eeq 
From part (I.a), we see that, if we just impose orthonormal factors, then $\bm{\mathcal H}_{\infty}= \lim_{n\to\infty}\mbf K=\bm\Upsilon_0$ (because of Assumption \ref{ass:sign_eval}(a)), which is a rotation, while from parts (II.a) or (III.a), we see that, if we also impose orthogonal loadings, then $\bm{\mathcal H}_{\infty}=\bm{\mathcal S}_0$. 

\begin{sidewaystable}[htbp]
\caption{Identifying constraints and their implications}\label{tab:ident}
\centering
\scriptsize{
\begin{tabular}{l | ll | l | l | l | l}
\hline
\hline
&&&&&&\\[-8pt]
& \multicolumn{2}{c|}{Constraints}&$n,T$&$\bm\lambda_i^\prime$ &$\mbf F_t$ &  $\wh{\mbf H}$   \\[1pt]
\hline
&&&&&&\\[-8pt]
I.a& $\bm\Gamma^F=\mbf I_r$ 
& $\bm\Lambda$  unrestricted 
& $n$ fixed
& $\mbf v_i^{C\prime}(\mbf M^{C})^{1/2}\mbf K^\prime$
& $\mbf K(\mbf M^{C})^{-1/2} \mbf V^{{C}\prime}\bm{C}_t$
& $ \wh{\mbf H}^\prime\wh{\mbf H}=\mbf I_r+ O_{\mathrm P}\l(\max\l(\frac 1n,\frac 1{\sqrt T}\r)\r)$\\[3pt]
&&& $n\to\infty$
& $ \bm p_{i0}^\prime (\bm V_0)^{1/2}\bm\Upsilon_0^\prime$
& ${\text{\tiny m.s.}\,} \bm\Upsilon_0(\bm V_0)^{-1/2}\bm W_{\infty}^\prime \bm C_{t,\infty}$
&\\[3pt]
\hline
&&&&&&\\[-8pt]
II.a& $\bm\Gamma^F=\mbf I_r$ 
& $\frac{\bm\Lambda^\prime\bm \Lambda}n$ diagonal 
& $n$ fixed
& $\mbf v_i^{C\prime}(\mbf M^{C})^{1/2}\bm S$
& $\bm S(\mbf M^{C})^{-1/2} \mbf V^{{C}\prime}\bm{C}_t$
& $\wh{\mbf H}=\wh{\bm J}+ O_{\mathrm P}\l(\max\l(\frac 1n,\frac 1{\sqrt T}\r)\r)$\\[3pt]
&&& $n\to\infty$
& $ \bm p_{i0}^\prime (\bm V_0)^{1/2}\bm{\mathcal S}_0$
& ${\text{\tiny m.s.}\,} \bm{\mathcal S}_0(\bm V_0)^{-1/2}\bm W_{\infty}^\prime \bm C_{t,\infty}$
&\\[3pt]
\hline
&&&&&&\\[-8pt]
III.a&$\bm\Gamma^F=\mbf I_r$ 
& $\bm\Sigma_\Lambda$  diagonal 
& $n\to\infty$
& $ \bm p_{i0}^\prime (\bm V_0)^{1/2}\bm{\mathcal S}_0$
& ${\text{\tiny m.s.}\,} \bm{\mathcal S}_0(\bm V_0)^{-1/2}\bm W_{\infty}^\prime \bm C_{t,\infty}$
& $\wh{\mbf H}=\wh{\bm J}+ O_{\mathrm P}\l(\max\l(\frac 1{\sqrt n},\frac 1{\sqrt T}\r)\r)$\\[3pt]
\hline
&&&&&&\\[-8pt]
IV.a& $\frac{ \bm F^\prime\bm F}T=\mbf I_r$
& $\bm\Lambda$  unrestricted 
& $n,T$ fixed
& $\wh{\mbf v}_i^{C\prime}(\wh{\mbf M}^C)^{1/2} \wh{\mbf Q}^\prime$
&  $\wh{\mbf Q}(\wh{\mbf M}^{C})^{-1/2} \wh{\mbf V}^{{C}\prime}\bm{C}_t$
& $\wh{\mbf H}^\prime\wh{\mbf H}=\mbf I_r+ O_{\mathrm P}\l(\max\Big(\frac 1n,\frac 1{ T}\Big)\r)$\\[3pt]
&&& $T\to\infty$
& $\mbf v_i^{C\prime}(\mbf M^{C})^{1/2}\mbf K^\prime$
& $\text{\tiny m.s.}\, \mbf K(\mbf M^{C})^{-1/2} \mbf V^{{C}\prime}\bm{C}_t$
&\\[3pt]
&&& $n,T\to\infty$
& $\bm p_{i0}^\prime (\bm V_0)^{1/2}\bm\Upsilon_0^\prime$
& ${\text{\tiny m.s.}\,} \bm\Upsilon_0(\bm V_0)^{-1/2}\bm W_{\infty}^\prime \bm C_{t,\infty}$
&\\[3pt]
\hline
&&&&&&\\[-8pt]
V.a& $\frac{ \bm F^\prime\bm F}T=\mbf I_r$
& $\frac{\bm\Lambda^\prime\bm \Lambda}n$ diagonal 
& $n,T$ fixed
& $\wh{\mbf v}_i^{C\prime}(\wh{\mbf M}^C)^{1/2} \wh{\bm S}$
&  $\wh{\bm S}(\wh{\mbf M}^{C})^{-1/2} \wh{\mbf V}^{{C}\prime}\bm{C}_t$
& $\wh{\mbf H}=\wh{\bm J}+ O_{\mathrm P}\l(\max\Big(\frac 1n,\frac 1{ T}\Big)\r)$\\[3pt]
&&&$T\to\infty$
& $\mbf v_i^{C\prime}(\mbf M^{C})^{1/2}\bm S$
& $\text{\tiny m.s.}\, \bm S(\mbf M^{C})^{-1/2} \mbf V^{{C}\prime}\bm{C}_t$
&\\[3pt]
&&& $n,T\to\infty$
& $ \bm p_{i0}^\prime (\bm V_0)^{1/2}\bm{\mathcal S}_0$
& ${\text{\tiny m.s.}\,} \bm{\mathcal S}_0(\bm V_0)^{-1/2}\bm W_{\infty}^\prime \bm C_{t,\infty}$
&\\[3pt]
\hline
&&&&&&\\[-8pt]
VI.a& $\frac{ \bm F^\prime\bm F}T=\mbf I_r$
& $\bm\Sigma_\Lambda$   diagonal 
& $n,T\to\infty$
& $\bm p_{i0}^\prime (\bm V_0)^{1/2}\bm{\mathcal S}_0$
& ${\text{\tiny m.s.}\,}\bm{\mathcal S}_0(\bm V_0)^{-1/2}\bm W_{\infty}^\prime \bm C_{t,\infty}$
& $\wh{\mbf H}=\wh{\bm J}+ O_{\mathrm P}\l(\max\l(\frac 1{\sqrt n},\frac 1{ T}\r)\r)$\\[3pt]
\hline
\hline
&&&&&&\\[-8pt]
I.b& $\bm\Sigma_\Lambda=\mbf I_r$ 
& $\bm F$ unrestricted 
& $n\to\infty$
& $\bm p_{i0}^{\prime}   \bm\Upsilon_1^\prime$
& ${\text{\tiny m.s.}\,} \bm \Upsilon_1 \bm W_{\infty}^\prime\bm C_{t,\infty}$
& $\wh{\mbf H}^\prime\wh{\mbf H}=\frac{\wh{\mbf M}^x}{n}+ O_{\mathrm P}\l(\max\l(\frac 1{\sqrt n},\frac 1{T}\r)\r)$\\[3pt]
\hline
&&&&&\\[-8pt]
II.b& $\bm\Sigma_\Lambda=\mbf I_r$ 
& $\frac{ \bm F^\prime\bm F}T$ diagonal
& $n,T\to\infty$
& $\text{\tiny m.s.}\, \bm p_{i0}^\prime \bm{\mathcal S}_0$
& $\text{\tiny m.s.}\, \bm {\mathcal S}_0 \bm W_{\infty}^\prime\bm C_{t,\infty}$
&$\wh{\mbf H}=\wh{\bm J}\l(\frac{\wh{\mbf M}^x}n\r)^{1/2}+ O_{\mathrm P}\l(\max\l(\frac 1{\sqrt n},\frac 1{T}\r)\r)$\\[3pt]
\hline
&&&&&\\[-8pt]
III.b& $\bm\Sigma_\Lambda=\mbf I_r$ 
& $\bm\Gamma^F$ diagonal
& $n\to\infty$
& $\bm p_{i0}^\prime \bm{\mathcal S}_0$
& $\text{\tiny m.s.}\, \bm {\mathcal S}_0 \bm W_{\infty}^\prime\bm C_{t,\infty}$
&$\wh{\mbf H}=\wh{\bm J}\l(\frac{\wh{\mbf M}^x}n\r)^{1/2}+ O_{\mathrm P}\l(\max\l(\frac 1{\sqrt n},\frac 1{\sqrt T}\r)\r)$\\[3pt]
\hline
&&&&&\\[-8pt]
IV.b& $\frac{\bm\Lambda^\prime\bm \Lambda}n=\mbf I_r$
& $\bm F$ unrestricted
& $n$ fixed
& $n \mbf v_i^{C\prime} (\mbf M^{C})^{-1/2}\mbf K^\prime(\bm\Gamma^F)^{1/2}$
& $ (\bm\Gamma^F)^{1/2}\mbf K(\mbf M^{C})^{-1/2}\mbf V^{C\prime}\bm C_t$
& $\wh{\mbf H}^\prime\wh{\mbf H}=\frac{\wh{\mbf M}^x}{n}+ O_{\mathrm P}\l(\max\Big(\frac 1n,\frac 1{ T}\Big)\r)$ \\[3pt]
&&& $n\to\infty$
& $\bm p_{i0}^{\prime}   \bm\Upsilon_1^\prime$
& $\text{\tiny m.s.}\,\bm \Upsilon_1 \bm W_{\infty}^\prime\bm C_{t,\infty}$
& \\[3pt]
\hline
&&&&&\\[-8pt]
V.b& $\frac{\bm\Lambda^\prime\bm \Lambda}n=\mbf I_r$
& $\frac{ \bm F^\prime\bm F}T$ diagonal
& $n,T$ fixed
& $\sqrt n \wh{\mbf v}_i^{C\prime} \wh{\bm S}$
& $\frac 1{\sqrt n}\wh{\bm S} \wh{\mbf V}^{C\prime}\bm C_t$
&$\wh{\mbf H}=\wh{\bm J}\l(\frac{\wh{\mbf M}^x}n\r)^{1/2}+ O_{\mathrm P}\l(\max\Big(\frac 1n,\frac 1{ T}\Big)\r)$\\[3pt]
&&& $T\to\infty$
& $\text{\tiny m.s.}\, \sqrt n {\mbf v}_i^{C\prime} {\bm S}$
& $\text{\tiny m.s.}\, \frac 1{\sqrt n}{\bm S} {\mbf V}^{C\prime}\bm C_t$
& \\[3pt]
&&& $n,T\to\infty$
& $\text{\tiny m.s.}\, \bm p_{i0}^\prime \bm{\mathcal S}_0$
& $\text{\tiny m.s.}\, \bm {\mathcal S}_0 \bm W_{\infty}^\prime\bm C_{t,\infty}$
& \\[3pt]
\hline
&&&&&\\[-8pt]
VI.b& $\frac{\bm\Lambda^\prime\bm \Lambda}n=\mbf I_r$
& $\bm\Gamma^F$ diagonal 
& $n$ fixed
& $\sqrt n  \mbf v_i^{C\prime}\bm S$
& $\frac 1{\sqrt n}\bm S \mbf V^{C\prime}\bm C_t$
&$\wh{\mbf H}=\wh{\bm J}\l(\frac{\wh{\mbf M}^x}n\r)^{1/2}+ O_{\mathrm P}\l(\max\l(\frac 1{ n},\frac 1{\sqrt T}\r)\r)$\\[3pt]
&&& $n\to\infty$
& $ \bm p_{i0}^\prime \bm{\mathcal S}_0$
& $\text{\tiny m.s.}\, \bm {\mathcal S}_0 \bm W_{\infty}^\prime\bm C_{t,\infty}$
& \\[3pt]
\hline
\hline
\end{tabular}
}

\begin{tabular}{p{.88\textwidth}}
$\mbf V^C$ normalized eigenvectors of $\bm\Lambda \bm\Gamma^F \bm\Lambda ^\prime$ with  rows $\mbf v_i^{C\prime}$, $i\in\mathbb N$, and eigenvalues $\mbf M^C$; 
$\mbf K$ normalized eigenvectors of $(\bm\Gamma^F)^{1/2}(n^{-1}\bm \Lambda^\prime\bm \Lambda) (\bm\Gamma^F)^{1/2}$ with eigenvalues $n^{-1}\mbf M^C$;\\
$\bm \Upsilon_0$ normalized eigenvectors of $(\bm\Gamma^F)^{1/2}\bm \Sigma_\Lambda (\bm\Gamma^F)^{1/2}$ with eigenvalues $\bm V_0$;
$\bm\Upsilon_1$  normalized eigenvectors of $(\bm \Sigma_\Lambda)^{1/2}\bm\Gamma^F(\bm \Sigma_\Lambda)^{1/2}$ with eigenvalues $\bm V_0$;\\
$\wh{\mbf V}^C$ normalized eigenvectors of $\bm\Lambda(T^{-1}\bm F^\prime\bm F) \bm\Lambda ^\prime$ with eigenvalues $\wh{\mbf M}^C$; 
$\wh{\mbf Q}$ normalized eigenvectors of  $(T^{-1}\bm F^\prime\bm F)^{1/2}(n^{-1}\bm \Lambda^\prime\bm \Lambda) (T^{-1}\bm F^\prime\bm F)^{1/2}$ with eigenvalues $n^{-1}\wh{\mbf M}^C$;\\ 
$\bm S$ is an $r\times r$ diagonal matrix with entries $\pm 1$ depending only on $n$; 
$\bm{\mathcal S}_0$  is an $r\times r$ diagonal matrix with entries $\pm 1$ independent of $n$ and $T$; 
$\wh{\bm S}$ and $\wh{\bm J}$ are $r\times r$ diagonal matrices with entries $\pm 1$ depending on $n$ and $T$;\\
$\bm p_{i0}:=\lim_{n\to\infty}\sqrt n \mbf v_i^C$, $i\in\mathbb N$; $\bm W_{\infty}:=\lim_{n\to\infty}n^{-1/2}\bm P_{n,\infty}$, with $\bm P_{n,\infty}:=(\bm p_{10}\cdots \bm p_{n0})^\prime$; $\bm{C}_{t,\infty}:=\text{\upshape m.s.-}\lim_{n\to\infty} n^{-1/2}\bm C_t$.
\end{tabular}
\end{sidewaystable}

As already remarked, Theorems \ref{th:CLTL} and \ref{th:CLTF} are not directly usable for inference, because, under Assumptions \ref{ass:common}-\ref{ass:rates} used to prove those results, the true loadings and factors are not identified, hence the presence of $\wh{\mbf H}$. The identifying restrictions studied in Proposition \ref{corol:K00} provide a way to address this problem. By restricting the true loadings and/or factors, they imply corresponding restrictions on the matrices linking the estimated and true quantities. The next result makes this implication explicit for $\wh{\mbf H}$, defined in \eqref{eq:acca} (see Appendix \ref{corol:H00proof} for a proof and Appendix \ref{app:calH} for the properties of $\bm{\mathcal H}$).

\begin{prop}\label{corol:H00}
Under Assumptions \ref{ass:common}, \ref{ass:idio}, \ref{ass:ind}, and \ref{ass:sign_eval},
with Assumption \ref{ass:common}(a) holding with rate $\sqrt n$
and Assumption \ref{ass:common}(c-ii) holding with rate $\sqrt T$, as $n,T\to\infty$,
\begin{compactenum}
\item [(I.a)] 
if $\bm\Gamma^F=\mbf I_r$ and
$\bm\Lambda$  is unrestricted, 
$\min(n,\sqrt T) \Vert \wh{\mbf H}^\prime\wh{\mbf H}-\mbf I_r\Vert=O_{\mathrm P}(1)$;

\item [(II.a)] 
if $\bm\Gamma^F=\mbf I_r$ and
$n^{-1}\bm\Lambda^\prime\bm \Lambda$  is diagonal for all $n\in\mathbb N$, 
$\min(n,\sqrt T)\Vert\wh{\mbf H}-\wh{\bm J}\Vert=O_{\mathrm P}(1)$;

\item [(III.a)] if 
$\bm\Gamma^F=\mbf I_r$ and
$\bm\Sigma_\Lambda$  is diagonal, 
$\min(\sqrt n,\sqrt T)\Vert\wh{\mbf H}-\wh{\bm J}\Vert=O_{\mathrm P}(1)$;

\item [(IV.a)]
if $T^{-1}\bm F^\prime\bm F=\mbf I_r$  for all $T\in\mathbb N$ and
$\bm\Lambda$ is unrestricted, 
$\min(n, T)\Vert \wh{\mbf H}^\prime\wh{\mbf H} -\mbf I_r \Vert = O_{\mathrm P}(1)$;

\item [(V.a)] 
if $T^{-1} \bm F^\prime \bm F=\mbf I_r$ and  
$n^{-1}\bm\Lambda^\prime\bm \Lambda$  is diagonal for all $n,T\in\mathbb N$, 
$\min(n, T)\Vert\wh{\mbf H}-\wh{\bm J}\Vert=O_{\mathrm P}(1)$; 

\item [(VI.a)] 
if $T^{-1} \bm F^\prime \bm F=\mbf I_r$  for all $T\in\mathbb N$  and  
$\bm\Sigma_\Lambda$  is diagonal,
$\min(\sqrt n, T)\Vert\wh{\mbf H}-\wh{\bm J}\Vert=O_{\mathrm P}(1)$; 

\item [(I.b)] 
if $\bm\Sigma_\Lambda=\mbf I_r$ 
and $\bm F$ is unrestricted, 
$\min(\sqrt n, T)\Vert \wh{\mbf H}{^\prime}\wh{\mbf H} - n^{-1}\wh{\mbf M}^x \Vert = O_{\mathrm P}(1)$;

\item [(II.b)] 
if  $\bm\Sigma_\Lambda=\mbf I_r$ and
$T^{-1}\bm F^\prime \bm F$ is diagonal for all $T\in\mathbb N$, 
$\min(\sqrt n, T)\Vert\wh{\mbf H}-\wh{\bm J}(n^{-1}\wh{\mbf M}^x)^{1/2}\Vert=O_{\mathrm P}(1)$;

\item [(III.b)] if 
$\bm\Sigma_\Lambda=\mbf I_r$ and
$\bm\Gamma^F$ is diagonal, 
$\min(\sqrt n, \sqrt T)\Vert\wh{\mbf H}-\wh{\bm J}(n^{-1}\wh{\mbf M}^x)^{1/2}\Vert=O_{\mathrm P}(1)$;

\item [(IV.b)] if
 $n^{-1}\bm\Lambda^\prime\bm \Lambda=\mbf I_r$  for all $n\in\mathbb N$ and
 $\bm F$ is unrestricted, 
$\min(n,T)\Vert  \wh{\mbf H}{^\prime}\wh{\mbf H} - n^{-1}\wh{\mbf M}^x \Vert = O_{\mathrm P}(1)$;

\item [(V.b)] if 
$n^{-1}\bm\Lambda^\prime\bm \Lambda=\mbf I_r$  and
$T^{-1} \bm F^\prime \bm F$ is diagonal for all $n,T\!\in\mathbb N$, 
$\min(n,T)\Vert\wh{\mbf H}-\wh{\bm J}(n^{-1}\wh{\mbf M}^x)^{1/2}\Vert=O_{\mathrm P}(1)$;

\item [(VI.b)] if 
$n^{-1}\bm\Lambda^\prime\bm \Lambda=\mbf I_r$ for all $n\in\mathbb N$ and
$\bm\Gamma^F$ is diagonal, 
$\min(n,\sqrt T)\Vert\wh{\mbf H}-\wh{\bm J}(n^{-1}\wh{\mbf M}^x)^{1/2}\Vert=O_{\mathrm P}(1)$;
\end{compactenum}
where $\wh{\bm J}$ is an $r\times r$ diagonal matrix with entries $\pm 1$, depending on $n$ and $T$. 
 
\end{prop}

As an immediate consequence of  Proposition \ref{corol:H00} notice that, 
by means of 
\eqref{eq:Htilde} in Appendix \ref{sec:cmppca},  and \eqref{eq:sameHsw}, \eqref{eq:sameHD}, and \eqref{eq:sameHD3} in Appendix \ref{sec:cmppcaCD}, we can also derive the properties of $\wt{\mbf H}$, $\wideparen {\mbf H}$, and $\bar{\mbf H}$, which are the matrices determining identification under approaches B1, A2, and B2, respectively. In particular, all statements for $\wh{\mbf H}$ hold also for $\wt{\mbf H}$ 
while the results for $\wh {\mbf H}$ under the constraints (a) in Proposition \ref{corol:H00} hold for $\wideparen {\mbf H}$ and $\bar{\mbf H}$ but under the constraints (b).

Although in some cases we impose non-asymptotic identifying conditions, we can derive only asymptotic results. Indeed, $\wh{\mbf H}$ depends on estimated quantities so its properties hold only asymptotically. The different convergence rates depend on the columns of $\bm\Lambda$ and $\bm F$ being exactly or just asymptotically orthogonal/orthonormal.


Under parts (II.a), (III.a), (V.a) and (VI.a), we see that $\wh{\mbf H}$ can be reduced to an asymptotically diagonal matrix of signs $\wh{\bm J}$. The sign indeterminacy, given by $\wh{\bm J}$, can always be fixed, for example, as follows. Let $\iota_j$ be the first index of the $j$th column of $\bm\Lambda$ such that $\lambda_{\iota_j j}\ne 0$, $j=1,\ldots, r$. Then, we assume $\lambda_{\iota_j j}> 0$, for $j=1,\ldots,r$. This condition implies that $\wh{\bm J}=\mbf I_r$ in Proposition \ref{corol:H00}.

\section{Implications for inference}\label{sec:ottavo}

The results in Proposition \ref{corol:H00} are directly applicable to  the  consistency  results proved in Propositions \ref{prop:L2} and \ref{prop:F2}. Specifically, under all (a) cases the PC estimators are consistent for the true loadings and factors as defined in Proposition \ref{corol:K00}. Indeed, after 
 setting $\wh{\bm J}=\mbf I_r$ as discussed in the previous section, by
 Propositions \ref{prop:L2} and \ref{prop:F2},
\begin{align}
\l\Vert \wh{\bm\lambda}_i-{\bm\lambda}_i \r\Vert&\le \l\Vert\wh{\bm\lambda}_i-\wh{\mbf H}^\prime{\bm\lambda}_i\r \Vert +\l \Vert\wh{\mbf H}^\prime-\mbf I_r\r\Vert\, \l\Vert{\bm\lambda}_i\r\Vert = O_{\mathrm P}\l(\max\l(\frac 1n,\frac 1{\sqrt T}\r)\r) + O_{\mathrm P}\l(\eta_{nT}\r),\label{eq:finita1}\\
\l\Vert\wh{\mbf F}_t- {\mbf F}_t\r\Vert &= \l\Vert \wh{\mbf F}_t-\wh{\mbf H}^{-1}{\mbf F}_t\r\Vert  +\l \Vert \wh{\mbf H}^{-1}-\mbf I_r\r \Vert\, \l\Vert\mbf F_t\r \Vert
= O_{\mathrm P}\l(\max\l(\frac 1{\sqrt n},\frac 1{ T}\r)\r) + O_{\mathrm P}\l(\eta_{nT}\r),\label{eq:finita2}
\end{align}
where $\eta_{nT}$ is given in Proposition \ref{corol:H00} and depends on the imposed constraints. 
 
 The magnitude of the remainder  $\eta_{nT}$ has important implications for inference. Under part (II.a) $\eta_{nT}=\max(n^{-1},T^{-1/2})$ this means that $ \sqrt T(\wh{\mbf H}^\prime-\mbf I_r) = O_{\mathrm P}(1)$ and the CLT for the loadings in Theorem \ref{th:CLTL} cannot hold, but, since $\sqrt n(\wh{\mbf H}^{-1}-\mbf I_r)= O_{\mathrm P}(\sqrt {n/T})$,  Theorem \ref{th:CLTF} for the factors can still hold provided $n/T\to 0$, as $n,T\to\infty$, which is a stronger constraint. Under part (VI.a) $\eta_{nT}=\max(n^{-1/2},T^{-1})$ which implies that Theorem \ref{th:CLTF}  for the factors cannot hold, but,
  since $\sqrt T(\wh{\mbf H}^{-1}-\mbf I_r)= O_{\mathrm P}(\sqrt {T/n})$, 
  Theorem \ref{th:CLTL}  for the loadings can still hold provided $T/n\to 0$, as $n,T\to\infty$. Under part (III.a) $\eta_{nT}=\max(n^{-1/2},T^{-1/2})$ and neither Theorem \ref{th:CLTL} nor Theorem \ref{th:CLTF} can hold and we conjecture that, if any asymptotic normality can be proved, the asymptotic covariance will be larger than the one in those theorems.
 
Still, under (II.a), (III.a) or (VI.a), and after setting the sign as discussed above, ${\text{P-lim}}_{n,T\to\infty} \wh{\mbf H}=\mbf I_r$.
So we can test hypothesis of the form $\text H_0: \mbf R \bm\lambda_i=\mbf 0_q$ with $\mbf R$ being $q\times r$ and full-rank for some $q\le r$, since, by Theorem \ref{th:CLTL}, under $\text H_0$, 
$
\sqrt T\, \mbf R\wh{\bm\lambda}_i\to_d\mathcal N(\mbf 0_q, \mbf R\bm\Theta_i^{\text{\tiny OLS}}\mbf R^\prime ),\nn
$
and we can immediately build a Wald statistic for $\text H_0$. Similarly, from Theorem \ref{th:CLTF} we can build pointwise $(1-\alpha)$-confidence intervals for the factors given by $\{\wh{F}_{jt}\pm z_{1-\alpha/2}{n^{-1/2} [{\bm \Pi}_t^{\text{\tiny OLS}}]_{jj}^{1/2}} \}$, $j=1,\ldots, r$.

However, to make inference for general hypotheses, we have to resort to the constraints in part (V.a), which are also  considered by \citet[Theorem 1]{baing13}. In this case, following again Theorems \ref{th:CLTL} and \ref{th:CLTF} and combining them with \eqref{eq:finita1} and \eqref{eq:finita2}, we obtain
\begin{align}
\sqrt T\l(\wh{\bm\lambda}_i-{\bm\lambda}_i\r )&=\sqrt T\l(\wh{\bm\lambda}_i-\wh{\mbf H}^\prime{\bm\lambda}_i \r) + o_{\mathrm P}(1)\to_d\mathcal N\l(\mbf 0_r,  
{\bm \Theta}_i^{\text{\tiny OLS}}
\r),\label{eq:CLTLesatto}\\
\sqrt n\l(\wh{\mbf F}_t- {\mbf F}_t\r) &=  \sqrt n\l(\wh{\mbf F}_t-\wh{\mbf H}^{-1}{\mbf F}_t\r)  + o_{\mathrm P}(1)\to_d\mathcal N\l(\mbf 0_r, {\bm \Pi}_t^{\text{\tiny OLS}}\r),\label{eq:CLTFesatto}
\end{align}
where we also used  the fact that $\Vert\bm\lambda_i\Vert = O(1)$, because of Assumption \ref{ass:common}(a), and $\Vert \mbf F_t\Vert = O_{\mathrm P}(1)$, because of Lemma \ref{lem:FTLN}(ii).

It is then clear that under condition (V.a), neither the  location nor the covariance of the asymptotic distribution depend on $\wh{\mbf H}$, or its limit, anymore. 
Still, the price to be paid is not irrelevant. Indeed, condition (V.a) is stated for a sample quantity, and, thus, in order for it  to hold, we should either consider the factors as a deterministic sequence or we should interpret the CLTs in \eqref{eq:CLTLesatto} and \eqref{eq:CLTFesatto} as conditional on a realization of the factors $\mbf F_1,\ldots, \mbf F_T$.  Under this interpretation, we can then impose condition (V.a) and use \eqref{eq:CLTLesatto} and \eqref{eq:CLTFesatto} to make inference not only on the loadings, but also on the factors since now $\mbf F_t$ is no longer random. 

We conclude by noticing that any inferential procedure will also require estimating ${\bm \Theta}_i^{\text{\tiny OLS}}$ and ${\bm \Pi}_t^{\text{\tiny OLS}}$, defined in Assumptions \ref{ass:CLT}(a) and \ref{ass:CLT}(b), respectively.
 Hereafter, let $\wh{e}_{it}:=x_{it}-\wh{\bm\lambda}_i^\prime \wh{\mbf F}_t$. Then, a consistent estimator for ${\bm \Theta}_i^{\text{\tiny OLS}}$ is the classical HAC estimator \citep{andrews91}:
\beq\label{eq:thetahatols}
\wh{\bm \Theta}_i^{\text{\tiny OLS}} := \l(\frac{\wh{\bm F}^\prime\wh{\bm F}}{T}\r)^{-1}
\l(\frac 1T \sum_{t=1}^T\sum_{s=1}^T\mathrm K\l(\frac{ t-s}{M_T}\r)
\wh{\mbf F}_t\wh{\mbf F}_s^\prime \wh e_{it}\wh e_{is}
\r)
\l(\frac{\wh{\bm F}^\prime\wh{\bm F}}{T}\r)^{-1},
\eeq
where $\mathrm K(\cdot) $ is any real kernel function on $\mathbb R$ with support $[-1,1]$, so that $\mathrm K(u)=0$ for $|u|>1$, with bandwidth $M_T$.

Concerning ${\bm \Pi}_t^{\text{\tiny OLS}}$, a possible estimator is the
time-series average of spatial HAC estimators \citep{kim2022robust}:
\beq\label{eq:pihatols}
\wh{\bm \Pi}^{\text{\tiny OLS}}:=\l(\frac{\wh{\bm\Lambda}^\prime\wh{\bm\Lambda}}{n}\r)^{-1}
\l(\frac 1n\sum_{i=1}^n\sum_{j=1}^n\mathrm K\l(\frac{d_{ij}}{d_n}\r) \wh{\bm\lambda}_i\wh{\bm\lambda}_j^\prime \frac 1T\sum_{t=1}^T \wh{e}_{it}\wh{e}_{jt}
\r)
\l(\frac{\wh{\bm\Lambda}^\prime\wh{\bm\Lambda}}{n}\r)^{-1},
\eeq
where $\mathrm K(\cdot) $ is any real kernel function on $\mathbb R$ with support $[-1,1]$, so that $\mathrm K(u)=0$ for $|u|>1$, with bandwidth $d_n$ and $d_{ij}$ is the correlation based distance between time series $i$ and $j$. 
 Note that 
by Assumption \ref{ass:idio}(b), $\E[e_{it}e_{jt}]$ does not depend on time, so ${\bm \Pi}^{\text{\tiny OLS}}_t$ is in fact  also independent of time and $\wh{\bm \Pi}^{\text{\tiny OLS}}$ can be considered as its estimator. If instead we allowed for $\E[e_{it}e_{jt}]$ to be time dependent, then $\wh{\bm \Pi}^{\text{\tiny OLS}}$ should be seen as an estimator of $T^{-1}\sum_{t=1}^T {\bm \Pi}^{\text{\tiny OLS}}_t$. 

Alternative estimators to \eqref{eq:pihatols}  are obtained by setting $\mathrm K(\cdot)=1$ therein and either summing only over a random subset of units \citep[see the cross-sectional HAC by][]{baing06} or summing  only over those units such that $T^{-1}\sum_{t=1}^T \wh{e}_{it}\wh{e}_{jt}$ is above certain threshold \citep[see the adaptive thresholding idiosyncratic cross-correlation by][]{fresoli2025idiosyncratic}.

Finally, for a formal proof of consistency of $\wh{\bm \Theta}_i^{\text{\tiny OLS}}$ and $\wh{\bm \Pi}^{\text{\tiny OLS}}$ we refer to the above references, while here we limit ourselves to notice that for such proofs to hold we should strengthen Assumptions \ref{ass:idio}(c-i) and \ref{ass:idio}(c-ii) by requiring the idiosyncratic components to have finite 8th order moments and summable 8th order cumulants, as in \citet[Assumptions C.1 and C.5]{bai03}.



\singlespacing
{{
\setlength{\bibsep}{.2cm}
\bibliographystyle{chicago}
\bibliography{BL_Biblio}
}}

\setcounter{section}{0}%
\setcounter{subsection}{0}
\setcounter{equation}{0}
\renewcommand{\theHsection}{supp.\Alph{section}}
\renewcommand{\theHsubsection}{supp.\Alph{section}.\arabic{subsection}}
\renewcommand{\theHequation}{supp.\Alph{section}.\arabic{equation}}
\renewcommand{\theHfigure}{supp.\Alph{section}.\arabic{figure}}
\renewcommand{\theHtable}{supp.\Alph{section}.\arabic{table}}
\gdef\thesection{ \Alph{section}}
\gdef\thesubsection{\Alph{section}.\arabic{subsection}}
\gdef\thefigure{\Alph{section}\arabic{figure}}
\gdef\theequation{\Alph{section}\arabic{equation}}
\gdef\varphible{\Alph{section}\arabic{table}}
\gdef\thefootnote{\Alph{section}\arabic{footnote}}

\newpage
\begin{center}
\textsc{\large{Principal Component Analysis \\ 
for High-Dimensional Approximate Factor Models in Time Series:\\ 
Assumptions, Asymptotic Theory, and Identification\\ \vskip .3cm
Supplementary material}}
\end{center}

\small

\section{Other equivalent approaches to Principal Component Analysis}\label{sec:otherPCA}

\paragraph{Approach A2.} 
If we impose $n^{-1} {{\bm L}^\prime {\bm L}}=\mbf I_r$, then by construction each column of $n^{-1/2}{\bm L}$ is normalized and, by definition of eigenvectors and eigenvalues, the value of the objective function in \eqref{eq:maxforni} must be the sum of the $r$ largest eigenvalues of $\wh{\bm\Gamma}^x=T^{-1}{\bm X^\prime\bm X}$ divided by $n$, i.e., it must give $n^{-1}\text{tr}({\wh{\mbf M}^x})$. So, our estimator $\wideparen{\bm\Lambda}$ must be such that $n^{-1/2}\wideparen{\bm\Lambda}
$ is the matrix of the corresponding normalized eigenvectors. Therefore, the PC estimators of the loadings and the factors (obtained by linear projection) are
\begin{align}
&\wideparen{\bm \Lambda}:=\wh{\mbf V}^x\sqrt n,\label{eq:estLsw}\\
&\wideparen{\bm F}:=\bm X \wideparen{\bm\Lambda}(\wideparen{\bm\Lambda}^\prime \wideparen{\bm\Lambda})^{-1}=
\frac{\bm X \wh{\mbf V}^x}{\sqrt n}.\label{eq:estFsw}
\end{align}
And, as expected, $T^{-1}{\wideparen{\bm F}^\prime\wideparen{\bm F}}=n^{-1}{\wh{\mbf M}^x}$, which  is diagonal. These are the estimators studied by \citet{stockwatson02JASA}, who prove only consistency although not with the sharpest possible rate.

\paragraph{Approach B2.} 
If we impose $T^{-1}{{\bm G}^\prime{\bm G}}$ to be diagonal, then each column of ${\bm G}
({\bm G}^\prime{\bm G})^{-1/2}$ is normalized and, by definition of eigenvectors and eigenvalues, the value of the objective function in \eqref{eq:maxbai} must be the sum of the $r$ largest eigenvalues of $\wt{\bm\Gamma}^x=n^{-1}{\bm X\bm X^\prime}$ divided by $T$,  i.e., it must give $T^{-1}\text{tr}(\wt{\mbf M}^x)$. 
So, our estimator $\bar{\bm F}$ must be such that $\bar{\bm F}
(\bar{\bm F}^\prime\bar{\bm F})^{-1/2}$ is the matrix of the corresponding normalized eigenvectors.
Therefore,   the PC estimators of the factors and the loadings  (obtained by linear projection) are
\begin{align}
&\bar{\bm F}:=\wt{\mbf V}^x(\wt{\mbf M}^x)^{1/2}, \label{eq:estFig}\\
&\bar{\bm \Lambda}
:=\bm X^\prime \bar{\bm F}(\bar{\bm F}^\prime \bar{\bm F})^{-1}
=\bm X^\prime \wt{\mbf V}^x(\wt{\mbf M}^x)^{-1/2}
.\label{eq:estLig}
\end{align}
And, as expected,   $n^{-1}{\bar{\bm \Lambda}^\prime\bar{\bm \Lambda}}=\mbf I_r$. This formulation is never found in the literature.

\paragraph{Equivalence.} Now because of \eqref{eq:evaleq} the objective function in \eqref{eq:maxforni} and \eqref{eq:maxbai} has always the same value, and given that approaches A2 and B2 are based on the same identification conditions, the solutions \eqref{eq:estLsw}-\eqref{eq:estFsw} and \eqref{eq:estFig}-\eqref{eq:estLig} must coincide, i.e.,  it must be that
$\wideparen{\bm F}=\bar{\bm F}$ and  $\wideparen{\bm \Lambda}=\bar{\bm \Lambda}$.
\setcounter{equation}{0}
\section{Asymptotic properties of approach B1} \label{sec:cmppca}

The PC estimators derived under approach A1 coincide with those studied by \citet{bai03}, indeed, as shown in Section \ref{sec:ABCD}, $\wt{\bm\Lambda}=\wh{\bm\Lambda}$ and $\wt{\bm F}=\wh{\bm F}$. The purpose of this section is then to show that, from the results in \citet{bai03} we can derive the same asymptotic expansions as in Theorems \ref{th:CLTL} and \ref{th:CLTF}, showing the relation between the PC and the unfeasible OLS estimators. These are more interpretable than the original results. 
Notice that, by virtue of Table \ref{tab:ass}, the results in  \citet{bai03} can  be proved under the same assumptions made in this paper so there is no need to prove them again here. Although the estimators are identical, in this section we keep using the notation $\wt{\bm \Lambda}$ and $\wt{\bm F}$ to highlight that their properties are derived using approach B1 rather than A1.

\paragraph{Consistency.} 
First of all notice that Proposition \ref{prop:L} still holds, and its proof using $\wt{\bm\Lambda}$ and $\wt{\bm F}$ is a special case of the proof by \citet[Theorem 4]{FLM13}, when no uniform bounds are computed. It follows the same steps as in \citet{bai03}, but it derives slower rates since fewer cross-moment assumptions are imposed.

Now, by definition of eigenvectors, it holds that
\beq\label{eq:EVECBAI}
\frac{\bm X\bm X^\prime}{nT}\wt{\bm F}=\wt{\bm F}\frac{\wt{\mbf M}^x}{T}.
\eeq
By using $\bm X=\bm F\bm\Lambda^\prime+\bm E$ and taking the $t$th row of \eqref{eq:EVECBAI},  for any $t=1,\ldots,T$,
\begin{align}
\wt{\mbf F}_t^\prime -  \mbf F_t^\prime \wt{\mbf H} &= \l(
\underbrace{\frac 1{nT} {\mbf F}_t^\prime \sum_{i=1}^n \sum_{s=1}^T  \bm\lambda_i  e_{is} \wt{\mbf F}_s^\prime}_{\text{(1.a)}}
+
\underbrace{\frac 1{nT} \sum_{i=1}^n  e_{it}\bm\lambda_i^\prime \sum_{s=1}^T {\mbf F}_s \wt{\mbf F}_s^\prime}_{\text{(1.b)}}
+\underbrace{\frac 1{nT} \sum_{i=1}^n \sum_{s=1}^T e_{is} e_{it}\wt{\mbf F}_s^\prime}_{\text{(1.c)}}
\r) \l(\frac{\wt {\mbf M}^x}{T}\r)^{-1},\label{eq:sviluppoLambdaBAI}
\end{align}
which is the expansion given in \citet[Equation (A.1)]{bai03} and where
\beq\label{eq:Htilde}
\wt{\mbf H}:=\frac{\bm\Lambda^\prime\bm\Lambda}{n}\frac{\bm F^\prime\wt{\bm F}}{T} \l(\frac{\wt {\mbf M}^x}{T}\r)^{-1}.
\eeq
Notice that $\Vert\wt{\mbf H}\Vert= O_{\mathrm P}(1)$ and $\Vert\wt{\mbf H}^{-1}\Vert= O_{\mathrm P}(1)$ because, as $n,T\to\infty$, all its terms tend to finite and invertible quantities (see Assumption \ref{ass:common}(a), 
Proposition \ref{prop:KKKbis}, and
Lemma \ref{lem:covarianze}(iii) jointly with Lemmas \ref{lem:Vzero}(i) and \ref{lem:Vzero}(iii)).

 Then, from \citet[Lemma A.2]{bai03}, $\Vert \text{(1.a)}\Vert= O_{\mathrm P}(\max(n^{-1},(nT)^{-1/2}) )$,
$\Vert \text{(1.b)}\Vert= O_{\mathrm P}( n^{-1/2})$, and
$\Vert \text{(1.c)}\Vert= O_{\mathrm P}(\max(n^{-1},(nT)^{-1/2}, T^{-1}) )$. Therefore, the estimated factors are consistent:
$$
\l\Vert \wh{\mbf F}_t-\wt{\mbf H}^{\prime}{\mbf F}_t \r\Vert = O_{\mathrm P} \l(\max\l(\frac 1{\sqrt n},\frac 1T\r)\r).
$$
This result coincides with Proposition \ref{prop:F2}, but when stated using $\wt{\mbf H}$ in place of $\wh{\mbf H}$. 

From \citet[Equation before (B.2)]{bai03}, for any $i=1,\ldots, n$ (recall that $\bm\varepsilon_i:=(e_{i1}\cdots e_{iT})^\prime$), 
\begin{align}
\wt{\bm\lambda}_i^\prime-{\bm\lambda}_i^\prime (\wt{\mbf H}^{-1})^\prime &=
\underbrace{\bm\lambda_i^\prime\frac{(\bm F-\wt{\bm F} \wt{\mbf H}^{-1})^\prime\wt{\bm F} }{T}}_{\text{(2.a)}}
+
\underbrace{\bm\varepsilon_i^\prime\frac{(\wt{\bm F}- \bm F\wt{\mbf H} )}{T}}_{\text{(2.b)}}
+
\underbrace{\frac{\bm\varepsilon_i^\prime\bm F\wt{\mbf H}}{T}}_{\text{(2.c)}},\label{eq:sviluppoFactorBAI}
\end{align}
which is  obtained from the linear projection of $\bm X^\prime$ onto the estimated factors $\wt{\bm F}$ in a similar way as we obtained  the linear projection of $\bm X$ onto the estimated loadings $\wh{\bm \Lambda}$ giving \eqref{eq:sviluppoFactor}. Then, from \citet[Lemmas B.3 and B.1]{bai03},
$\Vert \text{(2.a)}\Vert= O_{\mathrm P}(\max(n^{-1},(nT)^{-1/2},T^{-1}) )$,
$\Vert \text{(2.b)}\Vert= O_{\mathrm P}(\max(n^{-1},(nT)^{-1/2},T^{-1}) )$,
and $\Vert \text{(2.c)}\Vert= O_{\mathrm P}(T^{-1/2} )$. Therefore, the estimated loadings are consistent: 
$$
\l\Vert \wt{\bm\lambda}_i-\wt{\mbf H}^{-1}{\bm\lambda}_i\r\Vert = O_{\mathrm P} \l(\max\l(\frac 1{ n},\frac 1{\sqrt T}\r)\r).
$$
This result coincides with Proposition \ref{prop:L2}, but when stated using $\wt{\mbf H}$ in place of $\wh{\mbf H}$.

\paragraph{The role of $ \wt{\mbf H}$.} First, from \citet[Lemma B.3]{bai03}, 
\begin{align}
&\l\Vert \wt{\mbf H}- ({\bm F}^\prime {\bm F})^{-1}{\bm F}^\prime\wt{\bm F} \r\Vert = O_{\mathrm P}\l(\max\l(\frac 1n,\frac 1{\sqrt{nT}},\frac 1T\r)\r),\label{eq:prop9Htilde}
\end{align}
which is the analogue of Proposition \ref{prop:LLFF}(b). This immediately implies also the analogue of Propositions \ref{prop:LLFF}(d), \ref{prop:LLFF}(f), and \ref{prop:LLFF}(h). 

Second, following the same reasoning used to prove Proposition \ref{prop:HHAT}, but this time using \eqref{eq:prop9Htilde}, it follows that
\begin{align}
&\l\Vert \frac{\wt{\mbf M}^x}{T} -  \wt{\mbf H}^{-1}  \frac{\bm\Lambda^\prime\bm\Lambda}{n} (\wt{\mbf H}^{-1})^{\prime}\r\Vert =  O_{\mathrm P}\l(\max\l(\frac 1n,\frac 1{\sqrt{nT}},\frac 1T\r)\r),\label{eq:starp1BAI}
\end{align}
from which it follows that we must have
\begin{align}
&\l\Vert
\wt{\mbf H} -  \l(\frac{\bm F^\prime\bm F}{T}\r)^{-1/2} \wh{\mbf Q}
\r\Vert= O_{\mathrm P}\l(\max\l(\frac 1n,\frac 1{\sqrt{T}}\r)\r),\label{eq:sameQ}
\\
&\l\Vert
\wt{\mbf H}^{-1} -\wh{\mbf Q}^\prime \l(\frac{\bm F^\prime\bm F}{T}\r)^{1/2} 
\r\Vert= O_{\mathrm P}\l(\max\l(\frac 1n,\frac 1{\sqrt{T}}\r)\r),\label{eq:sameQinv}
\end{align}
where $\wh {\mbf Q}$ are the normalized eigenvectors of $(T^{-1}\bm F^\prime\bm F)^{1/2}(n^{-1}\bm \Lambda^\prime\bm \Lambda) (T^{-1}\bm F^\prime\bm F)^{1/2}$, which has $n^{-1}\wh{\mbf M}^C$ as eigenvalues, and these, in turn, are such that $n^{-1}\Vert \wh{\mbf M}^C-\wh{\mbf M}^x\Vert = O_{\mathrm P}(\max(n^{-1},T^{-1/2}))$ with
$T^{-1}\wt{\mbf M}^x=n^{-1}\wh{\mbf M}^x$. Results \eqref{eq:sameQ} and \eqref{eq:sameQinv} are the analogue of Proposition \ref{prop:HHAT}. Once again, we see that the loadings and the factors can be consistently estimated  up to: (i) a scale $(T^{-1}{\bm F^\prime\bm F})^{-1/2}$ and (ii) a rotation $\wh{\mbf Q}$. 

We can then derive a limit for $\wt{\mbf H}$ which depends only on population quantities. This can be done in two ways.
First, by Assumptions \ref{ass:common}(a) and \ref{ass:common}(c-ii), we also have $\Vert(T^{-1}\bm F^\prime\bm F)^{1/2}(n^{-1}\bm \Lambda^\prime\bm \Lambda) (T^{-1}\bm F^\prime\bm F)^{1/2}-(\bm\Gamma^F)^{1/2} \bm\Sigma_\Lambda(\bm\Gamma^F)^{1/2}\Vert \!=\!o_{\mathrm P}(1)$. Therefore, since the eigenvalues of $(\bm\Gamma^F)^{1/2} \bm\Sigma_\Lambda(\bm\Gamma^F)^{1/2}$ are distinct by Assumption \ref{ass:eval}, by Davis-Kahan theorem \citep[Corollary 1]{yu15}, the corresponding normalized eigenvectors satisfy:
$\Vert\wh{\mbf Q}-\bm\Upsilon_0\bm{\mathcal J}_0\Vert=o_{\mathrm P}(1)$, which, jointly with \eqref{eq:sameQ}  and  Assumption \ref{ass:common}(c-ii), implies
\beq
\l\Vert \wt{\mbf H}-  (\bm\Gamma^F)^{-1/2} \bm\Upsilon_0\bm{\mathcal J}_0\r\Vert = o_{\mathrm P}(1)
 \;\text{ and }\;
\l\Vert \wt{\mbf H}^{-1}-  \bm{\mathcal J}_0 \bm\Upsilon_0^\prime (\bm\Gamma^F)^{1/2}\r\Vert = o_{\mathrm P}(1), \label{eq:Htildelim2}
\eeq
which is the analogue of Proposition \ref{cor:semplice}.

Second, consider the following spectral decomposition: 
\beq\label{eq:U1V0U1}
(\bm\Sigma_\Lambda)^{1/2}\bm\Gamma^F(\bm\Sigma_\Lambda)^{1/2} =: \bm\Upsilon_1\bm V_0\bm\Upsilon_1^\prime,
\eeq
where $\bm\Upsilon_1$ is the $r\times r$ matrix having as columns the normalized eigenvectors of $(\bm\Sigma_\Lambda)^{1/2}\bm\Gamma^F(\bm\Sigma_\Lambda)^{1/2}$, and $\bm V_0$ is the $r\times r$ matrix of corresponding eigenvalues sorted in descending order, which coincide with those of 
$(\bm\Gamma^F)^{1/2}\bm\Sigma_\Lambda ( \bm\Gamma^F)^{1/2}$, and are given by $\bm V_0=\lim_{n\to\infty} n^{-1}\mbf M^{C}$. 
Then, from \citet[Proposition 1]{bai03}
\beq\label{eq:KKKBai}
\l\Vert \frac{\wt{\bm F}^\prime\bm F}{T}-\bm V_0^{1/2} \bm{\mathcal J}_1\bm\Upsilon_1^\prime (\bm\Sigma_\Lambda)^{-1/2}\r\Vert = o_{\mathrm P}(1),
\eeq
where $\bm{\mathcal J}_1$ is an $r \times r$ diagonal matrix with entries $\pm 1$. This matrix  is not present in the original proof since  the sign of the eigenvectors is implicitly fixed therein. Notice that the limiting quantity in \eqref{eq:KKKBai} is finite and invertible because of Assumption \ref{ass:common}(a) and Lemmas \ref{lem:Vzero}(i) and \ref{lem:Vzero}(iii). 
From \eqref{eq:Htilde}, \eqref{eq:KKKBai}, Assumption \ref{ass:common}(a), and Lemma \ref{lem:Vzero}(iv), 
\beq\label{eq:Htildelim}
\l\Vert \wt{\mbf H}-(\bm\Sigma_\Lambda)^{1/2} \bm\Upsilon_1 \bm{\mathcal J}_1 \bm V_0^{-1/2} \r\Vert = o_{\mathrm P}(1) \;\text{ and }\; \l\Vert \wt{\mbf H}^{-1}- \bm V_0^{1/2}\bm{\mathcal J}_1\bm\Upsilon_1^\prime (\bm\Sigma_\Lambda)^{-1/2} \r\Vert = o_{\mathrm P}(1).
\eeq
If Assumption \ref{ass:common}(a) holds with rate $\sqrt n$ and Assumption \ref{ass:common}(c-ii) holds with rate $\sqrt T$, then \eqref{eq:Htildelim2}, \eqref{eq:KKKBai}, and \eqref{eq:Htildelim} hold with rate $\min(\sqrt n,\sqrt T)$.

\paragraph{A new limit for $\wh{\mbf H}$.} From \eqref{eq:Htildelim2} and \eqref{eq:Htildelim} and uniqueness of the limits, 
\begin{align}
&
(\bm\Sigma_\Lambda)^{1/2} \bm\Upsilon_1 \bm{\mathcal J}_1 \bm V_0^{-1/2}
=(\bm\Gamma^F)^{-1/2} \bm\Upsilon_0\bm{\mathcal J}_0
,\label{eq:riunioni2}\\
&
\bm V_0^{1/2}
\bm{\mathcal J}_1 
\bm\Upsilon_1^\prime 
(\bm\Sigma_\Lambda)^{-1/2}
=
\bm{\mathcal J}_0\bm\Upsilon_0^\prime(\bm\Gamma^F)^{1/2} 
.\label{eq:riunioni}
\end{align}
And, from  Proposition \ref{prop:KKK}(b) and \eqref{eq:riunioni2}, it follows that we also have
\beq\label{eq:KKKBaiLL}
\l\Vert \frac{\wh{\bm \Lambda}^\prime\bm \Lambda}{n}- 
\bm V_0^{1/2}\bm{\mathcal J}_1\bm\Upsilon_1^\prime (\bm\Sigma_\Lambda)^{1/2}\r\Vert = o_{\mathrm P}(1).
\eeq
Therefore, from \eqref{eq:acca}, \eqref{eq:KKKBaiLL}, Assumption \ref{ass:common}(c-ii), and Lemma \ref{lem:Vzero}(iv),
\beq
\l\Vert\wh{\mbf H}-  \bm\Gamma^F(\bm\Sigma_\Lambda)^{1/2}\bm\Upsilon_1 \bm{\mathcal J}_1\bm V_0^{-1/2}\r\Vert= o_{\mathrm P}(1),
\label{eq:ennesimaespansioneAA}
\eeq
and, since from \eqref{eq:U1V0U1}, $\bm\Gamma^F(\bm\Sigma_\Lambda)^{1/2}=(\bm\Sigma_\Lambda)^{-1/2}\bm\Upsilon_1\bm V_0\bm\Upsilon_1^\prime$, from \eqref{eq:ennesimaespansioneAA} we obtain
\beq
\l\Vert\wh{\mbf H}-  (\bm\Sigma_\Lambda)^{-1/2}\bm\Upsilon_1  \bm{\mathcal J}_1\bm V_0^{1/2}\r\Vert= o_{\mathrm P}(1)
\;\text{ and }\;
\l\Vert\wh{\mbf H}^{-1}- \bm V_0^{-1/2} \bm{\mathcal J}_1\bm\Upsilon_1^\prime(\bm\Sigma_\Lambda)^{1/2}\r\Vert= o_{\mathrm P}(1). 
\label{eq:ennesimaespansione}
\eeq
Once again, if Assumption \ref{ass:common}(a) holds with rate $\sqrt n$ and Assumption \ref{ass:common}(c-ii) holds with rate $\sqrt T$, the rate in \eqref{eq:KKKBaiLL}-\eqref{eq:ennesimaespansione} is $\min(\sqrt n,\sqrt T)$. 


\paragraph{The relation between $\wt{\mbf H}$ and $\wh{\mbf H}$.}
Since $\wt{\bm F}=\wh{\bm F}$ from Proposition \ref{prop:LLFF}(b) and \eqref{eq:prop9Htilde}, we see that we must have 
\begin{align}
&\l\Vert\wt{\mbf H}-(\wh{\mbf H}^{-1})^\prime\r\Vert = O_{\mathrm P}\l(\max\l(\frac 1n,\frac 1{\sqrt{nT}},\frac 1T\r)\r)\;\text{and}\;
&\l\Vert\wt{\mbf H}^{-1}-\wh{\mbf H}^\prime \r\Vert = O_{\mathrm P}\l(\max\l(\frac 1n,\frac 1{\sqrt{nT}},\frac 1T\r)\r).\label{eq:sameH}
\end{align}
It follows that, by substituting \eqref{eq:sameH} into Proposition \ref{prop:LLFF}(a), we obtain its analogous:
\begin{align}
&\l\Vert  \wt{\mbf H}^{-1} - \wt{\bm\Lambda}^\prime{\bm\Lambda}(\bm\Lambda^\prime\bm\Lambda)^{-1}\r\Vert= O_{\mathrm P}\l(\max\l(\frac 1n,\frac 1{\sqrt{nT}},\frac 1T\r)\r),\label{eq:prop9HtildeL}
\end{align}
which immediately implies also the analogue of Propositions \ref{prop:LLFF}(c), \ref{prop:LLFF}(e), and \ref{prop:LLFF}(g).

\paragraph{Asymptotic normality.} From \eqref{eq:sviluppoLambdaBAI} and \citet[Theorem 1]{bai03}, if $\sqrt n/T\to 0$, as $n,T\to\infty$,
\begin{align}
\sqrt n&\l(\wt{\mbf F}_t-\wt{\mbf H}^{\prime}{\mbf F}_t\r) =\l(\frac{\wt {\mbf M}^x}{T} \r)^{-1} \frac{\wt{\bm F}^\prime \bm F}{T}\l(\frac 1{\sqrt n}\sum_{i=1}^n {\bm\lambda}_i e_{it}\r)+o_{\mathrm P}(1)= \wt{\mbf H}^\prime \l(\frac{\bm\Lambda^\prime\bm\Lambda}{n}\r)^{-1}\l(\frac 1{\sqrt n}\sum_{i=1}^n {\bm\lambda}_i e_{it}\r)+o_{\mathrm P}(1)\nn\\
& = \l\{ \underset{n,T\to\infty}{\text{P-lim}} \wt{\mbf H}^\prime \r\}
 \l(\frac{\bm\Lambda^\prime\bm\Lambda}{n}\r)^{-1}\l(\frac 1{\sqrt n}\sum_{i=1}^n \bm\lambda_i e_{it}\r)+o_{\mathrm P}(1)=  \bm V_0^{-1/2}\bm{\mathcal J}_1\bm\Upsilon_1^\prime (\bm\Sigma_\Lambda)^{1/2}  \sqrt n\l(\wh{\mbf F}_t^{\text{\tiny OLS}}-\mbf F_t\r)+ o_{\mathrm P}(1)\nn\\
&= \bm{\mathcal J}_0\bm\Upsilon_0^\prime  (\bm\Gamma^F)^{-1/2} \sqrt n\l(\wh{\mbf F}_t^{\text{\tiny OLS}}-\mbf F_t\r)+ o_{\mathrm P}(1).
\label{eq:CLTFBAI}
\end{align}
where the first line is the only asymptotic expansion reported by \citet[Theorem 1]{bai03}, and for the subsequent equalities we used also  \eqref{eq:Htilde}, \eqref{eq:Htildelim2}, and \eqref{eq:Htildelim}, or \eqref{eq:riunioni2}. The last expression in \eqref{eq:CLTFBAI} coincides with \eqref{eq:CLTF} in the proof of Theorem \ref{th:CLTF}, hence,  by Slutsky's theorem, 
\beq\label{eq:ANFBAI1}
\sqrt n\l(\wt{\mbf F}_t-\wt{\mbf H}^{\prime}{\mbf F}_t\r) \to_d\mathcal N\l(\mbf 0_r,   \bm\Upsilon_0^\prime  (\bm\Gamma^F)^{-1/2}
\bm\Pi_t^{\text{\tiny \upshape OLS}}
(\bm\Gamma^F)^{-1/2}\bm\Upsilon_0
\r),
\eeq
Moreover, using the second last line of \eqref{eq:CLTFBAI}, by Slutsky's theorem, we also have
\beq\label{eq:ANFBAI}
\sqrt n\l(\wt{\mbf F}_t-\wt{\mbf H}^{\prime}{\mbf F}_t\r) \to_d\mathcal N\l(\mbf 0_r,   \bm V_0^{-1/2}\bm\Upsilon_1^\prime (\bm\Sigma_\Lambda)^{1/2}\bm\Pi_t^{\text{\tiny \upshape OLS}}(\bm\Sigma_\Lambda)^{1/2}\bm\Upsilon_1 \bm V_0^{-1/2}
\r),
\eeq
 where $\bm\Pi_t^{\text{\tiny \upshape OLS}}=(\bm\Sigma_\Lambda)^{-1}\bm\Gamma_t(\bm\Sigma_\Lambda)^{-1}$ with $\bm\Gamma_t$ defined in Assumption \ref{ass:CLT}(b). Furthermore, let $\bm{\mathcal Q}:=\bm V_0^{1/2}\bm\Upsilon_1^\prime(\bm\Sigma_\Lambda)^{-1/2}$. Then, the asymptotic covariance matrix of $\wt{\mbf F}_t$ can also be written as $(\bm V_0)^{-1}\bm {\mathcal Q}\bm\Gamma_t\bm {\mathcal Q}^\prime (\bm V_0)^{-1}$, which is the expression given in \citet[Theorem 1]{bai03}. Notice that from \eqref{eq:ANFBAI} it follows that we can state Theorem \ref{th:CLTF} also as
 \beq\label{eq:ANFbis}
\sqrt n\l(\wh{\mbf F}_t-\wh{\mbf H}^{-1}{\mbf F}_t\r) \to_d\mathcal N\l(\mbf 0_r,   \bm V_0^{-1/2}\bm\Upsilon_1^\prime (\bm\Sigma_\Lambda)^{1/2}\bm\Pi_t^{\text{\tiny \upshape OLS}}(\bm\Sigma_\Lambda)^{1/2}\bm\Upsilon_1 \bm V_0^{-1/2}
\r).
\eeq


Similarly, from \eqref{eq:sviluppoFactorBAI} and \citet[Theorem 2]{bai03}, if $\sqrt T/n\to 0$, as $n,T\to\infty$,
\begin{align}
\sqrt T&\l(\wt{\bm\lambda}_i-\wt{\mbf H}^{-1}{\bm\lambda}_i\r) 
= \wt{\mbf H}^\prime \l(\frac 1{\sqrt T}\sum_{t=1}^T \mbf F_t e_{it}\r)+o_{\mathrm P}(1)=\l(\frac{\wt{\bm F}^\prime\wt{\bm F}}{T}\r)^{-1}  \wt{\mbf H}^\prime \l(\frac 1{\sqrt T}\sum_{t=1}^T \mbf F_t e_{it}\r)+o_{\mathrm P}(1)\nn\\
&=\l(\frac{\wt{\mbf H}^{\prime}{\bm F}^\prime{\bm F}\wt{\mbf H}}{T}\r)^{-1}  \wt{\mbf H}^\prime \l(\frac 1{\sqrt T}\sum_{t=1}^T \mbf F_t e_{it}\r)+o_{\mathrm P}(1)= \wt{\mbf H}^{-1}\l(\frac{{\bm F}^\prime{\bm F}}{T}\r)^{-1}\l(\frac 1{\sqrt T}\sum_{t=1}^T \mbf F_t e_{it}\r)+o_{\mathrm P}(1)\nn\\
&=\l\{ \underset{n,T\to\infty}{\text{P-lim}} \wt{\mbf H}^{-1} \r\}\l(\frac{{\bm F}^\prime{\bm F}}{T}\r)^{-1}\l(\frac 1{\sqrt T}\sum_{t=1}^T \mbf F_t e_{it}\r)+o_{\mathrm P}(1)=  \bm V_0^{1/2}\bm{\mathcal J}_1\bm\Upsilon_1^\prime (\bm\Sigma_\Lambda)^{-1/2} \sqrt T(\wh{\bm\lambda}_i^{\text{\tiny OLS}}-{\bm\lambda}_i) +o_{\mathrm P}(1)\nn\\
&= \bm{\mathcal J}_0\bm\Upsilon_0^\prime(\bm\Gamma^F)^{1/2}\sqrt T\l(\wh{\bm\lambda}_i^{\text{\tiny OLS}}-{\bm\lambda}_i\r) +o_{\mathrm P}(1)
,\label{eq:CLTLBAI}
\end{align}
where the first line is the only asymptotic expansion reported by \citet[proof of Theorem 2]{bai03}, and for the subsequent equalities we used also
the fact that $T^{-1}\wt{\mbf F}^\prime\wt{\mbf F}=\mbf I_r$ by definition, 
\eqref{eq:prop9Htilde} which implies $\Vert T^{-1}\wt{\bm F}^\prime\wt{\bm F}-T^{-1} \wt{\mbf H}^{\prime}{\bm F}^\prime{\bm F}\wt{\mbf H} \Vert=o_{\mathrm P}(1)$,
\eqref{eq:Htildelim2}, and \eqref{eq:Htildelim}, or \eqref{eq:riunioni}.
The last expression in \eqref{eq:CLTLBAI} coincides with \eqref{eq:CLTL} in the proof of Theorem \ref{th:CLTL}, hence, by Slutsky's theorem, 
\beq\label{eq:ANLBAI1}
\sqrt T\l(\wt{\bm\lambda}_i-\wt{\mbf H}^{-1}{\bm\lambda}_i\r) \to_d\mathcal N\l(\mbf 0_r,  \bm\Upsilon_0^\prime(\bm\Gamma^F)^{1/2} \bm\Theta_i^{\text{\tiny \upshape OLS}}(\bm\Gamma^F)^{1/2}\bm\Upsilon_0
\r).
\eeq
Moreover, using the second last line of \eqref{eq:CLTLBAI}, by Slutsky's theorem, we also have
\beq\label{eq:ANLBAI}
\sqrt T\l(\wt{\bm\lambda}_i-\wt{\mbf H}^{-1}{\bm\lambda}_i\r) \to_d\mathcal N\l(\mbf 0_r,  \bm V_0^{1/2}\bm\Upsilon_1^\prime (\bm\Sigma_\Lambda)^{-1/2} \bm\Theta_i^{\text{\tiny \upshape OLS}}(\bm\Sigma_\Lambda)^{-1/2}\bm\Upsilon_1\bm V_0^{1/2}
\r),
\eeq
where $\bm\Theta_i^{\text{\tiny \upshape OLS}}=(\bm\Gamma^F)^{-1}\bm\Phi_i(\bm\Gamma^F)^{-1}$ with  $\bm\Phi_i$ defined in Assumption \ref{ass:CLT}(a). Furthermore, from the second line of \eqref{eq:CLTLBAI}, using \eqref{eq:Htildelim}, we also have that the asymptotic covariance matrix of $\wt{\bm\lambda}_i$  can also be written as \linebreak $\bm V_0^{-1/2}\bm\Upsilon_1^\prime (\bm\Sigma_\Lambda)^{1/2}
\bm\Phi_i
(\bm\Sigma_\Lambda)^{1/2}\bm\Upsilon_1\bm V_0^{-1/2}$
and by letting $\bm {\mathcal Q}:=(\bm V_0)^{1/2}\bm\Upsilon_1^\prime(\bm\Sigma_\Lambda)^{-1/2}$ this is equivalent to
$(\bm {\mathcal Q}^{\prime})^{-1} \bm\Phi_i\bm {\mathcal Q}^{-1}\!$, which is the expression given in \citet[Theorem 2]{bai03}. Notice that from \eqref{eq:ANLBAI} it follows that we can state Theorem \ref{th:CLTL} also as
\beq\label{eq:ANLbis}
\sqrt T\l(\wh{\bm\lambda}_i-\wh{\mbf H}^{\prime}{\bm\lambda}_i\r) \to_d\mathcal N\l(\mbf 0_r,  \bm V_0^{1/2}\bm\Upsilon_1^\prime (\bm\Sigma_\Lambda)^{-1/2} \bm\Theta_i^{\text{\tiny \upshape OLS}}(\bm\Sigma_\Lambda)^{-1/2}\bm\Upsilon_1\bm V_0^{1/2}
\r).
\eeq
\setcounter{equation}{0}
\section{Asymptotic properties of approaches A2 and B2} \label{sec:cmppcaCD}

\subsection{Approach A2} \label{sec:cmppcaSW}

\paragraph{Consistency.} First of all, since $\wideparen{\bm\Lambda}$ and $\wideparen{\bm F}$, defined in \eqref{eq:estLsw} and \eqref{eq:estFsw}, depend only on the eigenvectors $\wh{\mbf V}^x$ but not on the eigenvalues, Proposition \ref{prop:L} still applies by using only Davis-Kahan theorem \citep[Corollary 1]{yu15}. In particular,
\beq\label{eq:prop1sw}
\l\Vert\frac{\wideparen{\bm\Lambda}-\bm\Lambda\bm{\mathcal H}}{\sqrt n}\r\Vert= O_{\mathrm P}\l(\max\l(\frac 1n,\frac 1{\sqrt T}\r)\r).
\eeq
Moreover, by definition of eigenvectors, we get
\beq\label{eq:EVECSW}
\frac{\bm X^\prime\bm X}{nT}\wideparen{\bm\Lambda}=\wideparen{\bm\Lambda}\frac{\wh{\mbf M}^x}{n}.
\eeq
By using $\bm X=\bm F\bm\Lambda^\prime+\bm E$ and taking the $i$th row of \eqref{eq:EVECSW},  for any $i=1,\ldots,n$,
\begin{align}
\wideparen{\bm\lambda}_i^\prime-{\bm\lambda}_i^\prime\wideparen{\mbf H}=&\,
\l(
\frac 1{nT}{\bm\lambda}_i^\prime\sum_{t=1}^T\sum_{j=1}^n\mbf F_t e_{jt}\wideparen{\bm\lambda}_j^\prime
+\frac 1{nT} \sum_{t=1}^T  e_{it}\mbf F_t^\prime\sum_{j=1}^n\bm\lambda_j\wideparen{\bm\lambda}_j^\prime
+\frac 1{nT} \sum_{t=1}^T\sum_{j=1}^n e_{it} e_{jt} \wideparen{\bm\lambda}_j^\prime
\r) \l(\frac{\wh{\mbf M}^x}{n}\r)^{-1},
\label{eq:start4sw}
\end{align}
which is equivalent to the expression in \eqref{eq:sviluppoLambda}, and where
\beq\label{eq:accasw}
\wideparen{\mbf H}:=
\frac{\bm F^\prime\bm F}{T}
\frac{\bm\Lambda^\prime\wideparen{\bm\Lambda}}{n}
\l(\frac{\wh{\mbf M}^x}{n}\r)^{-1}.
\eeq
Notice that $\Vert\wideparen{\mbf H}\Vert= O_{\mathrm P}(1)$ and $\Vert\wideparen{\mbf H}^{-1}\Vert= O_{\mathrm P}(1)$ because $\wideparen{\mbf H}$ is equivalent to $\wh{\mbf H}$ (see \eqref{eq:sameHsw} below) up to a scaling factor which is   asymptotically finite and  positive definite (see Lemma \ref{lem:covarianze}(iii) joint with Lemmas \ref{lem:Vzero}(i) and \ref{lem:Vzero}(iii)).
Similarly, by linear projection of $\bm X^\prime$ onto the estimated loadings $\wideparen{\bm \Lambda}$,
 following the same steps leading to \eqref{eq:sviluppoFactor}, we obtain the equivalent expression, for any $t=1,\ldots,T$,
\begin{align}
\wideparen{\mbf F}_t^\prime -\mbf F_t^\prime (\wideparen{\mbf H}^{-1})^\prime&=
\mbf F_t^\prime\frac{(\bm\Lambda-\wideparen{\bm\Lambda}\wideparen{\mbf H}^{-1})^\prime\wideparen{\bm\Lambda}}{n}+
\bm e_t^\prime \frac{(\wideparen{\bm\Lambda}-\bm\Lambda \wideparen{\mbf H})}{n}+
\bm e_t^\prime \frac{\bm\Lambda\wideparen{\mbf H}}{n}.\label{eq:FFF2sw}
\end{align}
Hence, from \eqref{eq:prop1sw}, \eqref{eq:start4sw}, and \eqref{eq:FFF2sw} we can prove consistency using Lemmas \ref{prop:load} and \ref{prop:factor}:
\begin{align}
&\l\Vert \wideparen{\bm\lambda}_i-\wideparen{\mbf H}^{\prime}{\bm\lambda}_i\r\Vert = O_{\mathrm P} \l(\max\l(\frac 1{ n},\frac 1{\sqrt T}\r)\r),\quad\l\Vert \wideparen{\mbf F}_t-\wideparen{\mbf H}^{-1}{\mbf F}_t \r\Vert = O_{\mathrm P} \l(\max\l(\frac 1{ \sqrt n},\frac 1{T}\r)\r),\nn
\end{align}
which are the analogue of Propositions \ref{prop:L2} and \ref{prop:F2}, but when stated using $\wideparen{\mbf H}$ in place of $\wh{\mbf H}$. 

\paragraph{The role of $ \wideparen{\mbf H}$.} 
From Lemma \ref{prop:factor}(a), which applies also in this case, we can prove that
\begin{align}
&\l\Vert  \wideparen{\mbf H} - (\bm\Lambda^\prime\bm\Lambda)^{-1}
{\bm\Lambda}^\prime\wideparen{\bm\Lambda}
\r\Vert=O_{\mathrm P}\l(\max\l(\frac 1n,\frac{1}{\sqrt{nT}},\frac 1T\r)\r),\label{eq:prop9HtildeLsw}
\end{align}
which is the analogue of Proposition \ref{prop:LLFF}(a), and the analogue of Propositions \ref{prop:LLFF}(c), \ref{prop:LLFF}(e),
and \ref{prop:LLFF}(g) follow immediately.
Moreover, following the same reasoning used to prove Proposition \ref{prop:HHAT}, but this time using \eqref{eq:prop9HtildeLsw}, it follows that
\begin{align}
&\l\Vert
\frac{\wh{\mbf M}^x}{n} - \wideparen{\mbf H}^{-1} \frac{\bm F^\prime\bm F}{T}  (\wideparen{\mbf H}^{-1})^{\prime}
\r\Vert=O_{\mathrm P}\l(\max\l(\frac 1n,\frac{1}{\sqrt{nT}},\frac 1T\r)\r).\label{eq:starop3sw}
\end{align}
from which it follows that we must have
\begin{align}
&\l\Vert \wideparen{\mbf H} -  \l(\frac{\bm\Lambda^\prime\bm\Lambda}{n}\r)^{-1/2} \wideparen {\mbf Q}\r\Vert = O_{\mathrm P}\l(\max\l(\frac 1n,\frac{1}{\sqrt{T}}\r)\r),\label{eq:sameQsw}\\
&\l\Vert \wideparen{\mbf H}^{-1} - \wideparen {\mbf Q}^\prime \l(\frac{\bm\Lambda^\prime\bm\Lambda}{n}\r)^{1/2}\r\Vert = O_{\mathrm P}\l(\max\l(\frac 1n,\frac{1}{\sqrt{T}}\r)\r) ,\label{eq:sameQinvsw}
\end{align}
where $\wideparen {\mbf Q}$ are the normalized eigenvectors of $(n^{-1}\bm \Lambda^\prime\bm \Lambda)^{1/2} (T^{-1}\bm F^\prime\bm F) (n^{-1}\bm \Lambda^\prime\bm \Lambda)^{1/2}$, which has $n^{-1}\wh{\mbf M}^C$ as eigenvalues, and these, in turn, are such that $n^{-1}\Vert \wh{\mbf M}^C-\wh{\mbf M}^x\Vert = O_{\mathrm P}(\max(n^{-1},T^{-1/2}))$. Results  \eqref{eq:sameQsw} and \eqref{eq:sameQinvsw} are  the analogue of Proposition \ref{prop:HHAT}.
Once again, we see that the loadings and the factors can be consistently estimated  up to: (i) a scale $(n^{-1}{\bm \Lambda^\prime\bm \Lambda})^{-1/2}$ and (ii) a rotation $\wideparen{\mbf Q}$. 

We can then derive a limit for $\wideparen{\mbf H}$ which depends only on population quantities.  By Assumptions \ref{ass:common}(a) and \ref{ass:common}(c-ii),  $\Vert
(n^{-1}\bm \Lambda^\prime\bm \Lambda)^{1/2} (T^{-1}\bm F^\prime\bm F) (n^{-1}\bm \Lambda^\prime\bm \Lambda)^{1/2}-
(\bm\Sigma_\Lambda)^{1/2}(\bm\Gamma^F)(\bm\Sigma_\Lambda)^{1/2}\Vert =o_{\mathrm P}(1)$, and by Assumption \ref{ass:eval} and Davis-Kahan theorem \citep[Corollary 1]{yu15}, the corresponding normalized eigenvectors satisfy:
$\Vert\wideparen{\mbf Q}-\bm\Upsilon_1\bm{\mathcal J}_1\Vert=o_{\mathrm P}(1)$. Hence, from \eqref{eq:sameQsw} and using also Assumption \ref{ass:common}(a),
\beq\label{eq:semplicesw}
\l\Vert\wideparen{\mbf H} - (\bm\Sigma_\Lambda)^{-1/2}\bm\Upsilon_1\bm{\mathcal J}_1\r\Vert=o_{\mathrm P}(1) \;\text{ and }\; 
\l\Vert\wideparen{\mbf H}^{-1} - \bm{\mathcal J}_1\bm\Upsilon_1^\prime  (\bm\Sigma_\Lambda)^{1/2}\r\Vert=o_{\mathrm P}(1),
\eeq
which is the analogue of \eqref{eq:ennesimaespansione}. If Assumption \ref{ass:common}(a) holds with rate $\sqrt n$ and Assumption \ref{ass:common}(c-ii) holds with rate $\sqrt T$, then \eqref{eq:semplicesw} holds with rate $\min(\sqrt n,\sqrt T)$.

\paragraph{The relation between $\wideparen{\mbf H}$, $\wh{\mbf H}$, and $\wt{\mbf H}$.} By definition, $n^{-1/2}\wideparen{\bm\Lambda} = \wh{\bm\Lambda} (\wh{\mbf M}^x)^{-1/2}=\wh{\mbf V}^x$ (see \eqref{eq:estL} and \eqref{eq:estLsw}). Thus, by comparing the definitions of $\wh{\mbf H}$ and $\wideparen{\mbf H}$ in \eqref{eq:acca} and \eqref{eq:accasw}, respectively, we see that
\beq
\wideparen{\mbf H} = \wh{\mbf H} \l(\frac{\wh{\mbf M}^x}{n} \r)^{-1/2}\;\text{ and }\qquad
\wideparen{\mbf H}^{-1} = \l(\frac{\wh{\mbf M}^x}{n} \r)^{1/2}\wh{\mbf H}^{-1} 
.\label{eq:sameHsw}
\eeq 
By \eqref{eq:sameH} and \eqref{eq:sameHsw}, we also have
\begin{align}
&\l\Vert\wideparen{\mbf H}-(\wt{\mbf H}^{-1})^\prime \l(\frac{\wh{\mbf M}^x}{n}\r)^{-1/2} \r\Vert = O_{\mathrm P}\l(\max\l(\frac 1n,\frac 1{\sqrt{nT}},\frac 1T\r)\r),\label{eq:sameHsw2}\\
&\l\Vert\wideparen{\mbf H}^{-1}-\l(\frac{\wh{\mbf M}^x}{n}\r)^{1/2}
\wt{\mbf H}^\prime\r\Vert = O_{\mathrm P}\l(\max\l(\frac 1n,\frac 1{\sqrt{nT}},\frac 1T\r)\r). \nn
\end{align}
It follows that, from \eqref{eq:sameHsw} and Proposition \ref{prop:LLFF}(b), 
\begin{align}
&\l\Vert \wideparen{\mbf H}^{-1}- \wideparen{\bm F}^\prime{\bm F} ({\bm F}^\prime {\bm F})^{-1}\r\Vert = O_{\mathrm P}\l(\max\l(\frac 1n,\frac 1{\sqrt{nT}},\frac 1T\r)\r),\label{eq:prop9Htildesw}
\end{align}
where we used also the fact that $\wideparen{\bm F}=\wh{\bm F}(n^{-1}\wh{\mbf M}^x)^{1/2}$. The analogue of Propositions \ref{prop:LLFF}(d), \ref{prop:LLFF}(f), and \ref{prop:LLFF}(h) follow immediately.
Last, from Proposition \ref{cor:semplice} and \eqref{eq:sameHsw}, by using also Lemma \ref{lem:Vzero}(iv), we get
\beq
\l\Vert \wideparen{\mbf H} -(\bm\Gamma^F)^{1/2}\bm\Upsilon_0\bm{\mathcal J}_0 \bm V_0^{-1/2}\r\Vert = o_{\mathrm P}(1)
\;\text{and }\;
\l\Vert \wideparen{\mbf H}^{-1} -
\bm V_0^{1/2}
\bm{\mathcal J}_0 
\bm\Upsilon_0^\prime
(\bm\Gamma^F)^{-1/2}
\r\Vert = o_{\mathrm P}(1).\label{eq:semplicesw2}
\eeq
Once again, if Assumption \ref{ass:common}(a) holds with rate $\sqrt n$ and Assumption \ref{ass:common}(c-ii) holds with rate $\sqrt T$, then
the rate in
 \eqref{eq:semplicesw2} is $\min(\sqrt n,\sqrt T)$. Note that by comparing \eqref{eq:semplicesw} and \eqref{eq:semplicesw2}, we prove again \eqref{eq:riunioni2} and \eqref{eq:riunioni}.

\paragraph{Asymptotic normality.} By noticing that Lemma \ref{prop:load} applies also to \eqref{eq:start4sw}, if $\sqrt T/n\to 0$, as $n,T\to\infty$, 
\begin{align}
\sqrt T(\wideparen{\bm\lambda}_i&-\wideparen{\mbf H}^{\prime}{\bm\lambda}_i) =  \l(\frac{\wh{\mbf M}^x}{n}\r)^{-1}\frac{ \wideparen{ \bm\Lambda}^\prime\bm\Lambda}{n} \l(\frac 1{\sqrt T}\sum_{t=1}^T \mbf F_t e_{it}\r)
+ o_{\mathrm {P}}(1)=\wideparen{\mbf H}^\prime \l(\frac{\bm F^\prime\bm F}{T}\r)^{-1} \l(\frac 1{\sqrt T}\sum_{t=1}^T \mbf F_t e_{it}\r)
+ o_{\mathrm {P}}(1)\nn\\
&=\l\{ \underset{n,T\to\infty}{\text{P-lim}} \wideparen{\mbf H}^\prime \r\}\l(\frac{{\bm F}^\prime{\bm F}}{T}\r)^{-1}\l(\frac 1{\sqrt T}\sum_{t=1}^T \mbf F_t e_{it}\r)+o_{\mathrm P}(1)=  \bm{\mathcal J}_1\bm\Upsilon_1^\prime (\bm\Sigma_\Lambda)^{-1/2} \sqrt T(\wh{\bm\lambda}_i^{\text{\tiny OLS}}-{\bm\lambda}_i) +o_{\mathrm P}(1)\nn\\
&=\bm V_0^{-1/2} \bm{\mathcal J}_0\bm\Upsilon_0^\prime(\bm\Gamma^F)^{1/2}\sqrt T(\wh{\bm\lambda}_i^{\text{\tiny OLS}}-{\bm\lambda}_i) +o_{\mathrm P}(1)
,\label{eq:CLTLSW}
\end{align}
where we used \eqref{eq:accasw}, \eqref{eq:semplicesw}, and \eqref{eq:semplicesw2}. Using the last two lines of \eqref{eq:CLTLSW}, by Slutsky's theorem, 
\begin{align}
&\sqrt T(\wideparen{\bm\lambda}_i-\wideparen{\mbf H}^{\prime}{\bm\lambda}_i) \to_d\mathcal N\l(\mbf 0_r,  \bm V_0^{-1/2} \bm\Upsilon_0^\prime(\bm\Gamma^F)^{1/2} \bm\Theta_i^{\text{\tiny \upshape OLS}}(\bm\Gamma^F)^{1/2}\bm\Upsilon_0\bm V_0^{-1/2}
\r),\label{eq:ANLSW1}\\
&\sqrt T(\wideparen{\bm\lambda}_i-\wideparen{\mbf H}^{\prime}{\bm\lambda}_i) \to_d\mathcal N\l(\mbf 0_r,  \bm\Upsilon_1^\prime (\bm\Sigma_\Lambda)^{-1/2} 
\bm\Theta_i^{\text{\tiny \upshape OLS}}
(\bm\Sigma_\Lambda)^{-1/2} \bm\Upsilon_1
\r),\label{eq:ANLSW2}
\end{align}
where $\bm\Theta_i^{\text{\tiny \upshape OLS}}=(\bm\Gamma^F)^{-1}\bm\Phi_i(\bm\Gamma^F)^{-1}$ with  $\bm\Phi_i$ defined in Assumption \ref{ass:CLT}(a). 

Similarly, by noticing that Lemma \ref{prop:factor} applies also to \eqref{eq:FFF2sw}, if $\sqrt n/T\to 0$, as $n,T\to\infty$,
\begin{align}
\sqrt n(\wideparen{\mbf F}_t&-\wideparen{\mbf H}^{-1}{\mbf F}_t )
= \wideparen{\mbf H}^\prime \l(\frac 1{\sqrt n}\sum_{i=1}^n \bm \lambda_i e_{it}\r)
+ o_{\mathrm {P}}(1)=  \l(\frac{\wideparen{\bm\Lambda}^\prime\wideparen{\bm\Lambda}}{n}\r)^{-1}\wideparen{\mbf H}^\prime \l(\frac 1{\sqrt n}\sum_{i=1}^n \bm \lambda_i e_{it}\r)
+ o_{\mathrm {P}}(1)\nn\\
&=  \l(\frac{\wideparen{\mbf H}^{\prime}{\bm \Lambda}^\prime{\bm \Lambda}\wideparen{\mbf H}}{n}\r)^{-1}\wideparen{\mbf H}^\prime \l(\frac 1{\sqrt n}\sum_{i=1}^n \bm \lambda_i e_{it}\r)
+ o_{\mathrm {P}}(1)= \wideparen{\mbf H}^{-1}\l(\frac{{\bm \Lambda}^\prime{\bm \Lambda}}n\r)^{-1}\l(\frac 1{\sqrt n}\sum_{i=1}^n \bm \lambda_i e_{it}\r)
+ o_{\mathrm {P}}(1)\nn\\
&= \l\{ \underset{n,T\to\infty}{\text{P-lim}} \wideparen{\mbf H}^{-1} \r\}
\l(\frac{{\bm \Lambda}^\prime{\bm \Lambda}}n\r)^{-1}\l(\frac 1{\sqrt n}\sum_{i=1}^n \bm \lambda_i e_{it}\r)+ o_{\mathrm {P}}(1)=  \bm{\mathcal J}_1\bm\Upsilon_1^\prime  (\bm\Sigma_\Lambda)^{1/2}\sqrt n(\wh{\mbf F}_t^{\text{\tiny OLS}}-\mbf F_t)+ o_{\mathrm P}(1)\nn\\
&= \bm V_0^{1/2} \bm{\mathcal J}_0\bm\Upsilon_0^\prime  (\bm\Gamma^F)^{-1/2} \sqrt n(\wh{\mbf F}_t^{\text{\tiny OLS}}-\mbf F_t)+ o_{\mathrm P}(1),\label{eq:CLTFSW}
\end{align}
where we used
the fact that $n^{-1}\wideparen{\bm\Lambda}^\prime\wideparen{\bm\Lambda}=\mbf I_r$ by definition, 
\eqref{eq:prop9HtildeLsw} which implies  $\Vert n^{-1}\wideparen{\bm \Lambda}^\prime\wideparen{\bm \Lambda}-n^{-1} \wideparen{\mbf H}^{\prime}{\bm \Lambda}^\prime{\bm \Lambda}\wideparen{\mbf H} \Vert=o_{\mathrm P}(1)$,
\eqref{eq:semplicesw}, and \eqref{eq:semplicesw2}. Using the last two lines of \eqref{eq:CLTFSW}, by Slutsky's theorem,
\begin{align}
&\sqrt n(\wideparen{\mbf F}_t-\wideparen{\mbf H}^{-1}{\mbf F}_t ) \to_d\mathcal N\l(\mbf 0_r, \bm V_0^{1/2} \bm\Upsilon_0^\prime  (\bm\Gamma^F)^{-1/2}\bm\Pi_t^{\text{\tiny \upshape OLS}}(\bm\Gamma^F)^{-1/2}\bm\Upsilon_0 \bm V_0^{1/2}\r),\label{eq:ANFSW1}\\
&\sqrt n(\wideparen{\mbf F}_t-\wideparen{\mbf H}^{-1}{\mbf F}_t ) \to_d\mathcal N\l(\mbf 0_r, \bm\Upsilon_1^\prime  (\bm\Sigma_\Lambda)^{1/2}\bm\Pi_t^{\text{\tiny \upshape OLS}} (\bm\Sigma_\Lambda)^{1/2} \bm\Upsilon_1\r),\label{eq:ANFSW2}
\end{align}
 where $\bm\Pi_t^{\text{\tiny \upshape OLS}}=(\bm\Sigma_\Lambda)^{-1}\bm\Gamma_t(\bm\Sigma_\Lambda)^{-1}$ with $\bm\Gamma_t$ defined in Assumption \ref{ass:CLT}(b). 

\subsection{Approach B2} \label{sec:cmppcaD}


\paragraph{Consistency.} 
Clearly, \eqref{eq:prop1sw}  still holds with the same $\bm{\mathcal H}$. 
Moreover, by definition of eigenvectors, we get
\beq\label{eq:EVECD}
\frac{\bm X\bm X^\prime}{nT}\bar{\bm F}=\bar{\bm F}\frac{\wt{\mbf M}^x}{T}.
\eeq
By using $\bm X=\bm F\bm\Lambda^\prime+\bm E$ and taking the $t$th row of \eqref{eq:EVECD}, for any $t=1,\ldots,T$,
\begin{align}
\bar{\mbf F}_t^\prime -  \mbf F_t^\prime \bar{\mbf H} &= \l(
\frac 1{nT} {\mbf F}_t^\prime \sum_{i=1}^n \sum_{s=1}^T  \bm\lambda_i  e_{is} \bar{\mbf F}_s^\prime
+
\frac 1{nT} \sum_{i=1}^n  e_{it}\bm\lambda_i^\prime \sum_{s=1}^T {\mbf F}_s \bar{\mbf F}_s^\prime
+
\frac 1{nT} \sum_{i=1}^n \sum_{s=1}^T e_{is} e_{it}\bar{\mbf F}_s^\prime
\r) \l(\frac{\wt {\mbf M}^x}{T}\r)^{-1},\label{eq:sviluppoLambdaD}
\end{align}
which is equivalent to the expression in \eqref{eq:sviluppoLambdaBAI}, and where 
\beq\label{eq:Hbar}
\bar{\mbf H}:=\frac{\bm\Lambda^\prime\bm\Lambda}{n}\frac{\bm F^\prime\bar{\bm F}}{T} \l(\frac{\wt {\mbf M}^x}{T}\r)^{-1}.
\eeq
Notice that $\Vert\bar{\mbf H}\Vert= O_{\mathrm P}(1)$ and $\Vert\bar{\mbf H}^{-1}\Vert= O_{\mathrm P}(1)$ because $\bar{\mbf H}$ is equivalent to $\wt{\mbf H}$ (see \eqref{eq:sameHD} below) up to a scaling factor which is  asymptotically finite and  positive definite (see Lemma \ref{lem:covarianze}(iii) joint with Lemmas \ref{lem:Vzero}(i) and \ref{lem:Vzero}(iii)).

From the linear projection of $\bm X^\prime$ onto the estimated factors $\bar{\bm F}$ we get, for any $i=1,\ldots,n$,
\begin{align}
\bar{\bm\lambda}_i^\prime-{\bm\lambda}_i^\prime (\bar{\mbf H}^{-1})^\prime &=
\l(\bm\lambda_i^\prime\frac{(\bm F-\bar{\bm F} \bar{\mbf H}^{-1})^\prime\bar{\bm F} }{T}
+
\bm\varepsilon_i^\prime\frac{(\bar{\bm F}- \bm F\bar{\mbf H} )}{T}
+
\frac{\bm\varepsilon_i^\prime\bm F\bar{\mbf H}}{T}\r) \l(\frac{\wt {\mbf M}^x}{T}\r)^{-1}
,\label{eq:sviluppoFactorD}
\end{align}
which is equivalent to the expression in \eqref{eq:sviluppoFactorBAI}, but for the scaling term $(T^{-1}\wt{\mbf M}^x)^{-1}$, which, however, is irrelevant being $O_{\mathrm P}(1)$ by Lemma \ref{lem:MO1}(iv). Hence, from \eqref{eq:sviluppoLambdaD} and \eqref{eq:sviluppoFactorD} 
we can prove consistency using the results in \citet[Lemmas A.2, B.1, and B.3]{bai03}, which hold by virtue of Table \ref{tab:ass}:
\begin{align}
&\l\Vert \bar{\bm\lambda}_i-\bar{\mbf H}^{-1}{\bm\lambda}_i\r\Vert = O_{\mathrm P} \l(\max\l(\frac 1{ n},\frac 1{\sqrt T}\r)\r),\quad\l\Vert \bar{\mbf F}_t-\bar{\mbf H}^{\prime}{\mbf F}_t \r\Vert = O_{\mathrm P} \l(\max\l(\frac 1{ \sqrt n},\frac 1{T}\r)\r),\nn
\end{align}
which are the analogue of Propositions \ref{prop:L2} and \ref{prop:F2}, but when stated using $\bar{\mbf H}$ in place of $\wh{\mbf H}$.

\paragraph{The role of $\bar{\mbf H}$.} By using the same arguments in \citet[Lemma B.3]{bai03}, it is straightforward to see that $\bar{\mbf H}$ satisfies:
\begin{align}
&\l\Vert \bar{\mbf H}- ({\bm F}^\prime {\bm F})^{-1}{\bm F}^\prime\bar{\bm F} \r\Vert = O_{\mathrm P}\l(\max\l(\frac 1n,\frac 1{\sqrt{nT}},\frac 1T\r)\r),\label{eq:prop9HtildeD}
\end{align}
which is the analogue of Proposition \ref{prop:LLFF}(b). This immediately implies also the analogue of Propositions \ref{prop:LLFF}(d), \ref{prop:LLFF}(f), and \ref{prop:LLFF}(h).
Moreover, following the same reasoning used to prove Proposition \ref{prop:HHAT}, but this time using \eqref{eq:prop9HtildeD}, it follows that
\begin{align}
&\l\Vert\bar{\mbf H} - \frac{\bm \Lambda^\prime\bm\Lambda}{n} (\bar{\mbf H}^{-1})^{\prime}\r\Vert =O_{\mathrm P}\l(\max\l(\frac 1n,\frac{1}{\sqrt{nT}},\frac 1T\r)\r),\label{eq:starop1D}\\
&\l\Vert\frac{\wt{\mbf M}^x}{T} - \bar{\mbf H}^{-1}\frac{\bm\Lambda^\prime\bm\Lambda}{n}\frac{\bm F^\prime\bm F}{T}  \bar{\mbf H}\r\Vert=O_{\mathrm P}\l(\max\l(\frac 1n,\frac{1}{\sqrt{nT}},\frac 1T\r)\r).\label{eq:starop2D}
\end{align}
And from \eqref{eq:starop1D} and \eqref{eq:starop2D}, 
\begin{align}
&\l\Vert
\frac{\wt{\mbf M}^x}{T} - \bar{\mbf H}^{-1} \frac{\bm\Lambda^\prime\bm\Lambda}{n}\frac{\bm F^\prime\bm F}{T} \frac{\bm\Lambda^\prime\bm\Lambda}{n} (\bar{\mbf H}^{-1})^{\prime}
\r\Vert=O_{\mathrm P}\l(\max\l(\frac 1n,\frac{1}{\sqrt{nT}},\frac 1T\r)\r),\label{eq:starop3D}
\end{align}
from which it follows that we must have
\begin{align}
&\l\Vert \bar{\mbf H} -  \l(\frac{\bm\Lambda^\prime\bm\Lambda}{n}\r)^{1/2} \wideparen {\mbf Q}\r\Vert = O_{\mathrm P}\l(\max\l(\frac 1n,\frac{1}{\sqrt{T}}\r)\r),\label{eq:sameQswD}\\
&\l\Vert \bar{\mbf H}^{-1} - \wideparen {\mbf Q}^\prime \l(\frac{\bm\Lambda^\prime\bm\Lambda}{n}\r)^{-1/2}\r\Vert = O_{\mathrm P}\l(\max\l(\frac 1n,\frac{1}{\sqrt{T}}\r)\r) ,\label{eq:sameQinvswD}
\end{align}
where we recall that $\wideparen {\mbf Q}$ are the normalized eigenvectors of $(n^{-1}\bm \Lambda^\prime\bm \Lambda)^{1/2} (T^{-1}\bm F^\prime\bm F) (n^{-1}\bm \Lambda^\prime\bm \Lambda)^{1/2}$,  which has $n^{-1}\wh{\mbf M}^C$ as eigenvalues, and these, in turn, are such that $n^{-1}\Vert \wh{\mbf M}^C-\wh{\mbf M}^x\Vert = O_{\mathrm P}(\max(n^{-1},T^{-1/2}))$ with
$T^{-1}\wt{\mbf M}^x=n^{-1}\wh{\mbf M}^x$. Results  \eqref{eq:sameQswD} and \eqref{eq:sameQinvswD} are  the analogue of Proposition \ref{prop:HHAT}.
Once again, we see that the loadings and the factors can be consistently estimated  up to: (i) a scale $(n^{-1}{\bm \Lambda^\prime\bm \Lambda})^{1/2}$ and (ii) a rotation $\wideparen{\mbf Q}$. 

We can then derive a limit for $\bar{\mbf H}$ which depends only on population quantities.  By Assumptions \ref{ass:common}(a) and \ref{ass:common}(c-ii),  $\Vert
(n^{-1}\bm \Lambda^\prime\bm \Lambda)^{1/2} (T^{-1}\bm F^\prime\bm F) (n^{-1}\bm \Lambda^\prime\bm \Lambda)^{1/2}-
(\bm\Sigma_\Lambda)^{1/2}(\bm\Gamma^F)(\bm\Sigma_\Lambda)^{1/2}\Vert =o_{\mathrm P}(1)$, by Assumption \ref{ass:eval} and Davis-Kahan theorem \citep[Corollary 1]{yu15}, the corresponding normalized eigenvectors satisfy:
$\Vert\wideparen{\mbf Q}-\bm\Upsilon_1\bm{\mathcal J}_1\Vert=o_{\mathrm P}(1)$. Hence, from \eqref{eq:sameQswD} and using also Assumption \ref{ass:common}(a),
\beq\label{eq:sempliceD}
\l\Vert\bar{\mbf H} - (\bm\Sigma_\Lambda)^{1/2}\bm\Upsilon_1\bm{\mathcal J}_1\r\Vert=o_{\mathrm P}(1) \;\text{ and }\; 
\l\Vert\bar{\mbf H}^{-1} - \bm{\mathcal J}_1\bm\Upsilon_1^\prime  (\bm\Sigma_\Lambda)^{-1/2}\r\Vert=o_{\mathrm P}(1),
\eeq
which is the analogue of \eqref{eq:ennesimaespansione}. If Assumption \ref{ass:common}(a) holds with rate $\sqrt n$ and Assumption \ref{ass:common}(c-ii) holds with rate $\sqrt T$, the rate in 
 \eqref{eq:sempliceD} is $\min(\sqrt n,\sqrt T)$.
 
\paragraph{The relation between $\bar{\mbf H}$, $\wh{\mbf H}$, $\wt{\mbf H}$, and $\wideparen{\mbf H}$.}  By definition, $T^{-1/2}\wt{\bm F} = \bar{\bm F} (\wt{\mbf M}^x)^{-1/2}=\wt{\mbf V}^x$ (see \eqref{eq:estFbai} and \eqref{eq:estFig}). Thus, by comparing the definitions of $\wt{\mbf H}$ and $\bar{\mbf H}$ in \eqref{eq:Htilde} and \eqref{eq:Hbar}, respectively, we see that
\beq
\bar{\mbf H} = \wt{\mbf H} \l(\frac{\wt{\mbf M}^x}{T} \r)^{1/2}\;\text{ and }\qquad
\bar{\mbf H}^{-1} = \l(\frac{\wt{\mbf M}^x}{T} \r)^{-1/2}\wt{\mbf H}^{-1} 
.\label{eq:sameHD}
\eeq 
By \eqref{eq:sameH} and \eqref{eq:sameHD}, we also have
\begin{align}
&\l\Vert\bar{\mbf H}-(\wh{\mbf H}^{-1})^\prime \l(\frac{\wt{\mbf M}^x}{T}\r)^{1/2} \r\Vert = O_{\mathrm P}\l(\max\l(\frac 1n,\frac 1{\sqrt{nT}},\frac 1T\r)\r),\label{eq:sameHD3}\\
&\l\Vert\bar{\mbf H}^{-1}-\l(\frac{\wt{\mbf M}^x}{T}\r)^{-1/2}
\wh{\mbf H}^\prime\r\Vert = O_{\mathrm P}\l(\max\l(\frac 1n,\frac 1{\sqrt{nT}},\frac 1T\r)\r). \nn
\end{align}
Moreover, from \eqref{eq:sameHD} and \eqref{eq:sameHsw2} and since $T^{-1}\wt{\mbf M}^x=n^{-1}\wh{\mbf M}^x$, we also have
\begin{align}
&\l\Vert\bar{\mbf H}-(\wideparen{\mbf H}^{-1})^\prime\r\Vert = O_{\mathrm P}\l(\max\l(\frac 1n,\frac 1{\sqrt{nT}},\frac 1T\r)\r),\label{eq:sameHD2}\\
&\l\Vert\bar{\mbf H}^{-1}-\wideparen{\mbf H}^\prime \r\Vert = O_{\mathrm P}\l(\max\l(\frac 1n,\frac 1{\sqrt{nT}},\frac 1T\r)\r).\nn
\end{align}
And, from \eqref{eq:sameHD} and \eqref{eq:prop9HtildeL}, it follows that
\begin{align}
&\l\Vert  \bar{\mbf H}^{-1} - \bar{\bm\Lambda}^\prime{\bm\Lambda}(\bm\Lambda^\prime\bm\Lambda)^{-1}\r\Vert= O_{\mathrm P}\l(\max\l(\frac 1n,\frac 1{\sqrt{nT}},\frac 1T\r)\r),\label{eq:prop9HtildeLD}
\end{align}
where we used also the fact that $\bar{\bm \Lambda}=\wt{\bm \Lambda}(T^{-1}\wt{\mbf M}^x)^{-1/2}$. This is the analogue of Proposition \ref{prop:LLFF}(a). The analogues of Propositions \ref{prop:LLFF}(c), \ref{prop:LLFF}(e), and \ref{prop:LLFF}(g) follow immediately.
Last, from \eqref{eq:Htildelim2} and \eqref{eq:sameHD}, by using also Lemma \ref{lem:Vzero}(ii), we get
\beq
\l\Vert \bar{\mbf H}-  (\bm\Gamma^F)^{-1/2} \bm\Upsilon_0\bm{\mathcal J}_0 \bm V_0^{1/2}\r\Vert = o_{\mathrm P}(1)
 \;\text{ and }\;
\l\Vert \bar{\mbf H}^{-1}-  \bm V_0^{-1/2}\bm{\mathcal J}_0 \bm\Upsilon_0^\prime (\bm\Gamma^F)^{1/2}\r\Vert = o_{\mathrm P}(1), \label{eq:sempliceD2}
\eeq
which is the analogue of Proposition \ref{cor:semplice}. Once again,
if Assumption \ref{ass:common}(a) holds with rate $\sqrt n$ and Assumption \ref{ass:common}(c-ii) holds with rate $\sqrt T$,
 the rate in 
 \eqref{eq:sempliceD2} is $\min(\sqrt n,\sqrt T)$. Note that by comparing \eqref{eq:sempliceD} and \eqref{eq:sempliceD2}, we prove again \eqref{eq:riunioni2} and \eqref{eq:riunioni}.

\paragraph{Asymptotic normality.}  From \eqref{eq:sviluppoLambdaD} and \citet[Theorem 1]{bai03}, if $\sqrt n/T\to 0$, as $n,T\to\infty$,
\begin{align}
\sqrt n(\bar{\mbf F}_t&-\bar{\mbf H}^{\prime}{\mbf F}_t) =\l(\frac{\wt {\mbf M}^x}{T} \r)^{-1} \frac{\bar{\bm F}^\prime \bm F}{T}\l(\frac 1{\sqrt n}\sum_{i=1}^n {\bm\lambda}_i e_{it}\r)+o_{\mathrm P}(1)= \bar{\mbf H}^\prime \l(\frac{\bm\Lambda^\prime\bm\Lambda}{n}\r)^{-1}\l(\frac 1{\sqrt n}\sum_{i=1}^n {\bm\lambda}_i e_{it}\r)+o_{\mathrm P}(1)\nn\\
& = \l\{ \underset{n,T\to\infty}{\text{P-lim}} \bar{\mbf H}^\prime \r\}
 \l(\frac{\bm\Lambda^\prime\bm\Lambda}{n}\r)^{-1}\l(\frac 1{\sqrt n}\sum_{i=1}^n \bm\lambda_i e_{it}\r)+o_{\mathrm P}(1)=  \bm{\mathcal J}_1\bm\Upsilon_1^\prime (\bm\Sigma_\Lambda)^{1/2}  \sqrt n(\wh{\mbf F}_t^{\text{\tiny OLS}}-\mbf F_t)+ o_{\mathrm P}(1)\nn\\
&=\bm V_0^{1/2} \bm{\mathcal J}_0\bm\Upsilon_0^\prime  (\bm\Gamma^F)^{-1/2} \sqrt n(\wh{\mbf F}_t^{\text{\tiny OLS}}-\mbf F_t)+ o_{\mathrm P}(1).
\label{eq:CLTFD}
\end{align}
where we used  \eqref{eq:Hbar}, \eqref{eq:sempliceD}, and \eqref{eq:sempliceD2}. The last expression in \eqref{eq:CLTFD} coincides with \eqref{eq:CLTFSW}, hence,  by Slutsky's theorem, 
\begin{align}
&\sqrt n(\bar{\mbf F}_t-\bar{\mbf H}^{\prime}{\mbf F}_t ) \to_d\mathcal N\l(\mbf 0_r, \bm V_0^{1/2} \bm\Upsilon_0^\prime  (\bm\Gamma^F)^{-1/2}\bm\Pi_t^{\text{\tiny \upshape OLS}}(\bm\Gamma^F)^{-1/2}\bm\Upsilon_0 \bm V_0^{1/2}\r),\label{eq:ANFD1}\\
&\sqrt n(\bar{\mbf F}_t-\bar{\mbf H}^{\prime}{\mbf F}_t ) \to_d\mathcal N\l(\mbf 0_r, \bm\Upsilon_1^\prime  (\bm\Sigma_\Lambda)^{1/2}\bm\Pi_t^{\text{\tiny \upshape OLS}} (\bm\Sigma_\Lambda)^{1/2} \bm\Upsilon_1\r),\label{eq:ANFD2}
\end{align}
 where $\bm\Pi_t^{\text{\tiny \upshape OLS}}=(\bm\Sigma_\Lambda)^{-1}\bm\Gamma_t(\bm\Sigma_\Lambda)^{-1}$ with $\bm\Gamma_t$ defined in Assumption \ref{ass:CLT}(b). 
 
 Similarly, from \eqref{eq:sviluppoFactorD} and \citet[Theorem 2]{bai03}, if $\sqrt T/n\to 0$, as $n,T\to\infty$,
\begin{align}
\sqrt T(\bar{\bm\lambda}_i&-\bar{\mbf H}^{-1}{\bm\lambda}_i) 
=\l(\frac{\wt {\mbf M}^x}{T} \r)^{-1} \bar{\mbf H}^\prime \l(\frac 1{\sqrt T}\sum_{t=1}^T \mbf F_t e_{it}\r)+o_{\mathrm P}(1)=\l(\frac{\bar{\bm F}^\prime\bar{\bm F}}{T}\r)^{-1}  \bar{\mbf H}^\prime \l(\frac 1{\sqrt T}\sum_{t=1}^T \mbf F_t e_{it}\r)+o_{\mathrm P}(1)\nn\\
&=\l(\frac{\bar{\mbf H}^{\prime}{\bm F}^\prime{\bm F}\bar{\mbf H}}{T}\r)^{-1}  \bar{\mbf H}^\prime \l(\frac 1{\sqrt T}\sum_{t=1}^T \mbf F_t e_{it}\r)+o_{\mathrm P}(1)= \bar{\mbf H}^{-1}\l(\frac{{\bm F}^\prime{\bm F}}{T}\r)^{-1}\l(\frac 1{\sqrt T}\sum_{t=1}^T \mbf F_t e_{it}\r)+o_{\mathrm P}(1)\nn\\
&=\l\{ \underset{n,T\to\infty}{\text{P-lim}} \bar{\mbf H}^{-1} \r\}\l(\frac{{\bm F}^\prime{\bm F}}{T}\r)^{-1}\l(\frac 1{\sqrt T}\sum_{t=1}^T \mbf F_t e_{it}\r)+o_{\mathrm P}(1)= \bm{\mathcal J}_1\bm\Upsilon_1^\prime (\bm\Sigma_\Lambda)^{-1/2} \sqrt T(\wh{\bm\lambda}_i^{\text{\tiny OLS}}-{\bm\lambda}_i) +o_{\mathrm P}(1)\nn\\
&=  \bm V_0^{-1/2}\bm{\mathcal J}_0\bm\Upsilon_0^\prime(\bm\Gamma^F)^{1/2}\sqrt T(\wh{\bm\lambda}_i^{\text{\tiny OLS}}-{\bm\lambda}_i) +o_{\mathrm P}(1)
,\label{eq:CLTLD}
\end{align}
where we used 
the fact that $T^{-1}\bar{\mbf F}^\prime\bar{\mbf F}=\mbf I_r$ by definition, 
\eqref{eq:prop9HtildeD} which implies $\Vert T^{-1}\bar{\bm F}^\prime\bar{\bm F}-T^{-1} \bar{\mbf H}^{\prime}{\bm F}^\prime{\bm F}\bar{\mbf H} \Vert=o_{\mathrm P}(1)$,
 \eqref{eq:sempliceD}, and \eqref{eq:sempliceD2}. The last expression in \eqref{eq:CLTLD} coincides with \eqref{eq:CLTLSW}, hence,  by Slutsky's theorem, 
\begin{align}
&\sqrt T(\bar{\bm\lambda}_i-\bar{\mbf H}^{-1}{\bm\lambda}_i) \to_d\mathcal N\l(\mbf 0_r,  \bm V_0^{-1/2} \bm\Upsilon_0^\prime(\bm\Gamma^F)^{1/2} \bm\Theta_i^{\text{\tiny \upshape OLS}}(\bm\Gamma^F)^{1/2}\bm\Upsilon_0\bm V_0^{-1/2}
\r),\label{eq:ANLD1}\\
&\sqrt T(\bar{\bm\lambda}_i-\bar{\mbf H}^{-1}{\bm\lambda}_i) \to_d\mathcal N\l(\mbf 0_r,  \bm\Upsilon_1^\prime (\bm\Sigma_\Lambda)^{-1/2} 
\bm\Theta_i^{\text{\tiny \upshape OLS}}
(\bm\Sigma_\Lambda)^{-1/2} \bm\Upsilon_1
\r),\label{eq:ANLD2}
\end{align}
where $\bm\Theta_i^{\text{\tiny \upshape OLS}}=(\bm\Gamma^F)^{-1}\bm\Phi_i(\bm\Gamma^F)^{-1}$ with  $\bm\Phi_i$ defined in Assumption \ref{ass:CLT}(a). 

\setcounter{equation}{0}
\section{Estimation of the covariance matrix}\label{sec:hannan}

To be able to apply PCA and, in particular, to prove Proposition \ref{prop:L}, we need a consistent estimator of the covariance matrix of $\{\mbf x_t\}$. In fact, 
for factor analysis, we do not need a consistency result for the whole matrix, but we just need to ensure that
\beq\label{eq:xxGx}
\frac 1n \l\Vert\wh{\bm\Gamma}^x-\bm\Gamma^x \r\Vert= O_{\mathrm P}\l(\frac 1 {\sqrt T}\r)\; \text{ or }\;
\frac 1n \l\Vert\wh{\bm\Gamma}^x-\bm\Gamma^x \r\Vert= O_{\mathrm {m.s.}}\l(\frac 1 {\sqrt T}\r).
\eeq
Thus, even if when $n>T$ the sample covariance matrix is not positive definite, this does not affect PCA nor its asymptotic properties. This is what distinguishes a factor model with pervasive factors, i.e., where the leading eigenvalues explode as $n\to\infty$, from the spiked covariance model studied in random matrix theory where all eigenvalues are bounded even when $n\to\infty$, thus corresponding to weak factors \citep{onatski2012asymptotics}. While in the former case, by allowing few eigenvalues to dominate, we are essentially considering a low-dimensional problem, in the latter case, we are effectively facing a high-dimensional problem as all $n$ eigenvalues are equally important, thus, PCA would require proving a much stronger consistency result where no rescaling is applied. See also \citet{hallin2022manfred} for a comparison between the two settings.

A proof of  \eqref{eq:xxGx}  is obtained directly from Assumption \ref{ass:common}(c-ii) for estimation of the covariance of the factors and from Assumption \ref{ass:idio}(c-ii) for estimation of the covariance of the idiosyncratic components (see Lemma \ref{lem:covarianze}(i-a)). Both these assumptions are high-level ones and it is natural to ask when  they are satisfied. 

In this section, we introduce four different sets of more primitive conditions on the data generating process of $\{\mbf F_t\}$ and $\{ e_{it}\}$, which ensure  that  \eqref{eq:xxGx} holds. Specifically, the assumed conditions ensure that (see Lemma \ref{lem:covarianze}(i-b) for a proof under Assumptions \ref{ass:Wold}, or \ref{ass:Hannan}, or \ref{ass:Wu}, below)
\beq\label{eq:A1c}
\l\Vert \frac 1 T\sum_{t=1}^T\l\{\mbf F_t\mbf F_t^\prime -\bm\Gamma^F\r\}\r\Vert = O_{\mathrm {P}}\l(\frac 1{\sqrt T}\r)\; \text{ or }\; \l\Vert \frac 1 T\sum_{t=1}^T\l\{\mbf F_t\mbf F_t^\prime -\bm\Gamma^F\r\}\r\Vert = O_{\mathrm {m.s.}}\l(\frac 1{\sqrt T}\r),
\eeq
and that, for all $i,j\in\mathbb N$,
\begin{align}
\l\vert \frac 1 T\sum_{t=1}^T \l\{ e_{it} e_{jt} -\E[ e_{it} e_{jt}]\r\}\r\vert =O_{\mathrm{P}}\l(\frac 1{\sqrt T}\r)\;&\text{ or }\;\l\vert \frac 1 T\sum_{t=1}^T \l\{ e_{it} e_{jt} -\E[ e_{it} e_{jt}]\r\}\r\vert =O_{\mathrm {m.s.}}\l(\frac 1{\sqrt T}\r),\label{eq:BBB}\\
\l\Vert\frac 1T\sum_{t=1}^T \mbf F_{t} e_{it}\r\Vert = O_{\mathrm {P}}\l(\frac 1{\sqrt T}\r)\;&\text{ or }\l\Vert\frac 1T\sum_{t=1}^T \mbf F_{t} e_{it}\r\Vert = O_{\mathrm {m.s.}}\l(\frac 1{\sqrt T}\r).\label{eq:Covzero}
\end{align}

\subsection{Estimation of the covariances of the factors and the idiosyncratic components}

\paragraph{Linear representation.}
Let us assume that $\{\mbf F_t\}$ admits the following linear representation:
\beq
\mbf F_t :=\sum_{j=1}^r \sum_{k=0}^\infty \mbf c_{kj} u_{j,t-k}= \mbf C(L) \mbf u_t,\label{eq:SDFM2R}
\eeq
where $\mbf C(L)=\sum_{k=0}^\infty \mbf C_k L^k$ is an $r\times r$ matrix of polynomials with coefficients $\mbf C_k:=(\mbf c_{k1}\cdots \mbf c_{kr})$.  We then characterize \eqref{eq:SDFM2R} by means of the following assumption.
\begin{ass}\label{ass:Wold}
$\{\mbf F_t\}$ satisfies \eqref{eq:SDFM2R} and it is such that:
\begin{compactenum}	
\item [(W1)] $\mbf {C}(z)=\sum_{k=0}^{\infty} \mbf{C}_{k}z^{k}$, where ${\mbf C}_{k}$ are $r\times r$ and $\sum_{k=0}^\infty \Vert \mbf C_k\Vert\le M_C$ for some finite positive real $M_C$;

 \item [(W2)] for all $t\in\mathbb Z$, $\E_{}[\mbf u_t]=\mbf 0_r$ and $\E_{}[\mbf u_t\mbf u_t^\prime]=\mbf I_r$;
 
 \item [(W3)] for all $t\in\mathbb Z$ and all $k\in\mathbb Z$ with $k\ne 0$, $\mbf u_t$ is independent of $\mbf u_{t-k}$;
 
 \item [(W4)] for all $t\in\mathbb Z$, $\E_{}[\Vert \mbf u_{t}\Vert^{4}]\le K_u$ for some finite positive real $K_u$ independent of $t$. 
 
\end{compactenum}
\end{ass}

Because of Assumption \ref{ass:Wold}(W1)-\ref{ass:Wold}(W4) and by using explicitly \eqref{eq:SDFM2R}, we can prove that \eqref{eq:A1c} holds (see Lemma \ref{lem:covarianzeF}(i)). 
Independence  of the innovations in Assumption \ref{ass:Wold}(W3) might be too much to ask, and, by relaxing it we could allow, for example,  for conditional heteroskedasticity in $\{\mbf F_t\}$.  This can be done by asking that for all $j_1,j_2,j_3,j_4=1,\ldots, r$, 
\beq\label{19}
\frac 1 {T} \sum_{t,s=1}^T \l\vert\E[u_{j_1t}u_{j_2t} u_{j_3s}u_{j_4s}]\r\vert \le C_u,\qquad \frac 1 {T} \sum_{t,s=1}^T \l\vert\E[u_{j_1t}u_{j_2t}]\E[u_{j_3s}u_{j_4s}]\r\vert \le C_u,
\eeq
for some finite positive real $C_u$ independent of $t$, $s$, $j_1$, $j_2$, $j_3$, and $j_4$.
Notice that the second condition in \eqref{19} holds trivially if $\{\mbf u_t\}$ is a white noise process. In this case, however, we should  change Assumption \ref{ass:Wold}(W4) by asking that for all $j_1,j_2,j_3,j_4=1,\ldots, r$, all $t_1,t_2,t_3,t_4=1,\ldots, T$ and all $T\in\mathbb N$, $\E[\vert u_{j_1t_1}u_{j_2t_2}u_{j_3t_3}u_{j_4t_4}\vert]\le K_u$ for some finite positive real $K_u$ independent of $j_1,j_2,j_3,j_4, t_1,t_2,t_3,t_4$, and $T$.  These changes would still ensure that $T^{-1}\sum_{t=1}^T \mbf u_t\mbf u_t^\prime$ is a  consistent estimator of the covariance matrix of $\{\mbf u_t\}$ (see also the arguments below for $\{\mbf F_t\}$). 

Likewise, we could assume a linear representation for the idiosyncratic components, i.e., for all $i\in\mathbb N$,
$
 e_{it}=\sum_{j=1}^n\sum_{k=0}^\infty \beta_{k,ij} \eta_{j,t-k}.
$
Then, letting $\bm\eta_t=(\eta_{1t}\cdots\eta_{nt})^\prime$, we can assume the analogue of Assumptions \ref{ass:Wold}(W1)-\ref{ass:Wold}(W4). Namely, 
{\it 
\begin{inparaenum}
\item [(WI1) ]For all $k\in\mathbb Z^+$ and all $i,j\in\mathbb N$, $\vert\beta_{k,ij}\vert\le \rho^k B_{ij}$ with $0\le \rho<1$, $\sum_{i=1, i\ne j}^n B_{ij}\le B$ and $\sum_{j=1, j\ne i}^n B_{ij}\le B$ for some finite positive real $B$ independent of $i$ and $j$. 
\item [(WI2)] For all $t\in\mathbb Z$ and all $n\in\mathbb N$, $\E[\bm\eta_{t}]=\mbf 0_n$ and $\E[\bm\eta_t\bm\eta_t^\prime]=\mbf I_n$.
\item [(WI3)] For all $t\in\mathbb Z$, all $n\in\mathbb N$, and all $k\in\mathbb Z$ with $k\ne 0$, $\bm\eta_t$ is independent of $\bm \eta_{t-k}$. 
\item [(WI4)] For all $t\in\mathbb Z$ and all $j\in\mathbb N$, $\E_{}[\vert \eta_{jt}\vert^{4}]\le K_\eta$ for some finite positive real $K_\eta$ independent of~$t$~and~$j$. 
\end{inparaenum}
}

Under (WI1)-(WI4) we could prove that \eqref{eq:BBB} still holds. Such a proof would follow the same argument as for the factors and it is therefore omitted. Note also that (WI1)-(WI4) imply also Assumption \ref{ass:idio}(b) which controls serial and cross-sectional idiosyncratic dependences  \citep[Propositions 1 and 4]{FHLZ17}.

\paragraph{Summability of fourth order cumulants.}
Alternatively, we could directly make assumptions on 4th order moments. So for the factors we assume.
\begin{ass}\label{ass:Hannan}
For all $i,j=1,\ldots, r$,
\begin{compactenum}
\item [(H1)] $T^{-1}\sum_{t,s=1}^T\l\vert \E[F_{it}F_{jt}F_{is}F_{js}]\r\vert \le C_F$;
\item [(H2)] $T^{-1}\sum_{t,s=1}^T\l\vert \E[F_{it}F_{jt}]\E[F_{is}F_{js}]\r\vert \le C_F$;
\item [(H3)] $T^{-1}\sum_{t,s=1}^T\l\vert \E[F_{it}F_{is}]\E[F_{jt}F_{js}]\r\vert \le C_F$;
\item [(H4)] $T^{-1}\sum_{t,s=1}^T\l\vert \E[F_{it}F_{js}]\E[F_{is}F_{jt}]\r\vert \le C_F$;
\end{compactenum}
for some finite positive real $C_F$ independent of $i$ and $j$. 
\end{ass}

Then, it is straightforward to see that \eqref{eq:A1c} holds (see Lemma 
\ref{lem:covarianzeF}(i)). Notice that, since $\E[F_{it}]=0$ by Assumption \ref{ass:common}(b), then
\[
\text{cum}_4(F_{it},F_{jt},F_{is},F_{js})=\E[F_{it}F_{jt}F_{is}F_{js}]-\E[F_{it}F_{jt}]\E[F_{is}F_{js}]-\E[F_{it}F_{is}]\E[F_{jt}F_{js}]-\E[F_{it}F_{js}]\E[F_{jt}F_{is}],
\]
and we are in fact assuming finite and summable 4th order cumulants, which is a necessary and sufficient condition for \eqref{eq:A1c} to hold (\citealp[pp. 209-211]{hannan}). This approach is a classical one and it is high-level in that it does not make any specific assumption on the dynamics of $\{\mbf F_t\}$ or $\{ e_{it}\}$ (see, e.g., \citealp{FGLR09}).

Regarding the idiosyncratic component, Assumption \ref{ass:idio}(c-ii) already implies, as a special case,  a condition on summability of 4th order moments along the time dimension as assumed in Assumption \ref{ass:Hannan} for the factors. Hence, it immediately follows that, for any $i,j=1,\ldots,n$ \eqref{eq:BBB} holds (see Lemma \ref{lem:covarianzeF}(ii)).

\paragraph{Physical or functional dependence.} Let $\mathcal F_{\nu,t}$ be the $\sigma$-field generated by $\{\bm \nu_s,\, s\le t\}$ where $\bm \nu_t=(\nu_{1t}\cdots \nu_{rt})^\prime$. Let also $g_{j,u}(\cdot)$ be a real-valued measurable function of $\mathcal F_{\nu,t}$ and define $\mbf G_u(\cdot)=(g_{1,u}(\cdot)\cdots g_{r,u}(\cdot))^\prime$. Then, by following \citet{wu05}, we assume that $\{\mbf F_t\}$ admits the following representation:
\begin{align}
\mbf F_t &:=\sum_{j=1}^r \sum_{k=0}^\infty \mbf c_{kj}^\prime u_{jt}=\mbf C(L)\mbf u_t,\label{eq:SDFM2RWu}\\
\mbf u_{t}&:= \mbf G_{u}(\mathcal F_{\nu,t}).\label{eq:SDFM2RWu2}
\end{align}
For any $\ell=1,\ldots, r$ and $t\in\mathbb Z$, define the element-wise functional dependence measure as
$$
\delta_{t,d,\ell}^u := \{\E[\Vert g_{\ell,u}(\mathcal F_{\nu,t})-g_{\ell,u}(\mathcal F_{\nu,t,\{0\}})\Vert^d ]\}^{1/d},
$$
where $\mathcal F_{\nu,t,\{0\}}$ is the $\sigma$-algebra generated by $\{\bm\nu_t,\ldots,\bm\nu_1,\bm\nu_0^*,\bm\nu_{-1},\ldots\}$ where $\bm\nu_0^*$ is an i.i.d.~copy of $\bm\nu_0$.
Finally, define the uniform dependence adjusted norm as:
\beq\label{eq:PHI}
\Phi_{d,\alpha}^u:=\max_{\ell=1,\ldots, r}\sup_{k\ge 0} (1+k)^\alpha \sum_{t=k}^\infty\delta_{t,d,\ell}^u.
\eeq
We characterize \eqref{eq:SDFM2RWu}-\eqref{eq:SDFM2RWu2} by means of the following assumption.
\begin{ass}\label{ass:Wu}
$\{\mbf F_t\}$ satisfies \eqref{eq:SDFM2RWu}-\eqref{eq:SDFM2RWu2} and it is such that:
\begin{compactenum}	
\item [(P1)] $\mbf {C}(z)=\sum_{k=0}^{\infty} \mbf{C}_{k}z^{k}$, where ${\mbf C}_{k}$ are $r\times r$ and $\sum_{k=0}^\infty \Vert \mbf C_k\Vert\le M_C$ for some finite positive real $M_C$;

 \item [(P2)] for all $t\in\mathbb Z$, $\E_{}[\bm \nu_t]=\mbf 0_r$, $\E_{}[\bm \nu_t\bm \nu_t^\prime]=\mbf I_r$, $\E[\mbf u_t]=\mbf 0_r$, and $\bm\Gamma^u:=\E[\mbf u_t\mbf u_t^\prime]$  is $r\times r$ positive definite and such that $\Vert \bm\Gamma^u\Vert\le M_u$ for some finite positive real $M_u$ independent of $t$;
 
 \item [(P3)] for all $t\in\mathbb Z$ and all $k\in\mathbb Z$ with $k\ne 0$, $\bm \nu_t$ is independent of $\bm \nu_{t-k}$ and $\E[\mbf u_t\mbf u_{t-k}^\prime]=\mbf 0_{r\times r}$;
 
 \item [(P4)] for all  $t\in\mathbb Z$ and all $\ell=1,\ldots, r$,  $\E\l[\l\{g_{\ell,u}(\mathcal F_{\nu,t})\r\}^q\r]\le K_u$ for some $q\ge 4$ and some finite positive real $K_u$ independent of $\ell$ and $t$;

 \item [(P5)] for all $d\le q$ with $q>4$, $\Phi_{d,\alpha}^u\le M_{u,d}$ for some finite positive real $M_{u,d}$ independent of $\alpha$. 
\end{compactenum}
\end{ass}

We can then prove that, under Assumption \ref{ass:Wu}, \eqref{eq:A1c} still holds (see Lemma \ref{lem:covarianzeF}(i)). This is the most recent approach, considered for estimation of the generalized dynamic factor models by, e.g., \citet{FHLZ17} and \citet{BCO22}.

Parts (P1)-(P3) are simply defining the Wold representation of $\{\mbf F_t\}$ as being driven by a weak white noise $\{\mbf u_t\}$ which in turn can be a function of an i.i.d. process $\{\bm\nu_t\}$. Moreover, from (P4) and (P5) it follows that we must have $\E[\vert u_{\ell t}\vert^{4+\epsilon}]\le K_u$ which is slightly stronger than Assumption \ref{ass:Wold}(W4). 

Finally (P5) deserves two comments. First, as seen from the definition of physical dependence \eqref{eq:PHI}, for (P5) to hold we must choose an appropriate value of $\alpha$, which, in principle, is constrained by the degree of serial dependence in  $\{\mbf u_t\}$. Now, since in (P3) we make the natural assumption of $\{\mbf u_t\}$ being a weak white noise, then no constraint on $\alpha$ is needed for (P5) to hold. And the same is true if we assumed a geometric decay of the lag-$k$ autocovariance of $\{\mbf u_t\}$, i.e., $\rho^k$ for some $0<\rho<1$. In general, we could even assume a hyperbolic decay of the lag-$k$ autocovariance of $\{\mbf u_t\}$, i.e., as $(1+k)^{-\varsigma}$, with $\varsigma>2$ and in that case (P5) would hold as long as we choose $\alpha\le \varsigma-1$ (see \citealp[Lemma C.1]{BCO22}). In all cases, since $q>4$, we can choose $\alpha>\frac 12-\frac 2q$, which is the condition needed to apply the results of \citet{zhang2021}.

Second, part (P5) is very general. It includes nonlinear models as threshold autoregressive models, exponential autoregressive models, and conditionally heteroskedastic models, e.g., univariate stationary GARCH(1,1) models, i.e., when  $g_{\ell,u}(\mathcal F_{\nu,t})=(\omega_\ell+\alpha_\ell u_{\ell,t-1}^2+\beta_\ell \sigma_{\ell,t-1}^2)^{1/2} \nu_{\ell t}$ with $\omega_\ell>0$, $\alpha_\ell,\beta_\ell\ge 0$ and $\alpha_\ell+\beta_\ell<1$ for all $\ell=1,\ldots, r$. For all these cases, it is shown in \citet[page 14152]{wu05} that Assumption \ref{ass:Wu}(P5) is satisfied. Obviously, the linear representation in \eqref{eq:SDFM2R} is a special case of \eqref{eq:SDFM2RWu}-\eqref{eq:SDFM2RWu2} when $\bm\nu_t\equiv \mbf u_t$ for all $t\in\mathbb Z$ and $g_{\ell,u}(\mathcal  F_{\nu,t})=u_{\ell t}$ for all $\ell=1,\ldots, r$.  Then, it is easy to see that Assumption \ref{ass:Wu}(P5) holds if we make Assumption \ref{ass:Wold}(W4), i.e., finite 4th moments of innovations. This can be shown using the same arguments in \citet[Proposition 5]{FHLZ17}, but applied to $\{\mbf F_t\}$.

Likewise, we could instead introduce an  $n$-dimensional process $\{\bm\omega_t\}$ such that $\{ e_{it}\}$ follows the representation: 
$ e_{it}=\sum_{j=1}^n\sum_{k=0}^\infty \beta_{k,ij} \eta_{j,t-k}$, with
$\eta_{jt}=g_{j,\eta}(\mathcal F_{\omega,t})$,
where $\mathcal F_{\omega,t}$ is the $\sigma$-field generated by $\{\bm \omega_s,\, s\le t\}$, and $g_{j,\eta}(\cdot)$ is a real-valued measurable function of $\mathcal F_{\omega,t}$. Define the uniform dependence adjusted norm as:
$
\Phi_{d,\alpha}^\eta:=\max_{\ell=1,\ldots, n}\sup_{k\ge 0} (1+k)^\alpha \sum_{t=k}^\infty \{\E[\Vert g_{\ell,\eta}(\mathcal F_{\omega,t})-g_{\ell,\eta}(\mathcal F_{\omega,t,\{0\}})\Vert^d ] \}^{1/d},
$
where $\mathcal F_{\omega,t,\{0\}}$ is the $\sigma$-algebra generated by $\{\bm\omega_t,\ldots,\bm\omega_1,\bm\omega_0^*,\bm\omega_{-1},\ldots\}$ where $\bm\omega_0^*$ is an i.i.d. copy of $\bm\omega_0$. Then, we could assume the analogue of Assumption \ref{ass:Wu}(P1)-\ref{ass:Wu}(P5). Namely,
{\it
\begin{inparaenum}
\item [(PI1) ]For all $k\in\mathbb Z^+$, all $i,j=1,\ldots, n$, and all $n\in\mathbb N$, $\vert\beta_{k,ij}\vert\le \rho^k B_{ij}$ with $0\le \rho<1$, $\sum_{i=1}^n B_{ij}\le B$ and $\sum_{j=1}^n B_{ij}\le B$ for some finite positive real $B$ independent of $i$ and $j$. 
\item [(PI2)] For all  $t\in\mathbb Z$ and all $n\in\mathbb N$,  $\E[\bm\omega_t]=\mbf 0_n$, $\E[\bm\omega_t\bm\omega_t^\prime]=\mbf I_n$, $\E[\bm\eta_{t}]=\mbf 0_n$, and $\bm\Gamma^\eta:=\E[\bm\eta_t\bm\eta_t^\prime]$ is positive definite and such that $\Vert \bm\Gamma^\eta\Vert\le M_\eta$ for some finite positive real $M_\eta$ independent of $t$.
\item [(PI3)]  For all $t\in\mathbb Z$, all $n\in\mathbb N$, and all $k\in\mathbb Z$ with $k\ne 0$, $\bm\omega_t$ is independent of $\bm \omega_{t-k}$~and\, $\E[\bm\eta_t\bm\eta_{t-k}^\prime]~=~\mbf 0_{n\times n}$. 
\item [(PI4)]  For all $t\in\mathbb Z$, all $j=1,\ldots, n$, and all $n\in\mathbb N$, $\E[\l\{g_{j,\eta}(\mathcal F_{\omega,t})\r\}^{q}]\le K_\eta$ for some $q\ge 4$ and some finite positive real $K_\eta$ independent of $t$ and $j$.
\item  [(PI5)] For all $d\le q$ with $q>4$, $\Phi_{d,\alpha}^\eta\le M_{\eta,d}$ for some finite positive real $M_{\eta,d}$ independent of $\alpha$. 
\end{inparaenum}
}
The same comments made for Assumption \ref{ass:Wu}(P5) in the case of the factors, apply also here, in particular notice that (PI4)-(PI5) imply $\E[\vert\eta_{j t}\vert^{4+\epsilon}]\le K_\eta$ which is slightly stronger than with (WI4). Under (PI1)-(PI5) we could prove that \eqref{eq:BBB} still holds. Such a proof would follow the same proof made for the factors and it is therefore omitted.

\paragraph{Mixing.} Finally, we can assume the following:
{\it \begin{inparaenum}
\item [(M1)] $\{\mbf F_t\}$ is strong mixing, i.e., $\alpha$-mixing, with exponentially decaying mixing coefficients 
$\alpha(T)\le \exp(-c_1T^{c_2})$,
for some finite positive reals $c_1$ and $c_2$ independent of $T$.
\item [(M2)] One of the following two conditions holds:
(M2a) $\E[\Vert\mbf F_t\Vert^{4}]\le K_F$ for some finite positive real $K_F$ independent of $t$. 
(M2b) For all $s>0$ and $j=1,\ldots, r$, $\mathrm P(\vert F_{jt}\vert>s)\le \exp(- bs ^k)$  for some finite positive reals $b$ and $k$ independent of $t$ and $j$. 
\end{inparaenum}
}

Notice that (M2b) is much stronger than (M2a), indeed it is equivalent to the Cram\'er condition:\linebreak $\sup_{m\ge 1} r^{-1/k} \l(\E\l[\l\vert F_{jt}\r\vert^m\r]\r)^{1/m} \le K_F^\prime$, for some finite positive real $K_F^\prime$ independent of $t$ and $j$ (\citealp[Section 2]{KC18}), which implies (M2a). The approach based on (M1) and (M2b) is considered in \citet{FLM13}.

From (M1)-(M2) it follows that  $\{\mbf F_t\}$ is ergodic (\citealp[Proposition 3.44]{white01}, and \citealp{rosenblatt1972}), and therefore  $\{\mbf F_t\mbf F_t^\prime\}$ is also ergodic (\citealp[Theorem 3.35]{white01}, and \citealp[ pp. 170, 182]{stout1974almost}), so that \eqref{eq:A1c} holds at least in probability. This is the approach followed, e.g., by \citet{FLM13}.

In the same spirit of (M1)-(M2), we could assume $\{\mbf F_t\}$ to admit the linear representation \eqref{eq:SDFM2R}, with the coefficients $\{\mbf C_k\}$ satisfying Assumption \ref{ass:Wold}(W1) and $\{\mbf u_t\}$ satisfying Assumptions \ref{ass:Wold}(W2)-\ref{ass:Wold}(W3), plus the integral Lipschitz condition:
{\it \begin{inparaenum}
\item [(M3)] for all $t\in\mathbb Z$, $\{\mbf u_t\}$ has pdf $f_{\mbf u_t}(\bm u)$ such that 
$\int_{\mathbb R^r}\l\vert f_{\mbf u_t}(\bm u+\bm v)-f_{\mbf u_t}(\bm v)\r\vert\le C_f\Vert \bm v\Vert
$, for any $\bm v\in\mathbb R^r$ and for some finite positive real $C_f$ independent of $t$.
\end{inparaenum}
}
Then, Assumptions \ref{ass:Wold}(W1)-\ref{ass:Wold}(W3) and (M3) imply that $\{\mbf F_t\}$ is strong mixing with exponentially decaying coefficients, i.e., (M1) holds (see \citealp[Theorem 3.1]{PT85}). Note that independence in Assumption \ref{ass:Wold}(W3) is not strictly necessary for having (M1) to hold, as we could allow for GARCH effects by assuming geometric ergodicity of $\{\mbf u_t\}$ instead. Indeed, geometric ergodicity implies $\beta$-mixing, which implies strong mixing (\citealp{FZ06}).

Likewise, we could assume $\{ e_{it}\}$ to be strong mixing, i.e., $\alpha$-mixing, with exponentially decaying mixing coefficients and that $\E[\vert e_{it}\vert^{4}]\le K_ e$ for some finite positive real $K_ e$ independent of $t$ and $i$. Or we could even assume exponentially decaying tails as in (M2b). This would imply ergodicity for $\{ e_{it} e_{jt}\}$, and, therefore, that \eqref{eq:BBB} holds at least in probability.

\subsection{Estimation of the covariance between factors and idiosyncratic components}\label{sec:covFXI}

From Assumption \ref{ass:ind} it follows that for all $t\in\mathbb Z$, all $i=1,\ldots n$, and all $n\in\mathbb N$, $\E[\mbf F_{t} e_{it}]=\mbf 0_r$. Moreover, as a consequence of Assumption \ref{ass:CLT}(a), we immediately have that \eqref{eq:Covzero} holds in probability.
It is then natural to ask if we can find primitive assumptions such that \eqref{eq:Covzero} or even Assumption \ref{ass:CLT}(a) holds.

Assume that one of the following holds:
{\it 
\begin{inparaenum}
\item [(A)] Assumption \ref{ass:Wold}(W1)-\ref{ass:Wold}(W4) and (WI1)-(WI4) hold with $\{\mbf u_t\}$ and $\{\bm\eta_t\}$ mutually independent. 
\item [(B)] Assumption \ref{ass:Wu}(P1)-\ref{ass:Wu}(P5) and (PI1)-(PI5) hold with $\{\bm \nu_t\}$ and $\{\bm\omega_t\}$ mutually independent. 
\end{inparaenum}
}
It is immediate to show that under (A) \eqref{eq:Covzero} holds, even in mean-square. 
Under (B) the same holds, just notice that because $\{\bm \nu_t\}$ and $\{\bm\omega_t\}$ are mutually independent then also $\{\mbf u_t\}$ and $\{\bm\eta_t\}$ are mutually independent and the uniform dependence adjusted norm of the process $\{(\mbf u_t^\prime,\bm\eta_t^\prime)^\prime\}$ is such that $\Phi_{d,\alpha}^{u\eta}\le \Phi_{d,\alpha}^{u}\Phi_{d,\alpha}^{\eta}\le M_{u,d}M_{\eta,d}$. Then, the results by \citet{zhang2021} can be applied to the estimation of the covariance of the process $\{\mbf F_t e_{it}\}$.

If we assume that $\{\mbf F_t\}$ and $\{ e_{it}\}$ are both strongly mixing with exponentially decaying coefficients and have exponentially decaying tails, then $\{\mbf F_t e_{it}\}$ is also strongly mixing (see, e.g., \citealp[Theorem 5.1.a]{bradley05}) with exponentially decaying coefficients and has exponentially decaying tails (see, e.g., \citealp[Lemma A2]{FLM11}), and by \citet[Theorem 1.4]{ibra62} the CLT for strongly mixing processes as in Assumption \ref{ass:CLT}(a) holds and so \eqref{eq:Covzero} holds.

\section{Asymptotic properties of \texorpdfstring{$\pmb{\mathcal H}$}{cal H} under identifying restrictions}\label{app:calH}

Here we are considering the properties of $\bm{\mathcal H}$ when imposing the restrictions in Proposition \ref{corol:K00}.
Recall from Proposition \ref{prop:L} that $\bm{\mathcal H}$ depends on $T$ only through a sign matrix $\mbf J$. This can be fixed, without loss of generality, by means of the following assumption.


\begin{ass}\label{ass:Jhat}
For all $n,T\in \mathbb N$
$[\wh{\mbf V}^{x\prime}\mbf V^C]_{jj}>0$ for all $j=1,\ldots, r$
\end{ass}

This, jointly with Lemma \ref{lem:covarianze}(iv), implies that $\wh{\mbf V}^{x}$ and $\mbf V^C$ coincide asymptotically as $n,T\to\infty$, so that $\mbf J=\mbf I_r$. 
We then derive the implications for $\bm{\mathcal H}$ as defined in Proposition \ref{prop:L} (see Appendix \ref{corol:K00Hproof} for a proof).\footnote{If we did not make Assumption \ref{ass:Jhat}, then $\bm S$ in Proposition \ref{corol:K00H}(II.a) would depend also on $T$, while convergence in Proposition \ref{corol:K00H}(III.a) would be in probability only, thus requiring also $T\to\infty$. 
Moreover, $\bm S$ and $\bm{\mathcal S}_0$ in Proposition \ref{corol:K00H} would still be diagonal matrices with entries $\pm 1$, but, in general, they would be different from $\bm S$ and $\bm{\mathcal S}_0$ in Proposition \ref{corol:K00}.}


\begin{prop}\label{corol:K00H}
Under Assumptions \ref{ass:common}, \ref{ass:sign_eval}, and \ref{ass:Jhat},
\begin{compactenum}
\item [(I.a)] if $\bm\Gamma^F=\mbf I_r$ and 
$\bm\Lambda$  is unrestricted, then, 
for all $n\in\mathbb N$, $\bm{\mathcal H}^\prime \bm{\mathcal H}=\mbf I_r$;

\item [(II.a)] if $\bm\Gamma^F=\mbf I_r$ and  $n^{-1}\bm\Lambda^\prime\bm \Lambda$  is diagonal for all $n\in\mathbb N$, then, for all $n\in\mathbb N$, $\bm{\mathcal H}=\bm S$;

\item [(III.a)] if $\bm\Gamma^F=\mbf I_r$ and 
$\bm\Sigma_\Lambda$  is diagonal, then, $\lim_{n\to\infty}\Vert\bm{\mathcal H}-\bm{\mathcal S}_0\Vert=0$; 
%

\end{compactenum}
where $\bm S$ is an $r\times r$ diagonal matrix with entries $\pm 1$ depending only on $n$, and $\bm{\mathcal S}_0$ is an $r\times r$ diagonal matrix with entries $\pm 1$ independent of $n$ and $T$.
\end{prop}





Here, we consider only the case of orthonormal factors and orthogonal loadings, which is consistent with the PC estimators considered under approach A.1 or, equivalently, B.1. The case of orthonormal loadings and orthogonal factors, consistent with approaches A.2 and B.2, can be easily derived from the proof of Proposition \ref{corol:K00}, hence it is omitted. Furthermore, for simplicity we also do not cover the cases in which we constrain $T^{-1}\bm F^\prime\bm F$ because $\bm{\mathcal H}$ does not depend on $T$.

Part (I.a) shows that if we restrict only the factors to be orthonormal, then $\bm{\mathcal H}$ is an orthogonal matrix. This is straightforward from its definition in Proposition \ref{prop:L}. 
Under parts (II.a) and (III.a) we see that $\bm{\mathcal H}$ can be reduced exactly or, at least, asymptotically to a diagonal matrix of signs. By comparing Propositions \ref{corol:K00} and \ref{corol:K00H}, from \eqref{eq:phil3}, we also see that $\bm{\mathcal H}_{\infty}=\lim_{n\to\infty}\bm{\mathcal H}$.




\setcounter{equation}{0}
\section{Proofs of main Propositions}\label{app:mainproof}
Throughout, recall that ``for all $n\in\mathbb N$'' always means ``for all $n>N$'' with $N$ defined in Assumption \ref{ass:common}(d).

\subsection{Proof of Proposition \ref{prop:L}} \label{prop:Lproof}
First, since $\bm\Gamma^C=\bm\Lambda \bm\Gamma^F\bm\Lambda^\prime$ has rank $r$, then
\beq\label{eq:effeuno}
\frac{\bm\Gamma^{C}}n = \mbf V^{C}\frac{\mbf M^{C}}n\mbf V^{{C}\prime}= \frac{\bm\Lambda}{\sqrt n}\bm\Gamma^F\frac{\bm\Lambda^\prime}{\sqrt n}. 
\eeq
Thus, the columns of $\frac{\mbf V^{C}(\mbf M^{C})^{1/2}}{\sqrt n}$ and the columns of $\frac{\bm\Lambda(\bm\Gamma^F)^{1/2}}{\sqrt n}$ must span the same space. 
So there exists an $r\times r$ invertible matrix $\mbf K$ such that
\beq\label{eq:kappaproj}
\bm\Lambda(\bm\Gamma^F)^{1/2}\mbf K= {\mbf V^{C}(\mbf M^{C})^{1/2}}.
\eeq
In Lemma \ref{lem:KO1} it is shown that $\mbf K$ is finite and invertible.
Let
%
\beq\label{eq:mcHbis}
\bm{\mathcal H}=(\bm\Gamma^F )^{1/2}\mbf K\mbf J,
\eeq
where $\mbf J$ is a diagonal $r\times r$ matrix with entries $\pm 1$ and independent of $n$ as defined in Lemma \ref{lem:covarianze}.
Therefore, from \eqref{eq:kappaproj} and \eqref{eq:mcHbis}, 
\beq\label{eq:drin}
\bm\Lambda\bm{\mathcal H}= {\mbf V}^{C} ( {\mbf M}^{C})^{1/2}\mbf J.
\eeq
Note that $\bm{\mathcal H}$ is finite and invertible because of Lemma \ref{lem:HO1bis}. 

Now, 
because of Lemmas \ref{lem:covarianze}(iii), \ref{lem:covarianze}(iv), \ref{lem:MO1}(i), using \eqref{eq:estL} and \eqref{eq:drin},
\begin{align}
\l\Vert\frac { \wh{\bm\Lambda} - \bm\Lambda\bm{\mathcal H}}{\sqrt n}\r\Vert=&\,
\l\Vert\wh{\mbf V}^x\l(\frac{\wh{\mbf M}^x}n\r)^{1/2} - {\mbf V}^{C} \l(\frac{ {\mbf M}^{C}}n\r)^{1/2} \mbf J\r\Vert=\ \l\Vert\wh{\mbf V}^x\l(\frac{\wh{\mbf M}^x}n\r)^{1/2} - {\mbf V}^{C}\mbf J \l(\frac{ {\mbf M}^{C}}n\r)^{1/2} \r\Vert\nn\\
\le&\, \l\Vert \wh{\mbf V}^x-{\mbf V}^{C}\mbf J\r\Vert\,\l\Vert\l(\frac{\mbf M^{C}}{n}\r)^{1/2} \r\Vert+ 
\l\Vert\frac 1 {\sqrt n}\l\{\l(\wh{\mbf M}^x\r)^{1/2}-\l(\mbf M^{C}\r)^{1/2}\r\}\r\Vert \,\l\Vert \mbf V^{C}\r\Vert\nn\\
&+ \l\Vert \wh{\mbf V}^x-{\mbf V}^{C}\mbf J\r\Vert\,
\l\Vert\frac 1 {\sqrt n}\l\{\l(\wh{\mbf M}^x\r)^{1/2}-\l(\mbf M^{C}\r)^{1/2}\r\}\r\Vert\nn\\
&= O_{\mathrm P}\l(\max\l(\frac 1 n,\frac 1{\sqrt T}\r)\r)+ O_{\mathrm P}\l(\max\l(\frac 1 {n^2},\frac 1{T}\r)\r), \nn
\end{align}
since $\Vert \mbf V^{C}\Vert=1$ because eigenvectors are normalized.
This proves part (a). 

For part (b), since from \eqref{eq:estL}, for any $i=1,\ldots, n$, 
\beq\label{eq:estLi}
\wh{\bm\lambda}_i^\prime = \wh{\mbf v}^{x\prime}_i(\wh{\mbf M}^x)^{1/2},
\eeq
then, because of Lemma \ref{lem:covarianze}(iii), \ref{lem:covarianzerighe}(ii), and \ref{lem:covarianzerighe}(iii), using  \eqref{eq:mcHbis} and \eqref{eq:estLi}, 
\begin{align}
\l\Vert\wh{\bm\lambda}_i^\prime - \bm\lambda_i^\prime\bm{\mathcal H}\r\Vert=&\,
\l\Vert\sqrt n\wh{\mbf v}^{x\prime}_i\l(\frac{\wh{\mbf M}^x}n\r)^{1/2} - \sqrt n{\mbf v}_i^{{C}\prime} \mbf J \l(\frac{ {\mbf M}^{C}}n\r)^{1/2}\r\Vert\nn\\
\le&\, \l\Vert \sqrt n\wh{\mbf v}_i^{x\prime}-\sqrt n{\mbf v}_i^{{C}\prime}\mbf J\r\Vert\,\l\Vert\l(\frac{\mbf M^{C}}{n}\r)^{1/2} \r\Vert+ \l\Vert\frac 1 {\sqrt n}\l\{\l(\wh{\mbf M}^x\r)^{1/2}-\l(\mbf M^{C}\r)^{1/2}\r\}\r\Vert\,
\l\Vert\sqrt n \mbf v_i^{{C}\prime}\r\Vert\nn\\
&+\l\Vert \sqrt n\wh{\mbf v}_i^{x\prime}-\sqrt n{\mbf v}_i^{{C}\prime}\mbf J\r\Vert
\,\l\Vert\frac 1 {\sqrt n}\l\{\l(\wh{\mbf M}^x\r)^{1/2}-\l(\mbf M^{C}\r)^{1/2}\r\}\r\Vert\nn\\
&= O_{\mathrm P}\l(\max\l(\frac 1{\sqrt n},\frac 1{\sqrt T}\r)\r)+ O_{\mathrm P}\l(\max\l(\frac 1 {n},\frac 1{T}\r)\r). \nn
\end{align}
This proves part (b).

For part (c), using \eqref{eq:estF}
\begin{align}
\wh{\bm F}&=\bm X\wh{\bm\Lambda}(\wh{\bm\Lambda}^\prime\wh{\bm\Lambda})^{-1}= \frac{\bm X\wh{\bm\Lambda}}{n}\l(\frac{\wh{\mbf M}^x}{n}\r)^{-1}= \frac{\bm X(\wh{\bm\Lambda}-\bm\Lambda\bm{\mathcal H}+\bm\Lambda\bm{\mathcal H})}{n}\l(\frac{\wh{\mbf M}^x}{n}\r)^{-1}\nn\\
&=
\frac{\bm F\bm\Lambda^\prime\bm\Lambda\bm{\mathcal H}}{n}\l(\frac{\wh{\mbf M}^x}{n}\r)^{-1}
+\l\{ \frac{\bm X(\wh{\bm\Lambda}-\bm\Lambda\bm{\mathcal H})}{n}+\frac{\bm  E \bm\Lambda\bm{\mathcal H}}{n}
\r\}
\l(\frac{\wh{\mbf M}^x}{n}\r)^{-1}.\label{eq:fattorisviluppati}
\end{align}
Now, from \eqref{eq:mcHbis}
\beq\label{eq:mcHinv}
\bm{\mathcal H}^{-1}=\mbf J\mbf K^{-1}(\bm\Gamma^F )^{-1/2}.
\eeq
which is finite and invertible because of Lemma \ref{lem:HO1bis}(ii). 
Moreover, from \eqref{eq:kappaproj},
$\mbf V^{C\prime}{\bm\Lambda(\bm\Gamma^F)^{1/2}}\mbf K= (\mbf M^{C})^{1/2},$
which is equivalent to 
\beq\label{eq:mcHbisinv}
(\mbf M^{C})^{-1/2} \mbf V^{C\prime}{\bm\Lambda}= \mbf K^{-1}(\bm\Gamma^F)^{1/2}.
\eeq
So from
\eqref{eq:mcHbis}, \eqref{eq:mcHinv}, and \eqref{eq:mcHbisinv}, we get
\begin{align}
\bm\Lambda^\prime\bm\Lambda\bm{\mathcal H}\l(\mbf M^{C}\r)^{-1}
& = \bm\Lambda^\prime\mbf V^{C}\l(\mbf M^{C}\r)^{-1/2}\mbf J=(\bm\Gamma^F)^{-1/2} (\mbf K^{-1})^\prime\mbf J = (\bm{\mathcal H}^{-1})^\prime.\label{eq:mcHinvbis}
\end{align}
Then, because of Lemma \ref{lem:covarianze}(iii), \ref{lem:FTLN}(i),  \ref{lem:MO1}(ii), \ref{lem:MO1}(iv), and \ref{lem:HO1bis}(i), by using \eqref{eq:mcHinvbis},
\begin{align}
\l\Vert \frac{\bm\Lambda^\prime\bm\Lambda\bm{\mathcal H}}{n}\l(\frac{\wh{\mbf M}^x}{n}\r)^{-1}-(\bm{\mathcal H}^{-1})^\prime\r\Vert&=\l\Vert \frac{\bm\Lambda^\prime\bm\Lambda\bm{\mathcal H}}{n}\l(\frac{\wh{\mbf M}^x}{n}\r)^{-1}-\frac{\bm\Lambda^\prime\bm\Lambda\bm{\mathcal H}}{n}\l(\frac{\mbf M^{C}}{n}\r)^{-1}\r\Vert\nn\\
&\le \l\Vert  \frac{\bm\Lambda^\prime\bm\Lambda}{n}\r\Vert\, \l\Vert \bm{\mathcal H}\r\Vert\,\l\Vert\l(\frac{\wh{\mbf M}^x}{n}\r)^{-1}-\l(\frac{\mbf M^{C}}{n}\r)^{-1}\r\Vert\nn\\
&\le \l\Vert  \frac{\bm\Lambda^\prime\bm\Lambda}{n}\r\Vert\, \l\Vert \bm{\mathcal H}\r\Vert\,
\l\Vert\l(\frac{\wh{\mbf M}^x}{n}\r)^{-1}\r\Vert\,
\l\Vert\frac{\wh{\mbf M}^x}{n}-\frac{\mbf M^{C}}{n}\r\Vert\,
\l\Vert\l(\frac{\mbf M^{C}}{n}\r)^{-1}\r\Vert\nn\\
&= O_{\mathrm P}\l(\max\l(\frac 1{ n}, \frac 1{\sqrt T}\r)\r).\label{eq:mcHinvbis2}
\end{align}

By using \eqref{eq:mcHinvbis2} in \eqref{eq:fattorisviluppati}, because of part (a), Lemma \ref{lem:FTLN}(ii), \ref{lem:FTLN}(iv), \ref{lem:LLN}(ii), and Lemma \ref{lem:MO1}(iv), and \ref{lem:HO1bis}(i)
\begin{align}
\l\Vert\frac{ \wh{\bm F} - \bm F(\bm{\mathcal H}^{-1})^\prime}{\sqrt T}\r\Vert
\le&\, 
\l\Vert
\frac{\bm\Lambda^\prime\bm\Lambda\bm{\mathcal H}}{n}\l(\frac{\wh{\mbf M}^x}{n}\r)^{-1}
-(\bm{\mathcal H}^{-1})^\prime
\r\Vert\,
\l\Vert
\frac{\bm F}{\sqrt T}
\r\Vert\nn\\
& +\l\{ \l\Vert\frac{(\wh{\bm\Lambda}-\bm\Lambda\bm{\mathcal H})}{\sqrt n}
\r\Vert
\,\l\Vert
\frac{\bm X}{\sqrt {nT}}
\r\Vert+\l\Vert
\frac{\bm  E \bm\Lambda}{n\sqrt T}
\r\Vert\, \Vert \bm{\mathcal H}\Vert
\r\}\cdot \l\Vert
\l(\frac{\wh{\mbf M}^x}{n}\r)^{-1}
\r\Vert\nn\\
=&\,O_{\mathrm P}\l(\max\l(\frac 1{ n},\frac 1{\sqrt T}\r)\r) +O_{\mathrm P}\l(\max\l(\frac 1n,\frac 1{\sqrt T}\r)\r)  
+O_{\mathrm P}\l(\frac 1{\sqrt n}\r).\nn
\end{align}
This proves part (c).

Similarly, for part (d), from \eqref{eq:estF} and \eqref{eq:fattorisviluppati}, for any $t=1,\ldots, T$, 
\begin{align}
\wh{\mbf F}_t&=(\wh{\bm\Lambda}^\prime\wh{\bm\Lambda})^{-1}\wh{\bm\Lambda}^\prime \mbf x_t=
\l(\frac{\wh{\mbf M}^x}{n}\r)^{-1} \frac{\bm{\mathcal H}^\prime\bm\Lambda^\prime\bm\Lambda\mbf F_t}{n}
+\l(\frac{\wh{\mbf M}^x}{n}\r)^{-1}\l\{ \frac{(\wh{\bm\Lambda}-\bm\Lambda\bm{\mathcal H})^\prime\mbf x_t}{n}+
\frac{\bm{\mathcal H}^\prime\bm\Lambda^\prime\bm e_t}{n}
\r\}.\label{eq:fattorisviluppatiriga}
\end{align}
Because of Assumption \ref{ass:common}(a) and Lemma \ref{lem:Gxi}(ii), 
\begin{align}
\max_{t=1,\ldots,T}\E\l[\l\Vert
\frac{\bm\Lambda'\bm e_t}{n}
\r\Vert^2
\r]&=\max_{t=1,\ldots,T} \frac 1{n^2}\sum_{k=1}^r \E\l[
\l(\sum_{i=1}^n  e_{it} \lambda_{ik}
\r)^2
\r] =\max_{t=1,\ldots,T} \frac 1{n^2}\sum_{k=1}^r \sum_{i=1}^n\sum_{j=1}^n \E\l[
  e_{it} e_{jt}\r] \lambda_{ik}\lambda_{jk}\nn\\
 &\le \frac {rM_\Lambda^2}{n^2}\max_{t=1,\ldots,T}\max_{k=1,\ldots, r}\sum_{i=1}^n\sum_{j=1}^n\vert \E\l[
  e_{it} e_{jt}\r]\vert 
 \le \frac {rM_\Lambda^2(C_e+M_ e)}{n},\label{eq:fatto2bis}
\end{align}
since $C_e$ and $M_ e$ are independent of $t$. This is a special case of Lemma \ref{lem:LLN}(ii).

Therefore, by using \eqref{eq:mcHinvbis2} and  \eqref{eq:fatto2bis} in \eqref{eq:fattorisviluppatiriga}, because of part (a), Lemma \ref{lem:FTLN}(ii), \ref{lem:FTLN}(iv), \ref{lem:MO1}(iv), \ref{lem:HO1bis}(i),
\begin{align}
\l\Vert\wh{\mbf F}_t - \bm{\mathcal H}^{-1}\mbf F_t\r\Vert\le&\,
\l\Vert
\l(\frac{\wh{\mbf M}^x}{n}\r)^{-1}\frac{\bm{\mathcal H}^\prime\bm\Lambda^\prime\bm\Lambda}{n}
-\bm{\mathcal H}^{-1}
\r\Vert\,\l\Vert
\mbf F_t
\r\Vert +\l\{ \l\Vert\frac{(\wh{\bm\Lambda}-\bm\Lambda\bm{\mathcal H})}{\sqrt n}
\r\Vert
\,\l\Vert
\frac{\mbf x_t}{\sqrt {n}}
\r\Vert+
\l\Vert
\frac{\ \bm\Lambda^\prime\bm e_t}{n}
\r\Vert\, \Vert \bm{\mathcal H}\Vert^2
\r\}\cdot \l\Vert
\l(\frac{\wh{\mbf M}^x}{n}\r)^{-1}
\r\Vert\nn\\
=&\,O_{\mathrm P}\l(\max\l(\frac 1{ n},\frac 1{\sqrt T}\r)\r) +O_{\mathrm P}\l(\max\l(\frac 1n,\frac 1{\sqrt T}\r)\r)  
+O_{\mathrm P}\l(\frac 1{\sqrt n}\r).\nn
\end{align}
This proves part (d) and it completes the proof. $\Box$

\subsection{Proof of Proposition \ref{prop:HHAT}}\label{prop:HHATproof} 
From \eqref{eq:acca} and the definition of $\wh{\bm\Lambda}$ in \eqref{eq:estL}
\begin{align}\label{eq:warwick1}
\l\Vert\wh{\mbf H} - \frac{\bm F^\prime\bm F}{T}(\wh{\mbf H}^{-1})^{\prime}\r\Vert
&=
\l\Vert \wh{\mbf H} - \frac{\bm F^\prime\bm F}{T}(\wh{\mbf H}^{-1})^{\prime} \frac{\wh{\bm\Lambda}^\prime\wh{\bm\Lambda}}{n}\l(\frac{\wh{\mbf M}^x}{n}\r)^{-1}\r\Vert\le
\l\Vert  \frac{\bm F^\prime\bm F}{T}\r\Vert\,
\l\Vert\frac{(\bm\Lambda-\wh{\bm\Lambda}\wh{\mbf H}^{-1})^\prime \wh{\bm\Lambda}}{n}\r\Vert \,\l\Vert\l(\frac{\wh{\mbf M}^x}{n}\r)^{-1}\r\Vert\nonumber\\
&=O_{\mathrm P}\l(\max\l(\frac 1n,\frac{1}{\sqrt{nT}},\frac 1T\r)\r),
\end{align}
by Proposition \ref{prop:LLFF}(c), and Lemmas  \ref{lem:LLN}(v) and \ref{lem:MO1}(iv). Similarly, from \eqref{eq:acca} it also follows that
\begin{align}\label{eq:warwick2}
\l\Vert\wh{\mbf H}- \frac{\bm F^\prime\bm F}{T} \frac{\bm\Lambda^\prime\bm\Lambda}{n} \wh{\mbf H} \l(\frac{\wh{\mbf M}^x}{n}\r)^{-1}\r\Vert&\le 
\l\Vert  \frac{\bm F^\prime\bm F}{T}\r\Vert\,
\l\Vert\frac{\bm\Lambda^\prime(\wh{\bm\Lambda}-\bm\Lambda\wh{\mbf H})}{n}\r\Vert\,
\l\Vert\l(\frac{\wh{\mbf M}^x}{n}\r)^{-1}\r\Vert\nonumber\\
&=O_{\mathrm P}\l(\max\l(\frac 1n,\frac{1}{\sqrt{nT}},\frac 1T\r)\r),
\end{align}
by Proposition \ref{prop:LLFF}(a), and Lemmas  \ref{lem:LLN}(v) and \ref{lem:MO1}(iv). Now, since $\Vert \wh{\mbf H}\Vert =O_{\mathrm P}(1)$ because of Lemma \ref{lem:HO1}(i), \eqref{eq:warwick1}  implies
\beq\label{eq:warwick3}
\l\Vert\wh{\mbf H}\wh{\mbf H}^{\prime} - \frac{\bm F^\prime\bm F}{T}\r\Vert =O_{\mathrm P}\l(\max\l(\frac 1n,\frac{1}{\sqrt{nT}},\frac 1T\r)\r),
\eeq
and, since $\Vert \frac{\wh{\mbf M}^x}n\Vert =O_{\mathrm P}(1)$ because of Lemma \ref{lem:MO1}(iii), \eqref{eq:warwick2} implies
\beq\label{eq:warwick4}
\l\Vert\wh{\mbf H} \frac{\wh{\mbf M}^x}{n} - \frac{\bm F^\prime\bm F}{T} \frac{\bm\Lambda^\prime\bm\Lambda}{n} \wh{\mbf H}\r\Vert=O_{\mathrm P}\l(\max\l(\frac 1n,\frac{1}{\sqrt{nT}},\frac 1T\r)\r).
\eeq
From \eqref{eq:warwick3} and \eqref{eq:warwick4} it follows that
\begin{align}
\l\Vert
\wh{\mbf H} \frac{\wh{\mbf M}^x}{n}\wh{\mbf H}^\prime - \frac{\bm F^\prime\bm F}{T} \frac{\bm\Lambda^\prime\bm\Lambda}{n}\frac{\bm F^\prime\bm F}{T} 
\r\Vert
&\le \l\Vert
\wh{\mbf H} \frac{\wh{\mbf M}^x}{n} - \frac{\bm F^\prime\bm F}{T} \frac{\bm\Lambda^\prime\bm\Lambda}{n} \wh{\mbf H}
\r\Vert\,\l\Vert \wh{\mbf H}^\prime\r\Vert + \l\Vert \frac{\bm F^\prime\bm F}{T}\r\Vert\,\l\Vert \frac{\bm\Lambda^\prime\bm\Lambda}{n} \r\Vert\,\l\Vert \frac{\bm F^\prime\bm F}{T}-\wh{\mbf H}\wh{\mbf H}^\prime\r\Vert\nn\\
&= O_{\mathrm P}\l(\max\l(\frac 1n,\frac{1}{\sqrt{nT}},\frac 1T\r)\r).\label{eq:warwick5}
\end{align}
again because of Lemmas \ref{lem:LLN}(iv), \ref{lem:LLN}(v), and \ref{lem:HO1}(i). And from \eqref{eq:warwick5}, since $\Vert \wh{\mbf H}\Vert =O_{\mathrm P}(1)$ because of Lemma \ref{lem:HO1}(i),  
\begin{align}
\l\Vert
\frac{\wh{\mbf M}^x}{n} - \wh{\mbf H}^{-1} \frac{\bm F^\prime\bm F}{T} \frac{\bm\Lambda^\prime\bm\Lambda}{n}\frac{\bm F^\prime\bm F}{T} (\wh{\mbf H}^{-1})^{\prime}
\r\Vert=O_{\mathrm P}\l(\max\l(\frac 1n,\frac{1}{\sqrt{nT}},\frac 1T\r)\r).\label{eq:madre}
\end{align}

Now, since by definition $\bm X=\bm F\bm\Lambda^\prime+\bm E$, we get
\begin{align}\label{eq:mistake3}
\l\Vert\frac{\bm X^\prime\bm X}{nT}-\frac{\bm\Lambda\bm F^\prime\bm F\bm\Lambda^\prime}{nT} \r\Vert&\le 
2 \l\Vert \frac{\bm F^\prime\bm E}{\sqrt nT}\r\Vert \,\l\Vert \frac{\bm\Lambda}{\sqrt n}\r\Vert+ \l\Vert \frac{\bm E^\prime\bm E}{nT}\r\Vert= O_{\mathrm P}\l(\frac 1{\sqrt T}\r)+O_{\mathrm P}\l(\max\l(\frac 1n,\frac 1{\sqrt T}\r)\r), 
\end{align}
by Lemmas \ref{lem:FTLN}(i), \ref{lem:LLN}(i), and \ref{lem:LLN}(iii). Moreover, the eigenvalues of $(\frac{\bm F^\prime\bm F}T)^{1/2}\frac{\bm \Lambda^\prime\bm \Lambda}n (\frac{\bm F^\prime\bm F}T)^{1/2}$ coincide with the 
eigenvalues of $(\frac{\bm \Lambda^\prime\bm \Lambda}n) \frac{\bm F^\prime\bm F}T$ which in turn coincide with the $r$ non-zero eigenvalues of 
$\frac {\bm\Lambda\bm F^\prime \bm F\bm\Lambda^\prime}{nT}$, collected into the $r\times r$ diagonal matrix $\frac{\wh{\mbf M}^C}{n}$. Thus, by \eqref{eq:mistake3} and Weyl's inequality \citep[Theorem 1]{MK04}:
\beq\label{eq:mistake}
 \l\Vert \frac{\wh{\mbf M}^x}n-\frac{\wh{\mbf M}^C}n\r\Vert\le \l\Vert\frac{\bm X^\prime\bm X}{nT}-\frac{\bm\Lambda\bm F^\prime\bm F\bm\Lambda^\prime}{nT} \r\Vert = O_{\mathrm P}\l(\max\l(\frac 1n,\frac 1{\sqrt T}\r)\r).
\eeq
%
Hence, from  \eqref{eq:madre} and \eqref{eq:mistake} 
\begin{align}
\l\Vert
\frac{\wh{\mbf M}^C}{n} - \wh{\mbf H}^{-1} \frac{\bm F^\prime\bm F}{T} \frac{\bm\Lambda^\prime\bm\Lambda}{n}\frac{\bm F^\prime\bm F}{T} (\wh{\mbf H}^{-1})^{\prime}
\r\Vert
&\le \l\Vert \frac{\wh{\mbf M}^x}{n} - \wh{\mbf H}^{-1} \frac{\bm F^\prime\bm F}{T} \frac{\bm\Lambda^\prime\bm\Lambda}{n}\frac{\bm F^\prime\bm F}{T} (\wh{\mbf H}^{-1})^{\prime}
\r\Vert+  \l\Vert \frac{\wh{\mbf M}^C}n-\frac{\wh{\mbf M}^x}n\r\Vert\nonumber\\
&=O_{\mathrm P}\l(\max\l(\frac 1n,\frac{1}{\sqrt{nT}},\frac 1T\r)\r)+O_{\mathrm P}\l(\max\l(\frac 1n,\frac 1{\sqrt T}\r)\r).\label{eq:mistake2}
\end{align}
Therefore,  it must be that 
\beq
\l\Vert\wh{\mbf H}^{-1} - \wh{\mbf Q}^\prime \l(\frac{\bm F^\prime\bm F}T\r)^{-1/2}\r\Vert =O_{\mathrm P}\l(\max\l(\frac 1n,\frac{1}{\sqrt{T}}\r)\r).\nn
\eeq
which proves part (b), while part (a) follows from the continuous mapping theorem and since $ (\wh{\mbf Q}^\prime)^{-1}= \wh{\mbf Q}$, we complete the proof. $\Box$

\subsection{Proof of Proposition \ref{th:CLTcent}}\label{app:CLTcentproof}

First consider part (a). Then, \eqref{eq:start} reads
\beq\label{eq:startcent}
\frac{(\bm X-\bar {\bm X})^\prime(\bm X-\bar {\bm X})}{nT}\wh{\bm\Lambda}=\wh{\bm\Lambda}\frac{\wh{\mbf M}^x}{n}.
\eeq 
Then, let $\bar{\bm F}$ be $T\times r$ with all $T$ rows equal to $\bar{\mbf F}^\prime$, $\bar{\bm E}$ be $T\times n$ with all $T$ rows equal to $\bar{\bm e}:=(\bar e_1\cdots \bar e_n)$, 
and 
\beq\label{eq:Hcent}
\wh{\mbf H}_c:=\frac{(\bm F-\bar{\bm F})^\prime(\bm F-\bar{\bm F})}{T}
\frac{\bm\Lambda^\prime\wh{\bm\Lambda}}{n}
\l(\frac{\wh{\mbf M}^x}{n}\r)^{-1}.
\eeq
Clearly $\Vert \wh{\mbf H}_c\Vert=O(1)$ and $\Vert \wh{\mbf H}_c^{-1}\Vert=O(1)$ because  the first term on the rhs of \eqref{eq:Hcent} is a consistent estimator of $\bm\Gamma^F=\E[\mbf F_t\mbf F_t^\prime]$ which is finite and positive definite by Assumption \ref{ass:common}(b). Indeed,
\begin{align}
\l\Vert \frac{(\bm F-\bar{\bm F})^\prime(\bm F-\bar{\bm F})}{T} -\bm\Gamma^F\r\Vert&=  \l\Vert \frac{\bm F^\prime \bm F}{T}- \frac{\bar{\bm F}^\prime \bar{\bm F}}{T} -\bm\Gamma^F\r\Vert
=\l\Vert \frac{\bm F^\prime \bm F}{T}- \frac 1T \sum_{t=1}^T \bar {\mbf F} \bar {\mbf F}^\prime-\bm\Gamma^F\r\Vert
=\l\Vert \frac{\bm F^\prime \bm F}{T}-  \bar {\mbf F} \bar {\mbf F}^\prime-\bm\Gamma^F\r\Vert\nn\\
&\le \l\Vert \frac{\bm F^\prime \bm F}{T}-\bm\Gamma^F\r\Vert + \Vert\bar {\mbf F} \Vert^2 = o_{\mathrm P}(1),\nn
\end{align}
by Assumption \ref{ass:common}(c-ii) which implies also that $\Vert \bar{\mbf F}\Vert =\Vert \bar{\mbf F}-\E[\mbf F_t]\Vert = o_{\mathrm P}(1)$, because $\E[\mbf F_t]=\mbf 0_r$ by  Assumption \ref{ass:common}(b).

Then, since, under Assumptions \ref{ass:center}(a) or \ref{ass:center}(b) or \ref{ass:center}(c), we always have
\beq\label{eq:profumo}
(\bm X-\bar {\bm X})= (\bm F-\bar {\bm F})\bm\Lambda^\prime+(\bm E-\bar {\bm E}),
\eeq 
from \eqref{eq:startcent} and \eqref{eq:Hcent} we get
\begin{align}
\wh{\bm\Lambda}-\bm\Lambda\wh{\mbf H}_c=&\, \l(\frac{\bm\Lambda(\bm F-\bar{\bm F})^\prime(\bm  E-\bar{\bm E})\wh{\bm\Lambda}}{nT}
+\frac{(\bm E-\bar{\bm E})^\prime(\bm F-\bar{\bm F})\bm\Lambda^\prime\wh{\bm\Lambda}}{nT}
+\frac{(\bm E-\bar{\bm E})^\prime(\bm E-\bar{\bm E})\wh{\bm\Lambda}}{nT}\r)\l(\frac{\wh{\mbf M}^x}{n}\r)^{-1}
.\label{eq:start4cent}
\end{align}
which is the analogue of \eqref{eq:start4}. It immediately follows that Proposition \ref{prop:L2} still holds. 
Indeed, all bounds in Lemma \ref{prop:load} still hold because the elements of $\bar{\bm F}$ and $\bar{\bm E}$ still satisfy Assumptions \ref{ass:common}(b), \ref{ass:common}(c), and \ref{ass:idio}. Furthermore, Propositions \ref{prop:KKKbis} and \ref{cor:semplice}(a) still hold, thus, $\Vert\wh{\mbf H}_c-  (\bm\Gamma^F)^{1/2}\bm\Upsilon_0 \bm{\mathcal J}_0   \Vert =o_{\mathrm P}(1)$. In particular,  Proposition \ref{prop:KKKbis} now holds for $\frac{\wh{\bm F}^\prime(\bm F-\bar{\bm F})}T$, because $\bar{\wh{\bm F}}=\mbf 0_{T\times r}$ by construction.
Therefore, \eqref{eq:finaleL} and \eqref{eq:CLTL} in the proof of Theorem \ref{th:CLTL} now give:
\begin{align*}
\sqrt T&\l(\wh{\bm\lambda}_i-\wh{\mbf H}_c^\prime{\bm\lambda}_i\r) 
=\bm{\mathcal J}_0\bm\Upsilon_0^\prime  (\bm\Gamma^F)^{1/2}  \l(\frac{(\bm F-\bar{\bm F})^\prime(\bm F-\bar{\bm F})}{T}\r)^{-1} \l(\frac 1{\sqrt T}\sum_{t=1}^T (\mbf F_t-\bar{\mbf F}) (e_{it}-\bar e_i)\r)+o_{\mathrm P}(1)\\
&=\bm{\mathcal J}_0\bm\Upsilon_0^\prime  (\bm\Gamma^F)^{1/2}  \l(\frac{(\bm F-\bar{\bm F})^\prime(\bm F-\bar{\bm F})}{T}\r)^{-1} \l(\frac 1{\sqrt T}\sum_{t=1}^T (\mbf F_t-\bar{\mbf F}) e_{it} - \frac 1{\sqrt T} \l\{\sum_{t=1}^T \mbf F_t- T\bar{\mbf F} \r\} \bar e_i \r)+o_{\mathrm P}(1)\\
&=\bm{\mathcal J}_0\bm\Upsilon_0^\prime  (\bm\Gamma^F)^{1/2} \sqrt T\l(\wh{\bm\lambda}_i^{\text{\tiny OLS}}-{\bm\lambda}_i\r) +o_{\mathrm P}(1).
\end{align*}
 So the asymptotic properties of the estimated loadings are unchanged. This proves part (a).

Turning to part (b).  Recall that, by Assumptions \ref{ass:Wold} or \ref{ass:Wu}, 
\[
\mbf F_t =\sum_{k=0}^\infty \mbf C_k \mbf u_{t-k}, \quad \sum_{k=0}^\infty \Vert \mbf C_k\Vert\le M_C, \quad \mbf u_t\sim i.i.d.(\mbf 0_r,\mbf I_r).
\]
Then, for any $\bm a\in\mathbb R^r$, by \citet[Theorem 7.1.2]{brockwell1991timeseries},
\beq\label{eq:CWOLD}
\sqrt T \bm a' \bar{\mbf F} = \frac 1{\sqrt T}\sum_{t=1}^T \bm a^\prime\mbf F_t \to_d\mathcal N\l(0, \bm a^\prime\bm\Omega_0 \bm a\r),
\eeq
where $\bm\Omega_0= \mbf C(1)\mbf C(1)^\prime$ with $\mbf C(1)=\sum_{k=0}^\infty \mbf C_k$, and note that $\Vert\bm\Omega_0\Vert\le  (\sum_{k=0}^\infty \Vert\mbf C_k\Vert)^2 \le M_C^2$. By \eqref{eq:CWOLD} by Cram\'er-Wold theorem it follows that
\beq\label{eq:triangolo2}
\sqrt T \bar{\mbf F}=\frac 1{\sqrt T}\sum_{t=1}^T \mbf F_t \to_d \mathcal N(\mbf 0_r,\bm\Omega_0),
\eeq
where we also notice that $\bm\Omega_0=\sum_{h=-\infty}^\infty \E[\mbf F_t\mbf F_{t-h}^\prime]$, indeed, 
\[
\sum_{h=-\infty}^\infty \E[\mbf F_t\mbf F_{t-h}^\prime] =
\sum_{h=-\infty}^\infty \sum_{k_1,k_2=0}^\infty \mbf C_{k_1}\E[\mbf u_{t-k_1}\mbf u_{t-h-k_2}^\prime]\mbf C_{k_2}^\prime
=\sum_{h=-\infty}^\infty \sum_{k_2=0}^\infty \mbf C_{h+k_2}\mbf C_{k_2}^\prime
=\mbf C(1)\mbf C(1)^\prime,
\]
since $\mbf C_k=\mbf 0_r$ if $k<0$.

Under Assumptions \ref{ass:common}(b) and \ref{ass:idio}(a) $\E[\mbf F_{t}]=\mbf 0_r$ and $\E[e_{it}]=0$. 
Thus, by \eqref{eq:triangolo2} and Assumptions \ref{ass:idio}(i) and \ref{ass:idio}(c-ii) 
\beq\label{eq:mediebelle}
\l\Vert \bar{\mbf F} \r\Vert=O_{\mathrm P}\l(\frac 1{\sqrt T}\r),\qquad \l\vert \bar e_i\r\vert = O_{\mathrm P}\l(\frac 1{\sqrt T}\r),
\eeq
where $\bar {e}_{i}:=T^{-1}\sum_{t=1}^T e_{it}$, $i\in\mathbb N$. From \eqref{eq:mediebelle} it follows that $\bar { x}_i$ is a $\sqrt T$-consistent estimator of $\E[x_{it}]$. 

By using \eqref{eq:profumo}, we immediately see that \eqref{eq:sviluppoFactor} is now replaced by
\begin{align}
\!\!\!\!\wh{\mbf F}_t^\prime -(\mbf F_t-\bar{\mbf F})^\prime (\wh{\mbf H}^{-1}_c)^\prime&=
\l(
(\mbf F_t-\bar{\mbf F})^\prime\frac{(\bm\Lambda-\wh{\bm\Lambda}\wh{\mbf H}^{-1}_c)^\prime\wh{\bm\Lambda}}{n}+
(\bm e_t-\bar{\bm e})^\prime \frac{(\wh{\bm\Lambda}-\bm\Lambda \wh{\mbf H}_c)}{n}+
(\bm e_t-\bar{\bm e})^\prime \frac{\bm\Lambda\wh{\mbf H}_c}{n}
\r)\l(\frac {\wh{\mbf M}^x}n\r)^{-1}.\label{eq:pave2}
\end{align}
Now, by Assumption \ref{ass:common}(a) and Lemma \ref{lem:Gxi}(i),
\begin{align}\label{eq:pave}
\E\l[\l\Vert \frac{\bar{\bm e}^\prime {\bm\Lambda}}n\r\Vert^2\r] &
=\E\l[\l\Vert \frac 1{nT} \sum_{t=1}^T{\bm e_t}^\prime {\bm\Lambda}\r\Vert^2\r]=\frac 1{n^2T^2}\sum_{j=1}^r \E\l[\l(\sum_{t=1}^T\sum_{i=1}^n \lambda_{ij} e_{it}\r)^2\r]\nn\\
&\le \frac{r}{n^2T^2}\max_{j=1,\ldots,r}\sum_{i=1}^n\sum_{k=1}^n \lambda_{ij}\lambda_{kj} \sum_{t=1}^T\sum_{s=1}^T\E[ e_{it} e_{ks}]\nn\\
&\le \frac{rM_\Lambda^2}{n^2T^2}\max_{j=1,\ldots,r}\sum_{i=1}^n\sum_{k=1}^n\sum_{t=1}^T\sum_{s=1}^T\l\vert \E[ e_{it} e_{ks}]\r\vert\nn\\
&\le \frac{rM_\Lambda^2M_{1e} }{nT}.
\end{align}
Therefore, from \eqref{eq:pave} which by Proposition \ref{prop:L2} implies also that $\Vert \frac{\bar{\bm e}^\prime ({\wh{\bm\Lambda}-\bm\Lambda\wh{\mbf H}})}n\Vert = o_{\mathrm P}(n^{-1/2}T^{-1/2})$,
and since  $\Vert \bar{\mbf F}\Vert = O_{\mathrm P}(T^{-1/2})$ by \eqref{eq:mediebelle},
we can apply Lemma \ref{prop:factor} to \eqref{eq:pave2} and get
\begin{align}
\wh{\mbf F}_t - \wh{\mbf H}^{-1}_c\mbf F_t 
&=-\wh{\mbf H}^{-1}_c\bar{\mbf F} 
+\l(\frac {\wh{\mbf M}^x}n\r)^{-1}\wh{\mbf H}_c^\prime \l(\frac 1{ n}\sum_{i=1}^n \bm \lambda_i e_{it}\r)
+ O_{\mathrm P}\l(\max\l(\frac 1n,\frac 1{\sqrt {nT}},\frac 1T\r)\r)\label{eq:pave3}\\
&=-\wh{\mbf H}^{-1}_c\l(\frac 1T\sum_{t=1}^T \mbf F_t\r)
+\wh{\mbf H}^{-1}_c
\l(\frac {\bm\Lambda^\prime\bm\Lambda}n\r)^{-1} 
\l(\frac 1{ n}\sum_{i=1}^n \bm \lambda_i e_{it}\r)
+ O_{\mathrm P}\l(\max\l(\frac 1n,\frac 1{\sqrt {nT}},\frac 1T\r)\r),\nn
\end{align}
where for the second term on the rhs we followed the same steps leading to \eqref{eq:finaleF}. Letting, hereafter,
$\delta_{nT}=\min(\sqrt n,\sqrt T)$,
from \eqref{eq:pave3} we immediately have
\begin{align}
\delta_{nT}\l(\wh{\mbf F}_t - \wh{\mbf H}^{-1}_c\mbf F_t \r)=&\,
- \frac{\delta_{nT}}{\sqrt T} \wh{\mbf H}^{-1}_c\l(\frac 1{\sqrt T}\sum_{t=1}^T \mbf F_t\r)+\frac{\delta_{nT}}{\sqrt n} \wh{\mbf H}^{-1}_c\l(\frac {\bm\Lambda^\prime\bm\Lambda}n\r)^{-1} 
\l(\frac 1{\sqrt n}\sum_{i=1}^n \bm \lambda_i e_{it}\r) + O_{\mathrm P}\l(\frac 1{\delta_{nT}}\r)\nn\\
&= - \frac{\delta_{nT}}{\sqrt T}  \bm \zeta_T + \frac{\delta_{nT}}{\sqrt n}\bm \xi_n + O_{\mathrm P}\l(\frac 1{\delta_{nT}}\r), \;\text{say.}\label{eq:truccolo}
\end{align}

Let us derive first the asymptotic distribution of $\bm \zeta_T$. 
Using  Proposition \ref{cor:semplice}(b), which still holds for $\wh{\mbf H}_c$,
\beq\label{eq:triangolo}
\bm \zeta_T=\sqrt T \wh{\mbf H}^{-1}_c\bar{\mbf F} = \l\{ \underset{n,T\to\infty}{\text{P-lim}} \wh{\mbf H}^{-1}_c \r\} \sqrt T \bar{\mbf F} +o_{\mathrm P}(1)=
\bm{\mathcal J}_0\bm\Upsilon_0^\prime  (\bm\Gamma^F)^{-1/2} \sqrt T \bar{\mbf F} +o_{\mathrm P}(1).
\eeq
Summing up, from \eqref{eq:triangolo} and \eqref{eq:triangolo2}, and Slutsky's theorem, it follows that
\beq\label{eq:truccoloT}
\bm \zeta_T \to_d \bm\zeta\sim \mathcal N(\mbf 0_r,\bm V_\zeta), \quad \bm V_\zeta=\bm\Upsilon_0^\prime  (\bm\Gamma^F)^{-1/2}
\bm\Omega_0
(\bm\Gamma^F)^{-1/2}
\bm\Upsilon_0. 
\eeq
Moreover, from Theorem \ref{th:CLTF}, and since Proposition \ref{cor:semplice}(b) holds also for $\wh{\mbf H}_c$,
\beq\label{eq:truccoloN}
\bm\xi_n \to_d \bm\xi\sim \mathcal N\l(\mbf 0_r, \bm V_\xi\r), \quad \bm V_\xi =\bm\Upsilon_0^\prime(\bm\Gamma^F)^{-1/2}\bm\Pi_t^{\text{\tiny \upshape OLS}}(\bm\Gamma^F)^{-1/2}\bm\Upsilon_0.
\eeq
%
%
%
%
Now $\bm\zeta$ and $\bm\xi$ are independent because of Assumption \ref{ass:ind}, the former being the limit of a sum over time, the latter being the limit of a sum over the cross-section.
Therefore, using the same arguments as those in \citet[proof of Theorem 3]{bai03},
we can prove that:
\beq
\l(\frac{\delta_{nT}^2}T\bm V_\zeta+\frac{\delta_{nT}^2}n\bm V_\xi
\r)^{-1/2}
\l(
-\frac{\delta_{nT}}{\sqrt T}\bm\zeta
+
\frac{\delta_{nT}}{\sqrt n}\bm\xi
\r)
\sim \mathcal N(\mbf 0_r,\mbf I_r).\label{eq:truccolo2}
\eeq
From \eqref{eq:truccolo}, \eqref{eq:truccoloT}, \eqref{eq:truccoloN}, and \eqref{eq:truccolo2}, and using Slutsky's theorem, 
\begin{align}
\l(
\frac{\bm V_\zeta}T+\frac{\bm V_\xi}n
\r)^{-1/2} 
\l(\wh{\mbf F}_t - \wh{\mbf H}^{-1}_c\mbf F_t \r)
&=
\l(\frac{\delta_{nT}^2}T\bm V_\zeta+\frac{\delta_{nT}^2}n\bm V_\xi
\r)^{-1/2} \delta_{nT}\l(\wh{\mbf F}_t - \wh{\mbf H}^{-1}_c\mbf F_t \r)\nn\\
&=
\l(\frac{\delta_{nT}^2}T\bm V_\zeta+\frac{\delta_{nT}^2}n\bm V_\xi
\r)^{-1/2} \l(
-\frac{\delta_{nT}}{\sqrt T}\bm\zeta_T
+
\frac{\delta_{nT}}{\sqrt n}\bm\xi_n
\r)+ O_{\mathrm P}\l(\frac 1{\delta_{nT}}\r)\nn\\
&\to_d \mathcal N\l(\mbf 0_r,\mbf I_r\r).\label{eq:magheggioratei2}
\end{align}
Then, notice that
\begin{align}\label{eq:magheggioratei}
\frac{\bm V_\zeta}T+\frac{\bm V_\xi}n &= \frac 1n \l(\bm V_\xi+\frac nT \bm V_\zeta  \r) = \frac{n+T}{nT}\frac{T}{n+T}\l(\bm V_\xi+\frac nT \bm V_\zeta  \r)=\frac{n+T}{nT} \frac 1{1+\frac nT}\l(\bm V_\xi+\frac nT \bm V_\zeta  \r).
\end{align}
By substituting \eqref{eq:magheggioratei} into \eqref{eq:magheggioratei2}
\begin{align}
\sqrt{\frac{nT}{n+T}}
\l\{\frac 1{1+\frac nT}\l(\bm V_\xi+\frac nT \bm V_\zeta  \r) 
\r\}^{-1/2}\l(\wh{\mbf F}_t - \wh{\mbf H}^{-1}_c\mbf F_t \r)\to_d \mathcal N\l(\mbf 0_r,\mbf I_r\r),\nn
\end{align}
which, by letting $\gamma=\lim_{n,T\to\infty} \frac nT$,  is equivalent to
\begin{align}
\sqrt{\frac{nT}{n+T}}
\l(\wh{\mbf F}_t - \wh{\mbf H}^{-1}_c\mbf F_t \r)\to_d \mathcal N\l(\mbf 0_r, \frac 1{1+\gamma}\l(\bm V_\xi+\gamma \bm V_\zeta  \r) \r).\nn
\end{align}
This proves part (b).

For part (c-i), by following the same reasoning leading to \eqref{eq:sviluppoFactor}: 
\begin{align}
\!\!\!\!\wh{\bm \mu}^\prime- (\bm\mu+\bar{\mbf F})^\prime(\wh{\mbf H}^{-1}_c)^\prime =
\bigg\{ \frac{\bm \mu^\prime(\bm\Lambda-\wh{\bm\Lambda}\wh{\mbf H}^{-1}_c)^\prime \wh{\bm\Lambda}}n
 +
\frac{\bar{\mbf F}^\prime(\bm\Lambda-\wh{\bm\Lambda}\wh{\mbf H}^{-1}_c)^\prime \wh{\bm\Lambda}}n
+
\frac{\bar{\mbf e}^\prime(\wh{\bm\Lambda}-\bm\Lambda\wh{\mbf H}_c)}{n}
+
\frac{\bar{\mbf e}^\prime\bm\Lambda\wh{\mbf H}_c}{n}
\bigg\}
\l(\frac {\wh{\mbf M}^x}n\r)^{-1}.\label{eq:simone}
\end{align}
Therefore, by using again \eqref{eq:pave} and since  $\Vert \bar{\mbf F}\Vert = O_{\mathrm P}(T^{-1/2})$ by  \eqref{eq:mediebelle} and $\Vert\bm \mu\Vert=O(1)$ by Assumption \ref{ass:center}(b),
we can apply Lemma \ref{prop:factor} to \eqref{eq:simone} and get
\begin{align}\label{eq:inoui}
\wh{\bm \mu}-\wh{\mbf H}^{-1}_c\bm\mu = \wh{\mbf H}^{-1}_c\bar{\mbf F} + O_{\mathrm P}\l(\max\l(\frac 1n,\frac 1{\sqrt {nT}},\frac 1T\r)\r).
\end{align}
So by \eqref{eq:triangolo} and \eqref{eq:truccoloT}
\[
\sqrt T\l(\wh{\bm \mu}-\wh{\mbf H}^{-1}_c\bm\mu\r)
\to_d \mathcal N\l(\mbf 0_r,\bm\Upsilon_0^\prime  (\bm\Gamma^F)^{-1/2}
\bm\Omega_0
(\bm\Gamma^F)^{-1/2}
\bm\Upsilon_0\r),
\]
which proves part (c-i).

For part (c-ii), by combining \eqref{eq:pave3} and \eqref{eq:inoui}, 
\begin{align}
\wh{\mbf F}_t-\wh{\mbf H}_c^{-1}\mbf F_t +\wh{\bm\mu}-\wh{\mbf H}_c^{-1}\bm\mu&=-\wh{\mbf H}^{-1}_c\bar{\mbf F}  +\wh{\mbf H}^{-1}_c
\l(\frac {\bm\Lambda^\prime\bm\Lambda}n\r)^{-1} 
\l(\frac 1{ n}\sum_{i=1}^n \bm \lambda_i e_{it}\r)
+\wh{\mbf H}^{-1}_c\bar{\mbf F} 
+ O_{\mathrm P}\l(\max\l(\frac 1n,\frac 1{\sqrt {nT}},\frac 1T\r)\r)\nn\\
&= \wh{\mbf H}^{-1}_c
\l(\frac {\bm\Lambda^\prime\bm\Lambda}n\r)^{-1} 
\l(\frac 1{ n}\sum_{i=1}^n \bm \lambda_i e_{it}\r)+O_{\mathrm P}\l(\max\l(\frac 1n,\frac 1{\sqrt {nT}},\frac 1T\r)\r).\label{eq:codex}
\end{align} 
Let  $\mbf G_t=\mbf F_t+\bm\mu$ and $\wh{\mbf G}_t^{\text{\tiny OLS}}$ obtained by regressing $(x_{1t}\cdots x_{nt})^\prime$ onto $\bm\Lambda$ thus coinciding with 
$\wh{\mbf F}_t^{\text{\tiny OLS}}$ when data has zero mean. Then, 
 since Proposition \ref{cor:semplice}(b) holds also for $\wh{\mbf H}_c$ and by Assumption \ref{ass:rates} $\sqrt n/T\to 0$ as $n,T\to\infty$, 
from \eqref{eq:codex} we get
\[
\sqrt n\l(\wh{\mbf F}_t-\wh{\mbf H}_c^{-1}\mbf F_t +\wh{\bm\mu}-\wh{\mbf H}_c^{-1}\bm\mu\r) = \bm{\mathcal J}_0\bm\Upsilon_0^\prime  (\bm\Gamma^F)^{-1/2} \sqrt n\l(\wh{\mbf G}_t^{\text{\tiny OLS}}-\mbf G_t\r)+ o_{\mathrm P}(1),
\] 
which coincides with \eqref{eq:CLTF} in the proof of Theorem \ref{th:CLTF} and so the estimated non-zero mean factors  have the asymptotic distribution given in Theorem \ref{th:CLTF}. This proves part (c-ii).

Finally, for part (d) which holds under Assumption \ref{ass:center}(c), by following the same reasoning leading to \eqref{eq:sviluppoFactor}: 
\begin{align}
\wh{\bm \mu}^\prime- (\bm\mu+\bar{\mbf F})^\prime(\wh{\mbf H}^{-1}_c)^\prime =&\,
\bigg\{ \frac{\bm \mu^\prime(\bm\Lambda-\wh{\bm\Lambda}\wh{\mbf H}^{-1}_c)^\prime \wh{\bm\Lambda}}n
 +
\frac{\bar{\mbf F}^\prime(\bm\Lambda-\wh{\bm\Lambda}\wh{\mbf H}^{-1}_c)^\prime \wh{\bm\Lambda}}n
+
\frac{\bm\alpha^\prime(\wh{\bm\Lambda}-\bm\Lambda\wh{\mbf H}_c)}{n}\nn\\
&+
\frac{\bm\alpha^\prime\bm\Lambda\wh{\mbf H}_c}{n}
+
\frac{\bar{\mbf e}^\prime(\wh{\bm\Lambda}-\bm\Lambda\wh{\mbf H}_c)}{n}
+
\frac{\bar{\mbf e}^\prime\bm\Lambda\wh{\mbf H}_c}{n}
\bigg\}
\l(\frac {\wh{\mbf M}^x}n\r)^{-1}.\label{eq:simonebiased}
\end{align}
First of all, notice that Proposition \ref{prop:L}(a) still holds. Indeed,  by definition the common component now is $\bm C_t=\bm\Lambda \mbf G_t$ where, however, $\E[\mbf G_t]=\bm\mu\ne \mbf 0_r$. Then, by definition of covariance matrix,
\beq
 \mbf V^C\mbf M^C\mbf V^{C\prime} = \bm\Lambda\E[(\mbf G_t-\bm\mu)(\mbf G_t-\bm\mu)^\prime]\bm\Lambda^\prime 
 = \bm\Lambda\E[\mbf F_t\mbf F_t^\prime]\bm\Lambda^\prime
 = \bm\Lambda\bm\Gamma^F\bm\Lambda^\prime,
\eeq
which coincides with \eqref{eq:effeuno} in the proof of Proposition \ref{prop:L}(a). So the definition of $\bm{\mathcal H}$ is unchanged. For the same reason, Lemmas \ref{lem:KO1}, \ref{lem:HO1bis}, and \ref{lem:HO1} still hold.

Now, from \eqref{eq:start4},
\begin{align}
\frac {(\wh{\bm\Lambda}-\bm\Lambda\wh{\mbf H}_c)^\prime \bm\alpha}n
=&\,  \l(\frac{\wh{\mbf M}^x}{n}\r)^{-1}\bm{\mathcal H}^\prime\frac{\bm\Lambda^\prime\bm\Lambda}{n}
\frac{\bm F^\prime\bm E\bm\alpha}{ nT}
+\l(\frac{\wh{\mbf M}^x}{n}\r)^{-1}\bm{\mathcal H}^\prime\frac{\bm\Lambda^\prime\bm E^\prime\bm F}{nT}\frac{\bm\Lambda^\prime\bm\alpha}{n} + \l(\frac{\wh{\mbf M}^x}{n}\r)^{-1}\bm{\mathcal H}^\prime \frac{\bm\Lambda^\prime\bm E^\prime\bm E\bm\alpha}{n^2T}\nn\\
&+\l(\frac{\wh{\mbf M}^x}{n}\r)^{-1}\frac{(\wh{\bm\Lambda}-\bm\Lambda\bm{\mathcal H})^\prime}{\sqrt n}\frac{\bm  E^\prime\bm F\bm\Lambda^\prime}{nT}\frac{\bm\alpha}{\sqrt n}
+\l(\frac{\wh{\mbf M}^x}{n}\r)^{-1}\frac{(\wh{\bm\Lambda}-\bm\Lambda\bm{\mathcal H})^\prime}{\sqrt n}\frac{\bm\Lambda\bm F^\prime\bm E}{nT}\frac{\bm\alpha}{\sqrt n}\nn\\
&+\l(\frac{\wh{\mbf M}^x}{n}\r)^{-1}\frac{(\wh{\bm\Lambda}-\bm\Lambda\bm{\mathcal H})^\prime}{\sqrt n}\frac{\bm E^\prime\bm  E}{nT}\frac{\bm\alpha}{\sqrt n}
= \bm I+\bm {II}+\bm {III}+\bm {IV}+\bm V+\bm {VI}, \;\text{say.} \label{eq:odeon}
\end{align}
Then, because of  \eqref{eq:oslo3} in the proof of Lemma \ref{lem:GxiBAI},
\begin{align}
\E\l[\l\Vert \frac{\bm F^\prime\bm E\bm\alpha}{ nT}\r\Vert^2\r] =
\E\l[\l\Vert\frac 1{nT}\sum_{t=1}^T\sum_{j=1}^n\mbf F_t e_{jt}{\alpha}_j \r\Vert^2\r] 
\le \E\l[\l\Vert\frac 1{nT}\sum_{t=1}^T\sum_{j=1}^n\mbf F_t e_{jt} \r\Vert^2\r] M_\alpha^2
\le \frac{r M_FM_{1 e}M_\alpha^2}{nT}.\label{eq:odeon2}
\end{align}
Therefore, by Lemmas \ref{lem:LLN}(iv), \ref{lem:MO1}(iv), \ref{lem:HO1}(i),  \ref{lem:HO1bis}(i),
and using 
\eqref{eq:2a3bis} in the proof of Lemma \ref{prop:factor} 
and
\eqref{eq:odeon2}, we get
\begin{align}
&\Vert \bm I\Vert \le \l\Vert \l(\frac{\wh{\mbf M}^x}{n}\r)^{-1}\r\Vert\,\l\Vert \bm{\mathcal H}\r\Vert\,\l\Vert\frac{\bm\Lambda^\prime\bm\Lambda}{n}\r\Vert\,
\l\Vert\frac{\bm F^\prime\bm E\bm\alpha }{ nT}\r\Vert\,\l\Vert \wh{\mbf H}\r\Vert=O_{\mathrm P}\l(\frac 1{\sqrt {nT}}\r),\label{eq:odeon3}\\
&\Vert \bm {II}\Vert\le  \l\Vert \l(\frac{\wh{\mbf M}^x}{n}\r)^{-1}\r\Vert\,\l\Vert \bm{\mathcal H}\r\Vert\,\l\Vert \frac{\bm\Lambda^\prime\bm E^\prime\bm F}{nT}\r\Vert \,\l\Vert \frac{\bm\Lambda}{\sqrt n} \r\Vert\,
\l\Vert\frac{\bm\alpha}{\sqrt n}\r\Vert\,
\l\Vert \wh{\mbf H}\r\Vert 
=O_{\mathrm P}\l(\frac 1{\sqrt {nT}}\r),\label{eq:odeon4}
\end{align}
since $\Vert\frac{\bm\alpha}{\sqrt n}\Vert^2 =\frac 1n\sum_{i=1}^n \alpha_i^2\le M_\alpha$ by Assumption \ref{ass:center}(c). Moreover, because of Lemmas \ref{lem:LLN}(iii), \ref{lem:MO1}(iv), \ref{lem:HO1}(i), and \ref{lem:HO1bis}(i),
\begin{align}\label{eq:odeon5}
\Vert \bm {III}\Vert \le  \l\Vert\l(\frac{\wh{\mbf M}^x}{n}\r)^{-1}\r\Vert\,\l\Vert\bm{\mathcal H}\r\Vert\, 
\l\Vert\frac{\bm\alpha}{\sqrt n}\r\Vert\,
\l\Vert\frac{\bm\Lambda^\prime\bm E^\prime\bm E }{n^{3/2}T}\r\Vert\,
\l\Vert\wh{\mbf H}\r\Vert=O_{\mathrm P}\l(\max\l(\frac 1n,\frac 1{\sqrt {nT}}\r)\r).
\end{align}
Similarly, because of Proposition \ref{prop:L}(a), Lemmas \ref{lem:FTLN}(i),  \ref{lem:LLN}(i),  \ref{lem:MO1}(iv), and \ref{lem:HO1}(i), 
\begin{align}
\Vert \bm {IV} \Vert &\le \l\Vert\l(\frac{\wh{\mbf M}^x}{n}\r)^{-1}\r\Vert\,\l\Vert \frac{\wh{\bm\Lambda}-\bm\Lambda\bm{\mathcal H}}{\sqrt n}\r\Vert \,\l\Vert \frac{\bm  E^\prime
\bm F}{\sqrt nT}\r\Vert\,
 \l\Vert \frac{\bm\Lambda }{\sqrt n}\r\Vert\, 
 \l\Vert \frac{\bm\alpha }{\sqrt n}\r\Vert\, 
\l\Vert \wh{\mbf H}\r\Vert=  O_{\mathrm P}\l(\max\l(\frac{1}{n\sqrt {T}},\frac 1{T}\r)\r), \label{eq:odeon6}\\ 
\Vert \bm V\Vert &\le \l\Vert\l(\frac{\wh{\mbf M}^x}{n}\r)^{-1}\r\Vert\,\l\Vert \frac{\wh{\bm\Lambda}-\bm\Lambda\bm{\mathcal H}}{\sqrt n}\r\Vert \,\l\Vert \frac{\bm  E^\prime\bm F\bm}{\sqrt nT}\r\Vert\, 
\l\Vert \frac{\bm\Lambda  }{\sqrt n}\r\Vert\, 
\l\Vert \frac{\bm\alpha  }{\sqrt n}\r\Vert\, 
\l\Vert \wh{\mbf H}\r\Vert=  O_{\mathrm P}\l(\max\l(\frac{1}{n\sqrt {T}},\frac 1{T}\r)\r),\label{eq:odeon7}
\end{align}
and, because of Proposition \ref{prop:L}(a), Lemmas \ref{lem:LLN}(iii), \ref{lem:MO1}(iv), and \ref{lem:HO1}(i),
\begin{align}\label{eq:odeon7bis}
\Vert \bm {VI}\Vert \le \l\Vert\l(\frac{\wh{\mbf M}^x}{n}\r)^{-1}\r\Vert\,\l\Vert \frac{\wh{\bm\Lambda}-\bm\Lambda\bm{\mathcal H}}{\sqrt n}\r\Vert \,\l\Vert \frac{\bm  E^\prime \bm E}{nT}\r\Vert\, \l\Vert \frac{\bm\alpha  }{\sqrt n}
\r\Vert\, \l\Vert \wh{\mbf H}\r\Vert
=  O_{\mathrm P}\l(\max\l(\frac 1{n^2},\frac{1}{n\sqrt {T}},\frac 1{T}\r)\r).
\end{align}
By using \eqref{eq:odeon3}, \eqref{eq:odeon4}, \eqref{eq:odeon5}, \eqref{eq:odeon6}, \eqref{eq:odeon7}, and \eqref{eq:odeon7bis} into \eqref{eq:odeon}
\beq\label{eq:odeon8}
\l\Vert \frac {(\wh{\bm\Lambda}-\bm\Lambda\wh{\mbf H})^\prime \bm\alpha}n \r\Vert = O_{\mathrm P}\l(\max\l(\frac 1 {n},\frac 1{\sqrt{nT}},\frac 1T\r)\r).
\eeq
Therefore, by using \eqref{eq:odeon8} and again \eqref{eq:pave} and since  $\Vert \bar{\mbf F}\Vert = O_{\mathrm P}(T^{-1/2})$ by  \eqref{eq:mediebelle} and $\Vert\bm \mu\Vert=O(1)$ by Assumption \ref{ass:center}(c),
we can apply Lemma \ref{prop:factor} to \eqref{eq:simonebiased} and get
\begin{align}
\wh{\bm \mu}^\prime- \bm\mu^\prime(\wh{\mbf H}^{-1})^\prime &=
\bar{\mbf F}^\prime(\wh{\mbf H}^{-1})^\prime+  \frac{\bm\alpha^\prime\bm\Lambda\wh{\mbf H}}{n}\l(\frac {\wh{\mbf M}^x}n\r)^{-1} +
  O_{\mathrm P}\l(\max\l(\frac 1n,\frac 1{\sqrt {nT}},\frac 1T\r)\r)\nn\\
&= \bar{\mbf F}^\prime(\wh{\mbf H}^{-1})^\prime+  {\bm\alpha^\prime\bm\Lambda  \l(\bm\Lambda^\prime\bm\Lambda\r)^{-1}
{\bm\Lambda}^\prime\wh{\bm\Lambda}
}\l(\wh{\bm\Lambda}^\prime \wh{\bm\Lambda} \r)^{-1} +
  O_{\mathrm P}\l(\max\l(\frac 1n,\frac 1{\sqrt {nT}},\frac 1T\r)\r)\nn\\
  &= \bar{\mbf F}^\prime(\wh{\mbf H}^{-1})^\prime+  \bm\alpha^\prime\bm\Lambda  \l(\bm\Lambda^\prime\bm\Lambda\r)^{-1}
(\wh{\mbf H}^{-1})^\prime +
  O_{\mathrm P}\l(\max\l(\frac 1n,\frac 1{\sqrt {nT}},\frac 1T\r)\r)\nn\\
  &=   \bm\alpha^\prime\bm\Lambda  \l(\bm\Lambda^\prime\bm\Lambda\r)^{-1}
(\wh{\mbf H}^{-1})^\prime +
  O_{\mathrm P}\l(\max\l(\frac 1n,\frac 1{\sqrt {T}}\r)\r),
\end{align}
by Propositions \ref{prop:LLFF}(a), \ref{prop:LLFF}(e), \eqref{eq:triangolo}, and \eqref{eq:truccoloT}.
  This proves part (d) and completes the proof. $\Box$
\subsection{Proof of Proposition \ref{corol:K00}}\label{corol:K00proof}

Throughout the proof, let
\bit
\item  [-] $\bm p_{i0}$ is such that $\lim_{n\to\infty} \Vert \sqrt n\mbf v_{i}^{C}-\bm p_{i0}\Vert=0$;
\item  [-] $\bm P_{n,\infty}=(\bm p_{01}\cdots \bm p_{0n})^{\prime}$ be the $n \times r$ matrix having as rows $\bm p_{i0}^{\prime}$, $i=1,\ldots,n$;
\item  [-] $\bm W_{\infty}=\lim_{n\to\infty} \frac{\bm P_{n,\infty}}{\sqrt n}$, which is $\infty\times r$;
\item  [-]$\bm C_{t,\infty}=\text{m.s.-}\lim_{n\to\infty} \frac{\bm{C}_{t}}{\sqrt n}$, which is an infinite dimensional process.
\eit

Then, since $\Vert \sqrt n\mbf v_{i}^{C\prime} \Vert = O(1)$ by Lemma \ref{lem:covarianzerighe}(ii),  $\Vert \bm p_{i0}^{\prime}\Vert=O(1)$, $\Vert \bm P_{n,\infty}\Vert=O(\sqrt n)$, so that $\Vert \bm W_{\infty}\Vert = O(1)$. And, since $\Vert\bm{C}_t\Vert =O_{\mathrm {m.s.}}(\sqrt n)$ by Lemmas \ref{lem:FTLN}(i) and \ref{lem:FTLN}(ii), then $\Vert\bm{C}_{t,\infty}\Vert =O_{\text{m.s.}}(1)$.

First of all, from Assumption \ref{ass:sign_eval}(a) it follows that, in \eqref{eq:ups000} in the proof of Proposition \ref{prop:KKK}, $\bm{\mathcal J}^*=\mbf I_r$, so that
\beq\label{eq:convenzione}
\lim_{n\to\infty}\Vert \mbf K-\bm\Upsilon_0\Vert = 0.
\eeq 
Moreover, from \eqref{eq:riunioni2},
\begin{align}
&
 \bm\Upsilon_0^\prime (\bm\Gamma^F)^{1/2}(\bm\Sigma_\Lambda)^{1/2} \bm\Upsilon_1  
=\bm{\mathcal J}_0\bm{\mathcal J}_1 \bm V_0^{1/2},\label{eq:convenzione2a}
\end{align}
since $\bm V_0$ and $\bm{\mathcal J}_1$ commute. So by Assumption \ref{ass:sign_eval}(a) and since $0<[\bm V_0]_{jj}<\infty$, $j=1,\ldots, r$, by Lemmas \ref{lem:Vzero}(i) and  \ref{lem:Vzero}(iii), from \eqref{eq:convenzione2a} we have 
\beq
\bm{\mathcal J}_0\bm{\mathcal J}_1 = \mbf I_r.\label{eq:convenzione2}
\eeq

We then consider all cases one-by-one.


\paragraph{I.a} In part (I.a) we impose $\bm\Gamma^F=\mbf I_r$ and leave $\bm\Lambda$ unrestricted. From \eqref{eq:kappaproj} in the proof of Proposition \ref{prop:L}, 
\beq\label{eq:kappaprojnew}
{\bm\Lambda}= \mbf V^{C}({\mbf M^{C}})^{1/2}\mbf K^{-1}= \mbf V^{C}({\mbf M^{C}})^{1/2}\mbf K^{\prime},
\eeq
since $\mbf K$ is orthogonal because of Lemma \ref{lem:KO1}(iii). By taking the $i$th row of \eqref{eq:kappaprojnew} we prove part (I.a.i) for fixed $n$, while for $n\to\infty$ the proof follows from the definition of $\bm p_{i0}$, Lemma \ref{lem:Vzero}(i), and \eqref{eq:convenzione}. 
Moreover, by linear projection of $\bm\Lambda$ onto $\bm{C}_t$ and using \eqref{eq:kappaprojnew} and since $\mbf K$ is orthogonal because of Lemma \ref{lem:KO1}(iii), for all $t\in\mathbb Z$,
\begin{align}
\mbf F_t &= \l(\bm\Lambda^\prime\bm\Lambda\r)^{-1}\bm\Lambda^\prime\bm{C}_t=\l((\mbf K^{-1})^\prime (\mbf M^{C})^{1/2}\mbf V^{{C}\prime}\mbf V^{C}(\mbf M^{C})^{1/2}\mbf K^{-1}\r)^{-1}(\mbf K^{-1})^\prime (\mbf M^{C})^{1/2}\mbf V^{{C}\prime}\bm{C}_t\nn\\
&= \l(\mbf K \mbf M^{C} \mbf K^{-1}\r)^{-1}\mbf K (\mbf M^{C})^{1/2}\mbf V^{{C}\prime}\bm{C}_t= \mbf K (\mbf M^{{C}})^{-1/2} \mbf V^{{C}\prime}\bm{C}_t.\nn
\end{align}
This proves part (I.a.ii) for fixed $n$, while for $n\to\infty$ the proof follows from the definition of $\bm W_{\infty}$ and $\bm C_{t,\infty}$, Lemma \ref{lem:Vzero}(i), and \eqref{eq:convenzione}.

\paragraph{II.a} 
 In part (II.a) we impose $\bm\Gamma^F=\mbf I_r$ and $\frac{\bm\Lambda^\prime\bm\Lambda}{n}$ to be diagonal for all $n\in\mathbb N$. 
 First, notice that
\beq\label{eq:LLprimo}
\bm\Gamma^{C}=\bm\Lambda\bm\Lambda^\prime,
\eeq
implying that the $r$ eigenvalues of $\bm\Gamma^{C}$ and of $\bm\Lambda^\prime\bm\Lambda$ coincide for all $n\in\mathbb N$, that is
\beq\label{eq:rotturaasyB}
 \frac{\bm\Lambda^\prime\bm\Lambda}{n} = \frac{\mbf M^{C}}{n}.
\eeq
 Second, from Lemma \ref{lem:KO1}(iii), 
\beq\label{eq:cirisiamo2BO}
 \frac{\mbf M^{C}}n  \mbf K^{\prime}=  \mbf K^\prime \frac{\bm\Lambda^\prime \bm\Lambda}n.
\eeq
From \eqref{eq:rotturaasyB} and \eqref{eq:cirisiamo2BO},
\beq\nn
\l(\frac{\mu^C_i}{n}-\frac{\mu^C_j}n\r)[\mbf K]_{ij}= 0, \quad i,j=1,\ldots, r.
\eeq
Hence, since by Assumption \ref{ass:sign_eval}(b), $\frac{\mu^C_i}n\ne \frac{\mu^C_j}n$ if $i\ne j$ and $0<\frac{\mu_j^C}{n}<\infty$, $j=1,\ldots, r$, for all $n\in\mathbb N$, by Lemmas \ref{lem:MO1}(i) and \ref{lem:MO1}(ii), and since $\mbf K$ has full-rank by Lemma \ref{lem:KO1}(ii), then each column of $\mbf K$ must have at least one non-zero entry, thus  $\mbf K$ must be diagonal. Therefore, since by Lemma \ref{lem:KO1}(iii) $\mbf K$ is also orthogonal, it must be that 
\beq\label{eq:KlimiteIBBO}
\mbf K=\bm S,
\eeq 
where $\bm S$ is diagonal with entries $\pm 1$ and depends only on $n$. From \eqref{eq:kappaprojnew} (which still holds) and \eqref{eq:KlimiteIBBO}, $\bm\Lambda=\mbf V^{C}({\mbf M^{C}})^{1/2}\bm S$ 
and, by taking the $i$th row, we prove part (II.a.i) for fixed $n$, while for $n\to\infty$ the proof is the same as that of part (III.a.i) below, because of Assumption \ref{ass:common}(a).
 Moreover, by linear projection of $\bm\Lambda$ onto $\bm{C}_t$, for all $t\in\mathbb Z$, 
$$
\mbf F_t = (\bm\Lambda^\prime\bm\Lambda)^{-1}\bm\Lambda^\prime\bm{C}_t= \bm S(\mbf M^{C})^{-1/2} \mbf V^{{C}\prime}\bm{C}_t, 
$$
which proves part (II.a.ii) for fixed $n$, while for $n\to\infty$ the proof is the same as that of part (III.a.ii) below, because of Assumption \ref{ass:common}(a).

\paragraph{III.a}
In part (III.a) we impose $\bm\Gamma^F=\mbf I_r$ and $\bm\Sigma_\Lambda$ to be diagonal. 
First, from \eqref{eq:LLprimo} (which still holds) we must have
\beq\label{eq:rotturaasy}
\lim_{n\to\infty} \l\Vert \frac{\bm\Lambda^\prime\bm\Lambda}{n} -  \frac{\mbf M^{C}}{n}\r\Vert =0,
\eeq
and, by Assumption \ref{ass:common}(a) and Lemma \ref{lem:Vzero}(i),  \eqref{eq:rotturaasy} is equivalent to
\beq\label{eq:stoppani}
\bm\Sigma_\Lambda=\bm V_0.
\eeq
Second, by definition the columns $\bm\Upsilon_0$ are eigenvectors of $(\bm\Gamma^F)^{1/2}\bm\Sigma_{\Lambda}(\bm\Gamma^F)^{1/2}=\bm\Sigma_{\Lambda}$, i.e.,
\beq\label{eq:stoppani2panda}
\bm\Sigma_\Lambda\bm\Upsilon_0^\prime= \bm\Upsilon_0^\prime\bm V_0.
\eeq
Therefore, because of \eqref{eq:stoppani} and  \eqref{eq:stoppani2panda}, 
\beq
\l([\bm V_0]_{ii}-[\bm V_0]_{jj}\r)  [ \bm\Upsilon_0]_{ij}= 0, \quad i,j=1,\ldots, r.
\eeq
Hence, since by Assumption \ref{ass:sign_eval}(b), which implies Assumption \ref{ass:eval}, $[\bm V_0]_{ii}\ne[\bm V_0]_{jj}$ for $i\ne j$, and since $0<[\bm V_0]_{jj}<\infty$ by Lemmas \ref{lem:Vzero}(i) and \ref{lem:Vzero}(iii),
$\bm\Upsilon_0$ must be diagonal, because $\bm\Upsilon_0$, being a matrix of eigenvectors, has full rank, thus it has at least one non-zero entry in each column. Therefore, since $\bm\Upsilon_0$ is orthogonal, it must be that 
\beq\label{eq:KlimiteIpanda}
\bm\Upsilon_0=\bm{\mathcal S}_0,
\eeq 
where $\bm{ S}_0$ is diagonal with entries $\pm 1$ and independent of $n$ and $T$. 
Furthermore, by \eqref{eq:KlimiteIpanda} and \eqref{eq:convenzione}, we also have
\beq\label{eq:KlimiteIunico}
\lim_{n\to\infty}\Vert \mbf K-\bm {\mathcal S}_0\Vert=\lim_{n\to\infty}\Vert \mbf K-\bm\Upsilon_0\Vert=0.
\eeq 
By taking the $i$th row of \eqref{eq:kappaprojnew} (which still holds), and by \eqref{eq:KlimiteIunico}, 
\begin{align}
\lim_{n\to\infty}\l\Vert {\bm\lambda}_i^\prime- \bm p_{i0}^{\prime}(\bm V_0)^{1/2} \bm{\mathcal S}_0\r\Vert \le&\,
\lim_{n\to\infty}\l\Vert {\bm\lambda}_i^\prime- \sqrt n\mbf v_{i}^{C\prime}\l(\frac{\mbf M^C}{n}\r)^{1/2} \mbf K^\prime\r\Vert\nn\\
&+\lim_{n\to\infty}\l\Vert  \sqrt n\mbf v_{i}^{C\prime}\l(\frac{\mbf M^C}{n}\r)^{1/2} \mbf K^\prime-\bm p_{i0}^{\prime}(\bm V_0)^{1/2} \bm{\mathcal S}_0\r\Vert=0,\label{eq:pi0}
\end{align}
because of Lemma \ref{lem:Vzero}(i). This proves part (III.a.i). 
Moreover, by linear projection of $\bm\Lambda$ onto $\bm{C}_t$, for all $t\in\mathbb Z$, 
$\mbf F_t = (\bm\Lambda^\prime\bm\Lambda)^{-1}\bm\Lambda^\prime\bm{C}_t$. By \eqref{eq:kappaprojnew} (which still holds),  \eqref{eq:rotturaasy}, \eqref{eq:KlimiteIunico}, and Lemma \ref{lem:Vzero}(i),
\begin{align}\label{eq:house3}
\lim_{n\to\infty}\l\Vert \frac{\bm\Lambda^\prime\bm\Lambda}n - \bm V_0\r\Vert\le&\, \lim_{n\to\infty}\l\Vert \frac{\bm\Lambda^\prime\bm\Lambda}n - \mbf K\frac{\mbf M^C}{n}\mbf K^\prime\r\Vert
+ \lim_{n\to\infty}  \l\Vert\mbf K\frac{\mbf M^C}{n}\mbf K^\prime-\bm V_0\r\Vert\nn\\
\le&\, \lim_{n\to\infty}\l\Vert \frac{\bm\Lambda^\prime\bm\Lambda}n - \frac{\mbf M^C}{n}\r\Vert
+ \lim_{n\to\infty}  \l\Vert\frac{\mbf M^C}{n}-\bm V_0\r\Vert\nn\\
& +\lim_{n\to\infty}\Vert \mbf K-\bm {\mathcal S}_0\Vert(1+\Vert \mbf K-\bm {\mathcal S}_0\Vert) \l\{ \l\Vert \frac{\mbf M^C}{n}\r\Vert+\l\Vert \bm V_0\r\Vert\r\}
=0.
\end{align}
 Then, by definition of $\bm p_{i0}$ and $\bm P_{n,\infty}$,
\begin{align}
\lim_{n\to\infty} \E\l[\l\Vert\frac{\mbf V^{C\prime} \bm C_t} {\sqrt n}-\frac{\bm P_{n,\infty}^\prime\bm C_{t }}n \r\Vert^2\r]\le &\,
\lim_{n\to\infty} \l\Vert \mbf V^{C\prime}-\frac{\bm P_{n,\infty}^\prime}{\sqrt n} \r\Vert^2_F\, \E\l[\l\Vert \frac{\bm C_t}{\sqrt n}\r\Vert^2\r]\nn\\
=&\, \lim_{n\to\infty}\frac 1n\sum_{i=1}^n \l\Vert\sqrt n\mbf v_{i}^C-\bm p_{i0} \r\Vert^2\, \E\l[\l\Vert \frac{\bm C_t}{\sqrt n}\r\Vert^2\r]\nn\\
\le&\, \lim_{n\to\infty}\max_{i=1,\ldots, n} \l\Vert\sqrt n\mbf v_{i}^C-\bm p_{i0} \r\Vert^2\, \E\l[\l\Vert \frac{\bm C_t}{\sqrt n}\r\Vert^2\r]\nn\\
=&\,o_{\mathrm{m.s}}(1),\label{eq:eataly}
\end{align}
since $\l\Vert\frac{\bm{C}_t}{\sqrt n}\r\Vert \le\l\Vert\frac{\bm\Lambda}{\sqrt n}\r\Vert\,\Vert \mbf F_t\Vert =O_{\mathrm {m.s.}}(1)$ because of Lemmas \ref{lem:FTLN}(i) and \ref{lem:FTLN}(ii).
Therefore, by \eqref{eq:kappaprojnew} (which still holds), \eqref{eq:KlimiteIunico}, \eqref{eq:house3}, \eqref{eq:eataly}, and Lemma \ref{lem:Vzero}(i),
\beq\label{eq:marcallo}
\text{m.s.-}\lim_{n\to\infty}\l\Vert \mbf F_t - \bm {\mathcal S}_0 \bm V_0^{-1/2} \frac{\bm P_{n,\infty}^\prime\bm C_{t }}n \r\Vert =
\text{m.s.-}\lim_{n\to\infty}\l\Vert \l(\frac{\bm\Lambda^\prime\bm\Lambda}n\r)^{-1} \frac{\bm\Lambda^\prime \bm C_t}n - \bm {\mathcal S}_0 \bm V_0^{-1/2} \frac{\bm P_{n,\infty}^\prime\bm C_{t }}n \r\Vert=0.
\eeq
Finally, by definition of $\bm W_{\infty}$ and $\bm C_{t,\infty}$, 
\beq\label{eq:marcallo2}
\text{m.s.-}\lim_{n\to\infty} \frac{\bm P_{n,\infty}^\prime \bm C_t}{n} =\text{m.s.-}\lim_{n\to\infty} \frac 1n\sum_{i=1}^n \bm p_{i0} C_{it} = \bm W_{\infty}^\prime \bm C_{t,\infty},
\eeq
and, by using \eqref{eq:marcallo2} into \eqref{eq:marcallo}, we prove part (III.a.ii). 

\paragraph{IV.a} In part (IV.a) we impose $\frac{\bm F^\prime \bm F}T=\mbf I_r$ for all $T
\in\mathbb N$ and leave $\bm\Lambda$ unrestricted. 
In this case, since $\wh{\bm\Gamma}^C =\wh{\mbf V}^{C}\wh{\mbf M}^{C}\wh{\mbf V}^{C\prime} = \bm\Lambda\bm\Lambda^\prime$, following analogous steps leading to \eqref{eq:kappaproj} in the proof of Proposition \ref{prop:L}, \eqref{eq:kappaprojnew} is replaced by
\beq\label{eq:kappaprojnewT}
{\bm\Lambda}= \wh{\mbf V}^{C}(\wh{\mbf M}^{C})^{1/2}\wh{\mbf Q}^\prime,
\eeq
where following similar steps leading to the proof of Lemma \ref{lem:KO1}(iii) we see that $\wh{\mbf Q}$ is the matrix of normalized eigenvectors of $(\frac{\bm F^\prime \bm F}T)^{1/2}\frac{\bm\Lambda^\prime\bm\Lambda}n(\frac{\bm F^\prime \bm F}T)^{1/2}=\frac{\bm\Lambda^\prime\bm\Lambda}n$. By taking the $i$th row of \eqref{eq:kappaprojnewT}, we prove part (IV.a.i) for fixed $n$ and $T$. Moreover, by linear projection of $\bm\Lambda$ onto $\bm{C}_t$, for all $t\in\mathbb Z$, 
$\mbf F_t = (\bm\Lambda^\prime\bm\Lambda)^{-1}\bm\Lambda^\prime\bm{C}_t$, and part (IV.a.ii) for fixed $n$ and $T$ follows directly from \eqref{eq:kappaprojnewT}. This holds only for $t=1,\ldots, T$ since $\wh{\mbf V}^{C}$ and $\wh{\mbf M}^{C}$ are obtained using only $T$ observations.
Then, $\lim_{T\to\infty} \Vert \bm\Gamma^F-T^{-1}\bm F^\prime \bm F\Vert=\Vert \bm\Gamma^F-\mbf I_r\Vert$ almost surely,
 because of Assumption \ref{ass:common}(c-ii), and since $\mathbf I_r$ does not depend on $T$ and so it is equal to its limit. So $\bm\Gamma^F=\mbf I_r$ almost surely and
 $\bm\Gamma^C=\bm\Lambda\bm\Lambda^\prime$ almost surely. Therefore, $\lim_{T\to\infty}\Vert\wh{\bm\Gamma}^C-{\bm\Gamma}^C\Vert=0$ almost surely and so their eigenvectors and eigenvalues coincide:
$\lim_{T\to\infty} \Vert \wh{\mbf V}^C-\mbf V^C\Vert=0$ and $\lim_{T\to\infty}  \frac 1n\Vert \wh{\mbf M}^C-\mbf M^C\Vert=0$  almost surely. The proof of part (IV.a) when $T\to\infty$ and $n$ is fixed follows immediately since $\lim_{T\to\infty}\Vert \wh{\mbf Q}-\mbf K\Vert =0$ almost surely.
The proof of part (IV.a) when $n,T\to\infty$ follows by simply taking the limit of the results for fixed $n$ and $T\to\infty$ and using the definitions of $\bm p_{i0}$, $\bm W_\infty$, and $\bm C_{t, \infty}$. 

\paragraph{V.a} In part (V.a) we impose $\frac{\bm F^\prime \bm F}T=\mbf I_r$ for all $T
\in\mathbb N$ and  $\frac{\bm\Lambda^\prime\bm\Lambda}n$ to be diagonal for all $n\in\mathbb N$. In this case \eqref{eq:rotturaasyB} and \eqref{eq:cirisiamo2BO} are replaced by
\begin{align}
&\frac{\bm\Lambda^\prime\bm\Lambda}{n} = \frac{\wh{\mbf M}^{C}}{n}, \label{eq:rotturaasyBT}\\
 &\frac{\wh{\mbf M}^{C}}n  \wh{\mbf Q}^{\prime}=  \wh{\mbf Q}^\prime \frac{\bm\Lambda^\prime \bm\Lambda}n.\label{eq:cirisiamo2BOT}
\end{align}
Hence, since $\lim_{T\to\infty} \frac 1n\Vert \wh{\mbf M}^C-\mbf M^C\Vert=0$ almost surely (see the proof of part (IV.a)), by Assumption \ref{ass:sign_eval}(b) $\frac{\wh{\mu}^C_i}n\ne \frac{\wh{\mu}^C_j}n$ if $i\ne j$ 
and $0<\lim_{T\to\infty} \frac{\wh{\mu}^C_j}n<\infty$, $j=1,\ldots, r$, by Lemmas \ref{lem:MO1}(i) and \ref{lem:MO1}(ii), almost surely and for all $n\in\mathbb N$,
 \eqref{eq:KlimiteIBBO} is replaced by
\beq\label{eq:KlimiteIBBOT}
\wh{\mbf Q}=\wh{\bm S},
\eeq 
where $\wh{\bm S}$ is diagonal with entries $\pm 1$ and depends on $n$ and $T$. Part (V.a.i) for fixed $n$ and $T$ then follows by taking the $i$th row of \eqref{eq:kappaprojnewT} (which still holds) and using \eqref{eq:KlimiteIBBOT}. Moreover, by linear projection of $\bm\Lambda$ onto $\bm{C}_t$, for all $t\in\mathbb Z$, 
$\mbf F_t = (\bm\Lambda^\prime\bm\Lambda)^{-1}\bm\Lambda^\prime\bm{C}_t$, and part (V.a.ii) for fixed $n$ and $T$ follows directly from \eqref{eq:kappaprojnewT} (which still holds) and using \eqref{eq:KlimiteIBBOT}. This holds only for $t=1,\ldots, T$ since $\wh{\mbf V}^{C}$ and $\wh{\mbf M}^{C}$ are obtained using only $T$ observations.
 Then, since $\lim_{T\to\infty}\Vert \wh{\mbf Q}-\mbf K\Vert =0$, there exists a matrix $\bm S$ diagonal with entries $\pm 1$ and dependent only on $n$ such that $\lim_{T\to\infty}\Vert \wh{\bm S}-\bm S\Vert=0$ almost surely, and the proof of part (IV.a) when $T\to\infty$ and $n$ is fixed follows immediately. The proof of part (V.a) when $n,T\to\infty$ follows by simply taking the limit of the results for fixed $n$ and $T\to\infty$ and using the definitions of $\bm p_{i0}$, $\bm W_\infty$, and $\bm C_{t, \infty}$. 

\paragraph{VI.a}  In part (VI.a) we impose $\frac{\bm F^\prime \bm F}T=\mbf I_r$ for all $T
\in\mathbb N$ and  $\bm\Sigma_\Lambda$ to be diagonal. In this case the proof is the same as the proof of part (V.a) when $n,T\to\infty$.


\paragraph{I.b}
In part (I.b) we  impose $\bm\Sigma_\Lambda=\mbf I_r$ and leave $\bm F$ unrestricted. Then, from \eqref{eq:kappaproj} in the proof of Proposition \ref{prop:L},  
\beq
\frac{\bm\Lambda^\prime\bm\Lambda}{n}(\bm\Gamma^F)^{1/2}\mbf K =\frac{\bm\Lambda^\prime}{\sqrt n} {\mbf V^{C}\l(\frac{\mbf M^{C}}n\r)^{1/2}},\nn
\eeq
which is equivalent to
\beq\label{eq:costume}
\l(\frac{\mbf M^{C}}n\r)^{-1/2}\mbf K^\prime(\bm\Gamma^F)^{1/2}\frac{\bm\Lambda^\prime\bm\Lambda}{n}=\frac{\mbf V^{C\prime}\bm\Lambda}{\sqrt n}.
\eeq
Since $\mbf V^{C}\mbf V^{C\prime}\bm\Lambda=\bm\Lambda$ because the columns of $\bm\Lambda$ and $\mbf V^C$ span the same space, from \eqref{eq:costume} we get
\beq\label{eq:costume2}
\frac{\bm\Lambda}{\sqrt n} = \mbf V^C\l(\frac{\mbf M^{C}}n\r)^{-1/2}\mbf K^\prime(\bm\Gamma^F)^{1/2}\frac{\bm\Lambda^\prime\bm\Lambda}{n}.
\eeq
By Assumption \ref{ass:common}(a), from \eqref{eq:costume2},
\beq\label{eq:costume3}
\lim_{n\to\infty} \l\Vert \frac{\bm\Lambda}{\sqrt n} - \mbf V^C\l(\frac{\mbf M^{C}}n\r)^{-1/2}\mbf K^\prime(\bm\Gamma^F)^{1/2}\r\Vert=0.
\eeq
Moreover, by Assumption \ref{ass:common}(a), and Lemmas \ref{lem:Vzero}(i) and \ref{lem:KO1}(iii),
\begin{align}
\lim_{n\to\infty} \l\Vert \mbf K^\prime (\bm\Gamma^F)^{1/2}\frac{\bm\Lambda^\prime\bm\Lambda}{n} (\bm\Gamma^F)^{1/2}\mbf K- \frac{\mbf M^C}{n} \r\Vert 
&=\lim_{n\to\infty} \l\Vert \mbf K^\prime \bm\Gamma^F\mbf K- \bm V_0 \r\Vert,\nn
\end{align}
which implies
\beq\label{eq:costume4}
\lim_{n\to\infty} \l\Vert \mbf K^\prime (\bm\Gamma^F)^{1/2} -\bm V_0 \mbf K^\prime(\bm\Gamma^F)^{-1/2} \r\Vert=0.
\eeq
By taking the $i$th row of \eqref{eq:costume3}, and using Lemma \ref{lem:Vzero}(i), \eqref{eq:convenzione} and \eqref{eq:costume4},
\beq\label{eq:costume5}
\lim_{n\to\infty} \l\Vert\bm\lambda_i^\prime - \bm p_{i0}^{C} (\bm V_0)^{1/2}  \bm\Upsilon_0^\prime(\bm\Gamma^F)^{-1/2} \r \Vert=
\lim_{n\to\infty} \l\Vert\bm\lambda_i^\prime - \sqrt n \mbf v_i^{C\prime} (\bm V_0)^{1/2} \mbf K^\prime(\bm\Gamma^F)^{-1/2} \r \Vert =0,
\eeq
where $\bm p_{i0}$ is defined in \eqref{eq:pi0}. Moreover, by \eqref{eq:convenzione2a} and \eqref{eq:convenzione2} we get
\beq\label{eq:costume6}
\lim_{n\to\infty}\l\Vert (\bm V_0)^{1/2}  \bm\Upsilon_0^\prime(\bm\Gamma^F)^{-1/2} - \bm{\mathcal J}_0\bm{\mathcal J}_1\bm\Upsilon_1^\prime\r\Vert=\lim_{n\to\infty}\l\Vert 
(\bm V_0)^{1/2}  \bm\Upsilon_0^\prime(\bm\Gamma^F)^{-1/2} - \bm\Upsilon_1^\prime\r\Vert=0.
\eeq
By using \eqref{eq:costume6} in \eqref{eq:costume5},
we prove part (I.b.i). Moreover, by linear projection of $\bm\Lambda$ onto $\bm{C}_t$, for all $t\in\mathbb Z$, 
$
\mbf F_t = (\bm\Lambda^\prime\bm\Lambda)^{-1}\bm\Lambda^\prime\bm{C}_t.
$
By Assumption \ref{ass:common}(a), $\lim_{n\to\infty}\Vert \frac{\bm\Lambda^\prime\bm\Lambda}n-\mbf I_r\Vert =0$. Therefore,  by \eqref{eq:riunioni}, \eqref{eq:costume3}, Lemma \ref{lem:Vzero}(i), and since, by definition, $\bm P_{n,\infty}=\sqrt n \mbf V^C$,
\begin{align}
\text{m.s.-}\lim_{n\to\infty}\l\Vert\mbf F_t -  
(\bm\Gamma^F)^{1/2}\mbf K(\bm V_0)^{-1/2}
\frac{\bm P_{n,\infty}^\prime\bm C_t}{n} \r\Vert&=
\text{m.s.-}\lim_{n\to\infty}\l\Vert \l(\frac{\bm\Lambda^\prime\bm\Lambda}n\r)^{-1} \frac{\bm\Lambda^\prime \bm C_t}n -  
(\bm\Gamma^F)^{1/2}\mbf K(\bm V_0)^{-1/2}
\frac{\bm P_{n,\infty}^\prime\bm C_t}{n} \r\Vert\nn\\
&=
\text{m.s.-}\lim_{n\to\infty}\l\Vert \l(\frac{\bm\Lambda^\prime\bm\Lambda}n\r)^{-1} \frac{\bm\Lambda^\prime \bm C_t}n -  
\bm \Upsilon_1
\frac{\bm P_{n,\infty}^\prime\bm C_t}{n} \r\Vert.\label{eq:marcallo3}
\end{align}
By definition of $\bm W_{\infty}$ and $\bm C_{t,\infty}$, from \eqref{eq:marcallo3} we prove part (I.b.ii).

\paragraph{III.b} In part (III.b) we impose $\bm\Sigma_\Lambda=\mbf I_r$ and 
$\bm\Gamma^F$ diagonal. First, notice that, since
$\bm\Gamma^{C}=\bm\Lambda\bm\Gamma^F\bm\Lambda^\prime$ has the same eigenvalues as $\bm\Lambda^\prime\bm\Lambda\bm\Gamma^F$, the $r$ eigenvalues of $\frac{\bm\Gamma^{C}}n$ and of $\bm\Gamma^F$ coincide asymptotically. Therefore, 
\beq\label{eq:rotturaasyFF}
\lim_{n\to\infty} \l\Vert \bm\Gamma^F -  \frac{\mbf M^{C}}{n}\r\Vert =0.
\eeq
Moreover, by Assumption \ref{ass:common}(a):
\beq\label{eq:biblioteca}
\lim_{n\to\infty} \l\Vert (\bm\Gamma^F)^{1/2}\frac{\bm\Lambda^\prime \bm\Lambda}n(\bm\Gamma^F)^{1/2} - \bm\Gamma^F \r\Vert=0,
\eeq
So from Lemmas \ref{lem:KO1}(i) and \ref{lem:KO1}(iii) and \eqref{eq:rotturaasyFF} and \eqref{eq:biblioteca} 
\beq\label{eq:biblioteca2}
 \lim_{n\to\infty}\l\Vert \frac{\mbf M^{C}}{n}\mbf K^\prime - \mbf K^\prime \frac{\mbf M^{C}}{n}\r\Vert \le  \lim_{n\to\infty}\l\Vert \frac{\mbf M^{C}}{n}\mbf K^\prime - \mbf K^\prime\bm\Gamma^F\r\Vert+ \lim_{n\to\infty}
 \l\Vert  \bm\Gamma^F -  \frac{\mbf M^{C}}{n} \r\Vert \,\Vert \mbf K\Vert=0.
\eeq
From \eqref{eq:biblioteca2}  (notice that $\mbf K$ depends on $n$)
\beq
\lim_{n\to\infty} \l(\frac{\mu^C_i}{n}-\frac{\mu^C_j}n\r)[\mbf K]_{ij}=\l([\bm V_0]_{ii}-[\bm V_0]_{jj}\r) \lim_{n\to\infty} [\mbf K]_{ij}= 0, \quad i,j=1,\ldots, r.
\eeq
Hence, since by Assumption \ref{ass:sign_eval}(b), which implies Assumption \ref{ass:eval}, $[\bm V_0]_{ii}\ne[\bm V_0]_{jj}$ for $i\ne j$, and $0<[\bm V_0]_{jj}<\infty$, $j=1,\ldots,r$, by Lemmas \ref{lem:Vzero}(i) and \ref{lem:Vzero}(iii), 
and since $\mbf K$ has full-rank as $n\to\infty$ by Lemma \ref{lem:KO1}(ii), then each column of $\lim_{n\to\infty} \mbf K$ must have at least one non-zero entry, thus  $\lim_{n\to\infty} \mbf K$ must be diagonal. Therefore, since by Lemma \ref{lem:KO1}(iii) $\mbf K$ is orthogonal for all $n\in\mathbb N$ and so also $\lim_{n\to\infty} \mbf K$ is orthogonal (see also \eqref{eq:ups} in the proof of Proposition \ref{prop:KKK}), it must be that \beq\label{eq:KlimiteIFF}
\lim_{n\to\infty}\Vert \mbf K-\bm{\mathcal S}_0\Vert=0,
\eeq 
where $\bm{\mathcal S}_0$ is diagonal with entries $\pm 1$ and independent of $n$. Then, from \eqref{eq:kappaproj} in the proof of Proposition \ref{prop:L}, 
\begin{align}
\lim_{n\to\infty}\l\Vert \frac{\bm\Lambda}{\sqrt n}-\mbf V^{C}\bm{\mathcal S}_0\r\Vert \le&\, 
\lim_{n\to\infty}  
\l\Vert
\mbf V^{C} \l(\frac{\mbf M^{C}}n\r)^{1/2} \mbf K^{-1} (\bm\Gamma^F)^{-1/2}  - \mbf V^{C}\bm{\mathcal S}_0
\r\Vert\nn\\
 \le&\,  \lim_{n\to\infty}  
\l\Vert
\mbf K^{-1}-\bm{\mathcal S}_0
\r\Vert\,
\l\Vert
\l(\frac{\mbf M^C}{n}\r)^{-1/2}
\r\Vert\,
\l\Vert
\l(\frac{\mbf M^C}{n}\r)^{1/2}
\r\Vert\nn\\
&+  \lim_{n\to\infty}
\l\Vert
(\bm\Gamma^F)^{-1/2}-
\l(\frac{\mbf M^C}{n}\r)^{-1/2}
\r\Vert\,
\l\Vert
\l(\frac{\mbf M^C}{n}\r)^{1/2}
\r\Vert\nn\\
&+\lim_{n\to\infty}
\l\Vert
\mbf K^{-1}-\bm{\mathcal S}_0
\r\Vert\,
\l\Vert
(\bm\Gamma^F)^{-1/2}-
\l(\frac{\mbf M^C}{n}\r)^{-1/2}
\r\Vert\,
\l\Vert
\l(\frac{\mbf M^C}{n}\r)^{1/2}
\r\Vert,\label{eq:biblioteca3}
\end{align}
by  \eqref{eq:rotturaasyFF} and \eqref{eq:KlimiteIFF}, Lemmas \ref{lem:MO1}(i) and \ref{lem:MO1}(ii),  and
since $\Vert \bm{\mathcal S}_0\Vert =1$ and
$\Vert \mbf V^{C}\Vert=1$ by definition of eigenvectors.  By taking the $i$th row of \eqref{eq:biblioteca3} and by definition of $\bm p_{i0}$, we prove part (III.b.i).  
Moreover, by linear projection of $\bm\Lambda$ onto $\bm{C}_t$, for all $t\in\mathbb Z$,  $\mbf F_t = (\bm\Lambda^\prime\bm\Lambda)^{-1}\bm\Lambda^\prime\bm{C}_t$. So from \eqref{eq:biblioteca3}  and since, by definition, $\bm P_{n,\infty}=\sqrt n \mbf V^C$,  
\begin{align}
\text{m.s.-}\lim_{n\to\infty}\l\Vert\mbf F_t -\bm{\mathcal S}_0
\frac{\bm P_{n,\infty}^\prime\bm C_t}{n} \r\Vert&=
\text{m.s.-}\lim_{n\to\infty}\l\Vert \l(\frac{\bm\Lambda^\prime\bm\Lambda}n\r)^{-1} \frac{\bm\Lambda^\prime \bm C_t}n -  
\bm{\mathcal S}_0
\frac{\bm P_{n,\infty}^\prime\bm C_t}{n} \r\Vert.\label{eq:marcallo4}
\end{align}
By definition of $\bm W_{\infty}$ and $\bm C_{t,\infty}$, from \eqref{eq:marcallo4} we prove part (III.b.ii).

\paragraph{II.b} In part (II.b) we impose $\bm\Sigma_\Lambda=\mbf I_r$ and  $\frac{\bm F^\prime\bm F}T$ diagonal for all $T\in\mathbb N$.
First, note that, since $\mbf K$ does not depend on $T$ then \eqref{eq:KlimiteIFF} still holds. 
Then, because of Assumption \ref{ass:common}(c-ii) (holding with rate $\sqrt T$),  $\l\Vert\frac{\bm F^\prime\bm F}T-\bm\Gamma^F \r\Vert = O_{\mathrm P}\l(\frac 1{\sqrt T}\r)$. Thus, by \eqref{eq:rotturaasyFF},
\beq\label{eq:rotturaasyFFP}
 \l\Vert \frac{\bm F^\prime\bm F}T -  \frac{\mbf M^{C}}{n}\r\Vert \le  \l\Vert \frac{\bm F^\prime\bm F}T -  \bm\Gamma^F\r\Vert + \l\Vert \bm\Gamma^F -  \frac{\mbf M^{C}}{n}\r\Vert = o_{\mathrm P}(1).
\eeq
This implies that the proof of part (II.b) is the same as for part (III.b) but with the difference that now \eqref{eq:biblioteca3} and \eqref{eq:marcallo4} hold in probability, as $n,T\to\infty$.

\paragraph{IV.b} In part (IV.b) we impose $\frac{\bm\Lambda^\prime\bm\Lambda}n=\mbf I_r$ for all $n\in\mathbb N$ and leave $\bm F$ unrestricted. From \eqref{eq:costume2} (which still holds)
\beq\label{eq:riesame}
\frac{\bm\Lambda}{\sqrt n} = \mbf V^C\l(\frac{\mbf M^{C}}n\r)^{-1/2}\mbf K^\prime(\bm\Gamma^F)^{1/2}.
\eeq
by taking the $i$th row of \eqref{eq:riesame} we prove part (IV.b.i) for fixed $n$, while for $n\to\infty$ the proof is the same as that of part (I.b.i) since \eqref{eq:riesame} is stronger than \eqref{eq:costume3},
and, by Lemma \ref{lem:KO1}(iii) and \eqref{eq:riesame}, we also have
$\mbf K^\prime \bm\Gamma^F\mbf K= \frac{\mbf M^C}{n}$, which still implies \eqref{eq:costume4}. Moreover, by linear projection of $\bm\Lambda$ onto $\bm{C}_t$, for all $t\in\mathbb Z$, $\mbf F_t = (\bm\Lambda^\prime\bm\Lambda)^{-1}\bm\Lambda^\prime\bm{C}_t$. So from \eqref{eq:riesame} we prove directly part (IV.b.ii) for fixed $n$, while for $n\to\infty$ the proof is the same as the one for part (I.b.ii) since \eqref{eq:riesame} still implies  \eqref{eq:marcallo3}.

\paragraph{VI.b} In part (VI.b) we impose $\frac{\bm\Lambda^\prime\bm\Lambda}{n}=\mbf I_r$ for all $n\in\mathbb N$ and $\bm\Gamma^F$ diagonal. 
First, notice that, since
$\bm\Gamma^{C}=\bm\Lambda\bm\Gamma^F\bm\Lambda^\prime$ has the same eigenvalues as $\bm\Lambda^\prime\bm\Lambda\bm\Gamma^F$, the $r$ eigenvalues of $\frac{\bm\Gamma^{C}}n$ and of $\bm\Gamma^F$ coincide. Therefore, 
\beq\label{eq:rugby}
 \bm\Gamma^F =  \frac{\mbf M^{C}}{n}.
\eeq
Moreover, from Lemma \ref{lem:KO1}(iii) 
\beq\label{eq:rugby2}
 \frac{\mbf M^{C}}{n}\mbf K^\prime = \mbf K^\prime\bm\Gamma^F.
\eeq
From \eqref{eq:rugby} and \eqref{eq:rugby2},
\beq
\l(\frac{\mu^C_i}{n}-\frac{\mu^C_j}n\r)[\mbf K]_{ij}= 0, \quad i,j=1,\ldots, r.\nn
\eeq
Hence, since by Assumption \ref{ass:sign_eval}(b) $\frac{\mu^C_i}n\ne \frac{\mu^C_j}n$ if $i\ne j$  and $0<\frac{\mu^C_j}n<\infty$, $j=1,\ldots, r$, for all $n\in\mathbb N$, by Lemmas \ref{lem:MO1}(i) and \ref{lem:MO1}(ii),
and since $\mbf K$ has full-rank by Lemma \ref{lem:KO1}(ii), then each column of $\mbf K$ must have at least one non-zero entry, thus  $\mbf K$ must be diagonal. Therefore, since by Lemma \ref{lem:KO1}(iii) $\mbf K$ is also orthogonal, it must be that 
\beq\label{eq:rugby3}
 \mbf K=\bm S,
\eeq 
where $\bm S$ is diagonal with entries $\pm 1$ and depends only on $n$. Then, from \eqref{eq:riesame} (which still holds) and using \eqref{eq:rugby} and \eqref{eq:rugby3}, 
\beq\label{eq:rugby5}
\frac{\bm\Lambda}{\sqrt n}
=  \mbf V^{C}\bm S.
\eeq
By taking the $i$th row of \eqref{eq:rugby5} we prove part (VI.b.i) for fixed $n$, while for $n\to\infty$, the proof is the same as that of part (III.b.i) since \eqref{eq:rugby2} and \eqref{eq:rugby3} are stronger than \eqref{eq:biblioteca2} and \eqref{eq:KlimiteIFF}, respectively. Clearly $\lim_{n\to\infty}\Vert \bm S-\bm{\mathcal S}_0\Vert=0$.
Moreover, by linear projection of $\bm\Lambda$ onto $\bm{C}_t$, for all $t\in\mathbb Z$,  $\mbf F_t = (\bm\Lambda^\prime\bm\Lambda)^{-1}\bm\Lambda^\prime\bm{C}_t$. So from \eqref{eq:rugby5} we prove directly part (VI.b.ii) for fixed $n$, while for $n\to\infty$ the proof is the same as that of part (III.b.ii) since \eqref{eq:rugby5} implies \eqref{eq:biblioteca3}.

\paragraph{V.b} In part (V.b)  we impose $\frac{\bm\Lambda^\prime\bm\Lambda}{n}=\mbf I_r$ for all $n\in\mathbb N$ and  $\frac{\bm F^\prime\bm F}T$ diagonal for all $T\in\mathbb N$. In this case, since $\wh{\bm\Gamma}^C =\wh{\mbf V}^{C}\wh{\mbf M}^{C}\wh{\mbf V}^{C\prime}=\bm\Lambda  \frac{\bm F^\prime\bm F}{T} \bm\Lambda^\prime$, then 
following analogous steps leading to \eqref{eq:kappaproj} in the proof of Proposition~\ref{prop:L}
\beq\label{eq:kappaprojnewTTT}
\frac{\bm\Lambda}{\sqrt n}= \wh{\mbf V}^{C}\l(\frac{\wh{\mbf M}^{C}} { n}\r)^{1/2}\wh{\mbf Q}^\prime\l( \frac{\bm F^\prime\bm F}{T}\r)^{-1/2},
\eeq
where following similar steps leading to the proof of Lemma \ref{lem:KO1}(iii) we see that $\wh{\mbf Q}$ is the matrix of normalized eigenvectors of $(\frac{\bm F^\prime \bm F}T)^{1/2}\frac{\bm\Lambda^\prime\bm\Lambda}n(\frac{\bm F^\prime \bm F}T)^{1/2}=\frac{\bm\Lambda^\prime\bm\Lambda}n$.
Moreover, $\frac{\wh{\bm\Gamma}^C}n$ has the same eigenvalues as $\frac{\bm F^\prime\bm F}{T}$ and \eqref{eq:rugby} is replaced by
\beq\label{eq:rugbyT}
 \frac{\bm F^\prime\bm F}{T} =  \frac{\wh{\mbf M}^{C}}{n},
\eeq
while \eqref{eq:rugby2} is replaced by
\beq\label{eq:rugby2T}
 \frac{\wh{\mbf M}^{C}}{n}\wh{\mbf Q}^\prime = \wh{\mbf Q}^\prime \frac{\bm F^\prime\bm F}{T} .
\eeq
From \eqref{eq:rugbyT} and \eqref{eq:rugby2T},
\beq
\l(\frac{\wh{\mu}^C_i}{n}-\frac{\wh{\mu}^C_j}n\r)[\wh{\mbf Q}]_{ij}= 0, \quad i,j=1,\ldots, r.\nn
\eeq
Hence, noticing that $\text{P}\l(\frac{\wh{\mu}^C_i}n- \frac{\wh{\mu}^C_j}n=0,\, i\ne j\r)=0$ since sample eigenvalues are continuous random variables,  and since by Assumption \ref{ass:sign_eval}(b) $\frac{\mu^C_i}n\ne \frac{\mu^C_j}n$ if $i\ne j$  and $0<\frac{\mu^C_j}n<\infty$, $j=1,\ldots, r$, for all $n\in\mathbb N$, by Lemmas \ref{lem:MO1}(i) and \ref{lem:MO1}(ii),
 \eqref{eq:rugby3} is replaced by
\beq\label{eq:rugby3T}
\wh{ \mbf Q}=\wh{\bm S},
\eeq 
where $\wh{\bm S}$ is diagonal with entries $\pm 1$ and depends on $n$ and $T$. By using \eqref{eq:rugbyT} and \eqref{eq:rugby3T} in  \eqref{eq:kappaprojnewTTT},
\beq\label{eq:rugby3TT}
\frac{\bm\Lambda}{\sqrt n}= \wh{\mbf V}^{C} \wh{\bm S}.
\eeq
Part (V.b.i) for fixed $n$ and $T$ then follows by taking the $i$th row of \eqref{eq:rugby3TT}. Moreover, by linear projection of $\bm\Lambda$ onto $\bm{C}_t$, for all $t\in\mathbb Z$, $\mbf F_t = (\bm\Lambda^\prime\bm\Lambda)^{-1}\bm\Lambda^\prime\bm{C}_t$, and part (V.b.ii) for fixed $n$ and $T$ follows directly from \eqref{eq:rugby3TT}. This holds only for $t=1,\ldots, T$ since $\wh{\mbf V}^{C}$ is obtained using only $T$ observations.
Then, because of Assumption \ref{ass:common}(c-ii),  $\l\Vert\frac{\bm F^\prime\bm F}T-\bm\Gamma^F \r\Vert = o_{\mathrm P}(1)$, and, thus  $\Vert \wh{\mbf V}^C-\mbf V^C\Vert=o_{\mathrm P}(1)$ (provided that their columns have the same sign which we can assume without loss of generality) and $\frac 1n\Vert \wh{\mbf M}^C-\mbf M^C\Vert =o_{\mathrm P}(1)$. Then, there exists a matrix $\bm S$ diagonal with entries $\pm 1$ and dependent only on $n$ such that $\Vert \wh{\bm S}-\bm S\Vert=o_{\mathrm P}(1)$
and the proof of part (V.b) when $T\to\infty$ and $n$ is fixed follows immediately. The proof of part (V.b) when $n,T\to\infty$ follows by simply taking the limit of the results for fixed $n$ and $T\to\infty$ and using the definitions of $\bm p_{i0}$, $\bm W_\infty$, and $\bm C_{t, \infty}$. This completes the proof. $\Box$

\subsection{Proof of Proposition \ref{corol:H00}}\label{corol:H00proof}


First, we prove all statements marked with (a) for $\wh{\mbf H}$ defined in \eqref{eq:acca}. Throughout, we use the fact that, from \eqref{eq:warwick1} in the proof of 
Proposition \ref{prop:HHAT}:
\beq\label{eq:tinta}
\l\Vert\wh{\mbf H} \wh{\mbf H}^\prime - \frac{\bm F^\prime\bm F}{T}\r\Vert = O_{\mathrm P}\l(\max\l(\frac 1{n},\frac 1{\sqrt{nT}},\frac 1T\r)\r).
\eeq

\paragraph{I.a} In part (I.a) we impose $\bm\Gamma^F=\mbf I_r$ and leave $\bm\Lambda$ unrestricted. Then, from \eqref{eq:tinta} and Assumption \ref{ass:common}(c-ii)
\beq
\l\Vert\wh{\mbf H} \wh{\mbf H}^\prime - \mbf I_r\r\Vert \le \l\Vert\wh{\mbf H} \wh{\mbf H}^\prime - \frac{\bm F^\prime\bm F}{T}\r\Vert + \l\Vert\frac{\bm F^\prime\bm F}{T} - \mbf I_r \r\Vert  = O_{\mathrm P}\l(\max\l(\frac 1{n},\frac 1{\sqrt{nT}},\frac 1T\r)\r)+O_{\mathrm P}\l(\frac 1{\sqrt T}\r),\label{eq:pipa}
\eeq
which proves part (I.a).


\paragraph{II.a}
In part (II.a) we impose $\bm\Gamma^F=\mbf I_r$ and $\frac{\bm\Lambda^\prime\bm\Lambda}{n}$ to be diagonal for all $n\in\mathbb N$.  Then, from \eqref{eq:warwick4} in the proof of Proposition \ref{prop:HHAT}
\begin{align}
\l\Vert
\frac{\wh{\mbf M}^x}{n}\wh{\mbf H}^\prime -\frac{\wh{\mbf H}^\prime \bm\Lambda^\prime\bm\Lambda}n
\r\Vert&\le
\l\Vert
\frac{\wh{\mbf M}^x}{n}\wh{\mbf H}^\prime -\frac{\wh{\mbf H}^\prime \bm\Lambda^\prime\bm\Lambda}n\frac{\bm F^\prime\bm F}{T}
\r\Vert
+\l\Vert\frac{\bm F^\prime\bm F}{T}-\mbf I_r \r\Vert\,\l\Vert \wh{\mbf H}\r\Vert
\, \l\Vert \frac{\bm\Lambda^\prime\bm\Lambda}{n}\r\Vert \nn\\
&=O_{\mathrm P}\l(\max\l(\frac 1{n},\frac 1{\sqrt{nT}},\frac 1{T}\r)\r) + O_{\mathrm P}\l(\frac 1{\sqrt T}\r),\label{eq:Hevecfixedn}
\end{align}
by Assumption \ref{ass:common}(c-ii), and Lemmas \ref{lem:LLN}(iv) and \ref{lem:HO1}(i).  
From \eqref{eq:Hevecfixedn} it follows that $\wh{\mbf H}$ is asymptotically equivalent to the matrix 
of eigenvectors of $\frac{\bm\Lambda^\prime\bm\Lambda}{n}$. Note that the eigenvectors are asymptotically normalized because of \eqref{eq:pipa} (which still holds).


Now, Proposition \ref{corol:K00}(II.a) implies:
\beq\label{eq:pipa2}
\frac{\bm\Lambda^\prime\bm\Lambda}{n} = \frac{\mbf M^C}{n},
\eeq
 which is diagonal. Then, by \eqref{eq:Hevecfixedn} and \eqref{eq:pipa2}
\beq\label{eq:pipa3}
\l(\frac{\wh{\mu}^x_i}{n}-\frac{{\mu}^C_j}n\r)[\wh{\mbf H}]_{ij} = O_{\mathrm P}\l(\max\l(\frac 1n,\frac 1{\sqrt T}\r)\r),\quad i,j=1,\ldots, r.
\eeq
Now, by Assumption \ref{ass:sign_eval}(b), $\frac{\mbf M^C}n$ has distinct entries sorted in descending order for all $n\in\mathbb N$. Moreover, since $\bm\Gamma^x = \bm\Gamma^C+\bm\Gamma^e$ by Assumption \ref{ass:ind}, by Weyl's inequality \citep[Theorem 1]{MK04} and Lemma \ref{lem:Gxi}(v),  for all $ i=1,\ldots, r$,
\begin{align}
\frac 1n \l\vert {\mu_i^x - \mu_i^C}\r\vert \le \frac 1n\l\Vert \bm\Gamma^x - \bm\Gamma^C \r\Vert = \frac1 n\l\Vert \bm\Gamma^e\r\Vert = \frac{\mu_1^e}n= O\l(\frac 1n\r),\label{eq:pipa7}
\end{align}
so the eigenvalues of $\frac{\bm\Gamma^x}n$ are also asymptotically distinct as $n\to\infty$ by Assumption \ref{ass:sign_eval}(b). It follows that $\mathrm P(\frac{\wh{\mu}^x_i}n= \frac{\wh{\mu}^x_j}n,\; i\ne j)\to 0$ as $n,T\to\infty$, because 
the sample eigenvalues are continuous random variables. Thus, by Lemma \ref{lem:covarianze}(iii), $\frac{\wh{\mbf M}^x}n$ has also asymptotically distinct entries as $n,T\to\infty$, which are sorted in descending order. 
Given that the entries of both $\frac{\mbf M^C}n$ and $\frac{\wh{\mbf M}^x}n$ are asymptotically distinct and sorted in descending order it must be that $\mathrm P(\frac{\wh{\mu}^x_i}n\ne \frac{{\mu}^C_j}n, i\ne j)\to 1$, as $n,T\to\infty$, and so $[\wh{\mbf H}]_{ij}=O_{\mathrm P}\l(\max\l(\frac 1n,\frac 1{\sqrt T}\r)\r)$. 
But since $\Vert\wh{\mbf H}^{-1}\Vert =O_{\mathrm P}(1)$ by Lemma \ref{lem:HO1}(ii), then it must also be that $\mathrm P([\wh{\mbf H}_{ii}]\ne 0)\to 1$ as $n,T\to\infty$ and, from \eqref{eq:pipa3}, $\frac{\wh{\mu}^x_i}n= \frac{{\mu}^C_i}n+O_{\mathrm P}\l(\max\l(\frac 1n,\frac 1{\sqrt T}\r)\r)$, for $i=1,\ldots, r$. Therefore, $\wh{\mbf H}$ is asymptotically diagonal, and since by \eqref{eq:pipa} (which still holds) $\wh{\mbf H}$ is also asymptotically orthogonal, i.e., it has eigenvalues $\pm 1$, it must be that 
\beq\nn
\l\Vert \wh{\mbf H}-\wh{\bm J}\r\Vert=O_{\mathrm P} \l(\max\l(\frac 1n,\frac 1{\sqrt T}\r)\r),
\eeq
where $\wh{\bm J}$ is an $r\times r$ diagonal matrix with entries $\pm 1$ which depends on $n$ and $T$. This proves part (II.a).

\paragraph{III.a}
In part (III.a) we impose $\bm\Gamma^F=\mbf I_r$ and $ \bm\Sigma_\Lambda$ to be diagonal. Then,  from \eqref{eq:warwick4} in the proof of Proposition \ref{prop:HHAT}
\begin{align}
\l\Vert
\frac{\wh{\mbf M}^x}{n}\wh{\mbf H}^\prime -\wh{\mbf H}^\prime \bm\Sigma_\Lambda
\r\Vert
&\le
\l\Vert
\frac{\wh{\mbf M}^x}{n}\wh{\mbf H}^\prime -\frac{\wh{\mbf H}^\prime \bm\Lambda^\prime\bm\Lambda}n\frac{\bm F^\prime\bm F}{T}
\r\Vert
+\l\Vert\frac{\bm F^\prime\bm F}{T}-\mbf I_r \r\Vert\,\l\Vert \wh{\mbf H}\r\Vert
\, \l\Vert \frac{\bm\Lambda^\prime\bm\Lambda}{n}\r\Vert +\l\Vert\frac{\bm \Lambda^\prime\bm \Lambda}{n}-\bm\Sigma_\Lambda \r\Vert\,\l\Vert \wh{\mbf H}\r\Vert
\, \l\Vert \frac{\bm F^\prime\bm F}{T}\r\Vert \nn\\
&=O_{\mathrm P}\l(\max\l(\frac 1{n},\frac 1{\sqrt{nT}},\frac 1{T}\r)\r) + O_{\mathrm P}\l(\frac 1{\sqrt T}\r)+ O\l(\frac 1{\sqrt n}\r),
\label{eq:Hevec}
\end{align}
by Assumptions \ref{ass:common}(a) and \ref{ass:common}(c-ii), and Lemmas \ref{lem:LLN}(iv), \ref{lem:LLN}(v), and \ref{lem:HO1}(i). From \eqref{eq:Hevec} it follows that $\wh{\mbf H}$ is asymptotically equivalent to the matrix 
of eigenvectors of $\bm\Sigma_\Lambda$. Note that the eigenvectors are asymptotically normalized because of \eqref{eq:pipa} (which still holds).

Now, since $\bm V_0$ is the matrix of eigenvalues of $\bm\Sigma_\Lambda\bm\Gamma^F$ and $\bm\Sigma_\Lambda$ is diagonal, it must be that
\beq\label{eq:pipa4}
\bm\Sigma_{\Lambda} = \bm V_0,
\eeq
because $\bm\Gamma^F=\mbf I_r$.
Then, by \eqref{eq:Hevec} and \eqref{eq:pipa4}
\beq\label{eq:pipa5}
\l(\frac{\wh{\mu}^x_i}{n}-[\bm V_0]_{jj}\r)[\wh{\mbf H}]_{ij} = O_{\mathrm P}\l(\max\l(\frac 1{\sqrt n},\frac 1{\sqrt T}\r)\r),\quad i,j=1,\ldots, r.
\eeq
Now, since Assumption \ref{ass:sign_eval}(b) implies Assumption \ref{ass:eval}, $\bm V_0$ has distinct entries sorted in descending order. Moreover, since $\bm\Gamma^x = \bm\Gamma^C+\bm\Gamma^e$ by Assumption \ref{ass:ind}, by Weyl's inequality \citep[Theorem 1]{MK04}, Lemmas \ref{lem:Gxi}(v) and \ref{lem:Vzero}(i),  for all $ i=1,\ldots, r$,
\begin{align}
\l\vert {\frac{\mu_i^x}n -[\bm V_0]_{ii}}\r\vert
&\le \frac 1n \l\vert {\mu_i^x - \mu_i^C}\r\vert + \l\vert {\frac{\mu_i^C}n -[\bm V_0]_{ii}}\r\vert
 \le \frac 1n\l\Vert \bm\Gamma^x - \bm\Gamma^C \r\Vert +  \l\vert {\frac{\mu_i^C}n -[\bm V_0]_{ii}}\r\vert\nn\\
 &= \frac1 n\l\Vert \bm\Gamma^e\r\Vert +  \l\vert {\frac{\mu_i^C}n -[\bm V_0]_{ii}}\r\vert = \frac{\mu_1^e}n+O\l(\frac 1{\sqrt n}\r)= O\l(\frac 1{ n}\r)+O\l(\frac 1{\sqrt n}\r),\label{eq:pipa8}
\end{align}
so the eigenvalues of $\frac{\bm\Gamma^x}n$ are also asymptotically distinct as $n\to\infty$ by Assumption \ref{ass:sign_eval}(b). It follows that $\mathrm P(\frac{\wh{\mu}^x_i}n= \frac{\wh{\mu}^x_j}n,\; i\ne j)\to 0$ as $n,T\to\infty$, because 
the sample eigenvalues are continuous random variables. Thus, by Lemmas \ref{lem:covarianze}(iii) and \ref{lem:Vzero}(i), $\frac{\wh{\mbf M}^x}n$ has also asymptotically distinct entries, as $n,T\to\infty$, which are sorted in descending order. 
Given that the entries of both $\bm V_0$ and $\frac{\wh{\mbf M}^x}n$ are asymptotically distinct and sorted in descending order it must be that
$\mathrm P(\frac{\wh{\mu}^x_i}n\ne [\bm V_0]_{jj}, i\ne j)\to 1$, as $n,T\to\infty$, and so $[\wh{\mbf H}]_{ij}=O_{\mathrm P}\l(\max\l(\frac 1{\sqrt n},\frac 1{\sqrt T}\r)\r)$. 
But since $\Vert\wh{\mbf H}^{-1}\Vert =O_{\mathrm P}(1)$ by Lemma \ref{lem:HO1}(ii), then it must also be that $\mathrm P([\wh{\mbf H}_{ii}]\ne 0)\to 1$ as $n,T\to\infty$ and, from \eqref{eq:pipa5}, $\frac{\wh{\mu}^x_i}n= [\bm V_0]_{jj}+O_{\mathrm P}\l(\max\l(\frac 1{\sqrt n},\frac 1{\sqrt T}\r)\r)$, for $i=1,\ldots, r$. Therefore, $\wh{\mbf H}$ is asymptotically diagonal, and since by \eqref{eq:pipa} (which still holds) $\wh{\mbf H}$ is also asymptotically orthogonal, i.e., it has eigenvalues $\pm 1$, it must be that 
\beq\nn
\l\Vert \wh{\mbf H}-\wh{\bm J}\r\Vert=O_{\mathrm P} \l(\max\l(\frac 1{\sqrt n},\frac 1{\sqrt T}\r)\r),
\eeq
where $\wh{\bm J}$ is an $r\times r$ diagonal matrix with entries $\pm 1$ which depends on $n$ and $T$. This proves part (III.a).

\paragraph{IV.a}
In part (IV.a) we impose $\frac{\bm F^\prime\bm F}T=\mbf I_r$  for all $T\in\mathbb N$ and leave
$\bm\Lambda$ unrestricted. Then, from \eqref{eq:tinta}
\begin{align}
\l\Vert \wh{\mbf H}\wh{\mbf H}^\prime- \mbf I_r\r\Vert = O_{\mathrm P}\l(\max\l(\frac 1{n},\frac 1{\sqrt{nT}}, \frac 1T\r)\r),\label{eq:pipa6}
\end{align}
which proves part (IV.a). 

\paragraph{V.a}
In part (V.a) we impose $\frac{\bm F^\prime\bm F}{T}=\mbf I_r$ and $\frac{\bm\Lambda^\prime\bm\Lambda}n$ to be diagonal for all $n,T\in\mathbb N$.
Then,  from \eqref{eq:warwick4} in the proof of Proposition \ref{prop:HHAT}, 
\begin{align}
&\l\Vert
\frac{\wh{\mbf M}^x}{n}\wh{\mbf H}^\prime -\frac{\wh{\mbf H}^\prime \bm\Lambda^\prime\bm\Lambda}n
\r\Vert
= O_{\mathrm P}\l(\max\l(\frac 1{n},\frac 1{\sqrt {nT}},\frac 1T\r)\r).\label{eq:HevecfixednT}
\end{align}
It follows that asymptotically the columns of $\wh{\mbf H}^\prime$ are eigenvectors of $\frac{\bm\Lambda^\prime\bm\Lambda}n$ and are asymptotically normalized because of \eqref{eq:pipa6} (which still holds). By following the same reasoning as in the proof of part (II.a), but using the rates in  \eqref{eq:HevecfixednT}, together with \eqref{eq:pipa7} and \eqref{eq:pipa6} (which still hold), we must have
\[
\l\Vert \wh{\mbf H}- \wh{\bm J}\r\Vert = O_{\mathrm P}\l(\max\l(\frac 1{n},\frac 1{\sqrt {nT}},\frac 1T\r)\r),
\]
which proves part (V.a).

\paragraph{VI.a}
In part (VI.a) we impose $\frac{\bm F^\prime\bm F}{T}=\mbf I_r$  for all $T\in\mathbb N$ and $\bm\Sigma_\Lambda$ to be diagonal.  Then,  from \eqref{eq:warwick4} in the proof of Proposition \ref{prop:HHAT}
\begin{align}
\l\Vert
\frac{\wh{\mbf M}^x}{n}\wh{\mbf H}^\prime -\wh{\mbf H}^\prime \bm\Sigma_\Lambda
\r\Vert
&\le
\l\Vert
\frac{\wh{\mbf M}^x}{n}\wh{\mbf H}^\prime -\frac{\wh{\mbf H}^\prime \bm\Lambda^\prime\bm\Lambda}n
\r\Vert
 +\l\Vert\frac{\bm \Lambda^\prime\bm \Lambda}{n}-\bm\Sigma_\Lambda \r\Vert\,\l\Vert \wh{\mbf H}\r\Vert
\, \l\Vert \frac{\bm F^\prime\bm F}{T}\r\Vert \nn\\
&=O_{\mathrm P}\l(\max\l(\frac 1{n},\frac 1{\sqrt{nT}},\frac 1{T}\r)\r) + O\l(\frac 1{\sqrt n}\r),
\label{eq:Hevecexact}
\end{align}
by Assumption \ref{ass:common}(a), and Lemmas \ref{lem:LLN}(v) and \ref{lem:HO1}(i). It follows that asymptotically the columns of $\wh{\mbf H}^\prime$ are eigenvectors of $\bm\Sigma_\Lambda$ and are asymptotically normalized because of \eqref{eq:pipa6} (which still holds). By following the same reasoning as in the proof of part (III.a), but using the rates in  \eqref{eq:Hevecexact}, together with \eqref{eq:pipa8} and \eqref{eq:pipa6} (which still hold), we must have
\[
\l\Vert \wh{\mbf H}- \wh{\bm J}\r\Vert = O_{\mathrm P}\l(\max\l(\frac 1{\sqrt n},\frac 1{ T}\r)\r),
\]
which proves part (VI.a).\\


We then prove all statements marked with (b) for $\bar{\mbf H}$ defined in \eqref{eq:Hbar} and by means of \eqref{eq:sameHD} and \eqref{eq:sameHD3}  we can then directly derive the properties of $\wh{\mbf H}$. Throughout, we use the fact that, from \eqref{eq:starop1D}:
\beq
\l\Vert\bar{\mbf H}\bar{\mbf H}^\prime - \frac{\bm \Lambda^\prime\bm\Lambda}{n} \r\Vert =O_{\mathrm P}\l(\max\l(\frac 1n,\frac{1}{\sqrt{nT}},\frac 1T\r)\r).\label{eq:tintaB}
\eeq

\paragraph{I.b}
In part (I.b) we leave $\bm F$ unrestricted and we  impose $\bm\Sigma_\Lambda=\mbf I_r$. Then,
 from \eqref{eq:tintaB} and Assumption \ref{ass:common}(a)
\begin{align}
\l\Vert\bar{\mbf H}\bar{\mbf H}^\prime -\mbf I_r \r\Vert&\le \l\Vert\bar{\mbf H}\bar{\mbf H}^\prime - \frac{\bm \Lambda^\prime\bm\Lambda}{n} \r\Vert + \l\Vert \frac{\bm \Lambda^\prime\bm \Lambda}{n}   - \mbf I_r \r\Vert =  O_{\mathrm P}\l(\max\l(\frac 1{n},\frac 1{\sqrt{nT}},\frac 1{T}\r)\r)+O\l(\frac 1{\sqrt n}\r),\label{eq:pipa9}
\end{align}
which proves  part (I.b).

\paragraph{II.b}
In part (II.b), we impose $\frac{\bm F^\prime\bm F}T$ to be diagonal for all $T\in\mathbb N$ and $\bm\Sigma_\Lambda=\mbf I_r$. Then, from \eqref{eq:starop2D}
\begin{align}
\l\Vert\frac{\wt{\mbf M}^x}{T} \bar{\mbf H}^\prime - \frac{ \bar{\mbf H}^\prime\bm F^\prime\bm F}{T} \r\Vert&\le 
\l\Vert\frac{\wt{\mbf M}^x}{T} \bar{\mbf H}^\prime - \frac{ \bar{\mbf H}^\prime\bm F^\prime\bm F}{T} \frac{\bm\Lambda^\prime\bm\Lambda}{n}\r\Vert
+\l\Vert\frac{\bm\Lambda^\prime\bm\Lambda}{n}-\mbf I_r \r\Vert\,\l\Vert \bar{\mbf H}\r\Vert
\, \l\Vert \frac{\bm F^\prime\bm F}{T}\r\Vert \nn\\
&=O_{\mathrm P}\l(\max\l(\frac 1n,\frac{1}{\sqrt{nT}},\frac 1T\r)\r) +O\l(\frac 1{\sqrt n}\r),\label{eq:starop2Dapp}
\end{align}
by Assumption \ref{ass:common}(a), Lemma \ref{lem:LLN}(v),   
where $\bar {\mbf H}$ is defined in \eqref{eq:Hbar} and it is such that $\Vert \bar{\mbf H}\Vert = O_{\mathrm P}(1)$ because of the same arguments leading to Lemma \ref{lem:HO1}(i). 
It follows that asymptotically the columns of $\bar{\mbf H}^\prime$ are eigenvectors of $\frac{\bm F^\prime\bm F}T$ and are asymptotically normalized because of \eqref{eq:pipa9} (which still holds).

Now, let $\wh{\mu}_j^C$, $j=1,\ldots, r$, be the $j$th largest eigenvalue of $\wh{\bm\Gamma}^C= \bm\Lambda \frac{\bm F^\prime\bm F}{T}\bm\Lambda^\prime$ which is also an eigenvalue of $\frac{\bm F^\prime\bm F}{T}{\bm\Lambda^\prime\bm\Lambda}$. Then, since $\frac{\bm F^\prime \bm F}{T}$ is diagonal, its entries coincide with its eigenvalues, and, by Assumption \ref{ass:common}(a), Lemma \ref{lem:LLN}(v), Weyl's inequality \citep[Theorem 1]{MK04}, and since $\bm\Sigma_\Lambda=\mbf I_r$, for all $j=1,\ldots, r$, we have
\begin{align}
\l\Vert  \l[\frac{\bm F^\prime \bm F}{T} \r]_{jj}- \frac{\wh{\mu}_j^C}n\r\Vert &=\l\Vert  \mu_j\l(\frac{\bm F^\prime \bm F}{T} \bm\Sigma_\Lambda\r)- \frac{\wh{\mu}_j^C}n\r\Vert \nn\\
&= \l\Vert  \mu_j\l(\frac{\bm F^\prime \bm F}{T}\l\{ \bm\Sigma_\Lambda-\frac{\bm\Lambda^\prime\bm\Lambda}n+\frac{\bm\Lambda^\prime\bm\Lambda}n\r\}\r)- \mu_j\l(\frac{\bm F^\prime \bm F}{T} \frac{\bm\Lambda^\prime\bm\Lambda}n\r) \r\Vert \nn\\
&\le \l\Vert \mu_1\l(\frac{\bm F^\prime \bm F}{T}\l\{\bm\Sigma_\Lambda-\frac{\bm\Lambda^\prime\bm\Lambda}{n}\r\} \r)+\mu_j\l(\frac{\bm F^\prime \bm F}{T} \frac{\bm\Lambda^\prime\bm\Lambda}n\r)-\mu_j\l(\frac{\bm F^\prime \bm F}{T} \frac{\bm\Lambda^\prime\bm\Lambda}n\r)
\r\Vert\nn\\
&= \l\Vert \frac{\bm F^\prime \bm F}{T}\l\{\bm\Sigma_\Lambda-\frac{\bm\Lambda^\prime\bm\Lambda}{n}\r\}\r\Vert\le \l\Vert \frac{\bm F^\prime \bm F}{T}\r\Vert\,   \l\Vert \bm\Sigma_\Lambda-\frac{\bm\Lambda^\prime\bm\Lambda}{n}\r\Vert= O\l(\frac 1{\sqrt n}\r).\label{eq:starop2DappBOH}
\end{align}
Note that this is an ordinary limit since $\frac{\bm F^\prime \bm F}{T}$ is diagonal for all $T\in\mathbb N$.
%
%
%
%
%
From \eqref{eq:starop2Dapp} and \eqref{eq:starop2DappBOH}, and since $\frac{\wt{\mbf M}^x}T=\frac{\wh{\mbf M}^x}n$,
\begin{align}\label{eq:menata}
\l(\frac{\wh{\mu}^x_i}{n}-\frac{\wh{\mu}^C_j}n\r)[\bar{\mbf H}]_{ij} &=
\l(\frac{\wh{\mu}^x_i}{n}-\l[\frac{\bm F^\prime\bm F}{T}\r]_{jj}\r)[\bar{\mbf H}]_{ij} + 
\l(\l[\frac{\bm F^\prime\bm F}{T}\r]_{jj}-\frac{\wh{\mu}^C_j}n\r)[\bar{\mbf H}]_{ij}\nn\\
&= O_{\mathrm P}\l(\max\l(\frac 1{\sqrt n},\frac 1{ T}\r)\r)+O\l(\frac 1{\sqrt n}\r),\quad i,j=1,\ldots, r.
\end{align}
Moreover, the eigenvalues of $\frac{\bm\Gamma^x}n$ are asymptotically distinct as $n\to\infty$ by \eqref{eq:pipa7} (which still holds) and Assumption \ref{ass:sign_eval}(b). Thus, by Lemma \ref{lem:covarianze}(iii), $\frac{\wh{\mbf M}^x}{n}$ has asymptotically distinct entries, $n,T\to\infty$, which are sorted in descending order. 
Similarly,  the eigenvalues of $\frac{\bm\Gamma^C}n$ are distinct for all $n\in\mathbb N$ by Assumption \ref{ass:sign_eval}(b). Thus, by Assumption \ref{ass:common}(c-ii), $\frac{\wh{\mbf M}^C}{n}$ has asymptotically distinct entries, as $T\to\infty$, which are sorted in descending order. Given that the entries of both $\frac{\wh{\mbf M}^C}n$ and $\frac{\wh{\mbf M}^x}n$ are asymptotically distinct and sorted in descending order it must be that
$\mathrm P(\frac{\wh{\mu}^x_i}n\ne \frac{\wh{\mu}^C_j}n, i\ne j)\to 1$, as $n,T\to\infty$, and so $[\bar{\mbf H}]_{ij}=O_{\mathrm P}\l(\max\l(\frac 1{\sqrt n},\frac 1{T}\r)\r)$. 
But since $\Vert\bar {\mbf H}^{-1}\Vert =O_{\mathrm P}(1)$ by Lemma \ref{lem:HO1}(ii), then it must also be that $\mathrm P([\wh{\mbf H}_{ii}]\ne 0)\to 1$ as $n,T\to\infty$ and, from \eqref{eq:menata}, $\frac{\wh{\mu}^x_i}n= \frac{\wh{\mu}^C_i}n+O_{\mathrm P}\l(\max\l(\frac 1{\sqrt n},\frac 1{T}\r)\r)$, for $i=1,\ldots, r$. Therefore, $\bar{\mbf H}$ is asymptotically diagonal, and since by \eqref{eq:pipa9} (which still holds) $\bar{\mbf H}$ is also asymptotically orthogonal, i.e., it has eigenvalues $\pm 1$, it must be that 
\beq\nn
\l\Vert \bar{\mbf H}- \wh{\bm J}\r\Vert = O_{\mathrm P}\l(\max\l(\frac 1{\sqrt n},\frac 1{ T}\r)\r), 
\eeq
which proves part (II.b).

\paragraph{III.b}
In part (III.b), we impose $\bm\Gamma^F$ to be diagonal and $\bm\Sigma_\Lambda=\mbf I_r$. 
Then, from \eqref{eq:starop2D}
\begin{align}
\l\Vert\frac{\wt{\mbf M}^x}{T} \bar{\mbf H}^\prime -  \bar{\mbf H}^\prime\bm\Gamma^F \r\Vert&\le 
\l\Vert\frac{\wt{\mbf M}^x}{T} \bar{\mbf H}^\prime - \frac{ \bar{\mbf H}^\prime\bm F^\prime\bm F}{T} \frac{\bm\Lambda^\prime\bm\Lambda}{n}\r\Vert
+\l\Vert\frac{\bm\Lambda^\prime\bm\Lambda}{n}-\mbf I_r \r\Vert\,\l\Vert \bar{\mbf H}\r\Vert
\, \l\Vert \frac{\bm F^\prime\bm F}{T}\r\Vert
+\l\Vert  \frac{\bm F^\prime\bm F}{T}-\bm\Gamma^F \r\Vert\,\l\Vert \bar{\mbf H}\r\Vert
\, \l\Vert\frac{\bm\Lambda^\prime\bm\Lambda}{n}\r\Vert
 \nn\\
&=O_{\mathrm P}\l(\max\l(\frac 1n,\frac{1}{\sqrt{nT}},\frac 1T\r)\r) +O\l(\frac 1{\sqrt n}\r)+O_{\mathrm P}\l(\frac 1{\sqrt T}\r),\label{eq:starop3Dapp}
\end{align}
by Assumptions \ref{ass:common}(a) and \ref{ass:common}(c-ii), Lemmas \ref{lem:LLN}(iv) and \ref{lem:LLN}(v),    
where $\bar {\mbf H}$ is defined in \eqref{eq:Hbar} and it is such that $\Vert \bar{\mbf H}\Vert = O_{\mathrm P}(1)$ because of the same arguments leading to Lemma \ref{lem:HO1}(i). 
It follows that asymptotically the columns of $\bar{\mbf H}^\prime$ are eigenvectors of $\bm\Gamma^F$ and are asymptotically normalized because of \eqref{eq:pipa9} (which still holds).

Now, since $\bm V_0$ is the matrix of eigenvalues of $\bm\Sigma_\Lambda\bm\Gamma^F$ and $\bm\Gamma^F$ is diagonal, it must be that
\beq\label{eq:pipa10}
\bm\Gamma^F = \bm V_0,
\eeq
because  $\bm\Sigma_\Lambda=\mbf I_r$. Then, by \eqref{eq:starop3Dapp} and \eqref{eq:pipa10}, and since $\frac{\wt{\mbf M}^x}T=\frac{\wh{\mbf M}^x}n$,
\beq
\l(\frac{\wh{\mu}^x_i}{n}-[\bm V_0]_{jj}\r)[\bar{\mbf H}]_{ij} = O_{\mathrm P}\l(\max\l(\frac 1{\sqrt n},\frac 1{\sqrt T}\r)\r),\quad i,j=1,\ldots, r,
\eeq
which coincides with \eqref{eq:pipa5}. Thus, we can follow the same steps in the proof of part (III.a), to show that $\bar{\mbf H}$ is asymptotically diagonal, and since by \eqref{eq:pipa9} (which still holds) $\bar{\mbf H}$ is also asymptotically orthogonal, i.e., it has eigenvalues $\pm 1$, it must be that 
\[
\l\Vert \bar{\mbf H}- \wh{\bm J}\r\Vert = O_{\mathrm P}\l(\max\l(\frac 1{\sqrt n},\frac 1{\sqrt T}\r)\r),
\]
which proves part (III.b).

\paragraph{IV.b}
In part (IV.b) we leave $\bm F$ unrestricted and impose
$\frac{\bm\Lambda^\prime \bm\Lambda}n=\mbf I_r$ for all $n\in\mathbb N$. Then,  from \eqref{eq:tintaB}
\begin{align}
\l\Vert \bar{\mbf H}\bar{\mbf H}^\prime-\mbf I_r\r\Vert = O_{\mathrm P}\l(\max\l(\frac 1{T},\frac 1{\sqrt{nT}},\frac 1{n}\r)\r),\label{eq:pipa12}
\end{align}
which proves part (IV.b).

\paragraph{V.b}
In part (V.b) we impose $\frac{\bm F^\prime\bm F}{T}$ to be diagonal and $\frac{\bm\Lambda^\prime\bm\Lambda}n=\mbf I_r$ for all $n,T\in\mathbb N$.
Then,  from \eqref{eq:starop2D}
\begin{align}
\l\Vert\frac{\wt{\mbf M}^x}{T} \bar{\mbf H}^\prime - \frac{ \bar{\mbf H}^\prime\bm F^\prime\bm F}{T} \r\Vert
&=O_{\mathrm P}\l(\max\l(\frac 1n,\frac{1}{\sqrt{nT}},\frac 1T\r)\r).\label{eq:HevecfixednTancora}
\end{align}
It follows that asymptotically the columns of $\bar{\mbf H}^\prime$ are eigenvectors of $\frac{\bm F^\prime\bm F}T$ and are asymptotically normalized because of \eqref{eq:pipa12} (which still holds). 

Now,  $\frac{\wh{\mbf M}^C}n$ are the eigenvalues of $\frac{\wh{\bm\Gamma}^C}n= \bm\Lambda \frac{\bm F^\prime\bm F}{nT}\bm\Lambda^\prime$, so they are also the eigenvalues of $\frac{\bm F^\prime\bm F}{T}\frac{\bm\Lambda^\prime\bm\Lambda}n=\frac{\bm F^\prime\bm F}{T}$. Therefore, for all $j=1,\ldots,r$,
\beq\label{eq:menatissima}
\l[\frac{\bm F^\prime \bm F}{T}\r]_{jj}= \frac{\wh{\mu}_j^C}n.
\eeq
From \eqref{eq:HevecfixednTancora} and \eqref{eq:menatissima}, and since $\frac{\wt{\mbf M}^x}T=\frac{\wh{\mbf M}^x}n$,
\begin{align}\label{eq:menata2}
\l(\frac{\wh{\mu}^x_i}{n}-\frac{\wh{\mu}^C_j}n\r)[\bar{\mbf H}]_{ij} &=
\l(\frac{\wh{\mu}^x_i}{n}-\l[\frac{\bm F^\prime\bm F}{T}\r]_{jj}\r)[\bar{\mbf H}]_{ij} 
= O_{\mathrm P}\l(\max\l(\frac 1{ n},\frac1{\sqrt{nT}},\frac 1{ T}\r)\r),\quad i,j=1,\ldots, r.
\end{align}
By following the same reasoning as in the proof of part (II.b), but using the rates in  \eqref{eq:menata2}, together with \eqref{eq:pipa12} (which still holds), we must have
\[
\l\Vert \bar{\mbf H}- \wh{\bm J}\r\Vert = O_{\mathrm P}\l(\max\l(\frac 1{n},\frac 1{\sqrt {nT}},\frac 1T\r)\r),
\]
which proves part (V.b).

\paragraph{VI.b}
In part (VI.b) we impose $\bm\Gamma^F$ to be diagonal and $\frac{\bm\Lambda^\prime\bm\Lambda}n=\mbf I_r$ for all $n\in\mathbb N$. Then, from \eqref{eq:starop2D}
\begin{align}
\l\Vert\frac{\wt{\mbf M}^x}{T} \bar{\mbf H}^\prime -  \bar{\mbf H}^\prime\bm\Gamma^F \r\Vert&\le 
\l\Vert\frac{\wt{\mbf M}^x}{T} \bar{\mbf H}^\prime - \frac{ \bar{\mbf H}^\prime\bm F^\prime\bm F}{T}\r\Vert
+\l\Vert  \frac{\bm F^\prime\bm F}{T}-\bm\Gamma^F \r\Vert\,\l\Vert \bar{\mbf H}\r\Vert
\, \l\Vert\frac{\bm\Lambda^\prime\bm\Lambda}{n}\r\Vert
 \nn\\
&=O_{\mathrm P}\l(\max\l(\frac 1n,\frac{1}{\sqrt{nT}},\frac 1T\r)\r) +O_{\mathrm P}\l(\frac 1{\sqrt T}\r),\label{eq:starop3Dappexact}
\end{align}
by Assumption \ref{ass:common}(c-ii), Lemma \ref{lem:LLN}(iv),    
where $\bar {\mbf H}$ is defined in \eqref{eq:Hbar} and it is such that $\Vert \bar{\mbf H}\Vert = O_{\mathrm P}(1)$ because of the same arguments leading to Lemma \ref{lem:HO1}(i). It follows that asymptotically the columns of $\bar{\mbf H}^\prime$ are eigenvectors of $\bm\Gamma^F$ and are asymptotically normalized because of \eqref{eq:pipa12} (which still holds). 

Now, since $\frac{\mbf M^C}{n}$ are the eigenvalues of $\frac{\bm\Gamma^C}n=\frac{\bm\Lambda\bm\Gamma^F\bm\Lambda^\prime}n$ they are also eigenvalues of $\bm\Gamma^F\frac{\bm\Lambda^\prime\bm\Lambda}{n}=\bm\Gamma^F$, which is diagonal. So it must be that
\beq\label{eq:menatissima2}
\bm\Gamma^F=\frac{\mbf M^C}{n}.
\eeq
From \eqref{eq:starop3Dappexact} and \eqref{eq:menatissima2}, and since $\frac{\wt{\mbf M}^x}T=\frac{\wh{\mbf M}^x}n$,
\begin{align}\label{eq:menata3}
\l(\frac{\wh{\mu}^x_i}{n}-\frac{{\mu}^C_j}n\r)[\bar{\mbf H}]_{ij} 
= O_{\mathrm P}\l(\max\l(\frac 1{ n},\frac1{\sqrt{T}}\r)\r),\quad i,j=1,\ldots, r.
\end{align}
By following the same reasoning as in the proof of part (I.a), but using the rates in  \eqref{eq:menata3}, together with \eqref{eq:pipa12} (which still holds), we must have
\beq\nn
\l\Vert \bar{\mbf H}- \wh{\bm J}\r\Vert = O_{\mathrm P}\l(\max\l(\frac 1{ n},\frac 1{\sqrt T}\r)\r),
\eeq
which proves part (VI.b). This completes the proof. $\Box$

\setcounter{equation}{0}
\section{Proofs of additional results}\label{sec:otherproofs}
Throughout, recall that ``for all $n\in\mathbb N$'' always means ``for all $n>N$'' with $N$ defined in Assumption \ref{ass:common}(d).

\subsection{Proof of Proposition \ref{corol:ovvio}}\label{corol:ovvioproof}
From Proposition \ref{prop:L}(a), and Lemmas \ref{lem:FTLN}(i) and \ref{lem:ranghi}, 
\begin{align}
\l\Vert \frac{\wh{\bm\Lambda}^\prime\bm\Lambda}{n}\l(\frac{\bm\Lambda^\prime\bm\Lambda}{n}\r)^{-1}-\bm {\mathcal H}^\prime\r\Vert
&\le\l\Vert \frac{\wh{\bm\Lambda}^\prime\bm\Lambda}{n}-\bm {\mathcal H}^\prime\frac{\bm\Lambda^\prime\bm\Lambda}{n}\r\Vert\, \l\Vert \l(\frac{\bm\Lambda^\prime\bm\Lambda}{n}\r)^{-1}\r\Vert\nn\\
&\le 
\l\Vert \frac{\wh{\bm\Lambda}^\prime-\bm {\mathcal H}^\prime\bm\Lambda^\prime}{\sqrt n}\r\Vert\,\l\Vert 
\frac{\bm\Lambda}{\sqrt n}\r\Vert\,  \l\Vert \l(\frac{\bm\Lambda^\prime\bm\Lambda}{n}\r)^{-1}\r\Vert = O_{\mathrm P}\l(\max\l(\frac 1 n,\frac 1{\sqrt T}\r)\r),\nn
\end{align}
which proves part (a). 

From Proposition \ref{prop:L}(c), Lemmas \ref{lem:FTLN}(ii),  \ref{lem:FFinv}, and \ref{lem:covarianzeF}(i)
\begin{align}
\l\Vert \frac{\wh{\bm F}^\prime\bm F}{T}\l(\frac{\bm F^\prime\bm F}{T}\r)^{-1}-\bm {\mathcal H}^{-1}\r\Vert
&\le\l\Vert \frac{\wh{\bm F}^\prime\bm F}{T}-\bm {\mathcal H}^{-1}\frac{\bm F^\prime\bm F}{T}\r\Vert\, \l\Vert \l(\frac{\bm F^\prime\bm F}{T}\r)^{-1}\r\Vert\nn\\
&\le 
\l\Vert \frac{\wh{\bm F}^\prime-\bm {\mathcal H}^{-1}\bm F^\prime}{\sqrt T}\r\Vert\,\l\Vert 
\frac{\bm F}{\sqrt T}\r\Vert\,  \l\Vert \l(\frac{\bm F^\prime\bm F}{T}\r)^{-1}\r\Vert = O_{\mathrm P}\l(\max\l(\frac 1 {\sqrt n},\frac 1{\sqrt T}\r)\r),\nn
\end{align}
which proves part (b).

Then, 
\begin{align}
\l\Vert \l(\frac{\wh{\bm\Lambda}^\prime\wh{\bm\Lambda}}{n}\r)^{-1}\frac{\wh{\bm\Lambda}^\prime\bm\Lambda}{n}-\bm {\mathcal H}^{-1}\r\Vert
&\le \l\Vert \l(\frac{\wh{\bm\Lambda}^\prime\wh{\bm\Lambda}}{n}\r)^{-1}\r\Vert\,
\l\Vert \frac{\wh{\bm\Lambda}^\prime\bm\Lambda}{n}-\frac{\wh{\bm\Lambda}^\prime\wh{\bm\Lambda}\bm {\mathcal H}^{-1}}{n}\r\Vert\nn\\
&=  \l\Vert \l(\frac{\wh{\mbf M}^x}{n}\r)^{-1}\r\Vert\,
\l\Vert
\frac{\wh{\bm\Lambda}^\prime(\bm\Lambda-\wh{\bm\Lambda}\bm{\mathcal H}^{-1})
}n
\r\Vert\nn\\
&\le   \l\Vert \l(\frac{\wh{\mbf M}^x}{n}\r)^{-1}\r\Vert\,
\l\Vert
\frac{\wh{\bm\Lambda}^\prime(\bm\Lambda\bm{\mathcal H}-\wh{\bm\Lambda})
}n
\r\Vert\,\l\Vert \bm{\mathcal H}^{-1}\r\Vert\nn\\
&\le  \l\Vert \l(\frac{\wh{\mbf M}^x}{n}\r)^{-1}\r\Vert
\l\{
\l\Vert
\frac{\bm{\mathcal H}^\prime{\bm\Lambda}^\prime(\bm\Lambda\bm{\mathcal H}-\wh{\bm\Lambda})
}n
\r\Vert
+
\l\Vert
\frac{(\bm\Lambda\bm{\mathcal H}-\wh{\bm\Lambda})^\prime(\bm\Lambda\bm{\mathcal H}-\wh{\bm\Lambda})
}n
\r\Vert
\r\}
\l\Vert \bm{\mathcal H}^{-1}\r\Vert\nn\\
&\le   \l\Vert \l(\frac{\wh{\mbf M}^x}{n}\r)^{-1}\r\Vert
\l\{
\l\Vert\bm{\mathcal H}\r\Vert\,
\l\Vert
\frac{\bm\Lambda}{\sqrt n}\r\Vert\,\l\Vert
\frac{\bm\Lambda\bm{\mathcal H}-\wh{\bm\Lambda}
}{\sqrt n}
\r\Vert
+
\l\Vert
\frac{\bm\Lambda\bm{\mathcal H}-\wh{\bm\Lambda}
}{\sqrt n}
\r\Vert^2
\r\}
\l\Vert \bm{\mathcal H}^{-1}\r\Vert\nn\\
&= O_{\mathrm P}\l(\max\l(\frac 1 n,\frac 1{\sqrt T}\r)\r),\nn
\end{align}
because of Proposition \ref{prop:L}(a), and Lemmas  \ref{lem:FTLN}(i), \ref{lem:MO1}(iv), \ref{lem:HO1bis}(i), and \ref{lem:HO1bis}(ii). This proves part (c).

Finally,
\begin{align}
\l\Vert \frac{{\bm F}^\prime\wh{\bm F}}{T}\l(\frac{\wh{\bm F}^\prime\wh{\bm F}}{T}\r)^{-1}-\bm {\mathcal H}\r\Vert&\le
\l\Vert
 \frac{{\bm F}^\prime\wh{\bm F}}{T} \frac{-\bm {\mathcal H}\wh{\bm F}^\prime\wh{\bm F}}{T}
\r\Vert\,
\l\Vert \l(\frac{\wh{\bm F}^\prime\wh{\bm F}}{T}\r)^{-1}\r\Vert\nn\\
&= 
\l\Vert
 \frac{({\bm F}-\wh{\bm F}\bm {\mathcal H}^\prime)^\prime\wh{\bm F}}{T}
\r\Vert\nn\\
&\le
\l\Vert
 \frac{({\bm F}(\bm {\mathcal H}^\prime)^{-1}-\wh{\bm F})^\prime\wh{\bm F}}{T}
\r\Vert\,\l\Vert \bm {\mathcal H}\r\Vert\nn\\
&\le
\l\{\l\Vert
 \frac{({\bm F}(\bm {\mathcal H}^\prime)^{-1}-\wh{\bm F})^\prime{\bm F}}{T}
\r\Vert
+
\l\Vert
 \frac{({\bm F}(\bm {\mathcal H}^\prime)^{-1}-\wh{\bm F})^\prime({\bm F}(\bm {\mathcal H}^\prime)^{-1}-\wh{\bm F})}{T}
\r\Vert
\r\}\l\Vert \bm {\mathcal H}\r\Vert\nn\\
&\le
\l\{\l\Vert
 \frac{\wh{\bm F}^\prime-\bm {\mathcal H}^{-1}{\bm F}^\prime}{\sqrt T}
\r\Vert\,\l\Vert \frac{\bm F}{\sqrt T}\r\Vert
+
\l\Vert
 \frac{\wh{\bm F}^\prime-\bm {\mathcal H}^{-1}{\bm F}^\prime}{\sqrt T}
\r\Vert^2
\r\}\l\Vert \bm {\mathcal H}\r\Vert\nn\\
&= O_{\mathrm P}\l(\max\l(\frac 1 {\sqrt n},\frac 1{\sqrt T}\r)\r),\nn
\end{align}
because of Proposition \ref{prop:L}(c), and Lemmas \ref{lem:FTLN}(ii) and \ref{lem:HO1bis}(i). This proves part (d) 
 and completes the proof. $\Box$

\subsection{Proof of Proposition \ref{prop:KKK}}\label{app:KKKc1}
Let $\bm V_0$ be the matrix of eigenvalues of $(\bm\Gamma^F)^{1/2}\bm\Sigma_\Lambda(\bm\Gamma^F)^{1/2}$ sorted in descending order denoted as $[\bm V_0]_{jj}$, $j=1,\ldots, r$,
and
$\bm\Upsilon_0$ be the matrix having as columns the corresponding normalized eigenvectors, so that
\beq\label{eq:UVU0app}
(\bm\Gamma^F)^{1/2}\bm\Sigma_\Lambda(\bm\Gamma^F)^{1/2} = \bm\Upsilon_0\bm V_0 \bm\Upsilon_0^\prime.
\eeq
Now, by Assumption \ref{ass:common}(a)
\beq
\lim_{n\to\infty} \l\Vert(\bm\Gamma^F)^{1/2}\frac{\bm\Lambda^\prime \bm\Lambda}n(\bm\Gamma^F)^{1/2}-(\bm\Gamma^F)^{1/2}\bm\Sigma_\Lambda(\bm\Gamma^F)^{1/2}\r\Vert=0.\label{eq:limB}
\eeq
Now, $[\bm V_0]_{jj} = \lim_{n\to\infty}\frac{\mu_j^C}{n}$ for all $j=1,\ldots, r$ by Lemma \ref{lem:Vzero}(i). Moreover, $[\bm V_0]_{jj}>[\bm V_0]_{j+1,j+1}$ for all $j=1,\ldots, r-1$ by Assumption \ref{ass:eval}. Then, since, as shown in Lemma \ref{lem:KO1}(iii), $\mbf K$ is the matrix of normalized eigenvectors of $(\bm\Gamma^F)^{1/2}\frac{\bm\Lambda^\prime \bm\Lambda}n(\bm\Gamma^F)^{1/2}$, from \citet[Corollary 1]{yu15},
\beq\label{eq:ups00}
\l\Vert\mbf K-\bm\Upsilon_0\bm{\mathcal J}\r\Vert \le \frac{2^{3/2}\sqrt r  \l\Vert(\bm\Gamma^F)^{1/2}\frac{\bm\Lambda^\prime \bm\Lambda}n(\bm\Gamma^F)^{1/2}-(\bm\Gamma^F)^{1/2}\bm\Sigma_\Lambda(\bm\Gamma^F)^{1/2}\r\Vert}
{\min\l\{\vert 
[\bm V_0]_{00}
-
[\bm V_0]_{11}
\vert,
\vert
[\bm V_0]_{rr}
-
[\bm V_0]_{r+1,r+1}
\vert
\r\}},
\eeq
where 
$[\bm V_0]_{00}=\infty$ and $[\bm V_0]_{r+1,r+1}=0$, 
and $\bm{\mathcal J}$ is an $r\times r$ diagonal matrix with entries $\pm 1$ and depending on $n$. By letting $\bm{\mathcal J}^*$ be the $r\times r$ diagonal matrix with entries $\pm 1$ and independent of $n$, such that $\lim_{n\to\infty}\Vert \bm{\mathcal J}^*-\bm{\mathcal J}\Vert =0$, 
by  \eqref{eq:limB} and \eqref{eq:ups00}, and since $0<[\bm V_0]_{jj}<\infty$, $=1,\ldots,r$ by Lemmas \ref{lem:Vzero}(i) and \ref{lem:Vzero}(iii),
we get 
\beq\label{eq:ups000}
 \lim_{n\to\infty}\l\Vert\mbf K-\bm\Upsilon_0\bm{\mathcal J}^*\r\Vert\le  \lim_{n\to\infty}\l\Vert\mbf K-\bm\Upsilon_0\bm{\mathcal J}\r\Vert+ 
 \lim_{n\to\infty}\l\Vert\bm{\mathcal J}-\bm{\mathcal J}^*\r\Vert\,\Vert\bm\Upsilon_0\Vert =0.
\eeq
Furthermore, let $\bm{\mathcal J}_0$ be the $r\times r$ diagonal matrix with entries $\pm 1$ such that  $\Vert \bm{\mathcal J}^*\mbf J-\bm{\mathcal J}_0\Vert=o_{\mathrm P}(1)$ as $n,T\to\infty$. Then, from \eqref{eq:ups000},
\beq\label{eq:ups}
\l\Vert\mbf K\mbf J-\bm\Upsilon_0\bm{\mathcal J}_0\r\Vert\le\l\Vert\mbf K-\bm\Upsilon_0\bm{\mathcal J}^*\r\Vert\,\Vert \mbf J\Vert+
\l\Vert\bm{\mathcal J}_0-\bm{\mathcal J}^*\mbf J\r\Vert\,\Vert\bm\Upsilon_0\Vert=o_{\mathrm P}(1).
\eeq
From \eqref{eq:mcHbis} in the proof of Proposition \ref{prop:L} and  \eqref{eq:ups} it follows that
\begin{align}
 \l\Vert \bm{\mathcal H}-(\bm\Gamma^F )^{1/2}\bm\Upsilon_0\bm{\mathcal J}_0\r\Vert&= \l\Vert(\bm\Gamma^F )^{1/2}\mbf K\mbf J-(\bm\Gamma^F )^{1/2}\bm\Upsilon_0\bm{\mathcal J}_0\r\Vert\nn\\
&\le \l\Vert\mbf K\mbf J-\bm\Upsilon_0\bm{\mathcal J}_0\r\Vert\,\Vert (\bm\Gamma^F )^{1/2}\Vert=o_{\mathrm P}(1),\label{eq:debosc}
\end{align}
by Assumption \ref{ass:common}(b).

By using Assumption \ref{ass:common}(a), \eqref{eq:UVU0app}, and \eqref{eq:debosc}, 
\begin{align}
\nn
\l\Vert\bm{\mathcal H}^\prime\frac{\bm\Lambda^\prime\bm\Lambda}{n} -\bm{\mathcal J}_0\bm\Upsilon_0^\prime(\bm\Gamma^F)^{1/2}\bm\Sigma_\Lambda\r\Vert&=
\l\Vert\bm{\mathcal H}^\prime\frac{\bm\Lambda^\prime\bm\Lambda}{n} -\bm{\mathcal J}_0\bm\Upsilon_0^\prime \bm\Upsilon_0\bm V_0 \bm\Upsilon_0^\prime(\bm\Gamma^F)^{-1/2}\r\Vert\nn\\
&=\l\Vert\bm{\mathcal H}^\prime\frac{\bm\Lambda^\prime\bm\Lambda}{n} -\bm V_0\bm{\mathcal J}_0\bm\Upsilon_0^\prime(\bm\Gamma^F)^{-1/2}\r\Vert=o_{\mathrm P}(1),\label{eq:exparta}
\end{align}
since the eigenvectors $\bm\Upsilon_0$ are orthonormal and $\bm{\mathcal J}_0$ is a diagonal matrix with entries $\pm 1$ so it commutes with any other diagonal $r\times r$ matrix.  

From \eqref{eq:exparta}, Proposition \ref{prop:L}(a), and Lemma \ref{lem:FTLN}(i), 
\begin{align}
\l\Vert \frac{\wh{\bm\Lambda}^\prime\bm\Lambda}{n}-\bm V_0\bm{\mathcal J}_0\bm\Upsilon_0^\prime(\bm\Gamma^F)^{-1/2}\r\Vert &\le
\l\Vert \frac{\wh{\bm\Lambda}^\prime\bm\Lambda}{n}-\bm {\mathcal H}^\prime\frac{\bm\Lambda^\prime\bm\Lambda}{n}\r\Vert+
\l\Vert \bm{\mathcal H}^\prime\frac{{\bm\Lambda}^\prime\bm\Lambda}{n}-\bm V_0\bm{\mathcal J}_0\bm\Upsilon_0^\prime(\bm\Gamma^F)^{-1/2}\r\Vert\nn\\
&\le \l\Vert \frac{\wh{\bm\Lambda}^\prime-\bm {\mathcal H}^\prime\bm\Lambda^\prime}{\sqrt n}\r\Vert\,\l\Vert 
\frac{\bm\Lambda}{\sqrt n}\r\Vert+
\l\Vert \bm{\mathcal H}^\prime\frac{{\bm\Lambda}^\prime\bm\Lambda}{n}-\bm V_0\bm{\mathcal J}_0\bm\Upsilon_0^\prime(\bm\Gamma^F)^{-1/2}\r\Vert = o_{\mathrm P}(1).\label{eq:parioli}
\end{align}
Note that $\bm V_0\bm{\mathcal J}_0\bm\Upsilon_0^\prime  (\bm\Gamma^F)^{-1/2}$ is finite and invertible because of Assumptions \ref{ass:common}(b) and  \ref{ass:common}(c-ii) and Lemmas \ref{lem:Vzero}(i) and \ref{lem:Vzero}(iii). 
Finally, if Assumption \ref{ass:common}(a) holds with rate $\sqrt n$, then \eqref{eq:limB}-\eqref{eq:ups000} hold all with rate $\sqrt n$, because, clearly, $\Vert\bm{\mathcal J}-\bm{\mathcal J}_1\Vert$ converges at the same rate at which $\Vert\frac{\bm\Lambda^\prime\bm\Lambda}{n} -\bm\Sigma_\Lambda\Vert$ converges. 
Likewise, if Assumption \ref{ass:common}(c-ii) holds with rate $\sqrt T$, Proposition \ref{prop:L}(a) holds with rate $\min(n,\sqrt T)$. Moreover, $\Vert\bm{\mathcal J}_1-\mbf J\Vert$ converges at the same rate at which $\frac 1n\Vert\wh{\bm\Gamma}^x -\bm\Gamma^C\Vert$ converges, i.e., $\min(n,\sqrt T)$ by Lemma \ref{lem:covarianze}(ii). 
Therefore, \eqref{eq:ups}-\eqref{eq:parioli} hold with rate $\min(\sqrt n,\sqrt T)$. This completes the proof. $\Box$

\subsection{Proof of Lemma \ref{prop:load}}\label{prop:loadproof}

From \eqref{eq:sviluppoLambda} and substituting on its right-hand-side $\wh{\bm\lambda}_j^\prime$ with $\bm\lambda_j^\prime\bm{\mathcal H}+\wh{\bm\lambda}_j^\prime-\bm\lambda_j^\prime\bm{\mathcal H}$, we get
\begin{align}
\wh{\bm\lambda}_i^\prime-{\bm\lambda}_i^\prime\wh{\mbf H}=&\,
\l(
\underbrace{\frac 1{nT}{\bm\lambda}_i^\prime\sum_{t=1}^T\sum_{j=1}^n\mbf F_t e_{jt}{\bm\lambda}_j^\prime}_{\text{(A)}}
+\underbrace{\frac 1{nT} \sum_{t=1}^T  e_{it}\mbf F_t^\prime\sum_{j=1}^n\bm\lambda_j\bm\lambda_j^\prime}_{\text{(B)}}
+\underbrace{\frac 1{nT} \sum_{t=1}^T\sum_{j=1}^n e_{it} e_{jt} \bm\lambda_j^\prime}_{\text{(C)}}
\r)\bm{\mathcal H} \l(\frac{\wh{\mbf M}^x}{n}\r)^{-1}\nn\\
&+
\l(
\underbrace{\frac 1{nT}{\bm\lambda}_i^\prime\sum_{t=1}^T\sum_{j=1}^n\mbf F_t e_{jt}(\wh{\bm\lambda}_j^\prime-\bm\lambda_j^\prime\bm{\mathcal H})}_{\text{(D)}}
+\underbrace{\frac 1{nT} \sum_{t=1}^T  e_{it}\mbf F_t^\prime\sum_{j=1}^n\bm\lambda_j(\wh{\bm\lambda}_j^\prime-\bm\lambda_j^\prime\bm{\mathcal H})}_{\text{(E)}}\r.\nn\\
&\;\;\l.+\underbrace{\frac 1{nT} \sum_{t=1}^T\sum_{j=1}^n e_{it} e_{jt} (\wh{\bm\lambda}_j^\prime-\bm\lambda_j^\prime\bm{\mathcal H})}_{\text{(F)}}\r) \l(\frac{\wh{\mbf M}^x}{n}\r)^{-1}.\label{eq:sviluppoLambda_app}
\end{align}

For part (a), notice that term (1.a) in \eqref{eq:sviluppoLambda} is given by $\text{A}\bm{\mathcal H}$ plus D in \eqref{eq:sviluppoLambda_app}. Start with A, for any $i=1,\ldots,n$, by Assumption \ref{ass:common}(a) 
\beq
\l\Vert\frac 1{nT}{\bm\lambda}_i^\prime\sum_{t=1}^T\sum_{j=1}^n\mbf F_t e_{jt}{\bm\lambda}_j^\prime\r\Vert \le \l\Vert \bm\lambda_i\r\Vert\,\l\Vert\frac 1{nT}\sum_{t=1}^T\sum_{j=1}^n\mbf F_t e_{jt}{\bm\lambda}_j^\prime \r\Vert
,\label{eq:1a}
\eeq
and by Lemma \ref{lem:GxiBAI}(v),
\beq
\E\l[\l\Vert\frac 1{nT}\sum_{t=1}^T\sum_{j=1}^n\mbf F_t e_{jt}{\bm\lambda}_j^\prime \r\Vert^2\r]
\le \frac{r^2 M_\Lambda^3 M_FM_{1 e}}{nT}.\label{eq:1aaaaa}
\eeq
So, by \eqref{eq:1a}, \eqref{eq:1aaaaa}, and Chebychev's inequality, term A in \eqref{eq:sviluppoLambda_app} is $O_{\mathrm P}(n^{-1/2}T^{-1/2})$. For term D,  for any $i=1,\ldots,n$, because of Assumption \ref{ass:common}(a)
\begin{align}
\l\Vert\frac 1{nT}{\bm\lambda}_i^\prime\sum_{t=1}^T\sum_{j=1}^n\mbf F_t e_{jt}(\wh{\bm\lambda}_j^\prime-{\bm\lambda}_j^\prime\bm{\mathcal H})\r\Vert &\le M_\Lambda\l\Vert\frac 1{nT}\sum_{t=1}^T\sum_{j=1}^n\mbf F_t e_{jt}(\wh{\bm\lambda}_j^\prime-{\bm\lambda}_j^\prime\bm{\mathcal H}) \r\Vert\nn\\
&= M_\Lambda\l\Vert\frac {\bm F^\prime\bm  E (\wh{\bm\Lambda}-\bm\Lambda\bm{\mathcal H} )}{nT} \r\Vert\le M_\Lambda \l\Vert\frac {\bm F^\prime\bm  E}{\sqrt n T} \r\Vert\, \l\Vert\frac{\wh{\bm\Lambda}-\bm\Lambda\bm{\mathcal H}}{\sqrt n}\r\Vert.\label{eq:1d}
\end{align}
Then, by using Lemma \ref{lem:LLN}(i) and Proposition \ref{prop:L}(a) in \eqref{eq:1d},  term D in \eqref{eq:sviluppoLambda_app} is $O_{\mathrm P}(\max(nT^{-1/2}, T^{-1}))$, hence, it is negligible with respect to term A. By noticing that $\Vert\bm{\mathcal H}\Vert= O(1)$ by Lemma \ref{lem:HO1bis}(i), we prove part (a).

For part (b), notice that term (1.b) in \eqref{eq:sviluppoLambda} is given by $\text{B}\bm{\mathcal H}$ plus E in \eqref{eq:sviluppoLambda_app}. Start with B, for any $i=1,\ldots,n$, because of Lemma \ref{lem:FTLN}(i),
\beq
\l\Vert\frac 1{nT} \sum_{t=1}^T  e_{it}\mbf F_t^\prime\sum_{j=1}^n\bm\lambda_j\bm\lambda_j^\prime\r\Vert=
\l\Vert\frac 1{nT} \sum_{t=1}^T  e_{it}\mbf F_t^\prime(\bm\Lambda^\prime\bm\Lambda)\r\Vert\le 
\l\Vert\frac 1{T} \sum_{t=1}^T  e_{it}\mbf F_t\r\Vert\,\l\Vert
\frac{\bm\Lambda}{\sqrt n}
\r\Vert^2\le \l\Vert\frac 1{T} \sum_{t=1}^T  e_{it}\mbf F_t\r\Vert M_\Lambda^2.\label{eq:1b}
\eeq
Then,  from \eqref{eq:oslo3} in the proof of Lemma \ref{lem:GxiBAI}, when setting $n=1$ therein,
\begin{align}
\E\l[\l\Vert\frac 1{T} \sum_{t=1}^T  e_{it}\mbf F_t\r\Vert^2\r]&\le\frac{M_{F e}}{T}.\label{eq:1b2}
\end{align}
By substituting \eqref{eq:1b2} into \eqref{eq:1b}, by Chebychev's inequality, term B in \eqref{eq:sviluppoLambda_app} is $O_{\mathrm P}(T^{-1/2})$. For term E, 
 for any $i=1,\ldots,n$,
\begin{align}
\l\Vert\frac 1{nT} \sum_{t=1}^T  e_{it}\mbf F_t^\prime\sum_{j=1}^n\bm\lambda_j(\wh{\bm\lambda}_j^\prime-{\bm\lambda}_j^\prime\bm{\mathcal H})\r\Vert&=
\l\Vert\frac 1{nT} \sum_{t=1}^T  e_{it}\mbf F_t^\prime\bm\Lambda^\prime(\wh{\bm\Lambda}-\bm\Lambda\bm{\mathcal H})\r\Vert\nn\\ 
&\le \l\Vert\frac 1{T} \sum_{t=1}^T  e_{it}\mbf F_t\r\Vert\,\l\Vert
\frac{\bm\Lambda}{\sqrt n}
\r\Vert \, \l\Vert\frac{\wh{\bm\Lambda}-\bm\Lambda\bm{\mathcal H}}{\sqrt n}\r\Vert .\label{eq:1e}
\end{align}
By \eqref{eq:1b2} (with Chebychev's inequality), Lemma \ref{lem:FTLN}(i), and part (a) of Proposition \ref{prop:L}, from \eqref{eq:1e}, term E in \eqref{eq:sviluppoLambda_app} is $O_{\mathrm P}(\max(nT^{-1/2}, T^{-1}))$, hence, it is negligible with respect to term B. By noticing that $\Vert\bm{\mathcal H}\Vert= O(1)$ by Lemma \ref{lem:HO1bis}(i), we prove part (b).

For part (c), notice that term (1.c) in \eqref{eq:sviluppoLambda} is given by $\text{C}\bm{\mathcal H}$ plus F in \eqref{eq:sviluppoLambda_app}. Start with C, for any $i=1,\ldots,n$,  because of Assumption \ref{ass:common}(a),
\begin{align}
\l\Vert\frac 1{nT} \sum_{t=1}^T\sum_{j=1}^n e_{it} e_{jt} \bm\lambda_j^\prime\r\Vert &=\l\{\sum_{k=1}^r \l(\frac 1{nT} \sum_{t=1}^T\sum_{j=1}^n e_{it} e_{jt} \lambda_{jk}\r)^2\r\}^{1/2}\le \sqrt rM_\Lambda\l\vert\frac 1{nT} \sum_{t=1}^T\sum_{j=1}^n e_{it} e_{jt} \r\vert\nn\\
&\le\sqrt r M_\Lambda\l\{ \l\vert\frac 1{nT} \sum_{t=1}^T\sum_{j=1}^n\l\{ e_{it} e_{jt}-\E[ e_{it} e_{jt}]\r\} \r\vert+\l\vert\frac 1{nT} \sum_{t=1}^T\sum_{j=1}^n\E[ e_{it} e_{jt}] \r\vert\r\}.\label{eq:1c}
\end{align}
Then, by Assumption \ref{ass:idio}(b),
\begin{align}
\l\vert\frac 1{nT} \sum_{t=1}^T\sum_{j=1}^n\E[ e_{it} e_{jt}] \r\vert\le 
\frac 1{nT}\sum_{t=1}^T\sum_{j=1}^n\l\vert\E[ e_{it} e_{jt}]\r\vert\le 
\max_{t=1,\ldots,T}\frac 1n\sum_{j=1}^n\l\vert\E[ e_{it} e_{jt}]\r\vert\le\frac 1n \sum_{j=1}^n M_{ij} \le \frac{M_ e}n,\label{eq:1c2}
\end{align}
since $M_ e$ is independent of $i$ and $t$. Moreover, by Assumption \ref{ass:idio}(c), 
\beq\label{eq:1c3}
\E\l[\l\vert\frac 1{nT} \sum_{t=1}^T\sum_{j=1}^n\l\{ e_{it} e_{jt}-\E[ e_{it} e_{jt}]\r\} \r\vert^2\r]\le \frac{K_ e}{nT}.
\eeq
By substituting \eqref{eq:1c2} and \eqref{eq:1c3} into \eqref{eq:1c},  by Chebychev's inequality,  term C in \eqref{eq:sviluppoLambda_app} is\linebreak $O_{\mathrm P}(\max(n^{-1}, n^{-1/2}T^{-1/2}))$. For term F, for any $i=1,\ldots,n$, let $\bm\varepsilon_i=( e_{i1}\cdots  e_{iT})^\prime$, then
\begin{align}
\l\Vert\frac 1{nT} \sum_{t=1}^T\sum_{j=1}^n e_{it} e_{jt} (\wh{\bm\lambda}_j^\prime-{\bm\lambda}_j^\prime\bm{\mathcal H})\r\Vert &= \l\Vert\frac{\bm\varepsilon_i^\prime\bm E\l(\wh{\bm\Lambda}-\bm\Lambda\bm{\mathcal H}\r)}{nT} \r\Vert \le \l\Vert\frac{\bm\varepsilon_i^\prime\bm E}{\sqrt n T}\r\Vert\, \l\Vert\frac{\wh{\bm\Lambda}-\bm\Lambda\bm{\mathcal H}}{\sqrt n}\r\Vert.\label{eq:1f}
\end{align}
Then, by the $C_r$-inequality with $r=2$,
\begin{align}
\l\Vert\frac{\bm\varepsilon_i^\prime\bm E}{\sqrt n T}\r\Vert^2&=
\l\Vert\frac 1 {\sqrt n T} \sum_{t=1}^T  e_{it} \bm e_t^\prime\r\Vert^2
 =\frac 1{n} \sum_{j=1}^n \l(\frac 1T\sum_{t=1}^T  e_{it} e_{jt}\r)^2\nn\\
& =\frac 1{n} \sum_{j=1}^n \l(\frac 1T\sum_{t=1}^T \l\{ e_{it} e_{jt}-\E[ e_{it} e_{jt}]\r\}+\frac 1T\sum_{t=1}^T \E[ e_{it} e_{jt}]\r)^2\nn\\
&\le\frac 2{n} \sum_{j=1}^n\l\{  \l(\frac 1T\sum_{t=1}^T \l\{ e_{it} e_{jt}-\E[ e_{it} e_{jt}]\r\}\r)^2+\l(\frac 1T\sum_{t=1}^T \E[ e_{it} e_{jt}]\r)^2\r\}.\label{eq:1f2}
\end{align}
By taking the expectation of \eqref{eq:1f2}, and because of Assumption \ref{ass:idio}(c) and Lemma \ref{lem:Gxi}(v),
\begin{align}
\E\l[\l\Vert\frac{\bm\varepsilon_i^\prime\bm E}{\sqrt n T}\r\Vert^2\r]&=\frac 2{n} \sum_{j=1}^n \E\l[
\l(\frac 1T\sum_{t=1}^T \l\{ e_{it} e_{jt}-\E[ e_{it} e_{jt}]\r\}\r)^2
\r]+\frac 2{n} \sum_{j=1}^n \l(\frac 1T\sum_{t=1}^T \E[ e_{it} e_{jt}]\r)^2\nn\\
&\le 2\max_{j=1,\ldots, n}\E\l[
\l(\frac 1T\sum_{t=1}^T \l\{ e_{it} e_{jt}-\E[ e_{it} e_{jt}]\r\}\r)^2
\r]+\frac{2}{n}\sum_{i=1}^n\l(\frac 1T\sum_{t=1}^T \E[ e_{it} e_{jt}]\r)^2\nn\\
&\le \frac{2K_ e}{T} + \frac{2}{nT^2}\sum_{i=1}^n\sum_{t=1}^T \E[ e_{it} e_{jt}]\sum_{s=1}^T\E[ e_{is} e_{js}]\nn\\
&\le \frac{2K_ e}{T} + \max_{t=1,\ldots ,T}\frac{2}{n}\sum_{i=1}^n \l\vert\E[ e_{it} e_{jt}]\r\vert \max_{s=1,\ldots,T}\max_{i,j=1,\ldots, n} \E[ e_{is} e_{js}]\nn\\
&\le \frac{2K_ e}{T} + \frac{2}n \sum_{i=1}^n M_{ij} \max_{i,j=1,\ldots, n} \bm\varepsilon_i^\prime\bm\Gamma^ e\bm\varepsilon_j\le  \frac{2K_ e}{T}+ \frac{2M_ e}{n}\Vert \bm\Gamma^ e\Vert\le \frac{2K_ e}{T}+ \frac{2M_ e M_{2 e}}{n}, \label{eq:1f3}
\end{align}
since $K_ e$ is independent of $j$ and $M_ e$ is independent of $i$, $j$, $t$, and $s$ and where $\bm\varepsilon_i$ is an $n$-dimensional vector with one in the $i$th entry and zero elsewhere.  
By substituting \eqref{eq:1f3} and part (a) of Proposition \ref{prop:L} into \eqref{eq:1f},  term F in \eqref{eq:sviluppoLambda_app} is $O_{\mathrm P}(\max(n^{-3/2}, n^{-1/2}T^{-1/2}, T^{-1}))$, hence, it is negligible with respect to term C. By noticing that $\Vert\bm{\mathcal H}\Vert= O(1)$ by Lemma \ref{lem:HO1bis}(i), we prove part (c) and complete the proof. $\Box$

\subsection{Proof of Proposition \ref{prop:L2}}\label{prop:L2proof}
Part (a), is proved  by applying Lemma 
\ref{prop:load} to \eqref{eq:sviluppoLambda} and since  $\Vert (\frac{\wh{\mbf M}^x}{n})^{-1} \bm{\mathcal H}^\prime \Vert = O_{\mathrm {P}}(1)$, because of Lemma \ref{lem:MO1}(iv) and \ref{lem:HO1bis}(i).

For part (b), from \eqref{eq:start4}
\begin{align}
\l\Vert\frac{\wh{\bm\Lambda}-\bm\Lambda\wh{\mbf H}}{\sqrt n}\r\Vert\le&\,
\l(\l\Vert\frac{\bm F^\prime\bm  E}{\sqrt nT}\r\Vert\,\l\Vert\frac{\bm\Lambda}{\sqrt n}\r\Vert^2
+\l\Vert \frac{\bm E^\prime\bm F}{\sqrt {n}T}\r\Vert\,\l\Vert \frac{\bm\Lambda^\prime{\bm\Lambda}}{n}\r\Vert
+\l\Vert\frac{\bm E^\prime\bm  E\bm\Lambda}{n^{3/2}T}\r\Vert\r)\l\Vert\bm{\mathcal H} \r\Vert \,\l\Vert\l(\frac{\wh{\mbf M}^x}{n}\r)^{-1}\r\Vert\nn\\
&+
\l(\l\Vert\frac{\bm F^\prime\bm  E}{\sqrt nT}\r\Vert\, \l\Vert\frac{\bm\Lambda}{\sqrt n}\r\Vert
+\l\Vert\frac{\bm E^\prime\bm F}{\sqrt nT}\r\Vert\, \l\Vert\frac{\bm\Lambda}{\sqrt n}\r\Vert
+\l\Vert\frac{\bm E^\prime\bm  E}{nT}\r\Vert\r)\l\Vert\frac{\wh{\bm\Lambda}-\bm\Lambda\bm{\mathcal H}}{\sqrt n}\r\Vert \l\Vert\l(\frac{\wh{\mbf M}^x}{n}\r)^{-1}\r\Vert,\nn
\end{align}
and the proof of part (b) follows from 
Proposition \ref{prop:L}(a) and Lemma
\ref{lem:FTLN}(i),
\ref{lem:LLN}(i),
\ref{lem:LLN}(iii),
\ref{lem:LLN}(iv),
\ref{lem:MO1}(iv),
\ref{lem:HO1bis}(i). This completes the proof. $\Box$

\subsection{Proof of Lemma \ref{prop:factor}}\label{prop:factorproof}
For part (a):
\begin{align}
\frac{\wh{\bm\Lambda}^\prime(\bm\Lambda-\wh{\bm\Lambda}\wh{\mbf H}^{-1})\mbf F_t}{n}&=-\frac{\wh{\bm\Lambda}^\prime(\wh{\bm\Lambda}-\bm\Lambda\wh{\mbf H})\wh{\mbf H}^{-1}\mbf F_t}{n}\nn\\
&=
-\l\{
\frac{(\wh{\bm\Lambda}-\bm\Lambda\wh{\mbf H})^\prime\bm\Lambda\wh{\mbf H}}{n}+ \frac{(\wh{\bm\Lambda}-\bm\Lambda\wh{\mbf H})^\prime (\wh{\bm\Lambda}-\bm\Lambda\wh{\mbf H})}{n}\r\}\wh{\mbf H}^{-1}\mbf F_t, \label{eq:2a0}
\end{align}
Then, from \eqref{eq:2a0}
\begin{align}
\l\Vert \frac{\wh{\bm\Lambda}^\prime(\bm\Lambda-\wh{\bm\Lambda}\wh{\mbf H}^{-1})\mbf F_t}{n}\r\Vert&\le 
\l\{\l\Vert
\frac{(\wh{\bm\Lambda}-\bm\Lambda\wh{\mbf H})^\prime\bm\Lambda\wh{\mbf H}}{n}\r\Vert+\l\Vert \frac{(\wh{\bm\Lambda}-\bm\Lambda\wh{\mbf H})^\prime (\wh{\bm\Lambda}-\bm\Lambda\wh{\mbf H})}{n}\r\Vert\r\}\, \l\Vert\wh{\mbf H}^{-1}\r\Vert\, \Vert\mbf F_t\Vert\nn\\
&=\l\{\Vert \bm I \Vert  +\Vert  \bm{II}\Vert \r\}\, \l\Vert\wh{\mbf H}^{-1}\r\Vert\, \Vert\mbf F_t\Vert
,\;\text{say.}\label{eq:2a}
\end{align}
First, consider $\bm I$ in \eqref{eq:2a}. From \eqref{eq:start4} 
\begin{align}
\bm I=&\,\frac {(\wh{\bm\Lambda}-\bm\Lambda\wh{\mbf H})^\prime \bm\Lambda \wh{\mbf H}}n\nn\\
=&\,  \l(\frac{\wh{\mbf M}^x}{n}\r)^{-1}\bm{\mathcal H}^\prime\frac{\bm\Lambda^\prime\bm\Lambda}{n}
\frac{\bm F^\prime\bm E\bm\Lambda \wh{\mbf H}}{ nT}
+\l(\frac{\wh{\mbf M}^x}{n}\r)^{-1}\bm{\mathcal H}^\prime\frac{\bm\Lambda^\prime\bm E^\prime\bm F}{nT}\frac{\bm\Lambda^\prime\bm\Lambda \wh{\mbf H}}{n} + \l(\frac{\wh{\mbf M}^x}{n}\r)^{-1}\bm{\mathcal H}^\prime \frac{\bm\Lambda^\prime\bm E^\prime\bm E\bm\Lambda \wh{\mbf H}}{n^2T}\nn\\
&+\l(\frac{\wh{\mbf M}^x}{n}\r)^{-1}\frac{(\wh{\bm\Lambda}-\bm\Lambda\bm{\mathcal H})^\prime}{\sqrt n}\frac{\bm  E^\prime\bm F\bm\Lambda^\prime}{nT}\frac{\bm\Lambda \wh{\mbf H}}{\sqrt n}
+\l(\frac{\wh{\mbf M}^x}{n}\r)^{-1}\frac{(\wh{\bm\Lambda}-\bm\Lambda\bm{\mathcal H})^\prime}{\sqrt n}\frac{\bm\Lambda\bm F^\prime\bm E}{nT}\frac{\bm\Lambda \wh{\mbf H}}{\sqrt n}\nn\\
&+\l(\frac{\wh{\mbf M}^x}{n}\r)^{-1}\frac{(\wh{\bm\Lambda}-\bm\Lambda\bm{\mathcal H})^\prime}{\sqrt n}\frac{\bm E^\prime\bm  E}{nT}\frac{\bm\Lambda \wh{\mbf H}}{\sqrt n}
= \bm I_a+\bm I_b+\bm I_c+\bm I_d+\bm I_e+\bm I_f, \;\text{say.} \label{eq:2a3}
\end{align}
Then, because of  \eqref{eq:1a} in the proof of Lemma \ref{prop:load},
\begin{align}
\E\l[\l\Vert \frac{\bm F^\prime\bm E\bm\Lambda}{ nT}\r\Vert^2\r] =
\E\l[\l\Vert\frac 1{nT}\sum_{t=1}^T\sum_{j=1}^n\mbf F_t e_{jt}{\bm\lambda}_j^\prime \r\Vert^2\r] 
\le \frac{r^2 M_\Lambda^3 M_FM_{1 e}}{nT}.\label{eq:2a3bis}
\end{align}
Therefore, by Lemmas \ref{lem:LLN}(iv), \ref{lem:MO1}(iv), \ref{lem:HO1}(i),  \ref{lem:HO1bis}(i),
and using \eqref{eq:2a3bis} , we get
\begin{align}
\Vert \bm I_a\Vert \le \l\Vert \l(\frac{\wh{\mbf M}^x}{n}\r)^{-1}\r\Vert\,\l\Vert \bm{\mathcal H}\r\Vert\,\l\Vert\frac{\bm\Lambda^\prime\bm\Lambda}{n}\r\Vert\,
\l\Vert\frac{\bm F^\prime\bm E\bm\Lambda }{ nT}\r\Vert\,\l\Vert \wh{\mbf H}\r\Vert=O_{\mathrm P}\l(\frac 1{\sqrt {nT}}\r),\label{eq:2aIa}\\
\Vert \bm I_b\Vert\le  \l\Vert \l(\frac{\wh{\mbf M}^x}{n}\r)^{-1}\r\Vert\,\l\Vert \bm{\mathcal H}\r\Vert\,\l\Vert \frac{\bm\Lambda^\prime\bm E^\prime\bm F}{nT}\r\Vert \,\l\Vert \frac{\bm\Lambda^\prime\bm\Lambda}n \r\Vert\,\l\Vert \wh{\mbf H}\r\Vert 
=O_{\mathrm P}\l(\frac 1{\sqrt {nT}}\r).\label{eq:2aIb}
\end{align}
Moreover, because of Lemmas \ref{lem:FTLN}(i), \ref{lem:LLN}(iii), \ref{lem:MO1}(iv), \ref{lem:HO1}(i), and \ref{lem:HO1bis}(i),
\begin{align}\label{eq:2aIc}
\Vert \bm I_c\Vert \le  \l\Vert\l(\frac{\wh{\mbf M}^x}{n}\r)^{-1}\r\Vert\,\l\Vert\bm{\mathcal H}\r\Vert\, 
\l\Vert\frac{\bm\Lambda}{\sqrt n}\r\Vert\,
\l\Vert\frac{\bm\Lambda^\prime\bm E^\prime\bm E }{n^{3/2}T}\r\Vert\,
\l\Vert\wh{\mbf H}\r\Vert=O_{\mathrm P}\l(\max\l(\frac 1n,\frac 1{\sqrt {nT}}\r)\r).
\end{align}
Similarly, because of Proposition \ref{prop:L}(a), Lemmas \ref{lem:FTLN}(i),  \ref{lem:LLN}(i),  \ref{lem:LLN}(iv), \ref{lem:MO1}(iv), and \ref{lem:HO1}(i), 
\begin{align}
\Vert \bm I_d\Vert &\le \l\Vert\l(\frac{\wh{\mbf M}^x}{n}\r)^{-1}\r\Vert\,\l\Vert \frac{\wh{\bm\Lambda}-\bm\Lambda\bm{\mathcal H}}{\sqrt n}\r\Vert \,\l\Vert \frac{\bm  E^\prime\bm F}{\sqrt nT}\r\Vert\, \l\Vert \frac{\bm\Lambda^\prime \bm\Lambda }{n}
\r\Vert\, \l\Vert \wh{\mbf H}\r\Vert=  O_{\mathrm P}\l(\max\l(\frac{1}{n\sqrt {T}},\frac 1{T}\r)\r), \label{eq:2aldd}\\ 
\Vert \bm I_e\Vert &\le \l\Vert\l(\frac{\wh{\mbf M}^x}{n}\r)^{-1}\r\Vert\,\l\Vert \frac{\wh{\bm\Lambda}-\bm\Lambda\bm{\mathcal H}}{\sqrt n}\r\Vert \,\l\Vert \frac{\bm  E^\prime\bm F\bm}{\sqrt nT}\r\Vert\, \l\Vert \frac{\bm\Lambda  }{\sqrt n}
\r\Vert^2\, \l\Vert \wh{\mbf H}\r\Vert=  O_{\mathrm P}\l(\max\l(\frac{1}{n\sqrt {T}},\frac 1{T}\r)\r),\label{eq:2aIde}
\end{align}
and, because of Proposition \ref{prop:L}(a), Lemmas \ref{lem:FTLN}(i), \ref{lem:LLN}(iii), \ref{lem:MO1}(iv), and \ref{lem:HO1}(i),
\begin{align}\label{eq:2aIf}
\Vert \bm I_f\Vert \le \l\Vert\l(\frac{\wh{\mbf M}^x}{n}\r)^{-1}\r\Vert\,\l\Vert \frac{\wh{\bm\Lambda}-\bm\Lambda\bm{\mathcal H}}{\sqrt n}\r\Vert \,\l\Vert \frac{\bm  E^\prime \bm E}{nT}\r\Vert\, \l\Vert \frac{\bm\Lambda  }{\sqrt n}
\r\Vert\, \l\Vert \wh{\mbf H}\r\Vert
=  O_{\mathrm P}\l(\max\l(\frac 1{n^2},\frac{1}{n\sqrt {T}},\frac 1{T}\r)\r).
\end{align}
By using \eqref{eq:2aIa}, \eqref{eq:2aIb}, \eqref{eq:2aIc}, \eqref{eq:2aldd}, \eqref{eq:2aIde}, and \eqref{eq:2aIf} into \eqref{eq:2a3}
\beq\label{eq:2a4}
\Vert\bm{I}\Vert = O_{\mathrm P}\l(\max\l(\frac 1 {n},\frac 1{\sqrt{nT}},\frac 1T\r)\r).
\eeq
Second, consider $\bm {II}$ in \eqref{eq:2a}. From Proposition \ref{prop:L2}(b),
\beq\label{eq:2a5}
\Vert\bm{II}\Vert  \le \frac 1 {n}\l\Vert \wh{\bm\Lambda}-{\bm\Lambda}\wh{\mbf H}\r\Vert^2 
= O_{\mathrm P}\l(\max\l(\frac{1}{n^2},\frac 1{T}\r)\r).
\eeq
And, by using \eqref{eq:2a4} and \eqref{eq:2a5} in \eqref{eq:2a}, because of Lemma \ref{lem:FTLN}(ii), and Lemma \ref{lem:HO1}(i), we prove part (a).

For part (b), from \eqref{eq:sviluppoLambda},
\begin{align}
\frac{\bm e_t^\prime (\wh{\bm\Lambda}-\bm\Lambda \wh{\mbf H}) }{n}&=
\frac{\bm e_t^\prime\bm\Lambda\bm F^\prime\bm  E{\bm\Lambda}}{n^2T}\bm{\mathcal H} \l(\frac{\wh{\mbf M}^x}{n}\r)^{-1}
+\frac{\bm e_t^\prime\bm E^\prime\bm F\bm\Lambda^\prime{\bm\Lambda}}{n^2T}\bm{\mathcal H} \l(\frac{\wh{\mbf M}^x}{n}\r)^{-1}
+\frac{\bm e_t^\prime\bm E^\prime\bm  E{\bm\Lambda}}{n^2T}\bm{\mathcal H} \l(\frac{\wh{\mbf M}^x}{n}\r)^{-1}\nn\\
&+\frac{\bm e_t^\prime\bm\Lambda\bm F^\prime\bm  E}{n^2T}(\wh{\bm\Lambda}-\bm\Lambda\bm{\mathcal H}) \l(\frac{\wh{\mbf M}^x}{n}\r)^{-1}
+\frac{\bm e_t^\prime\bm E^\prime\bm F\bm\Lambda^\prime}{n^2T}(\wh{\bm\Lambda}-\bm\Lambda\bm{\mathcal H}) \l(\frac{\wh{\mbf M}^x}{n}\r)^{-1}
+\frac{\bm e_t^\prime\bm E^\prime\bm  E}{n^2T}(\wh{\bm\Lambda}-\bm\Lambda\bm{\mathcal H}) \l(\frac{\wh{\mbf M}^x}{n}\r)^{-1}\nn\\
&= \bm {III}_a+\bm {III}_b+\bm {III}_c+\bm {III}_d+\bm {III}_e+\bm {III}_f, \;\text{say.} \label{eq:2b}
\end{align}
Then, because of Lemma \ref{lem:FTLN}(i), \ref{lem:FTLN}(iii), \ref{lem:MO1}(iv), and \ref{lem:HO1bis}(i),
and using \eqref{eq:2a3bis} in the proof of part (a)
\beq
\l\Vert \bm {III}_a\r\Vert \le \l\Vert\frac{\bm e_t}{\sqrt n}\r\Vert \,\l\Vert\frac{\bm\Lambda}{\sqrt n} \r\Vert\, \l\Vert \frac{\bm F^\prime\bm E\bm\Lambda}{nT}\r\Vert \,\l\Vert\bm{\mathcal H}\r\Vert\,\l\Vert \l(\frac{\wh{\mbf M}^x}{n}\r)^{-1}\r\Vert =O_{\mathrm P}\l(\frac 1{\sqrt {nT}}\r).\label{eq:IIIa}
\eeq
Now, by Assumptions  \ref{ass:common}(b), \ref{ass:idio}(c-ii), \ref{ass:ind}, and Lemma \ref{lem:Gxi}(i) 
\begin{align}
\E\l[\l\Vert \frac{\bm e_t^\prime\bm E^\prime \bm F}{nT}\r\Vert^2\r]&=\E\l[\l\Vert
\frac1{nT}\sum_{i=1}^n\sum_{s=1}^T  e_{it} e_{is}  \mbf F_{s}^\prime
\r\Vert^2\r]=\frac 1{n^2T^2} \sum_{k=1}^r\sum_{i_1=1}^n\sum_{i_2=1}^n \sum_{s_1=1}^T\sum_{s_2=1}^T\E\l[ e_{i_1t} e_{i_2t} e_{i_1s_1} e_{i_2s_2}F_{ks_1}F_{ks_2}\r]\nn\\
&\le\frac {r}{n^2T^2}\max_{k=1,\ldots, r}
\sum_{i_1=1}^n\sum_{i_2=1}^n \sum_{s_1=1}^T\sum_{s_2=1}^T\E\l[ e_{i_1t} e_{i_2t} e_{i_1s_1} e_{i_2s_2}\r]\,\E\l[F_{ks_1}F_{ks_2}\r]\nn\\
&\le r
\l\{\max_{k=1,\ldots, r}\max_{s_1,s_2=1,\ldots, T}\E\l[F_{ks_1}F_{ks_2}\r]\r\}\nn\\
&\cdot\l\{\frac {1}{n^2T^2}\sum_{i_1=1}^n\sum_{i_2=1}^n \sum_{s_1=1}^T\sum_{s_2=1}^T \l(\E\l[ e_{i_1t} e_{i_2t} e_{i_1s_1} e_{i_2s_2}\r]-\E\l[ e_{i_1t} e_{i_2t}\r]\,\E\l[ e_{i_1s_1} e_{i_2s_2}\r]
\r)\r.\nn\\
&\;\;\l.+\frac {1}{n^2T^2}\sum_{i_1=1}^n\sum_{i_2=1}^n \sum_{s_1=1}^T\sum_{s_2=1}^T
\E\l[ e_{i_1t} e_{i_2t}\r]\,\E\l[ e_{i_1s_1} e_{i_2s_2}\r]\r\}\nn\\
&\le  r\l\{\max_{k=1,\ldots, r} \bm\eta_k^\prime\bm\Gamma^F\bm\eta_k\r\}\nn\\
&\cdot \l\{\frac {1}{n^2T^2} \sum_{i_1=1}^n\sum_{i_2=1}^n\sum_{s_1=1}^T\sum_{s_2=1}^T \l\{\E\l[ e_{i_1t} e_{i_2t} e_{i_1s_1} e_{i_2s_2}\r]-\E\l[ e_{i_1t} e_{i_2t}\r]\,\E\l[ e_{i_1s_1} e_{i_2s_2}\r]
\r\}\r.\nn\\
&\;\;\l.+\max_{i_1,i_2=1,\ldots, n} \l\vert\E\l[ e_{i_1t} e_{i_2t}\r] \r\vert \frac {1}{n^2T^2}\sum_{i_1=1}^n\sum_{i_2=1}^n\sum_{s_1=1}^T\sum_{s_2=1}^T
\l\vert\E\l[ e_{i_1s_1} e_{i_2s_2}\r]\r\vert \r\}\nn\\
&\le \frac{rM_F K_ e}{nT}+\frac{rM_FM_ e M_{3 e}}{nT},\label{eq:IIIb1}
\end{align}
since $K_ e$,  $M_ e$, and $M_{3 e}$ are independent of $i_1$ and $i_2$, and  $M_F$ is independent of $s$, and where $\bm\eta_k$ is an $r$-dimensional vector with one in the $k$th entry and zero elsewhere. Notice that in \eqref{eq:IIIb1} we can use Assumption \ref{ass:idio}(c-ii) since 
\[
\sum_{i_1=1}^n\sum_{i_2=1}^n\sum_{s_1=1}^T\sum_{s_2=1}^T \l\{\E\l[ e_{i_1t} e_{i_2t} e_{i_1s_1} e_{i_2s_2}\r]-\E\l[ e_{i_1t} e_{i_2t}\r]\,\E\l[ e_{i_1s_1} e_{i_2s_2}\r]
\r\}=\E\l[\l(\sum_{i=1}^n\sum_{s=1}^T\l\{ e_{it} e_{is}-\E[ e_{it} e_{is}] \r\}\r)^2\r].
\]
So by Lemma \ref{lem:LLN}(iv), \ref{lem:MO1}(iv), \ref{lem:HO1bis}(i), and using \eqref{eq:IIIb1}, we get
\beq
\l\Vert \bm {III}_b \r\Vert \le  \l\Vert\frac{\bm e_t^\prime\bm E^\prime \bm F}{nT}\r\Vert\,\l\Vert \frac{\bm\Lambda^\prime\bm\Lambda}{n}\r\Vert\, \l\Vert\bm{\mathcal H}\r\Vert\,\l\Vert \l(\frac{\wh{\mbf M}^x}{n}\r)^{-1}\r\Vert=O_{\mathrm P}\l(\frac 1{\sqrt {nT}}\r).\label{eq:IIIb}
\eeq
Similarly, because of Lemma  \ref{lem:FTLN}(iii), \ref{lem:LLN}(iii), \ref{lem:MO1}(iv), and \ref{lem:HO1bis}(i),
\beq
\l\Vert \bm {III}_c \r\Vert \le\l\Vert \frac{\bm e_t}{\sqrt n}\r\Vert\,\l\Vert\frac{\bm E^\prime\bm E\bm\Lambda}{n^{3/2}T}\r\Vert
\l\Vert\bm{\mathcal H}\r\Vert\,\l\Vert \l(\frac{\wh{\mbf M}^x}{n}\r)^{-1}\r\Vert=O_{\mathrm P}\l(\max\l(\frac 1n,\frac 1{\sqrt {nT}}\r)\r).\label{eq:IIIc}
\eeq
Then, by Proposition \ref{prop:L}(a), Lemma \ref{lem:LLN}(i),
\ref{lem:MO1}(iv), and \ref{lem:HO1bis}(i),
and using  \eqref{eq:fatto2bis} in the proof of Proposition \ref{prop:L}
\begin{align}
\l\Vert \bm {III}_d \r\Vert \le
\l\Vert\frac{\bm e_t^\prime\bm\Lambda}{ n}\r\Vert\,
\l\Vert\frac{\bm F^\prime\bm E}{\sqrt n T}\r\Vert\,\l\Vert\frac{\wh{\bm\Lambda}-\bm\Lambda\bm{\mathcal H}}{\sqrt n}\r\Vert
\l\Vert\bm{\mathcal H}\r\Vert\,\l\Vert \l(\frac{\wh{\mbf M}^x}{n}\r)^{-1}\r\Vert = O_{\mathrm P}\l(\max\l(\frac1{n^{3/2}\sqrt T},\frac 1{\sqrt nT}\r)\r).\label{eq:IIId}
\end{align}
Similarly, by Proposition \ref{prop:L}(a), Lemma \ref{lem:FTLN}(i), \ref{lem:MO1}(iv), and \ref{lem:HO1bis}(i), and using \eqref{eq:IIIb1},
 \begin{align}
\l\Vert \bm {III}_e \r\Vert \le\l\Vert \frac{\bm e_t^\prime\bm E^\prime \bm F}{nT}\r\Vert\, \l\Vert\frac{\bm\Lambda}{\sqrt n}\r\Vert\, \l\Vert\frac{\wh{\bm\Lambda}-\bm\Lambda\bm{\mathcal H}}{\sqrt n}\r\Vert
\l\Vert\bm{\mathcal H}\r\Vert\,\l\Vert \l(\frac{\wh{\mbf M}^x}{n}\r)^{-1}\r\Vert= O_{\mathrm P}\l(\max\l(\frac1{n^{3/2}\sqrt T},\frac 1{\sqrt nT}\r)\r).\label{eq:IIIe}
\end{align}
And, finally, by Proposition \ref{prop:L}(a), Lemma \ref{lem:FTLN}(iii) , and Lemma  \ref{lem:LLN}(iii), \ref{lem:MO1}(iv), and \ref{lem:HO1bis}(i),
 \begin{align}
\l\Vert \bm {III}_f \r\Vert \le\l\Vert\frac{\bm e_t}{\sqrt n}\r\Vert\,\l\Vert\frac{\bm E^\prime\bm E}{nT}\r\Vert\, \l\Vert\frac{\wh{\bm\Lambda}-\bm\Lambda\bm{\mathcal H}}{\sqrt n}\r\Vert
\l\Vert\bm{\mathcal H}\r\Vert\,\l\Vert \l(\frac{\wh{\mbf M}^x}{n}\r)^{-1}\r\Vert= O_{\mathrm P}\l(\max\l(\frac 1{n^2},\frac1{n\sqrt T},\frac 1{T}\r)\r).\label{eq:IIIf}
\end{align}
Hence, by  using \eqref{eq:IIIa}, \eqref{eq:IIIb}, \eqref{eq:IIIc}, \eqref{eq:IIId}, \eqref{eq:IIIe}, and \eqref{eq:IIIf} in \eqref{eq:2b}, we prove part (b). 

Last, for part (c), by Lemma \ref{lem:HO1}(i), using  \eqref{eq:fatto2bis} in the proof of Proposition \ref{prop:L},
\begin{align}
\l \Vert \frac{\bm e_t^\prime\bm\Lambda\wh{\mbf H}}{n} \r\Vert\le 
\l\Vert \frac{\bm e_t^\prime\bm\Lambda}{n} \r\Vert\,
\l\Vert
\wh{\mbf H}
\r\Vert = O_{\mathrm P}\l(\frac 1{\sqrt n}\r).
\end{align}
Notice that this is a special case of Lemma \ref{lem:LLN}(ii). This completes the proof. $\Box$
\subsection{Proof of Proposition \ref{prop:F2}}\label{prop:F2proof}
Part (a) is proved by applying Lemma \ref{prop:factor}  to \eqref{eq:sviluppoFactor} and since $\l\Vert \l(\frac{\wh{\mbf M}^x}{n}\r)^{-1}\r\Vert = O_{\mathrm {P}}(1)$ because of Lemma \ref{lem:MO1}(iv).

For part (b), from \eqref{eq:FFF2} and using \eqref{eq:2a} in the proof of Lemma \ref{prop:factor}
\begin{align}
\l\Vert\frac{ \wh{\bm F}-{\bm F}(\wh{\mbf H}^{-1})^\prime}{\sqrt T}\r\Vert&\le\l\{ \l\Vert\frac{\bm F}{\sqrt T}\r\Vert\, \l\Vert\frac{(\bm\Lambda-\wh{\bm\Lambda}\wh{\mbf H}^{-1})^\prime\wh{\bm\Lambda}}{n}\r\Vert +
\l\Vert \frac{\bm  E(\wh{\bm\Lambda}-\bm\Lambda \wh{\mbf H})}{n\sqrt T}\r\Vert+
\l\Vert \frac{\bm E\bm\Lambda}{n\sqrt T}\r\Vert\,\l\Vert \wh{\mbf H}\r\Vert\r\}
\l\Vert\l(\frac {\wh{\mbf M}^x}n\r)^{-1}\r\Vert.\label{eq:prop5}
\end{align}
Then, from \eqref{eq:2a4} and \eqref{eq:2a5} in the proof of Lemma \ref{prop:factor},
\begin{align}
\l\Vert\frac{(\bm\Lambda-\wh{\bm\Lambda}\wh{\mbf H}^{-1})^\prime\wh{\bm\Lambda}}{n}\r\Vert 
&\le \l\Vert\frac{(\bm\Lambda-\wh{\bm\Lambda}\wh{\mbf H}^{-1})^\prime{\bm\Lambda}\wh{\mbf H}}{n}\r\Vert 
+\l\Vert\frac{(\bm\Lambda-\wh{\bm\Lambda}\wh{\mbf H}^{-1})^\prime(\wh{\bm\Lambda}-\bm\Lambda\wh{\mbf H})}{n}\r\Vert \nn\\
&= \Vert \bm I\Vert +\Vert \bm {II}\Vert
= O_{\mathrm P}\l(\max\l(\frac 1n,\frac 1{\sqrt {nT}},\frac 1 T\r)\r),\label{eq:UNOF}
\end{align}
where $\bm I$ and $\bm{II}$ are defined in \eqref{eq:2a} in the proof of Lemma \ref{prop:factor}. Moreover,  from \eqref{eq:2b} in the proof of Lemma \ref{prop:factor}
\begin{align}
 \frac{\bm  E(\wh{\bm\Lambda}-\bm\Lambda \wh{\mbf H})}{n\sqrt T}&=
 \frac{\bm  E \bm\Lambda\bm F^\prime\bm  E{\bm\Lambda}}{n^2T^{3/2}}\bm{\mathcal H} \l(\frac{\wh{\mbf M}^x}{n}\r)^{-1}
+\frac{\bm  E \bm E^\prime\bm F\bm\Lambda^\prime{\bm\Lambda}}{n^2T^{3/2}}\bm{\mathcal H} \l(\frac{\wh{\mbf M}^x}{n}\r)^{-1}
+\frac{\bm  E \bm E^\prime\bm  E{\bm\Lambda}}{n^2T^{3/2}}\bm{\mathcal H} \l(\frac{\wh{\mbf M}^x}{n}\r)^{-1}\nn\\
&+\frac{\bm  E\bm\Lambda\bm F^\prime\bm  E}{n^2T^{3/2}}(\wh{\bm\Lambda}-\bm\Lambda\bm{\mathcal H}) \l(\frac{\wh{\mbf M}^x}{n}\r)^{-1}
+\frac{\bm  E\bm E^\prime\bm F\bm\Lambda^\prime}{n^2T^{3/2}}(\wh{\bm\Lambda}-\bm\Lambda\bm{\mathcal H}) \l(\frac{\wh{\mbf M}^x}{n}\r)^{-1}
+\frac{\bm  E\bm E^\prime\bm  E}{n^2T^{3/2}}(\wh{\bm\Lambda}-\bm\Lambda\bm{\mathcal H}) \l(\frac{\wh{\mbf M}^x}{n}\r)^{-1}\nn\\
&= \bm {IV}_a+\bm {IV}_b+\bm {IV}_c+\bm {IV}_d+\bm {IV}_e+\bm {IV}_f, \;\text{say.} \label{eq:DUEF}
\end{align}
Then, because of Lemma \ref{lem:FTLN}(i), \ref{lem:FTLN}(iii), and Lemma \ref{lem:MO1}(iv), \ref{lem:HO1bis}(i),
and using \eqref{eq:2a3bis} in the proof of Lemma~\ref{prop:factor}
\beq
\l\Vert \bm {IV}_a\r\Vert \le \l\Vert\frac{\bm E}{\sqrt {nT}}\r\Vert \,\l\Vert\frac{\bm\Lambda}{\sqrt n} \r\Vert\, \l\Vert \frac{\bm F^\prime\bm E\bm\Lambda}{nT}\r\Vert \,\l\Vert\bm{\mathcal H}\r\Vert\,\l\Vert \l(\frac{\wh{\mbf M}^x}{n}\r)^{-1}\r\Vert =O_{\mathrm P}\l(\frac 1{\sqrt {nT}}\r).\label{eq:IVa}
\eeq
Now, by Assumptions \ref{ass:common}(b), \ref{ass:idio}(c-ii), \ref{ass:ind}, and Lemma \ref{lem:Gxi}(i) 
(see also \eqref{eq:IIIb1} in the proof of Lemma \ref{prop:factor}),
letting $\bm\varepsilon_{i}=( e_{i1}\cdots  e_{iT})^\prime$, 
\begin{align}
\E\l[\l\Vert \frac{\bm E\bm E^\prime \bm F}{nT^{3/2}}\r\Vert^2\r]&=\E\l[\l\Vert
\frac1{nT^{3/2}}\sum_{i=1}^n\sum_{s=1}^T \bm\varepsilon_{i} e_{is}  \mbf F_{s}^\prime
\r\Vert^2\r]=\frac 1{n^2T^3} \sum_{t=1}^T\sum_{k=1}^r\sum_{i_1=1}^n\sum_{i_2=1}^n \sum_{s_1=1}^T\sum_{s_2=1}^T\E\l[ e_{i_1t} e_{i_2t} e_{i_1s_1} e_{i_2s_2}F_{ks_1}F_{ks_2}\r]\nn\\
&\le\frac {r}{n^2T^2}\max_{k=1,\ldots, r}\max_{t=1,\ldots, T}
\sum_{i_1=1}^n\sum_{i_2=1}^n \sum_{s_1=1}^T\sum_{s_2=1}^T\E\l[ e_{i_1t} e_{i_2t} e_{i_1s_1} e_{i_2s_2}\r]\,\E\l[F_{ks_1}F_{ks_2}\r]\nn\\
&\le r
\l\{\max_{k=1,\ldots, r}\max_{s_1,s_2=1,\ldots, T}\E\l[F_{ks_1}F_{ks_2}\r]\r\}\nn\\
&\cdot\l\{\frac {1}{n^2T^2}\max_{t=1,\ldots, T}\sum_{i_1=1}^n\sum_{i_2=1}^n \sum_{s_1=1}^T\sum_{s_2=1}^T \l(\E\l[ e_{i_1t} e_{i_2t} e_{i_1s_1} e_{i_2s_2}\r]-\E\l[ e_{i_1t} e_{i_2t}\r]\,\E\l[ e_{i_1s_1} e_{i_2s_2}\r]
\r)\r.\nn\\
&\;\;\l.+\frac {1}{n^2T^2}\max_{t=1,\ldots, T}\sum_{i_1=1}^n\sum_{i_2=1}^n \sum_{s_1=1}^T\sum_{s_2=1}^T
\E\l[ e_{i_1t} e_{i_2t}\r]\,\E\l[ e_{i_1s_1} e_{i_2s_2}\r]\r\}\nn\\
&\le  r\l\{\max_{k=1,\ldots, r} \bm\eta_k^\prime\bm\Gamma^F\bm\eta_k\r\}\nn\\
&\cdot \l\{\frac {1}{n^2T^2}\max_{t=1,\ldots, T} \sum_{i_1=1}^n\sum_{i_2=1}^n\sum_{s_1=1}^T\sum_{s_2=1}^T \l\{\E\l[ e_{i_1t} e_{i_2t} e_{i_1s_1} e_{i_2s_2}\r]-\E\l[ e_{i_1t} e_{i_2t}\r]\,\E\l[ e_{i_1s_1} e_{i_2s_2}\r]
\r\}\r.\nn\\
&\;\;\l.+\max_{i_1,i_2=1,\ldots, n}\max_{t=1,\ldots, T} \l\vert\E\l[ e_{i_1t} e_{i_2t}\r] \r\vert \frac {1}{n^2T^2}\sum_{i_1=1}^n\sum_{i_2=1}^n\sum_{s_1=1}^T\sum_{s_2=1}^T
\l\vert\E\l[ e_{i_1s_1} e_{i_2s_2}\r]\r\vert \r\}\nn\\
&\le \frac{rM_F K_ e}{nT}+\frac{rM_FM_ e M_{3 e}}{nT},\label{eq:IVb1}
\end{align}
since $K_ e$,  $M_ e$, and $M_{3 e}$ are independent of $i_1$, $i_2$ and $t$, and  $M_F$ is independent of $s$, and where $\bm\eta_k$ is an $r$-dimensional vector with one in the $k$th entry and zero elsewhere. 
So by Lemma \ref{lem:LLN}(iv), \ref{lem:MO1}(iv), \ref{lem:HO1bis}(i), and using \eqref{eq:IVb1} , we get
\beq
\l\Vert \bm {IV}_b \r\Vert \le  \l\Vert\frac{\bm E\bm E^\prime \bm F}{nT^{3/2}}\r\Vert\,\l\Vert \frac{\bm\Lambda^\prime\bm\Lambda}{n}\r\Vert\, \l\Vert\bm{\mathcal H}\r\Vert\,\l\Vert \l(\frac{\wh{\mbf M}^x}{n}\r)^{-1}\r\Vert=O_{\mathrm P}\l(\frac 1{\sqrt {nT}}\r).\label{eq:IVb}
\eeq
Similarly, because of Lemma  \ref{lem:FTLN}(i), \ref{lem:FTLN}(iii), \ref{lem:LLN}(iii), \ref{lem:MO1}(iv), and \ref{lem:HO1bis}(i),
\beq
\l\Vert \bm {IV}_c \r\Vert \le\l\Vert \frac{\bm E}{\sqrt {nT}}\r\Vert\,\l\Vert\frac{\bm E^\prime\bm E\bm\Lambda}{n^{3/2}T}\r\Vert
\,\l\Vert\bm{\mathcal H}\r\Vert\,\l\Vert \l(\frac{\wh{\mbf M}^x}{n}\r)^{-1}\r\Vert=O_{\mathrm P}\l(\max\l(\frac 1n,\frac 1{\sqrt {nT}}\r)\r).\label{eq:IVc}
\eeq
Moreover, by Proposition \ref{prop:L}(a), Lemma \ref{lem:LLN}(i), \ref{lem:LLN}(ii), and Lemma \ref{lem:MO1}(iv) and \ref{lem:HO1bis}(i)
\begin{align}
\l\Vert \bm {IV}_d \r\Vert \le
\l\Vert\frac{\bm E\bm\Lambda}{\sqrt {nT}}\r\Vert\,
\l\Vert\frac{\bm F^\prime\bm E}{\sqrt n T}\r\Vert\,\l\Vert\frac{\wh{\bm\Lambda}-\bm\Lambda\bm{\mathcal H}}{\sqrt n}\r\Vert
\l\Vert\bm{\mathcal H}\r\Vert\,\l\Vert \l(\frac{\wh{\mbf M}^x}{n}\r)^{-1}\r\Vert = O_{\mathrm P}\l(\max\l(\frac1{n\sqrt T},\frac 1T\r)\r).\label{eq:IVd}
\end{align}
Similarly, by Proposition \ref{prop:L}(a), Lemma \ref{lem:FTLN}(i), \ref{lem:MO1}(iv), and \ref{lem:HO1bis}(i), and using \eqref{eq:IVb1},
 \begin{align}
\l\Vert \bm {IV}_e \r\Vert \le\l\Vert \frac{\bm E\bm E^\prime \bm F}{nT^{3/2}}\r\Vert\, \l\Vert\frac{\bm\Lambda}{\sqrt n}\r\Vert\, \l\Vert\frac{\wh{\bm\Lambda}-\bm\Lambda\bm{\mathcal H}}{\sqrt n}\r\Vert
\l\Vert\bm{\mathcal H}\r\Vert\,\l\Vert \l(\frac{\wh{\mbf M}^x}{n}\r)^{-1}\r\Vert= O_{\mathrm P}\l(\max\l(\frac1{n^{3/2}\sqrt T},\frac 1{\sqrt nT}\r)\r).\label{eq:IVe}
\end{align}
And, finally, by Proposition \ref{prop:L}(a), Lemma \ref{lem:FTLN}(iii) , and Lemma  \ref{lem:LLN}(iii), \ref{lem:MO1}(iv), and \ref{lem:HO1bis}(i),
 \begin{align}
\l\Vert \bm {IV}_f \r\Vert \le\l\Vert\frac{\bm E}{\sqrt {nT}}\r\Vert\,\l\Vert\frac{\bm E^\prime\bm E}{nT}\r\Vert\, \l\Vert\frac{\wh{\bm\Lambda}-\bm\Lambda\bm{\mathcal H}}{\sqrt n}\r\Vert
\l\Vert\bm{\mathcal H}\r\Vert\,\l\Vert \l(\frac{\wh{\mbf M}^x}{n}\r)^{-1}\r\Vert= O_{\mathrm P}\l(\max\l(\frac 1{n^2},\frac1{n\sqrt T},\frac 1{T}\r)\r).\label{eq:IVf}
\end{align}
By using \eqref{eq:IVa}, \eqref{eq:IVb}, \eqref{eq:IVc}, \eqref{eq:IVd}, \eqref{eq:IVe}, and \eqref{eq:IVf}
in \eqref{eq:DUEF} we get
\beq
\l\Vert \frac{\bm  E(\wh{\bm\Lambda}-\bm\Lambda \wh{\mbf H})}{n\sqrt T}\r\Vert = O_{\mathrm P}\l(\max\l(\frac 1{n},\frac 1{\sqrt{nT}},\frac 1T\r)\r).\label{eq:TREF} 
\eeq
The proof of part (b) follows from Lemma \ref{lem:FTLN}(ii), \ref{lem:LLN}(ii), \ref{lem:HO1}(i), and \ref{lem:MO1}(iv), and using \eqref{eq:UNOF} and \eqref{eq:TREF} in \eqref{eq:prop5}.
This completes the proof. $\Box$

\subsection{Proof of Proposition \ref{prop:sumF}}\label{prop:sumFproof}
From \eqref{eq:sviluppoFactor} and the $C_r$-inequality with $r=2$, 
\begin{align}
\frac 1T&\sum_{t=1}^T \l\Vert \wh{\mbf F}_t-\wh{\mbf H}^{-1}{\mbf F}_t  \r\Vert^2 \label{eq:aue}\\
&\le 3  \l\{\frac 1T\sum_{t=1}^T \l\Vert
\mbf F_t^\prime\frac{(\bm\Lambda-\wh{\bm\Lambda}\wh{\mbf H}^{-1})^\prime\wh{\bm\Lambda}}{n}\r\Vert^2+
 \frac 1T\sum_{t=1}^T\l\Vert \bm e_t^\prime \frac{(\wh{\bm\Lambda}-\bm\Lambda \wh{\mbf H})}{n}\r\Vert^2+
 \frac 1T\sum_{t=1}^T\l\Vert \bm e_t^\prime \frac{\bm\Lambda\wh{\mbf H}}{n}\r\Vert^2 \r\}
\l\Vert\l(\frac {\wh{\mbf M}^x}n\r)^{-1}\r\Vert^2.\nn
\end{align}
Consider the first term on the rhs of \eqref{eq:aue}
\begin{align}
\frac 1T\sum_{t=1}^T \l\Vert
\mbf F_t^\prime\frac{(\bm\Lambda-\wh{\bm\Lambda}\wh{\mbf H}^{-1})^\prime\wh{\bm\Lambda}}{n}\r\Vert^2&\le
\frac 1T\sum_{t=1}^T \l\Vert
\mbf F_t\r\Vert^2\,\l\Vert \frac{(\bm\Lambda-\wh{\bm\Lambda}\wh{\mbf H}^{-1})^\prime\wh{\bm\Lambda}}{n}\r\Vert^2\nn\\
&= \l\Vert \frac{\bm F}{\sqrt T}\r\Vert^2 \,\l\Vert \frac{(\bm\Lambda-\wh{\bm\Lambda}\wh{\mbf H}^{-1})^\prime\wh{\bm\Lambda}}{n}\r\Vert^2\nn\\
&\le \l\Vert \frac{\bm F}{\sqrt T}\r\Vert^2  2\l\{ \Vert \bm I\Vert^2 +\Vert\bm {II}\Vert^2\r\}\nn\\
&=  O_{\mathrm P}\l(\max\l(\frac 1 {n^2},\frac 1{{nT}},\frac 1{T^2}\r)\r),\label{eq:aue1}
\end{align}
where $\bm I$ and $\bm{II}$ are defined in \eqref{eq:2a} and we used the $C_r$-inequality with $r=2$, \eqref{eq:2a4} and \eqref{eq:2a5} in the proof of Lemma \ref{prop:factor} and Lemma \ref{lem:FTLN}(ii).

Then, consider the second term on the rhs of \eqref{eq:aue}. From \eqref{eq:2b} in the proof of Lemma \ref{prop:factor} and the $C_r$-inequality with $r=2$
 \begin{align}
  \frac 1T\sum_{t=1}^T\l\Vert \bm e_t^\prime \frac{(\wh{\bm\Lambda}-\bm\Lambda \wh{\mbf H})}{n}\r\Vert^2&\le  6 \frac 1T\sum_{t=1}^T\l\{ \Vert \bm {III}_a\Vert^2+\Vert\bm {III}_b\Vert^2+\Vert\bm {III}_c\Vert^2+\Vert\bm {III}_d\Vert^2+\Vert\bm {III}_e\Vert^2+\Vert\bm {III}_f\Vert^2\r\}.\label{eq:aue2}
\end{align}
Then, from \eqref{eq:IIIa} in the proof of Lemma \ref{prop:factor}
 \begin{align}
  \frac 1T\sum_{t=1}^T\l\Vert \bm {III}_a\r\Vert^2& \le   \frac 1T\sum_{t=1}^T\l\Vert\frac{\bm e_t}{\sqrt n}\r\Vert^2 \,\l\Vert\frac{\bm\Lambda}{\sqrt n} \r\Vert^2\, \l\Vert \frac{\bm F^\prime\bm E\bm\Lambda}{nT}\r\Vert^2 \,\l\Vert\bm{\mathcal H}\r\Vert^2\,\l\Vert \l(\frac{\wh{\mbf M}^x}{n}\r)^{-1}\r\Vert^2\nn\\
  &=  \l \Vert \frac{\bm  E} {\sqrt {nT}}\r\Vert^2  \,\l\Vert\frac{\bm\Lambda}{\sqrt n} \r\Vert^2\, \l\Vert \frac{\bm F^\prime\bm E\bm\Lambda}{nT}\r\Vert^2 \,\l\Vert\bm{\mathcal H}\r\Vert^2\,\l\Vert \l(\frac{\wh{\mbf M}^x}{n}\r)^{-1}\r\Vert^2=O_{\mathrm P}\l(\frac 1{nT}\r),\label{eq:aue2a}
\end{align}
 because of Lemma \ref{lem:FTLN}(i), \ref{lem:FTLN}(iii), \ref{lem:MO1}(iv), \ref{lem:HO1bis}(i),
and using \eqref{eq:2a3bis} in the proof of Lemma \ref{prop:factor} joint with Markov's inequality.
Similarly, from \eqref{eq:IIIb} in the proof of Lemma \ref{prop:factor}
 \begin{align}
  \frac 1T\sum_{t=1}^T\l\Vert \bm {III}_b \r\Vert^2 &\le   \frac 1T\sum_{t=1}^T\l\Vert \frac{\bm e_t}{\sqrt n}\r\Vert^2\, \l\Vert\frac{\bm E^\prime \bm F}{\sqrt nT}\r\Vert^2\,\l\Vert \frac{\bm\Lambda^\prime\bm\Lambda}{n}\r\Vert^2\, \l\Vert\bm{\mathcal H}\r\Vert^2\,\l\Vert \l(\frac{\wh{\mbf M}^x}{n}\r)^{-1}\r\Vert^2\nn\\
  &= \l \Vert \frac{\bm  E} {\sqrt {nT}}\r\Vert^2\, \l\Vert\frac{\bm E^\prime \bm F}{\sqrt nT}\r\Vert^2\,\l\Vert \frac{\bm\Lambda^\prime\bm\Lambda}{n}\r\Vert^2\, \l\Vert\bm{\mathcal H}\r\Vert^2\,\l\Vert \l(\frac{\wh{\mbf M}^x}{n}\r)^{-1}\r\Vert^2
  =O_{\mathrm P}\l(\frac 1{T}\r),\label{eq:aue2b}
\end{align}
 because of Lemma  \ref{lem:FTLN}(iii),  \ref{lem:LLN}(i), \ref{lem:LLN}(iv), \ref{lem:MO1}(iv), \ref{lem:HO1bis}(i). And  from \eqref{eq:IIIc} in the proof of Lemma \ref{prop:factor}
 \beq
  \frac 1T\sum_{t=1}^T\l\Vert \bm {III}_c \r\Vert^2 \le \l\Vert \frac{\bm E}{\sqrt {nT}}\r\Vert^2\,\l\Vert\frac{\bm E^\prime\bm E\bm\Lambda}{n^{3/2}T}\r\Vert^2
\l\Vert\bm{\mathcal H}\r\Vert^2\,\l\Vert \l(\frac{\wh{\mbf M}^x}{n}\r)^{-1}\r\Vert^2=O_{\mathrm P}\l(\max\l(\frac 1{n^2},\frac 1{ {nT}}\r)\r),\label{eq:aue2c}
\eeq
because of Lemma  \ref{lem:FTLN}(iii), \ref{lem:LLN}(iii), \ref{lem:MO1}(iv), and \ref{lem:HO1bis}(i). Following similar reasoning from \eqref{eq:IIId} and \eqref{eq:IIIe} in the proof of Lemma \ref{prop:factor},
\begin{align}
  \frac 1T\sum_{t=1}^T\l\Vert \bm {III}_d \r\Vert^2& \le
  \l\Vert \frac{\bm E}{\sqrt {nT}}\r\Vert^2\,
\l\Vert\frac{\bm\Lambda}{\sqrt  n}\r\Vert^2\,
\l\Vert\frac{\bm F^\prime\bm E}{\sqrt n T}\r\Vert^2\,\l\Vert\frac{\wh{\bm\Lambda}-\bm\Lambda\bm{\mathcal H}}{\sqrt n}\r\Vert^2
\l\Vert\bm{\mathcal H}\r\Vert^2\,\l\Vert \l(\frac{\wh{\mbf M}^x}{n}\r)^{-1}\r\Vert^2\nn\\
& = O_{\mathrm P}\l(\max\l(\frac1{n^2 T},\frac 1{T^2}\r)\r), \label{eq:aue2d}
\end{align}
and also 
 \begin{align}
  \frac 1T\sum_{t=1}^T\l\Vert \bm {III}_e \r\Vert^2 &\le
  \l\Vert \frac{\bm E}{\sqrt{nT}}\r\Vert^2\,
  \l\Vert \frac{\bm E^\prime \bm F}{\sqrt nT}\r\Vert^2\, \l\Vert\frac{\bm\Lambda}{\sqrt n}\r\Vert^2\, \l\Vert\frac{\wh{\bm\Lambda}-\bm\Lambda\bm{\mathcal H}}{\sqrt n}\r\Vert^2
\l\Vert\bm{\mathcal H}\r\Vert^2\,\l\Vert \l(\frac{\wh{\mbf M}^x}{n}\r)^{-1}\r\Vert^2\nn\\
& = O_{\mathrm P}\l(\max\l(\frac1{n^2 T},\frac 1{T^2}\r)\r), \label{eq:aue2e}
\end{align}
because of Proposition \ref{prop:L}(a) and Lemma \ref{lem:FTLN}(i), \ref{lem:FTLN}(iii), \ref{lem:LLN}(i), \ref{lem:MO1}(iv), and \ref{lem:HO1bis}(i). 
And, finally, from \eqref{eq:IIIf} in the proof of Lemma \ref{prop:factor}, 
 \begin{align}
  \frac 1T\sum_{t=1}^T\l\Vert \bm {III}_f \r\Vert^2& \le
  \l\Vert\frac{\bm E}{\sqrt {nT}}\r\Vert^2\,\l\Vert\frac{\bm E^\prime\bm E}{nT}\r\Vert^2\, \l\Vert\frac{\wh{\bm\Lambda}-\bm\Lambda\bm{\mathcal H}}{\sqrt n}\r\Vert^2
\l\Vert\bm{\mathcal H}\r\Vert^2\,\l\Vert \l(\frac{\wh{\mbf M}^x}{n}\r)^{-1}\r\Vert^2\nn\\
&= O_{\mathrm P}\l(\max\l(\frac 1{n^4},\frac1{n^2 T},\frac 1{T^2}\r)\r),\label{eq:aue2f}
\end{align}
because of  Proposition \ref{prop:L}(a), Lemma \ref{lem:FTLN}(iii), \ref{lem:LLN}(iii), \ref{lem:MO1}(iv), and \ref{lem:HO1bis}(i). By substituting 
\eqref{eq:aue2a},
\eqref{eq:aue2b},
\eqref{eq:aue2c},
\eqref{eq:aue2d},
\eqref{eq:aue2e}, and
\eqref{eq:aue2f} into \eqref{eq:aue2}
\beq
  \frac 1T\sum_{t=1}^T\l\Vert \bm e_t^\prime \frac{(\wh{\bm\Lambda}-\bm\Lambda \wh{\mbf H})}{n}\r\Vert^2= O_{\mathrm P}\l(\max\l(\frac 1{n^2},\frac 1T\r)\r).\label{eq:aue2F}
\eeq

Last, consider the third term on the rhs of \eqref{eq:aue}. We have
\beq
 \frac 1T\sum_{t=1}^T\l\Vert \bm e_t^\prime \frac{\bm\Lambda\wh{\mbf H}}{n}\r\Vert^2 \le \l\Vert  \frac{\bm E\bm\Lambda}{n\sqrt T}\r\Vert^2\,\Vert\wh{\mbf H}\Vert^2 = O_{\mathrm P}\l(\frac 1n\r),\label{eq:aue3}
\eeq
because of Lemma \ref{lem:LLN}(ii) and \ref{lem:HO1}(i).
By substituting \eqref{eq:aue1}, \eqref{eq:aue2F}, and \eqref{eq:aue3} into \eqref{eq:aue} and using Lemma \ref{lem:MO1}(iv), we complete the proof. $\Box$
\subsection{Proof of Proposition \ref{prop:LLFF}}\label{prop:LLFFproof}
For part (a), 
\begin{align}
\l\Vert  \wh{\mbf H} - \l(\frac{\bm\Lambda^\prime\bm\Lambda}n\r)^{-1}
\frac{{\bm\Lambda}^\prime\wh{\bm\Lambda}}n
\r\Vert
&\le 
\l\Vert  \l(\frac{\bm\Lambda^\prime\bm\Lambda}n\r)^{-1}\r\Vert\, \l\Vert  \frac{\bm\Lambda^\prime\bm\Lambda \wh{\mbf H}}n  -
\frac{{\bm\Lambda}^\prime\wh{\bm\Lambda}}n
\r\Vert= \l\Vert  \l(\frac{\bm\Lambda^\prime\bm\Lambda}n\r)^{-1}\r\Vert\, \l\Vert \frac {{\bm\Lambda}^\prime(\wh{\bm\Lambda}-\bm\Lambda \wh{\mbf H})}{n}\r\Vert\nn\\
&=  O_{\mathrm P}\l(\max\l(\frac 1n,\frac 1{\sqrt{nT}},\frac 1T\r)\r),\nn
\end{align}
because of Lemma \ref{lem:ranghi} and \eqref{eq:2a3} and \eqref{eq:2a4} in the proof of Lemma \ref{prop:factor}, where we can drop $\wh{\mbf H}$ since $\Vert \wh{\mbf H}\Vert=O_{\mathrm P}(1)$ by Lemma \ref{lem:HO1}(i). This proves part (a). 

For part (b), 
\begin{align}
 \l\Vert\wh{\mbf H}^{-1}-\frac{\wh{\bm F}^\prime{\bm F}}T\l(\frac{\bm F^\prime\bm F}T\r)^{-1}\r\Vert&\le
 \l\Vert\frac{\wh{\mbf H}^{-1}\bm F^\prime\bm F}T-\frac{\wh{\bm F}^\prime{\bm F}}T\r\Vert \,\l\Vert \l(\frac{\bm F^\prime\bm F}T\r)^{-1}\r\Vert\nn\\
 &= \l\Vert \frac{(\wh{\bm F}-\bm F(\wh{\mbf H}^{-1})^\prime)^\prime {\bm F}^\prime}{T}\r\Vert
 \,\l\Vert \l(\frac{\bm F^\prime\bm F}T\r)^{-1}\r\Vert.\label{eq:terronski2}
\end{align}
Then, from \eqref{eq:FFF2},
\begin{align}
\l\Vert\frac{\bm F^\prime(\wh{\bm F}-\bm F (\wh{\mbf H}^{-1})^\prime) }{T}\r\Vert
&=
\l\{\l\Vert
\frac{\bm F^\prime\bm F(\bm\Lambda-\wh{\bm\Lambda}\wh{\mbf H}^{-1})^\prime\wh{\bm\Lambda}}{nT}\r\Vert+
\l\Vert\frac{\bm F^\prime\bm  E(\wh{\bm\Lambda}-\bm\Lambda \wh{\mbf H})}{nT}\r\Vert+
\l\Vert\frac{\bm F^\prime\bm E\bm\Lambda\wh{\mbf H}}{nT}
\r\Vert\r\}
\l\Vert\l(\frac {\wh{\mbf M}^x}n\r)^{-1}\r\Vert\nn\\
&=\l\{ \Vert \bm{\mathcal I}\Vert+ \Vert \bm{\mathcal {II}}\Vert+ \Vert \bm{\mathcal {III}}\Vert\r\} \l\Vert\l(\frac {\wh{\mbf M}^x}n\r)^{-1}\r\Vert, \;\text{say.}\label{eq:sayyou}
\end{align}
First, consider $\bm{\mathcal I}$ in \eqref{eq:sayyou}. We have
\begin{align}
\Vert \bm{\mathcal I}\Vert&\le 
\l\Vert
\frac{\bm F^\prime\bm F}T\r\Vert\,\l\Vert\frac{(\bm\Lambda-\wh{\bm\Lambda}\wh{\mbf H}^{-1})^\prime\wh{\bm\Lambda}}{n}\r\Vert\le\l\Vert
\frac{\bm F^\prime\bm F}T\r\Vert\l\{
\l\Vert\frac{(\bm\Lambda-\wh{\bm\Lambda}\wh{\mbf H}^{-1})^\prime{\bm\Lambda}\wh{\mbf H}}{n}\r\Vert\,
+\l\Vert\frac{(\bm\Lambda-\wh{\bm\Lambda}\wh{\mbf H}^{-1})^\prime(\wh{\bm\Lambda}-\bm\Lambda\wh{\mbf H})}{n}\r\Vert\r\}
\nn\\
&=\l\Vert
\frac{\bm F^\prime\bm F}T\r\Vert\l\{\Vert \bm I \Vert  +\Vert  \bm{II}\Vert \r\},\nn
\end{align}
where $\bm I$ and $\bm {II}$ are defined in \eqref{eq:2a}  in the proof of Lemma \ref{prop:factor}. Then, because of \eqref{eq:2a4} and \eqref{eq:2a5} in the proof of Lemma \ref{prop:factor} and Lemma \ref{lem:LLN}(v), 
\beq
\Vert \bm{\mathcal I}\Vert = O_{\mathrm P}\l(\max\l(\frac 1n,\frac 1{\sqrt{nT}},\frac 1T\r)\r).\label{eq:sayme1}
\eeq
Second, consider $\bm{\mathcal{II}}$ in \eqref{eq:sayyou}. We have
\begin{align}
\Vert \bm{\mathcal {II}}\Vert&\le \l\Vert\frac{\bm F^\prime\bm  E}{\sqrt n T}\r\Vert\,\l\Vert\frac{(\wh{\bm\Lambda}-\bm\Lambda \wh{\mbf H})}{\sqrt n}\r\Vert =O_{\mathrm P}\l(\max\l(\frac1{n\sqrt T},\frac1{\sqrt n T}, \frac 1{T} \r)\r),\label{eq:sayme2}
\end{align}
because of Lemma \ref{lem:LLN}(i) and Proposition \ref{prop:L2}(b).
Last, consider $\bm{\mathcal{III}}$ in \eqref{eq:sayyou}. We have
\begin{align}\label{eq:sayme3}
\Vert \bm{\mathcal {III}}\Vert&\le \l\Vert\frac{\bm F^\prime\bm E\bm\Lambda}{nT}
\r\Vert\,\Vert \wh{\mbf H}\Vert = O_{\mathrm P}\l(\frac 1{\sqrt{nT}}\r),
\end{align}
because of \eqref{eq:1a} in the proof of Lemma \ref{prop:load} and Lemma \ref{lem:HO1}(i). By substituting \eqref{eq:sayme1}, \eqref{eq:sayme2}, and \eqref{eq:sayme3} in \eqref{eq:sayyou}, because of Lemma \ref{lem:MO1}(iv), 
\beq\label{eq:terronski1}
\l\Vert \frac{(\wh{\bm F}-\bm F(\wh{\mbf H}^{-1})^\prime)^\prime {\bm F}^\prime}{T}\r\Vert= O_{\mathrm P}\l(\max\l(\frac 1n,\frac 1{\sqrt{nT}},\frac 1T\r)\r).
\eeq
Hence, by substituting \eqref{eq:terronski1} into \eqref{eq:terronski2}, because of Lemma \ref{lem:FFinv}, we prove
part (b).

For part (c), using the definition of $\wh{\bm\Lambda}$ in \eqref{eq:estL} which implies $\wh{\bm\Lambda}^\prime\wh{\bm\Lambda}=\wh{\mbf M}^x$, 
\begin{align}
\l\Vert  \wh{\mbf H}^{-1} - \l(\frac{\wh{\bm\Lambda}^\prime\wh{\bm\Lambda}}n\r)^{-1}
\frac{\wh{\bm\Lambda}^\prime{\bm\Lambda}}n
\r\Vert
&\le 
\l\Vert \l(\frac{\wh{\bm\Lambda}^\prime\wh{\bm\Lambda}}n\r)^{-1} \r\Vert\,
\l\Vert \frac{\wh{\bm\Lambda}^\prime\wh{\bm\Lambda} \wh{\mbf H}^{-1} }n - \frac{\wh{\bm\Lambda}^\prime{\bm\Lambda}}n\r\Vert= 
\l\Vert \l(\frac{\wh{\mbf M}^x}n\r)^{-1} \r\Vert\,
\l\Vert \frac{\wh{\bm\Lambda}^\prime(\wh{\bm\Lambda} \wh{\mbf H}^{-1}-{\bm\Lambda})}n\r\Vert\nn\\
&\le 
\l\Vert \l(\frac{\wh{\mbf M}^x}n\r)^{-1} \r\Vert\,
\l\Vert \frac{\wh{\bm\Lambda}^\prime(\wh{\bm\Lambda} -{\bm\Lambda}\wh{\mbf H})}n\r\Vert\,\l\Vert \wh{\mbf H}^{-1}\r\Vert \nn\\
&= O_{\mathrm P}\l(\max\l(\frac 1n,\frac 1{\sqrt {nT}},\frac 1 T\r)\r),\nn
\end{align}
because of Lemmas \ref{lem:MO1}(iv) and \ref{lem:HO1}(ii), 
and \eqref{eq:UNOF} in the proof of Proposition \ref{prop:F2}. This proves part (c).

For part (d), using the definition of $\wh{\bm F}$ in \eqref{eq:estF} which implies $T^{-1}\wh{\bm F}^\prime\wh{\bm F}= \mbf I_r$, 
\begin{align}
\l\Vert \wh{\mbf H} -  \frac{{\bm F}^\prime\wh{\bm F}}T\l(\frac{\wh{\bm F}^\prime\wh{\bm F}}T\r)^{-1}\r\Vert
&\le \l\Vert \l(\frac{\wh{\bm F}^\prime\wh{\bm F}}T\r)^{-1}\r\Vert\,
\l\Vert \frac{ \wh{\mbf H} \wh{\bm F}^\prime\wh{\bm F}}T- \frac{{\bm F}^\prime\wh{\bm F}}T\r\Vert= \l\Vert \frac{ ( \wh{\bm F} \wh{\mbf H}^\prime- {\bm F})^\prime \wh{\bm F}}T\r\Vert\nn\\
&\le  \l\Vert \frac{ ( \wh{\bm F} - {\bm F}(\wh{\mbf H}^{-1})^\prime)^\prime \wh{\bm F}}T\r\Vert\, \l\Vert\wh{\mbf H}\r\Vert\nn\\
&\le 
\l\{
\l\Vert \frac{ ( \wh{\bm F} - {\bm F}(\wh{\mbf H}^{-1})^\prime)^\prime {\bm F}}T\r\Vert
+\l\Vert \frac{ ( \wh{\bm F} - {\bm F}(\wh{\mbf H}^{-1})^\prime)^\prime ( \wh{\bm F} - {\bm F}(\wh{\mbf H}^{-1})^\prime)}T\r\Vert
\r\} \l\Vert\wh{\mbf H}\r\Vert
\nn\\
&\le \l\Vert\wh{\mbf H}\r\Vert\, 
\l\{
\l\Vert \frac{ ( \wh{\bm F} - {\bm F}(\wh{\mbf H}^{-1})^\prime)^\prime {\bm F}}T\r\Vert
+\l\Vert \frac{ \wh{\bm F} - {\bm F}(\wh{\mbf H}^{-1})^\prime}{\sqrt T}\r\Vert^2
\r\}
\nn\\
&= O_{\mathrm P}\l(\max\l(\frac 1n,\frac 1{\sqrt {nT}},\frac 1 T\r)\r),\label{eq:remp}
\end{align}
because of Lemma \ref{lem:HO1}(i), \eqref{eq:terronski1}, and Proposition \ref{prop:F2}(b). This proves part (d). 

Parts (e) and (f) follow trivially from parts (a) and (b) and the continuous mapping theorem, respectively. 

For part (g), from part (e) and using the same reasoning for proving \eqref{eq:sayme1}, we also have
\begin{align}
 \l\Vert  \frac{(\wh{\mbf H}^{-1})^\prime\wh{\bm\Lambda}^\prime\wh{\bm\Lambda}\wh{\mbf H}^{-1}}n-\frac{\bm\Lambda^\prime\bm\Lambda}n\r\Vert
 &\le 
  \l\Vert  \frac{(\wh{\mbf H}^{-1})^\prime\wh{\bm\Lambda}^\prime\bm\Lambda}n-\frac{\bm\Lambda^\prime\bm\Lambda}n\r\Vert
  + \l\Vert  \frac{(\wh{\mbf H}^{-1})^\prime\wh{\bm\Lambda}^\prime(\wh{\bm\Lambda}\wh{\mbf H}^{-1}-\bm\Lambda)}n
  \r\Vert\nn\\
 &\le 
  \l\Vert (\wh{\mbf H}^{-1})^\prime -\frac{\bm\Lambda^\prime\bm\Lambda}n\l(\frac{\wh{\bm\Lambda}^\prime\bm\Lambda}n\r)^{-1}\r\Vert\,
  \l\Vert \frac{\wh{\bm\Lambda}^\prime\bm\Lambda}n\r\Vert
  + \l\Vert  \frac{(\wh{\mbf H}^{-1})^\prime\wh{\bm\Lambda}^\prime(\wh{\bm\Lambda}\wh{\mbf H}^{-1}-\bm\Lambda)}n
  \r\Vert\nn\\ 
 &= O_{\mathrm P}\l(\max\l(\frac 1n,\frac 1{\sqrt{nT}},\frac 1{\sqrt T}\r)\r).\nn
 \end{align}
This proves part (g).

For part (h), from part (f) and using the same reasoning for proving \eqref{eq:remp}, we also have
\begin{align}
\l\Vert \frac{\wh{\mbf H} \wh{\bm F}^\prime\wh{\bm F}\wh{\mbf H}^\prime}T-  \frac{{\bm F}^\prime{\bm F}}T\r\Vert &\le 
 \l\Vert \frac{\wh{\mbf H} \wh{\bm F}^\prime{\bm F}}T-  \frac{{\bm F}^\prime{\bm F}}T\r\Vert
 +\l\Vert\frac{ \wh{\mbf H} \wh{\bm F}^\prime(\wh{\bm F}\wh{\mbf H}^\prime-{\bm F})}T\r\Vert\nn\\
 &\le  
 \l\Vert \wh{\mbf H} -  \frac{{\bm F}^\prime{\bm F}}T\l(\frac{\wh{\bm F}^\prime{\bm F}}T\r)^{-1}\r\Vert\,
 \l\Vert\frac{ \wh{\bm F}^\prime{\bm F}}T\r\Vert
 +\l\Vert\frac{ \wh{\mbf H} \wh{\bm F}^\prime(\wh{\bm F}\wh{\mbf H}^\prime-{\bm F})}T\r\Vert\nn\\
 &= O_{\mathrm P}\l(\max\l(\frac 1n,\frac 1{\sqrt{nT}},\frac 1{\sqrt T}\r)\r).\nn
\end{align}
This proves part (h) and completes the proof. $\Box$

\subsection{Proof of Proposition \ref{prop:KKKbis}}\label{prop:KKKbisproof}
First, from \eqref{eq:acca}, 
\beq\label{eq:torino}
\frac{\bm\Lambda^\prime\wh{\bm\Lambda}}{n}=\l(\frac{\bm F^\prime\bm F}{T}\r)^{-1}\wh{\mbf H}\, \frac{\wh{\mbf M}^x}{n}.
\eeq
So from \eqref{eq:torino}, Propositions \ref{prop:KKK} (under conditions (B), which hold because of Assumption \ref{ass:ind}),  \ref{prop:LLFF}(f), Lemmas \ref{lem:FFinv}, \ref{lem:MO1}(i), \ref{lem:MO1}(ii), \ref{lem:Vzero}(ii), and \ref{lem:HO1}(i)
\begin{align}
\l\Vert
(\bm\Gamma^F)^{-1/2}\bm\Upsilon_0 \bm{\mathcal J}_0 -  \l(\frac{\wh{\bm F}^\prime{\bm F}}T\r)^{-1}
\r\Vert\le&\,
\l\Vert
(\bm\Gamma^F)^{-1/2}\bm\Upsilon_0 \bm{\mathcal J}_0 \bm V_0-  \l(\frac{\wh{\bm F}^\prime{\bm F}}T\r)^{-1}\bm V_0
\r\Vert\,
\l\Vert (\bm V_0)^{-1}\r\Vert
\label{eq:torino2}
\\
=&\,\l\Vert
(\bm\Gamma^F)^{-1/2}\bm\Upsilon_0 \bm{\mathcal J}_0\bm V_0  -\l(\frac{\bm F^\prime\bm F}{T}\r)^{-1}\frac{\bm F^\prime\bm F}{T}  \l(\frac{\wh{\bm F}^\prime{\bm F}}T\r)^{-1}
 \bm V_0
\r\Vert\,
\l\Vert (\bm V_0)^{-1}\r\Vert\nn\\
\le&\, \l\{
\l\Vert
(\bm\Gamma^F)^{-1/2}\bm\Upsilon_0 \bm{\mathcal J}_0\bm V_0  -\l(\frac{\bm F^\prime\bm F}{T}\r)^{-1}\wh{\mbf H}\, \frac{\wh{\mbf M}^x}{n}
\r\Vert\r.\nn\\
& +\l\Vert \l(\frac{\bm F^\prime\bm F}{T}\r)^{-1}\r\Vert \l\Vert \wh{\mbf H}\r\Vert
\,
\l\Vert
\bm V_0-\frac{\wh{\mbf M}^x}{n}
\r\Vert
 \nn\\
&+\l.\l\Vert \l(\frac{\bm F^\prime\bm F}{T}\r)^{-1}\r\Vert
\l\Vert
\frac{\bm F^\prime\bm F}{T}  \l(\frac{\wh{\bm F}^\prime{\bm F}}T\r)^{-1} -\wh{\mbf H}
\r\Vert\,
\l\Vert
 \bm V_0
\r\Vert
\r\} \l\Vert (\bm V_0)^{-1}\r\Vert\nn\\
=&\, \l\{
\l\Vert
(\bm\Gamma^F)^{-1/2}\bm\Upsilon_0 \bm{\mathcal J}_0\bm V_0  - \frac{\bm\Lambda^\prime\wh{\bm\Lambda}}{n}
\r\Vert + \l\Vert \l(\frac{\bm F^\prime\bm F}{T}\r)^{-1}\r\Vert \l\Vert \wh{\mbf H}\r\Vert
\,
\l\Vert
\bm V_0-\frac{\wh{\mbf M}^x}{n}
\r\Vert
 \r.\nn\\
&+\l. \l\Vert \l(\frac{\bm F^\prime\bm F}{T}\r)^{-1}\r\Vert
\l\Vert
\frac{\bm F^\prime\bm F}{T}  \l(\frac{\wh{\bm F}^\prime{\bm F}}T\r)^{-1} -\wh{\mbf H}
\r\Vert\,
\l\Vert
 \bm V_0
\r\Vert
\r\} \l\Vert (\bm V_0)^{-1}\r\Vert = o_{\mathrm P}(1).\nn
\end{align}
So from \eqref{eq:torino2} and the continuous mapping theorem
\beq
\l\Vert \frac{\wh{\bm F}^\prime\bm F}{T}-\bm{\mathcal J}_0\bm\Upsilon_0^\prime  (\bm\Gamma^F)^{1/2}\r\Vert = o_{\mathrm P}(1).\label{eq:parioli2}
\eeq
Finally, if Assumption \ref{ass:common}(a) holds with rate $\sqrt n$ and Assumption \ref{ass:common}(c-ii) with rate $\sqrt T$, then Proposition \ref{prop:KKK} and Lemma \ref{lem:Vzero}(ii) hold with rate $\min(\sqrt n,\sqrt T)$, so also \eqref{eq:parioli2} holds with the same. Note that $\bm{\mathcal J}_0\bm\Upsilon_0^\prime  (\bm\Gamma^F)^{1/2}$ is finite and invertible because of Assumptions \ref{ass:common}(b) and \ref{ass:common}(c-ii).
This completes the proof. $\Box$

\subsection{Proof of Proposition \ref{cor:semplice}}\label{cor:sempliceproof}
For part (a) from Propositions \ref{prop:LLFF}(d) and \ref{prop:KKKbis},
\begin{align}
\l\Vert\wh{\mbf H}^\prime -  \bm{\mathcal J}_0\bm\Upsilon_0^\prime  (\bm\Gamma^F)^{1/2} \r\Vert &\le 
\l\Vert\wh{\mbf H}^\prime - \frac{\wh{\bm F}^\prime\bm F}{T} \r\Vert +
\l\Vert \frac{\wh{\bm F}^\prime\bm F}{T}- \bm{\mathcal J}_0\bm\Upsilon_0^\prime  (\bm\Gamma^F)^{1/2}\r\Vert = o_{\mathrm P}(1),
\end{align}
and the leading rate is clearly the same as in Proposition \ref{prop:KKKbis}.
Alternatively, the proof follows from the definition of $\wh{\mbf H}$ in \eqref{eq:acca}, Assumption \ref{ass:common}(c-ii), Lemma \ref{lem:Vzero}(iv), and Proposition \ref{prop:KKK} (under conditions (B), which hold because of Assumption \ref{ass:ind}).  In this case the rate follows from Proposition \ref{prop:KKK}.

Part (b) follows from part (a) and the continuous mapping theorem, or alternatively, from Propositions \ref{prop:LLFF}(b) and \ref{prop:KKKbis},
\begin{align}
\l\Vert\wh{\mbf H}^{-1} -  \bm{\mathcal J}_0\bm\Upsilon_0^\prime  (\bm\Gamma^F)^{-1/2} \r\Vert 
&\le
\l\Vert\wh{\mbf H}^{-1} -  \frac{\wh{\bm F}^\prime\bm F}{T}\l(\frac{\bm F^\prime\bm F}T\r)^{-1}\r\Vert + \l\Vert  \frac{\wh{\bm F}^\prime\bm F}{T} -  \bm{\mathcal J}_0\bm\Upsilon_0^\prime  (\bm\Gamma^F)^{-1/2} \frac{\bm F^\prime\bm F}T\r\Vert\,\l\Vert
\l(\frac{\bm F^\prime\bm F}T\r)^{-1}\r\Vert\nn\\
&\le
\l\Vert\wh{\mbf H}^{-1} -  \frac{\wh{\bm F}^\prime\bm F}{T}\l(\frac{\bm F^\prime\bm F}T\r)^{-1}\r\Vert + \Bigg\{ \l\Vert  \frac{\wh{\bm F}^\prime\bm F}{T} -  \bm{\mathcal J}_0\bm\Upsilon_0^\prime  (\bm\Gamma^F)^{1/2} \r\Vert\nn\\
& +\l\Vert \bm{\mathcal J}_0\bm\Upsilon_0^\prime  (\bm\Gamma^F)^{-1/2}\r\Vert\, \l\Vert \frac{\bm F^\prime\bm F}T-\bm\Gamma^F\r\Vert
\Bigg\}\l\Vert
\l(\frac{\bm F^\prime\bm F}T\r)^{-1}\r\Vert\nn\\
&= O_{\mathrm P}\l(\max\l(\frac 1n,\frac 1{\sqrt{nT}},\frac 1T\r)\r)+o_{\mathrm P}(1),\label{eq:upswarwickroma}
\end{align}
where we used also Assumption \ref{ass:common}(c-ii), Lemma \ref{lem:FFinv}, and the fact that $\Vert \bm{\mathcal J}_0\bm\Upsilon_0^\prime  (\bm\Gamma^F)^{-1/2}\Vert = O_{\mathrm P}(1)$ again because of Assumption \ref{ass:common}(b) and since $\Vert \bm{\mathcal J}_0\Vert =\Vert \bm\Upsilon_0\Vert =1$. The rate in \eqref{eq:upswarwickroma} follows from Proposition \ref{prop:KKKbis} and Assumption \ref{ass:common}(c-ii).
This completes the proof. $\Box$

\subsection{Proof of Proposition \ref{corol:K00H}}\label{corol:K00Hproof}

First, using Assumption \ref{ass:Jhat} into \eqref{eq:DK} in the proof of Lemma \ref{lem:covarianze}, it follows that, for all $T\in\mathbb N$,
\beq\label{eq:convenzione3}
\mbf J=\mbf I_r.
\eeq
Recall the definition of $\bm{\mathcal H}$ in Proposition \ref{prop:L}, from \eqref{eq:convenzione3} we have
\beq\label{eq:HGKJ}
\bm{\mathcal H}=(\bm\Gamma^F)^{1/2}\mbf K.
\eeq
Then, we consider all cases one-by-one.

\paragraph{I.a} In part (I.a) we impose $\bm\Gamma^F=\mbf I_r$ and leave $\bm\Lambda$ unrestricted. 
In this case, from \eqref{eq:HGKJ}, 
$\bm{\mathcal H}=\mbf K$. So, for all $n\in\mathbb N$, $\bm{\mathcal H}$ is orthogonal  because $\mbf K$ is orthogonal by Lemma \ref{lem:KO1}(iii). 
This proves part (I.a). 

\paragraph{II.a} In part (II.a) we impose $\bm\Gamma^F=\mbf I_r$ and $\frac{\bm\Lambda^\prime\bm\Lambda}{n}$ to be diagonal for all $n\in\mathbb N$. 
First, from Lemma \ref{lem:KO1}(iii), 
\beq\label{eq:cirisiamo2}
 \frac{\mbf M^{C}}n \mbf K^{\prime}=  \mbf K^\prime \frac{\bm\Lambda^\prime \bm\Lambda}n.
\eeq
Then, from \eqref{eq:rotturaasyB} in the proof of Proposition \ref{corol:K00} and \eqref{eq:cirisiamo2}, we see that we must have 
\beq\nn
\l(\frac{\mu^C_i}{n}-\frac{\mu^C_j}n\r)[\mbf K]_{ij}= 0, \quad i,j=1,\ldots, r.
\eeq
Hence, since by Assumption \ref{ass:sign_eval}(b) $\frac{\mu^C_i}n\ne \frac{\mu^C_j}n$ if $i\ne j$  and $0<\frac{\mu^C_j}n<\infty$, $j=1,\ldots, r$, for all $n\in\mathbb N$, by Lemmas \ref{lem:MO1}(i) and \ref{lem:MO1}(ii),
and since $\mbf K$ has full-rank by Lemma \ref{lem:KO1}(ii), then each column of $\mbf K$ must have at least one non-zero entry, thus  $\mbf K$ must be diagonal. 
Therefore, since by Lemma \ref{lem:KO1}(iii) $\mbf K$ is also orthogonal, it must be that 
\beq\label{eq:KlimiteIB}
\mbf K=\bm S,
\eeq 
where $\bm S$ is diagonal with entries $\pm 1$ and depends only on $n$. In this case, from \eqref{eq:HGKJ}, 
$\bm{\mathcal H}=\mbf K$ and from \eqref{eq:KlimiteIB}
we prove part (II.a).

\paragraph{III.a}
In part (III.a) we impose $\bm\Gamma^F=\mbf I_r$ and $\bm\Sigma_\Lambda$ to be diagonal. 
From Proposition \ref{prop:KKK} recall that the columns of $\bm\Upsilon_0\bm{\mathcal J}_0$ are eigenvectors of $(\bm\Gamma^F)^{1/2}\bm\Sigma_{\Lambda}(\bm\Gamma^F)^{1/2}=\bm\Sigma_{\Lambda}$, i.e.,
\beq\label{eq:stoppani2}
\bm\Sigma_\Lambda \bm{\mathcal J}_0\bm\Upsilon_0^\prime= \bm{\mathcal J}_0\bm\Upsilon_0^\prime\bm V_0.
\eeq
Therefore, because of \eqref{eq:stoppani} in the proof of Proposition \ref{corol:K00} and  \eqref{eq:stoppani2}, 
\beq
\l([\bm V_0]_{ii}-[\bm V_0]_{jj}\r)  [ \bm\Upsilon_0\bm{\mathcal J}_0]_{ij}= 0, \quad i,j=1,\ldots, r.
\eeq
Hence, since by Assumption \ref{ass:sign_eval}(b), which implies Assumption \ref{ass:eval}, $[\bm V_0]_{ii}\ne[\bm V_0]_{jj}$ for $i\ne j$, and $0<[\bm V_0]_{jj}<\infty$ by Lemmas \ref{lem:Vzero}(i) and \ref{lem:Vzero}(iii), 
$\bm\Upsilon_0\bm{\mathcal J}_0$ must be diagonal, because $\bm\Upsilon_0$, being a matrix of eigenvectors, has full rank, thus it has at least one non-zero entry in each column. Therefore, since $\bm\Upsilon_0\bm{\mathcal J}_0$ is orthogonal, it must be that 
\beq\label{eq:KlimiteI}
\bm\Upsilon_0\bm{\mathcal J}_0=\bm{\mathcal S}_0,
\eeq 
where $\bm{ \mathcal S}_0$ is diagonal with entries $\pm 1$ and independent of $n$ and $T$. 
In this case, from \eqref{eq:HGKJ}, 
$\bm{\mathcal H}=\mbf K$ . Thus, by
 \eqref{eq:KlimiteI} and \eqref{eq:ups} in the proof of Proposition \ref{prop:KKK},
\beq\nn
\lim_{n\to\infty}\Vert \bm{\mathcal H}-\bm {\mathcal S}_0\Vert=\lim_{n\to\infty} \Vert \bm{\mathcal H}-\bm\Upsilon_0\bm{\mathcal J}_0\Vert= \lim_{n\to\infty} \Vert \mbf K-\bm\Upsilon_0\bm{\mathcal J}_0\Vert=o(1),
\eeq 
where we also used \eqref{eq:convenzione3}, which implies that this limit holds independently of $T$ and, therefore, is an ordinary one.
This proves part (III.a) and
 completes the proof. $\Box$ 

\setcounter{equation}{0}
\section{Auxiliary results}\label{sec:lemma}
Throughout, recall that ``for all $n\in\mathbb N$'' always means ``for all $n>N$'' with $N$ defined in Assumption \ref{ass:common}(d).

\begin{lem}\label{lem:Gxi}
Under Assumptions \ref{ass:common} and \ref{ass:idio}:
\begin{compactenum} 
\item [(i)] for all $n,T\in\mathbb N$, $\frac 1{nT}\sum_{i,j=1}^n\sum_{t,s=1}^T \vert\E_{}[ e_{it} e_{js}]\vert \le M_{1 e}$, for some finite positive real $M_{1 e}$ independent of $n$ and $T$;
\item [(ii)] for all $n\in\mathbb N$ and $t\in\mathbb Z$, $\frac 1{n}\sum_{i,j=1}^n \vert\E_{}[ e_{it} e_{jt}]\vert \le M_{2 e}$, for some finite positive real $M_{2 e}$ independent of $n$ and $t$;
\item [(iii)] for all $i\in\mathbb N$ and $T\in\mathbb N$, $\!\frac 1{T}\sum_{t,s=1}^T \vert\E_{}[ e_{it} e_{is}]\vert \le M_{3 e}$, $\!$for some finite positive real $M_{3 e}$ independent of $i$ and $T$;
\item [(iv)] 
for all $j=1,\ldots,r$, $\underline C_j\!\le \lim_{n\to\infty} \frac{\mu_{j}^{C}}n\le\! \overline C_j$, $\!\!$
for some finite positive reals $\underline C_j$ and $\overline C_j$;
\item [(v)] for all $n\in\mathbb N$, $\mu_{1}^ e  \le M_{2 e}$, where $M_{2 e}$ is defined in part (ii);
\item [(vi)] for all $j=1,\ldots,r$, $\underline C_j\le \lim_{n\to\infty} \frac{\mu_{j}^x}n\le \overline C_j$, and for all $n\in\mathbb N$, $\mu_{r+1}^x \le M_{2 e}$, where $M_{2 e}$ is defined in part (ii).
\end{compactenum}
\end{lem}

\noindent{\bf Proof.} Using Assumptions \ref{ass:idio}(a) and \ref{ass:idio}(b), 
\begin{align}
\frac 1{nT}\sum_{i,j=1}^n\sum_{t,s=1}^T \vert\E_{}[ e_{it} e_{js}]\vert &=\frac 1{n}\sum_{i,j=1}^n\sum_{k=-(T-1)}^{T-1} \l(1-\frac{\vert k\vert}{T}\r) \vert\E_{}[ e_{it} e_{j,t-k}]\vert\nn\\
&\le \frac 1n \sum_{i=1}^n\sum_{k=-\infty}^{\infty} \rho^{\vert k\vert} \E[ e_{it}^2]+ \max_{i=1,\ldots,n} \sum_{j=1, j\ne i}^n\sum_{k=-\infty}^{\infty} \rho^{\vert k\vert} M_{ij} \nn\\
&\le \frac{C_ e(1+\rho)}{1-\rho}+ \frac{M_ e(1+\rho)}{1-\rho} .\nn
\end{align}
Similarly, 
\begin{align}
\frac 1{n}\sum_{i,j=1}^n \vert\E_{}[ e_{it} e_{jt}]\vert &= \frac 1n\sum_{i=1}^n \E[ e_{it}^2]+\max_{i=1,\ldots, n}\sum_{j=1, j\ne i}^n M_{ij} \le C_ e+ M_ e,\nn
\end{align}
and 
\begin{align}
\frac 1{T}\sum_{t,s=1}^T \vert\E_{}[ e_{it} e_{is}]\vert &= \sum_{k=-(T-1)}^{T-1} \l(1-\frac{\vert k\vert}{T}\r) \vert\E_{}[ e_{it} e_{i,t-k}]\vert\le \sum_{k=-\infty}^{\infty} \rho^{\vert k\vert} \E[ e_{it}^2] \le \frac{C_ e (1+\rho)}{1-\rho}.\nn
\end{align}
Defining $M_{1 e}=\frac{(C_ e+M_ e)(1+\rho)}{1-\rho}$, $M_{2 e}=C_ e+M_ e$, and $M_{3 e}=\frac{C_ e(1+\rho)}{1-\rho}$ we prove parts (i), (ii), and (iii).

For part (iv), by \citet[Theorem 7]{MK04}, for all $j=1,\ldots, r$, 
\beq\label{eq:KUMAR}
\mu_j(\bm\Gamma^F) \lim_{n\to\infty}\frac{\mu_r(\bm\Lambda^\prime\bm\Lambda)}n \le
\lim_{n\to\infty} \frac{\mu_{j}^{C}}n \le 
\mu_1(\bm\Gamma^F) \lim_{n\to\infty} \frac{\mu_j(\bm\Lambda^\prime\bm\Lambda)}n,
\eeq
and since, for all $j=1,\ldots, r$, $\lim_{n\to\infty}\frac{\mu_j(\bm\Lambda^\prime\bm\Lambda)}n=\mu_j(\bm\Sigma_\Lambda)$, the proof follows from  Assumptions \ref{ass:common}(a) and \ref{ass:common}(b).
%

For part (v), by Assumption \ref{ass:idio}(b) and the Circle theorem by \citet[Satz II]{gershgorin1931}:
\begin{align}
\Vert\bm\Gamma^ e\Vert\le \max_{i=1,\ldots,n}\sum_{j=1}^n \vert \E[ e_{it} e_{jt}]\vert \le  \max_{i=1,\ldots, n} \E[ e_{it}^2]+ \max_{i=1,\ldots,n}\sum_{j=1,j\ne i}^n \vert \E[ e_{it} e_{jt}]\vert 
\le C_ e+M_{ e}.\nn
\end{align}
Part (vi) follows from parts (iv) and (v) and Weyl's inequality \citep[Theorem 1]{MK04}. This completes the proof. $\Box$

\begin{lem}\label{lem:GxiBAI}
Under Assumption \ref{ass:idio}:
\begin{compactenum}
\item [(i)]  for all $n\in\mathbb N$ and $T\in\mathbb N$, $\frac 1{T}\sum_{t,s=1}^T \l \vert\frac 1n \sum_{i=1}^n \E_{}[ e_{it} e_{is}]\r\vert \le M_{3 e}$, where $M_{3 e}$ is a finite positive real independent of $n$ and $T$, defined in Lemma \ref{lem:Gxi}(iii);
\item [(ii)] for all $t=1,\ldots, T$ and all $n,T\in\mathbb N$, $\sum_{s=1}^T  \l \vert\frac 1n \sum_{i=1}^n \E_{}[ e_{it} e_{is}]\r\vert\le M_{3 e}$, where $M_{3 e}$ is a finite positive real independent of $n$ and $T$, defined in Lemma \ref{lem:Gxi}(iii);
\item [(iii)] for all $i=1,\ldots, n$, all $t=1,\ldots, T$, and all $n,T\in\mathbb N$, $\sum_{j=1}^n \vert \E[ e_{it}  e_{jt}]  \vert\le M_{2 e}$,where $M_{2 e}$ is a finite positive real independent of $n$ and $T$, defined in Lemma \ref{lem:Gxi}(ii);
\item [(iv)] for all $t=1,\ldots, T$ and all $n,T\in\mathbb N$, $\E[\Vert\frac 1{\sqrt {n T} }\sum_{s=1}^T\sum_{k=1}^n \mbf F_{s} \{e_{ks}e_{kt}-\E[e_{ks}e_{kt}] \}
\Vert^2] \le M_{4e}$, for some finite positive real $M_{4 e}$ independent of $t$, $n$, and $T$;
\item [(v)] for all $n,T\in\mathbb N$,  $\E[\Vert\frac 1{\sqrt{nT}}\sum_{t=1}^T\sum_{j=1}^n\mbf F_t{\bm\lambda}_j^\prime  e_{jt} \Vert^2] 
\le M_{5e}$,  for some finite positive real $M_{5 e}$ independent of $n$ and $T$;
\item [(vi)]  for all $j=1,\ldots, n$ and all $n,T\in\mathbb N$,  
$\E[\Vert\frac 1{\sqrt {nT}}\sum_{s=1}^T\sum_{k=1}^n \bm\lambda_k
\{
e_{ks}e_{js}-\E[e_{ks}e_{js}]
\}
\Vert^2]\le M_{6e}$, for some finite positive real $M_{6 e}$ independent of $j$, $n$, and $T$.
\end{compactenum}
\end{lem}

\noindent{\bf Proof.} For part (i), from Assumption \ref{ass:idio}(b)
\begin{align}
\frac 1{T}\sum_{t,s=1}^T \l\vert\frac 1n \sum_{i=1}^n \E_{}[ e_{it} e_{is}]\r\vert &\le \frac 1{T}\sum_{t,s=1}^T \frac 1n \sum_{i=1}^n \l\vert\E_{}[ e_{it} e_{is}]\r\vert\le \sum_{k=-(T-1)}^{T-1} \l(1-\frac{|k|}{T}\r) \frac 1n \sum_{i=1}^n \vert \E_{}[ e_{it} e_{i,t-k}] \vert \nn\\
&\le \frac 1n \sum_{i=1}^n M_{ii} \sum_{k=-\infty}^{\infty}  \rho^{|k|}\le\frac {C_{ e}  (1+\rho)}{1-\rho}= M_{3 e}.\nn
\end{align}

For part (ii),  from Assumption \ref{ass:idio}(b)
\begin{align}
\sum_{s=1}^T  \l \vert\frac 1n \sum_{i=1}^n \E_{}[ e_{it} e_{is}]\r\vert&\le 
\sum_{s=1}^T \frac 1n \sum_{i=1}^n \l \vert \E_{}[ e_{it} e_{is}]\r\vert\le \sum_{s=1}^T \frac 1n \sum_{i=1}^n M_{ii} \rho^{|t-s|}\nn\\
&\le C_ e \sum_{s=1}^\infty \rho^{|t-s|} = C_ e \l\{ \sum_{s=1}^t \rho^{t-s} + \sum_{s=t+1}^\infty \rho^{s-t}\r\}\nn\\
&= C_ e\l\{ \sum_{k=0}^{t-1} \rho^k+\sum_{k=1}^\infty \rho^k\r\}= C_ e\l\{\frac{1-\rho^t}{1-\rho}+ \frac{\rho}{1-\rho}\r\}\nn\\
&= \frac{C_ e(1+\rho-\rho^t)}{1-\rho}\le \frac {C_{ e}  (1+\rho)}{1-\rho}= M_{3 e}. \nn
\end{align}

For part (iii),  from Assumption \ref{ass:idio}(b)
\begin{align}
\sum_{j=1}^n \vert \E[ e_{it}  e_{jt}]  \vert&\le M_{ii} + \sum_{j=1, j\ne i}^n M_{ij} \le C_ e +M_ e=M_{2 e}.\nn
\end{align}

For part (iv), by Assumptions \ref{ass:common}(b), \ref{ass:idio}(c-ii), and \ref{ass:ind}
\begin{align}
\E&\l[\l\Vert\frac 1{\sqrt {n T} }\sum_{s=1}^T\sum_{k=1}^n \mbf F_{s} \l\{e_{ks}e_{kt}-\E[e_{ks}e_{kt}] \r\}
\r\Vert^2\r] \nn\\
&= \frac 1{nT} \sum_{j=1}^r \sum_{s,u=1}^T \sum_{k,i=1}^n \E\l[F_{js}F_{ju}\l\{ e_{ks}e_{kt}e_{iu}e_{it}-\E[e_{ks}e_{kt}]\E[e_{iu}e_{it}] \r\} \r] \nn\\
&\le \l\{r \max_{j=1,\ldots,r} \max_{s,u=1,\ldots, T} \l\vert \E\l[F_{js}F_{ju}\r]\r\vert \r\}
\l\{\frac 1{nT}
\sum_{s,u=1}^T \sum_{k,i=1}^n  \E\l[ e_{ks}e_{kt}e_{iu}e_{it}-\E[e_{ks}e_{kt}]\E[e_{iu}e_{it}] \r] \r\}\nn\\
&\le \l\{r \max_{j=1,\ldots,r}  \E\l[F_{js}^2\r] \r\}
\l\{\E\l[\l\vert\frac 1{\sqrt{nT}} \sum_{k=1}^n\sum_{s=1}^T\l\{{e}_{ks}{e}_{kt}-\E[{e}_{ks}{e}_{kt}]\r\} \r\vert^2\r] \r\}
\le rM_F K_e,\nn
\end{align}
for some finite positive real $M_F$ independent of $j$ and $s$, and since, by Cauchy-Schwarz inequality, $ \l\vert \E\l[F_{js}F_{ju}\r]\r\vert \le  \E[F_{js}^2]$ for all $s,u\in\mathbb Z$. Defining $M_{4e}=rM_F K_e$, we prove part (iv).

For part (v), by Assumption \ref{ass:ind} and Lemma \ref{lem:Gxi}(i) (using similar steps as in part (iv)),
\begin{align}\label{eq:oslo3}
\E\l[\l\Vert\frac 1{ \sqrt {nT}}\sum_{t=1}^T\sum_{j=1}^n\mbf F_t e_{jt} \r\Vert^2\r]&= \frac 1{nT}\sum_{\ell=1}^r\E\l[\l(\sum_{t=1}^T\sum_{j=1}^n F_{\ell t} e_{jt}\r)^2\r]\nn\\
&\le\frac r{nT}\max_{\ell=1,\ldots, r}\sum_{j=1}^n\sum_{i=1}^n\sum_{t=1}^T\sum_{s=1}^T \E[F_{\ell t} e_{jt}F_{\ell s} e_{is}]\nn\\
&\le\frac r{nT}\max_{\ell=1,\ldots, r}\sum_{j=1}^n\sum_{i=1}^n\sum_{t=1}^T\sum_{s=1}^T \E[F_{\ell t}F_{\ell s}]\,\E[ e_{jt} e_{is}]\nn\\
&\le \l\{r \max_{\ell=1,\ldots, r}\max_{t,s=1,\ldots, T} \vert\E[F_{\ell t}F_{\ell s}]\vert\r\}\l\{\frac 1{nT} \sum_{j=1}^n\sum_{i=1}^n\sum_{t=1}^T\sum_{s=1}^T\vert\E[ e_{jt} e_{is}]\vert \r\}\nn\\
&\le  \l\{r \max_{\ell=1,\ldots,r}\E[F_{\ell s}^2] \r\}
\frac{(C_ e+M_ e)(1+\rho)}{(1-\rho)}\le r M_FM_{1 e}.
\end{align}
So,  from Assumption \ref{ass:common}(a) and \eqref{eq:oslo3}
\begin{align}
\E\l[\l\Vert\frac 1{\sqrt{nT}}\sum_{t=1}^T\sum_{j=1}^n\mbf F_t{\bm\lambda}_j^\prime e_{jt} \r\Vert^2\r] 
&\le \E\l[\l\Vert\frac 1{\sqrt{nT}}\sum_{t=1}^T\sum_{j=1}^n\mbf F_t {\bm\lambda}_j^\prime e_{jt} \r\Vert_F^2\r] \nn\\
&= \frac 1{nT} \sum_{\ell,k=1}^r \E\l[  \l( \sum_{t=1}^T\sum_{j=1}^n F_{\ell t} \lambda_{jk} e_{jt}  \r)^2\r]
\le r^2 M_\Lambda^2 M_FM_{1 e}.\nn
\end{align}
Defining $M_{5 e}=r^2 M_\Lambda^2 M_FM_{1 e}$, with $M_{1e}=\frac{(C_ e+M_ e)(1+\rho)}{1-\rho}$, we prove part (v). 

For part (vi), because of Assumptions \ref{ass:common}(a) and \ref{ass:idio}(c-ii), 
\begin{align}
\E&\l[\l\Vert\frac 1{\sqrt {nT}}\sum_{s=1}^T\sum_{k=1}^n \bm\lambda_k
\l\{
e_{ks}e_{js}-\E[e_{ks}e_{js}]
\r\}
\r\Vert^2\r]\nn\\
&= \frac 1{nT}\sum_{\ell=1}^r \sum_{s,u=1}^T \sum_{k,i=1}^n \lambda_{k\ell }\lambda_{i\ell }\E\l[
e_{ks}e_{js}e_{iu}e_{ju}-\E[e_{ks}e_{js}]\E[e_{iu}e_{ju}]
\r]\nn\\
&\le M_\Lambda^2 \l\{\frac 1{nT}\E\l[\l\vert \sum_{s=1}^T \sum_{k=1}^n \l\{e_{ks}e_{js}-\E[e_{ks}e_{js}]\r\}\r\vert^2\r]\r\}\le M_{\Lambda}^2 K_e.\nn
\end{align}
Defining $M_{6 e}= M_{\Lambda}^2 K_e$ we prove part (vi). This completes the proof. $\Box$

\begin{lem}\label{lem:FTLN}
Under Assumptions \ref{ass:common} and \ref{ass:idio}, together with Assumption~\ref{eq:ortho}:
\begin{compactenum}[(i)]
\item for all $n\in\mathbb N$, $\Vert\frac{\bm\Lambda}{\sqrt n}\Vert=O(1)$;
\item for all $t=1,\ldots, T$ and all $T\in\mathbb N$, $\Vert \mbf F_t\Vert = O_{\mathrm {m.s.}}(1)$ and $\Vert\frac{\bm F}{\sqrt T}\Vert=O_{\mathrm {m.s.}}(1)$;
\item for all $t=1,\ldots, T$ and all $n,T\in\mathbb N$, $\Vert\frac{\bm  e_t}{\sqrt {n}}\Vert=O_{\mathrm {m.s.}}(1)$ and $\Vert\frac{\bm  E}{\sqrt {nT}}\Vert=O_{\mathrm {m.s.}}(1)$;
\item for all $t=1,\ldots, T$ and all $n,T\in\mathbb N$, $\Vert\frac{\mbf x_t}{\sqrt {n}}\Vert=O_{\mathrm {m.s.}}(1)$ and $\Vert\frac{\bm X}{\sqrt {nT}}\Vert=O_{\mathrm {m.s.}}(1)$.
\end{compactenum}
\end{lem}

\noindent
\textbf{Proof.} By Assumption \ref{ass:common}(a), 
\beq
\sup_{n\in\mathbb N}\l\Vert\frac{\bm\Lambda}{\sqrt n}\r\Vert^2\le \sup_{n\in\mathbb N}\l\Vert\frac{\bm\Lambda}{\sqrt n}\r\Vert^2_F =\sup_{n\in\mathbb N}\frac 1n\sum_{j=1}^r\sum_{i=1}^n \lambda_{ij}^2 \le\sup_{n\in\mathbb N} \max_{i=1,\ldots, n} \Vert\bm\lambda_i\Vert^2\le M_\Lambda^2,\nn
\eeq
since $M_\Lambda$ is independent of $i$. This proves part (i).

By Assumption \ref{ass:common}(b), which holds for all $T\in\mathbb N$ because of stationarity (see also part (i) of Lemma \ref{lem:covarianze}),
\begin{align}\label{eq:vettoreF}
\sup_{T\in\mathbb N}\max_{t=1,\ldots ,T}\E[\Vert\mbf F_t\Vert^2]&=
\sup_{T\in\mathbb N} \max_{t=1,\ldots ,T}\sum_{j=1}^r \E[F_{jt}^2]\le
r\sup_{T\in\mathbb N}\max_{t=1,\ldots ,T}\max_{j=1,\ldots,r} \E[F_{jt}^2] \nn\\
&\le r\sup_{T\in\mathbb N}\max_{t=1,\ldots ,T}\max_{j=1,\ldots,r} \bm\eta_j^\prime\bm\Gamma^F\bm\eta_j\le r \Vert\bm\Gamma^F\Vert\le
r M_F,
\end{align}
since $M_F$ is independent of $t$ and where $\bm\eta_j$ is an $r$-dimensional vector with one in the $j$th entry and zero elsewhere. Therefore, from \eqref{eq:vettoreF}:
\begin{align}
\sup_{T\in\mathbb N}\E\l[\l\Vert\frac{\bm F}{\sqrt T}\r\Vert^2\r]\le\sup_{T\in\mathbb N} \E\l[\l\Vert\frac{\bm F}{\sqrt T}\r\Vert^2_F\r] =\sup_{T\in\mathbb N}\frac 1T\sum_{j=1}^r\sum_{t=1}^T \E[ F_{jt}^2] \le\sup_{T\in\mathbb N} \max_{t=1,\ldots, T} \E[\Vert\mbf F_t\Vert^2]
\le r M_F.\nn
\end{align}
This proves part (ii). 

Then, because of Lemma \ref{lem:Gxi}(v), which holds for all $n\in\mathbb N$ and holds also for all $T\in\mathbb N$ by stationarity, which is implied by Assumption \ref{ass:idio}(b),
\begin{align}\label{eq:vettorexi}
\sup_{n,T\in\mathbb N}\max_{t=1,\ldots, T}\E\l[\l\Vert\frac{\bm e_t}{\sqrt n}\r\Vert^2\r]&=\sup_{n,T\in\mathbb N}\max_{t=1,\ldots, T}\frac 1n\sum_{i=1}^n \E[ e_{it}^2]
\le \sup_{n,T\in\mathbb N}\max_{t=1,\ldots, T}\max_{i=1,\ldots,n} \E[ e_{it}^2]\nn\\
&= \sup_{n\in\mathbb N}\max_{i=1,\ldots, n} \bm\varepsilon_i^\prime\bm\Gamma^ e\bm\varepsilon_i\le  \sup_{n\in\mathbb N}\Vert\bm\Gamma^ e\Vert= \sup_{n\in\mathbb N}\mu_1^ e\le M_{2 e}.
\end{align}
since $M_ e$ is independent of $i$ and $t$ and where $\bm\varepsilon_i$ is an $n$-dimensional vector with one in the $i$th entry and zero elsewhere. Therefore, from \eqref{eq:vettorexi}
\begin{align}
\sup_{n,T\in\mathbb N}\E\l[\l\Vert\frac{\bm  E}{\sqrt {nT}}\r\Vert^2\r]\le\sup_{n,T\in\mathbb N} \E\l[\l\Vert\frac{\bm  E}{\sqrt {nT}}\r\Vert^2_F\r] &=\sup_{n,T\in\mathbb N}\frac 1{nT}\sum_{i=1}^n\sum_{t=1}^T \E[ e_{it}^2] \le\sup_{n,T\in\mathbb N} \max_{t=1,\ldots, T}\max_{i=1,\ldots, n} \E[ e_{it}^2]\le M_ e.\nn
\end{align}
This proves part (iii). 

Finally, as a consequence of parts (ii) and (iii), Assumption \ref{ass:common}(a), and Assumption~\ref{eq:ortho}:
\begin{align}\label{eq:vettoreX}
\sup_{n,T\in\mathbb N}\max_{t=1,\ldots, T}\E\l[\l\Vert\frac{\mbf x_t}{\sqrt n}\r\Vert^2\r]
&=\sup_{n,T\in\mathbb N}\max_{t=1,\ldots, T}\frac 1n\sum_{i=1}^n \E[x_{it}^2]\le\sup_{n,T\in\mathbb N}\max_{t=1,\ldots, T}\max_{i=1,\ldots, n} \E[x_{it}^2]\nn\\
&=\sup_{n,T\in\mathbb N}\max_{t=1,\ldots, T}\max_{i=1,\ldots, n} \l\{\bm\lambda_i^\prime \bm\Gamma^F  \bm\lambda_i+\bm\varepsilon_i^\prime\bm\Gamma^ e\bm\varepsilon_i\r\}\nn\\
&\le\sup_{n\in\mathbb N}\max_{i=1,\ldots, n} \Vert\bm\lambda_i\Vert^2 \, \Vert\bm\Gamma^F\Vert +\sup_{n\in\mathbb N}\Vert\bm\Gamma^ e\Vert\nn\\
&\le M_\Lambda^2 M_F +M_ e= M_X,\;\text{say,}
\end{align}
where $M_X$ is independent of $i$, $t$, $n$, and $T$. Therefore, from part (i) and \eqref{eq:vettoreX}:
\begin{align}
\sup_{n,T\in\mathbb N}\E\l[\l\Vert\frac{\bm X}{\sqrt {nT}}\r\Vert^2\r]\le\sup_{n,T\in\mathbb N} \max_{t=1,\ldots, T}\max_{i=1,\ldots, n} \E[x_{it}^2]\le M_X.\nn
\end{align}
This proves part (iv) and it completes the proof. $\Box$

\begin{lem}\label{lem:LLN}
Under Assumptions \ref{ass:common}, \ref{ass:idio}, and \ref{ass:ind}:
\begin{compactenum}[(i)]
\item for all $n,T\in\mathbb N$, $\sqrt T\l\Vert\frac{\bm F^\prime\bm  E}{\sqrt n T} \r\Vert = O_{\mathrm {m.s.}}(1)$;
\item for all $n,T\in\mathbb N$, $\sqrt n\l\Vert\frac{\bm  E\bm\Lambda}{n \sqrt T} \r\Vert = O_{\mathrm {m.s.}}(1)$;
\item for all $n,T\in\mathbb N$, $\min(n,\sqrt T)\l\Vert\frac{\bm  E^\prime\bm  E}{nT} \r\Vert = O_{\mathrm {m.s.}}(1)$ and $\min(n,\sqrt {nT})\l\Vert\frac{\bm\Lambda^\prime\bm  E^\prime\bm  E}{n^{3/2}T} \r\Vert = O_{\mathrm {m.s.}}(1)$;
\item for all $n\in\mathbb N$, $\l\Vert\frac{\bm \Lambda^\prime\bm \Lambda}{n} \r\Vert = O(1)$;
\item for all $T\in\mathbb N$, $\l\Vert\frac{\bm F^\prime\bm F}{T} \r\Vert = O_{\mathrm {m.s.}}(1)$.
\end{compactenum}
\end{lem}

\noindent
\textbf{Proof.}  For part (i), by Assumption \ref{ass:ind} and Lemma \ref{lem:Gxi}(iii) 
\begin{align}
\E\l[ \l\Vert\frac {\bm F^\prime\bm  E}{\sqrt n T} \r\Vert^2\r] &=\E\l[ \l\Vert\frac 1{\sqrt nT}\sum_{t=1}^T \mbf F_t\bm e_t^\prime \r\Vert^2\r]\le \E\l[ \l\Vert\frac 1{\sqrt nT}\sum_{t=1}^T \mbf F_t\bm e_t^\prime \r\Vert_F^2\r]\nn\\
&=\frac 1{nT^2}\sum_{j=1}^r\sum_{i=1}^n \E\l[\l(\sum_{t=1}^T F_{jt} e_{it}\r)^2\r]\nn\\
&\le\frac r{T^2}\max_{j=1\ldots, r} \max_{i=1\ldots, n} \sum_{t=1}^T\sum_{s=1}^T \E[ e_{it}F_{jt} e_{is}F_{js}]\nn\\
&=\frac r{T^2}\max_{j=1\ldots, r} \max_{i=1\ldots, n} \sum_{t=1}^T\sum_{s=1}^T \E[F_{jt}F_{js}]\, \E[ e_{it} e_{is}]\nn\\
&\le \l\{\frac{r}{T}\max_{j=1,\ldots, r}\max_{t,s=1,\ldots, T} \vert\E[F_{jt}F_{js}]\vert\r\}
\l\{\frac 1T  \sum_{t=1}^T\sum_{s=1}^T\l\vert\E[ e_{it} e_{is}]\r\vert\r\}\nn\\
&\le \frac r T \max_{j=1,\ldots, r}\max_{t,s=1,\ldots, T} \E[F_{jt}^2] \frac{M_ e(1+\rho)}{1-\rho}\le\frac{rM_F M_{3 e}}T,\nn
\end{align}
since $M_F$ is independent of $t$ and $s$ and $M_{3 e}$ is independent of $i$. This proves part (i).

For part (ii), let $\bm\varepsilon_i=( e_{i1}\cdots  e_{iT})^\prime$, then by
\begin{align}
\E\l[ \l\Vert\frac {\bm  E\bm\Lambda}{ n\sqrt T} \r\Vert^2\r] &=\E\l[ \l\Vert\frac 1{ n\sqrt T}\sum_{i=1}^n \bm\lambda_i\bm\varepsilon_i^\prime \r\Vert^2\r]\le \E\l[ \l\Vert\frac 1{ n\sqrt T}\sum_{i=1}^n \bm\lambda_i\bm\varepsilon_i^\prime \r\Vert^2_F\r]\nn\\
&=\frac 1{n^2T}\sum_{j=1}^r \sum_{t=1}^T\E\l[\l(\sum_{i=1}^n \lambda_{ij} e_{it}\r)^2\r]\nn\\
&\le \frac{r}{n^2}\max_{j=1,\ldots,r}\max_{t=1,\ldots, T}\sum_{i=1}^n\sum_{k=1}^n\lambda_{ij}\lambda_{kj}\E[ e_{it} e_{kt}]\nn\\
&\le \frac{rM_\Lambda^2}{n^2}\max_{j=1,\ldots,r}\max_{t=1,\ldots, T}\sum_{i=1}^n\sum_{k=1}^n\l\vert \E[ e_{it} e_{kt}]\r\vert\nn\\
&\le \frac{rM_\Lambda^2M_ e }{n},\nn
\end{align}
since $M_\Lambda$ is independent of $i$ and $k$ and $M_{ e}$ is independent of $t$. This proves part (ii).

For part (iii), first notice that, by Lemmas \ref{lem:Gxi}(v) and \ref{lem:covarianzeF}(ii)
\beq
\l\Vert\frac{\bm  E^\prime\bm  E}{nT} \r\Vert \le \l\Vert\frac{\bm  E^\prime\bm  E}{nT}-\frac{\bm\Gamma^ e}{n} \r\Vert +\l\Vert\frac{\bm\Gamma^ e}{n}\r\Vert
= \l\Vert\frac{\bm  E^\prime\bm  E}{nT}-\frac{\bm\Gamma^ e}{n} \r\Vert +\frac{\mu_1^ e}{n}=O_{\mathrm P}\l(\frac 1{\sqrt T}\r)+O\l(\frac 1{n}\r).
\eeq
Similarly, by Lemmas \ref{lem:FTLN}(i) and \ref{lem:Gxi}(v) 
\beq
\l\Vert\frac{\bm\Lambda^\prime\bm  E^\prime\bm  E}{n^{3/2}T} \r\Vert \le \l\Vert\frac{\bm\Lambda^\prime\bm  E^\prime\bm  E}{n^{3/2}T}-\frac{\bm\Lambda^\prime\bm\Gamma^ e}{n^{3/2}} \r\Vert +\l\Vert \frac{\bm\Lambda}{\sqrt n}\r\Vert\,\l\Vert\frac{\bm\Gamma^ e}{n}\r\Vert= 
\l\Vert\frac{\bm\Lambda^\prime\bm  E^\prime\bm  E}{n^{3/2}T}-\frac{\bm\Lambda^\prime\bm\Gamma^ e}{n^{3/2}} \r\Vert +O\l(\frac 1{n}\r).\label{eq:rifacciamo2}
\eeq
Then, because of Assumptions \ref{ass:common}(a) and \ref{ass:idio}(c-ii), 
\begin{align}
\E\l[\l\Vert\frac{\bm\Lambda^\prime\bm  E^\prime\bm  E}{n^{3/2}T}-\frac{\bm\Lambda^\prime\bm\Gamma^ e}{n^{3/2}} \r\Vert ^2\r]&\le 
\E\l[\l\Vert\frac{\bm\Lambda^\prime\bm  E^\prime\bm  E}{n^{3/2}T}-\frac{\bm\Lambda^\prime\bm\Gamma^ e}{n^{3/2}} \r\Vert ^2_F\r]\nn\\
&= \sum_{k=1}^r\sum_{j=1}^n\E\l[\l\vert\frac 1{n^{3/2}T}
\sum_{i=1}^n  \sum_{t=1}^T \l\{\lambda_{ik} e_{it} e_{jt}-\lambda_{ik}\E[ e_{it} e_{jt}]\r\}
\r\vert^2
\r]\nn\\
&\le \frac{rM_\Lambda n}{n^2T}
\max_{j=1,\ldots, n}
\E\l[\l\vert\frac 1{\sqrt{nT}}
\sum_{i=1}^n  \sum_{t=1}^T \l\{ e_{it} e_{jt}-\E[ e_{it} e_{jt}]\r\}
\r\vert^2
\r]\le \frac{rM_\Lambda K_ e}{nT},\label{eq:rifacciamo}
\end{align}
since $K_ e$ is independent of $j$. By using \eqref{eq:rifacciamo} into \eqref{eq:rifacciamo2},
we prove part (iii). 

Part (iv) follows from Lemma \ref{lem:FTLN}(i) since
\[
\l\Vert\frac{\bm \Lambda^\prime\bm \Lambda}{n} \r\Vert\le \l\Vert\frac{\bm \Lambda}{\sqrt n} \r\Vert^2\le M_\Lambda^2.
\]
Likewise, part (v) follows from Lemma \ref{lem:FTLN}(ii) since
\[
\l\Vert\frac{\bm F^\prime\bm F}{T} \r\Vert\le \l\Vert\frac{\bm F}{\sqrt T} \r\Vert^2\le r M_F.
\]
This completes the proof. $\Box$

\begin{lem}\label{lem:ranghi} 
Under Assumption  \ref{ass:common}, as $n\to\infty$,
 $\l\Vert \l(\frac{\bm\Lambda^\prime\bm\Lambda}{n}\r)^{-1}\r\Vert = O(1)$.
\end{lem}

\noindent
\textbf{Proof.}
First, by Assumption \ref{ass:common}(a) and Weyl's inequality \citep[Theorem 1]{MK04},
\beq\label{eq:scemopiu}
\lim_{n\to\infty}\l\vert \mu_r\l(\frac{\bm \Lambda^\prime \bm \Lambda}n\r)- \mu_r(\bm\Sigma_\Lambda)\r\vert\le \lim_{n\to\infty} \l\Vert \frac{\bm\Lambda^\prime\bm\Lambda}{n} - \bm\Sigma_\Lambda\r\Vert =0.
\eeq
 Then, using a first-order Taylor expansion and by \eqref{eq:scemopiu},
\begin{align}
\l\Vert\l(\frac{\bm \Lambda^\prime\bm \Lambda}{n} \r)^{-1}\r\Vert & = \frac 1{\mu_r(n^{-1}\bm \Lambda^\prime \bm \Lambda)}=  \frac 1{1+\frac{\mu_r(n^{-1}\bm \Lambda^\prime \bm \Lambda)- \mu_r(\bm\Sigma_\Lambda)}{\mu_r(\bm\Sigma_\Lambda)}}\frac 1{\mu_r(\bm\Sigma_\Lambda)}\nonumber\\
&= \frac 1{\mu_r(\bm\Sigma_\Lambda)} \l\{1 - \frac{\mu_r(n^{-1}\bm \Lambda^\prime \bm \Lambda)- \mu_r(\bm\Sigma_\Lambda)}{\mu_r(\bm\Sigma_\Lambda)} \r\} + O\l(\vert \mu_r(n^{-1}\bm \Lambda^\prime \bm \Lambda)- \mu_r(\bm\Sigma_\Lambda)\vert^2\r)\nn\\
&\le\frac 1{\mu_r(\bm\Sigma_\Lambda)}+\frac{ \mu_r(\bm\Sigma_\Lambda)-\mu_r(n^{-1}\bm \Lambda^\prime \bm \Lambda)}{\mu_r^2(\bm\Sigma_\Lambda)}+ O\l(\vert \mu_r(n^{-1}\bm \Lambda^\prime \bm \Lambda)- \mu_r(\bm\Sigma_\Lambda)\vert^2\r)\nn\\
&\le\frac 1{c_\Lambda}+\frac{ \l\vert \mu_r(\bm\Sigma_\Lambda)-\mu_r(n^{-1}\bm \Lambda^\prime \bm \Lambda)\r\vert }{c_\Lambda^2}+ O\l(\vert \mu_r(n^{-1}\bm \Lambda^\prime \bm \Lambda)- \mu_r(\bm\Sigma_\Lambda)\vert^2\r)\nn\\
&\le \frac 1{c_\Lambda}+o(1)=O(1),\nn
\end{align}
for some finite positive real $c_\Lambda$ independent of $n$, as implied by Assumption \ref{ass:common}(a).
This completes the proof. $\Box$
\begin{lem}\label{lem:covarianzeF}
$\,$
\begin{compactenum}[(i)]
\item Under either Assumption \ref{ass:Wold}, or \ref{ass:Hannan}, or \ref{ass:Wu} in Appendix \ref{sec:hannan}, for all $T\in\mathbb N$, $\sqrt T\l\Vert\frac{\bm F^\prime\bm F}{T} -\bm\Gamma^F\r\Vert = O_{\mathrm {m.s.}}(1)$;
\item Under Assumption \ref{ass:idio}, for all $n,T\in\mathbb N$, $\sqrt T \l\Vert\frac{\bm  E^\prime\bm  E}{nT}-\frac{\bm\Gamma^ e}{n} \r\Vert= O_{\mathrm {m.s.}}(1)$.
\end{compactenum}
\end{lem}

\noindent
\textbf{Proof.} 
Because of Assumption \ref{ass:Wold}, for any $i,j=1,\ldots, r$, letting $\gamma_{ij}^F$ be the $(i,j)$th entry of $\bm\Gamma^F$,
\begin{align}
\E&\l[\l\vert\frac 1T \sum_{t=1}^T \l\{F_{it}F_{jt}-\gamma_{ij}^F\r\} \r\vert^2\r]= \frac 1{T^2} \sum_{t,s=1}^T \l\{\E[F_{it}F_{jt}F_{is}F_{js}]- \E[F_{it}F_{jt}]^2\r\}\nn\\
&= \frac 1{T^2} \sum_{t,s=1}^T\l\{\sum_{k_1,k_2,k_3,k_4=0}^\infty\sum_{h_1,h_2,h_3,h_4=1}^r C_{k_1,ih_1} C_{k_2,jh_2}C_{k_3,ih_3} C_{k_4,jh_4}
\E\l[ u_{h_1,t-k_1} u_{h_2,t-k_2} u_{h_3,s-k_3} u_{h_4,s-k_4}
\r]\r.\nn\\
&\;\;\l. -\l(\sum_{k_1,k_2=0}^\infty\sum_{h_1,h_2=1}^r C_{k_1,ih_1} C_{k_2,jh_2} \E\l[u_{h_1,t-k_1} u_{h_2,t-k_2} \r]
\r)^2\r\}\nn\\
&\le  \frac {M_C^4}{T^2}  \l\{\sum_{t=1}^T\sum_{h=1}^r \E[u_{ht}^4]+\sum_{t,s=1}^T\sum_{h_1=1}^r \E[u_{h_1t}^2]\sum_{h_3=1}^r \E[u_{h_3s}^2]-\sum_{t,s=1}^T \l(\sum_{h=1}^r \E[u_{h,t} ^2]\r)^2\r\}\le \frac {M_C^4 r K_u}{T}. \nn
\end{align}
Then,
\begin{align}
\E\l[\l\Vert\frac{\bm F^\prime\bm F}{T}-\bm\Gamma^F\r\Vert^2\r]&=
\E\l[\l\Vert \frac 1T\sum_{t=1}^T \mbf F_t\mbf F_t^\prime-\bm\Gamma^F\r\Vert^2\r]=\sum_{i,j=1}^r \E\l[\l\vert\frac 1T \sum_{t=1}^T \l\{F_{it}F_{jt}-\gamma_{ij}^F\r\} \r\vert^2\r]\nn\\
&\le r^2\max_{i,j=1,\ldots, r}\E\l[\l\vert\frac 1T \sum_{t=1}^T \l\{F_{it}F_{jt}-\gamma_{ij}^F\r\} \r\vert^2\r]\le \frac {r^3 M_C^4  K_u}{T},\nn
\end{align}
since $M_C$ and $K_u$ do not depend on $i$ and $j$. This proves part (i) under Assumption \ref{ass:Wold}.

Alternatively, because of Assumption \ref{ass:Hannan},
\begin{align}
\E\l[\l\Vert\frac{\bm F^\prime\bm F}{T}-\bm\Gamma^F\r\Vert^2\r]&=
\E\l[\l\Vert \frac 1T\sum_{t=1}^T \mbf F_t\mbf F_t^\prime-\bm\Gamma^F\r\Vert^2\r]\le 
\E\l[\l\Vert \frac 1T\sum_{t=1}^T \l\{\mbf F_t\mbf F_t^\prime-\bm\Gamma^F\r\}\r\Vert^2_F\r]\nn\\
&=\frac 1{T^2} \sum_{j=1}^r\sum_{k=1}^r\E\l[\l(\sum_{t=1}^T \l\{F_{jt}F_{kt}-\E[F_{jt}F_{kt}] \r\}\r)^2\r]\nn\\
&=\frac {r^2}{T^2} \max_{j,k=1,\ldots,r}\sum_{t=1}^T\sum_{s=1}^T
\E\l[\l\{F_{jt}F_{kt}-\E[F_{jt}F_{kt}] \r\}\l\{F_{js}F_{ks}-\E[F_{js}F_{ks}] \r\}\r]\nn\\
&=\frac {r^2}{T^2} \max_{j,k=1,\ldots,r}\sum_{t=1}^T\sum_{s=1}^T
\l\{\E\l[F_{jt}F_{kt}F_{js}F_{ks}\r]-\E\l[F_{jt}F_{kt}\r]\E\l[F_{js}F_{ks}\r]\r\}\nn\\
&\le \frac {r^2}{T^2} \max_{j,k=1,\ldots,r}\l\{\sum_{t=1}^T\sum_{s=1}^T
\l\vert \E\l[F_{jt}F_{kt}F_{js}F_{ks}\r]\r\vert +\sum_{t=1}^T\sum_{s=1}^T\l\vert \E\l[F_{jt}F_{kt}\r]\E\l[F_{js}F_{ks}\r]\r\vert\r\}\nn\\
&\le \frac {r^2C_F}{T},\nn
\end{align}
since $C_F$ is independent of $j$ and $k$. This proves part (i) under Assumption \ref{ass:Hannan}.

Let us now consider the case in which we make Assumption \ref{ass:Wu}. For any $i,j=1,\ldots, r$,
\begin{align}
\frac 1T \sum_{t=1}^T \l\{F_{it}F_{jt}-\gamma_{ij}^F\r\} &  =
 \sum_{k,k'=0}^\infty\sum_{\ell,\ell'=1}^r C_{k,i\ell} C_{k',j\ell'} \l(
  \frac 1T \sum_{t=1}^T \l\{u_{\ell,t-k}u_{\ell',t-k'}-
  \E\l[u_{\ell,t-k}u_{\ell',t-k'}\r]\r\}
 \r).\label{eq:CCuu}
\end{align}
For simplicity let
\[
\mathcal C_{kk',\ell\ell',T}=
  \frac 1T \sum_{t=1}^T \l\{u_{\ell,t-k}u_{\ell',t-k'}-
  \E\l[u_{\ell,t-k}u_{\ell',t-k'}\r]\r\}
\]
Then, from \eqref{eq:CCuu}, because of Assumption \ref{ass:Wu}(P1), and Minkowski's inequality
\begin{align}
\l\{\E\l[\l\vert \frac 1T \sum_{t=1}^T \l\{F_{it}F_{jt}-\gamma_{ij}^F\r\} \r\vert^2\r]\r\}^{1/2}&=
\l\{
\E\l[\l\vert
 \sum_{k,k'=0}^\infty\sum_{\ell,\ell'=1}^r C_{k,i\ell} C_{k',j\ell'}\mathcal C_{kk',\ell\ell',T}
\r\vert^2\r]
\r\}^{1/2}\nn\\
&\le  \sum_{k,k'=0}^\infty\sum_{\ell,\ell'=1}^r \l\{\E\l[\l\vert
C_{k,i\ell} C_{k',j\ell'}\mathcal C_{kk',\ell\ell',T}
\r\vert^2
\r]\r\}^{1/2}\nn\\
&=\sum_{k,k'=0}^\infty\sum_{\ell,\ell'=1}^r\l\vert C_{k,i\ell}\r\vert \,\l\vert C_{k',j\ell'}\r\vert \l\{\E\l[\l\vert
\mathcal C_{kk',\ell\ell',T}
\r\vert^2
\r]\r\}^{1/2}\nn\\
&\le M_C^2 r^2 \sup_{k,k'\in\mathbb Z^+}\max_{\ell,\ell'=1,\ldots, r} \l\{\E\l[\l\vert
\mathcal C_{kk',\ell\ell',T}
\r\vert^2
\r]\r\}^{1/2}.\label{eq:fufix}
\end{align}
Now, since as explained in the text we can always choose $\alpha>\frac 12-\frac 2q$, 
then, from \citet[Proposition 3.3]{zhang2021},  for any $\ell,\ell'=1,\ldots, r$, $k,k'\in\mathbb Z^+$,
 and $z>0$,
(see also \citealp[Lemma C.4 and Remark C.1]{BCO22})
\begin{align}
\mathrm P\l( \l\vert \mathcal C_{kk',\ell\ell',T}\r\vert>z\r)&\le \frac{C_{q,\alpha}T (\Phi_{q,\alpha}^u)^q}{(Tz)^{q/2}}+ C_0\exp\l(-\frac{Tz^2}{C_\alpha(\Phi_{4,\alpha}^{u})^4}\r)\nn\\
&\le 
\frac{C_{q,\alpha}T M_{u,q}^q}{(Tz)^{q/2}}+ C_0\exp\l(-\frac{Tz^2}{C_\alpha M_{u,4}^4}\r)\nn\\
&= \frac{C_{1,q,\alpha}}{z^{q/2} T^{q/2-1}}+C_0 \exp\l(-\frac{Tz^2}{C_{2,\alpha}}\r)
,\label{eq:ZWapp}
\end{align}
where $q>4$, $C$ is a finite positive real independent of $T$, $i$, $j$, $\alpha$, and $q$, while $M_{u,q}$, $C_{q,\alpha}$, $C_\alpha$, $C_{1,q,\alpha}$, and $C_{2,\alpha}$ are finite positive reals depending only on their subscripts. So, as expected, from \eqref{eq:ZWapp}, $\vert\mathcal C_{kk',\ell\ell',T}\vert=O_{\mathrm P}\l(\max\l(T^{-1/2}, T^{2/q-1}\r)\r)=O_{\mathrm P}(T^{-1/2})$ since $q>4$.

Moreover, from \eqref{eq:ZWapp} it follows that
\begin{align}
\E\l[\l\vert \mathcal C_{kk',\ell\ell',T}\r\vert^2\r]&=\E\l[\int_0^\infty\mathbb I\l(\l\vert \mathcal C_{kk',\ell\ell',T}\r\vert^2>y\r) \mathrm dy\r]=\int_0^\infty \mathrm P\l(\l\vert \mathcal C_{kk',\ell\ell',T}\r\vert^2>y\r)\mathrm dy\nn\\
&= \int_0^{r_T} \mathrm P\l(\l\vert \mathcal C_{kk',\ell\ell',T}\r\vert^2>y\r)\mathrm dy+
\int_{r_T}^\infty \mathrm P\l(\l\vert \mathcal C_{kk',\ell\ell',T}\r\vert^2>y\r)\mathrm dy\nn\\
&= \int_0^{r_T} \mathrm P\l(\l\vert \mathcal C_{kk',\ell\ell',T}\r\vert>\sqrt y\r)\mathrm dy+
\int_{r_T}^\infty \mathrm P\l(\l\vert \mathcal C_{kk',\ell\ell',T}\r\vert>\sqrt y\r)\mathrm dy\nn\\
&\le r_T + \frac{C_{1,q,\alpha}}{T^{q/2-1}}\int_{r_T^2}^\infty \frac{1}{y^{q/4} }\mathrm d y
+C_0  \int_{r_T}^\infty \exp\l(-\frac{Ty}{C_{2,\alpha}}\r)\mathrm d y \nn\\
&= r_T- \frac{C_{1,q,\alpha}}{(q/4-1)T^{q/2-1}}\l.\frac{1}{y^{q/4-1}}\r\vert_{r_T}^\infty
- \frac{C_0 C_{2,\alpha}}{T} \l. \exp\l(-\frac{Ty}{C_{2,\alpha}}\r)\r\vert_{r_T}^\infty\nn\\
&= r_T+ \frac{C_{1,q,\alpha}}{(q/4-1)T^{q/2-1}r_T^{q/4-1}}
+ \frac{C_0 C_{2,\alpha}}{T}  \exp\l(-\frac{Tr_T}{C_{2,\alpha}}\r).\label{eq:ZWapp2}
\end{align}
Then, by choosing $r_T=\frac 1T$ from \eqref{eq:ZWapp2} and since $q>4$, we get
\begin{align}
\E\l[\l\vert \mathcal C_{kk',\ell\ell',T}\r\vert^2\r] &\le \frac 1T +  \frac{C_{1,q,\alpha}}{(q/4-1)T^{q/2-1}T^{1-q/4}}+  \frac{C_0 C_{2,\alpha}}{T}  \exp\l(-\frac{1}{C_{2,\alpha}}\r)\nn\\
&= \frac 1T\l(1+C_0 C_{2,\alpha}  \exp\l(-\frac{1}{C_{2,\alpha}}\r) \r)+\frac{C_{1,q,\alpha}}{(q/4-1)T^{q/4}}\nn\\
&\le\frac 1T\l(1+C_0 C_{2,\alpha}  \exp\l(-\frac{1}{C_{2,\alpha}}\r) +\frac{C_{1,q,\alpha}}{(q/4-1)}\r) \nn\\
&\le \frac{C_{u,\alpha,q}}{T}, \;\text {say,}\label{eq:ZWapp3}
\end{align}
where $C_{u,\alpha,q}$ is independent of $k$, $k'$, $\ell$, and $\ell'$. 

By substituting \eqref{eq:ZWapp3} into \eqref{eq:fufix}, we get
\beq
\E\l[\l\vert \frac 1T \sum_{t=1}^T \l\{F_{it}F_{jt}-\gamma_{ij}^F\r\} \r\vert^2\r]\le \frac{M_C^4 r^4 C_{u,\alpha,q}}{T} .\label{eq:CCuu3}
\eeq
Finally, from \eqref{eq:CCuu3}
\begin{align}
\E\l[\l\Vert\frac{\bm F^\prime\bm F}{T}-\bm\Gamma^F\r\Vert^2\r]&=
\E\l[\l\Vert \frac 1T\sum_{t=1}^T \mbf F_t\mbf F_t^\prime-\bm\Gamma^F\r\Vert^2\r]=\sum_{i,j=1}^r \E\l[\l\vert\frac 1T \sum_{t=1}^T \l\{F_{it}F_{jt}-\gamma_{ij}^F\r\} \r\vert^2\r]\nn\\
&\le r^2\max_{i,j=1,\ldots, r}\E\l[\l\vert\frac 1T \sum_{t=1}^T \l\{F_{it}F_{jt}-\gamma_{ij}^F\r\} \r\vert^2\r]\le \frac{M_C^4 r^6 C_{u,\alpha,q}}{T},\nn
\end{align}
since $M_C$ and $C_{u,\alpha,q}$ do not depend on $i$ and $j$. This proves part (i) under Assumption \ref{ass:Wu}.

For part (ii),  by Assumption \ref{ass:idio}(c), letting $\gamma_{ij}^ e$ be the $(i,j)$th entry of $\bm\Gamma^ e$,
\begin{align}
\E\l[\l\Vert\frac{\bm  E^\prime\bm  E}{nT}-\frac{\bm\Gamma^ e}{n} \r\Vert ^2\r]&=\E\l[\l\Vert \frac 1{nT} \sum_{t=1}^T\bm e_t\bm e_t^\prime -\frac{\bm\Gamma^ e}n \r\Vert^2\r]\le  \E\l[\l\Vert \frac 1{nT} \sum_{t=1}^T\l\{\bm e_t\bm e_t^\prime -\bm\Gamma^ e\r\} \r\Vert^2_F\r]\nn\\
&=\frac 1{n^2T^2}\sum_{i,j=1}^n\E\l[\l(\sum_{t=1}^T \l\{ e_{it} e_{jt}-\gamma_{ij}^ e\r\}\r)^2\r]\nn\\
&\le \max_{i,j=1,\ldots, n}\frac 1{T^2} \E\l[\l(\sum_{t=1}^T \l\{ e_{it} e_{jt}-\gamma_{ij}^ e\r\}\r)^2\r]
\le \frac {K_ e}{T},\nn
\end{align}
since $K_ e$ is independent of $i$ and $j$. This completes the proof. $\Box$

\begin{lem}\label{lem:FFinv}
As $T\to \infty$,
\begin{compactenum}
\item [(i-a)] under Assumption \ref{ass:common},  
$\l\Vert\l(\frac{\bm F^\prime\bm F}{T} \r)^{-1}\r\Vert = O_{\mathrm {P}}(1)$;
\item [(i-b)] under either Assumption \ref{ass:Wold}, or \ref{ass:Hannan}, or \ref{ass:Wu} in Appendix \ref{sec:hannan}, $\l\Vert\l(\frac{\bm F^\prime\bm F}{T} \r)^{-1}\r\Vert = O_{\mathrm {m.s.}}(1)$.
\end{compactenum}
\end{lem}

\noindent
\textbf{Proof.}
First, by Assumption \ref{ass:common}(c-ii) and Weyl's inequality \citep[Theorem 1]{MK04},
\beq\label{eq:midlands}
\l\vert \mu_r\l(\frac{\bm F^\prime \bm F}T\r)- \mu_r(\bm\Gamma^F)\r\vert \le \l\Vert \frac{\bm F^\prime \bm F}T-\bm\Gamma^F \r\Vert = O_{\mathrm P}\l(\frac 1{\sqrt T}\r).
\eeq
Then, by using a first-order Taylor expansion and by \eqref{eq:midlands},
\begin{align}
\l\Vert\l(\frac{\bm F^\prime\bm F}{T} \r)^{-1}\r\Vert & = \frac 1{\mu_r(T^{-1}\bm F^\prime \bm F)}=  \frac 1{1+\frac{\mu_r(T^{-1}\bm F^\prime \bm F)- \mu_r(\bm\Gamma^F)}{\mu_r(\bm\Gamma^F)}}\frac 1{\mu_r(\bm\Gamma^F)}\nonumber\\
&= \frac 1{\mu_r(\bm\Gamma^F)} \l\{1 - \frac{\mu_r(T^{-1}\bm F^\prime \bm F)- \mu_r(\bm\Gamma^F)}{\mu_r(\bm\Gamma^F)} \r\} + O_{\mathrm P}\l(\frac 1T\r)\nn\\
&\le\frac 1{\mu_r(\bm\Gamma^F)}+\frac{ \mu_r(\bm\Gamma^F)-\mu_r(T^{-1}\bm F^\prime \bm F)}{\mu_r^2(\bm\Gamma^F)}+ O_{\mathrm P}\l(\frac 1T\r)\nn\\
&\le\frac 1{m_F}+\frac{ \l\vert \mu_r(\bm\Gamma^F)-\mu_r(T^{-1}\bm F^\prime \bm F)\r\vert }{m_F^2}+ O_{\mathrm P}\l(\frac 1T\r)\nn\\
&\le\frac 1{m_F}+O_{\mathrm P}\l(\frac 1{\sqrt T}\r)=O_{\mathrm P}(1),\nn
\end{align}
for some finite positive real $m_F$ independent of $T$, as implied by Assumption \ref{ass:common}(b). This proves part (i-a). As for part (i-b) it suffices to note that, in this case, \eqref{eq:midlands} holds in mean-square because of Lemma \ref{lem:covarianzeF}(i).
This completes the proof. $\Box$

\begin{lem}\label{lem:covarianze}
Under Assumptions \ref{ass:common} through \ref{ass:eval} and Assumption~\ref{eq:ortho}:
\begin{compactenum}
\item [(i-a)] if also Assumption \ref{ass:ind} holds and Assumption \ref{ass:common}(c-ii) holds with rate $\sqrt T$, then, for all $n,T\in\mathbb N$,  $\frac {\sqrt T} n\Vert \wh{\bm\Gamma}^x-\bm\Gamma^x\Vert = O_{\mathrm {P}}(1)$;
\item [(i-b)] if either Assumption \ref{ass:Wold} or \ref{ass:Hannan} or \ref{ass:Wu} in Appendix \ref{sec:hannan} hold, then, for all $n,T\in\mathbb N$,  $\frac {\sqrt T} n\Vert \wh{\bm\Gamma}^x-\bm\Gamma^x\Vert = O_{\mathrm {m.s.}}(1)$;
\item [(ii)] if also either the conditions in (i-a) or in (i-b) hold, then,
for all $n,T\in\mathbb N$,  $\frac {\min(n,\sqrt T)} n\Vert \wh{\bm\Gamma}^x-\bm\Gamma^{C}\Vert = O_{\mathrm {m.s.}}(1)$;
\item [(iii)] if also either the conditions in (i-a) or in (i-b) hold, then, for all $n,T\in\mathbb N$,  $\frac  {\min(n,\sqrt T)} n \Vert \wh{\mbf M}^x-\mbf M^{C}\Vert=O_{\mathrm {m.s.}}(1)$;
\item [(iv)] if also either the conditions in (i-a) or in (i-b) hold, then, for all $T\in\mathbb N$ and as $n\to\infty$, ${\min(n,\sqrt T)} \Vert\wh{\mbf V}^x-\mbf V^{C}\mbf J\Vert = O_{\mathrm {m.s.}}(1)$, where $\mbf J$ is $r\times r$ diagonal with entries $\pm 1$ and independent of $n$.
\end{compactenum}
\end{lem}

\noindent
\textbf{Proof.} 
For part (i-a), from Assumption \ref{ass:common}(c-ii), and Lemmas \ref{lem:covarianzeF}(ii), and \ref{lem:FTLN}(i): 
\begin{align}
\l\Vert\frac  1n\l( \wh{\bm\Gamma}^x-\bm\Gamma^x\r)\r\Vert&\le\l\Vert \frac 1n\l\{ \bm\Lambda\l( \frac 1T\sum_{t=1}^T \mbf F_t\mbf F_t^\prime-\bm\Gamma^F\r)\bm\Lambda^\prime 
 + \frac 1T\sum_{t=1}^T\bm e_t\bm e_t^\prime-\bm\Gamma^ e\r\}\r\Vert
+\l\Vert\frac 2{nT}\sum_{t=1}^T\bm e_t\mbf F_t^\prime \r\Vert 
 \nn\\
 &\le \l\Vert\frac{\bm\Lambda}{\sqrt n}\r\Vert^2\,\l\Vert \frac 1T\sum_{t=1}^T \mbf F_t\mbf F_t^\prime-\bm\Gamma^F\r\Vert + 
 \l\Vert \frac 1n\l( \frac 1T\sum_{t=1}^T\bm e_t\bm e_t^\prime-\bm\Gamma^ e\r)\r\Vert +\l\Vert\frac 2{nT}\sum_{t=1}^T\bm e_t\mbf F_t^\prime \r\Vert \nn\\
 & = O_{\mathrm P}\l(\frac 1{\sqrt T}\r)+O_{\mathrm P}\l(\frac 1{\sqrt {nT}}\r). \nn
\end{align}
where we also used  either Lemma \ref{lem:LLN}(i), if Assumption \ref{ass:ind} holds, or the moment conditions \eqref{eq:oslo2}.

For part (i-b), if we make either Assumption \ref{ass:Wold} or \ref{ass:Hannan} or \ref{ass:Wu},  from Lemma \ref{lem:covarianzeF}(i), \ref{lem:covarianzeF}(ii) and \ref{lem:FTLN}(i) and the $C_r$-inequality with $r=2$,
\begin{align}
\E\l[\l\Vert\frac  1n\l( \wh{\bm\Gamma}^x-\bm\Gamma^x\r)\r\Vert^2\r]
& =
 \E\l[\l\Vert \frac 1n\l\{ \bm\Lambda\l( \frac 1T\sum_{t=1}^T \mbf F_t\mbf F_t^\prime-\bm\Gamma^F\r)\bm\Lambda^\prime 
 + \frac 1T\sum_{t=1}^T\bm e_t\bm e_t^\prime-\bm\Gamma^ e\r\}\r\Vert^2
 \r]\nn\\
& \le 
2\E\l[\l\Vert \frac 1n\l\{ \bm\Lambda\l( \frac 1T\sum_{t=1}^T \mbf F_t\mbf F_t^\prime-\bm\Gamma^F\r)\bm\Lambda^\prime 
\r\}\r\Vert^2\r]
 +2\E\l[\l\Vert \frac 1n\l( \frac 1T\sum_{t=1}^T\bm e_t\bm e_t^\prime-\bm\Gamma^ e\r)\r\Vert^2
 \r]\nn\\
&\le 2 \l\Vert\frac{\bm\Lambda}{\sqrt n}\r\Vert^4 \, \E\l[\l\Vert \frac 1T\sum_{t=1}^T \mbf F_t\mbf F_t^\prime-\bm\Gamma^F\r\Vert^2\r]+2 \E\l[\l\Vert \frac 1n\l( \frac 1T\sum_{t=1}^T\bm e_t\bm e_t^\prime-\bm\Gamma^ e\r)\r\Vert^2
 \r]\nn\\
&\le 2 \frac{M_\Lambda^4 K_F+K_ e}{T},\nn
\end{align}
where $K_F=r^3M_C^4K_u$ under Assumption \ref{ass:Wold}, or $K_F=r^2 C_F$ under Assumption \ref{ass:Hannan}, or $K_F=M_C^4 r^6 C_{u,\alpha,q}$ under Assumption \ref{ass:Wu}. This proves part (i).

Part (ii) follows from
\begin{align}
\E\l[\l\Vert\frac  1n\l( \wh{\bm\Gamma}^x-\bm\Gamma^{C}\r)\r\Vert^2\r] &\le \E\l[\l\Vert\frac  1n\l( \wh{\bm\Gamma}^x-\bm\Gamma^x\r)\r\Vert^2\r] +\l\Vert\frac  {\bm\Gamma^ e}n\r\Vert^2 \le  \frac{M_\Lambda^4K_F+K_ e}{T}+
\frac{\mu_1^ e}{n^2}\le \frac{M_\Lambda^4K_F+K_ e}{T}+\frac{M_{2 e}}{n^2}\nn\\
&\le M_1\max \l(\frac 1{n^2},\frac 1T\r),\;\text{say,}\nn
\end{align}
 because of part (i) and Lemma \ref{lem:Gxi}(v) and where $M_1$ is a finite positive real independent of $n$ and $T$.
 
For part (iii), for any $j=1,\ldots, r$, because of Weyl's inequality \citep[Theorem 1]{MK04} and part (ii), it holds that
\beq\label{eq:weyl}
\vert\wh\mu_j^x-\mu_j^{C} \vert\le \mu_1(\wh{\bm\Gamma}^x-\bm\Gamma^{C})
=\l\Vert\wh{\bm\Gamma}^x-\bm\Gamma^{C}\r\Vert.
\eeq
Hence, from \eqref{eq:weyl} and part (ii),
\begin{align}
\E\l[ \l\Vert\frac 1n\l( \wh{\mbf M}^x-\mbf M^{C}\r)\r\Vert^2\r] &\le  
\E\l[ \l\Vert\frac 1n\l( \wh{\mbf M}^x-\mbf M^{C}\r)\r\Vert^2_F\r] =
\frac 1{n^2} \sum_{j=1}^r \E\l[\l(\wh\mu_j^x-\mu_j^{C} \r)^2\r]\nn\\
&\le \frac{r}{n^2} \E\l[\l(\wh\mu_1^x-\mu_1^{C} \r)^2\r]\le r\E\l[\l\Vert\frac 1n\l(\wh{\bm\Gamma}^x-\bm\Gamma^{C}\r)\r\Vert^2\r]\le \frac{rM_\Lambda^4K_F+K_ e}{T}+\frac{rM_ e}{n^2}\nn\\
&\le rM_1\max\l(\frac 1{n^2},\frac 1T\r).
\end{align}
This proves part (iii).

For part (iv), start by noticing that, by Assumption \ref{ass:eval}, $\lim_{n\to\infty}n^{-1} \mu_j^C> \lim_{n\to\infty}n^{-1} \mu_{j+1}^C$ for all $j=1,\ldots, r-1$.
Then, we can apply \citet[Corollary 1]{yu15}, which is a special case of Davis-Kahan Theorem, as $n\to\infty$. So from part (ii) and Lemma \ref{lem:Gxi}(iv) 
there exists an $r\times r$ diagonal matrix $\mbf J$  with entries $\pm 1$ such that
%
%
%
%
\begin{align}
\lim_{n\to\infty} \min(n^2,T) \E\l[\Vert\wh{\mbf V}^x-\mbf V^{C}\mbf J\Vert^2\r]&\le \lim_{n\to\infty} \frac
{\min(n^2,T)  2^3 \frac r{n^2} \E\l[\l\Vert\wh{\bm\Gamma}^x-\bm\Gamma^{C} \r\Vert^2\r]}
{\frac 1{n^2}\l\{\min(\vert \mu_{0}^{C}-\mu_{1}^{C} \vert,\vert \mu_{r}^{C}-\mu_{r+1}^{C} \vert)\r\}^2}\nn\\
& \le \frac{8rM_1}{\underline C_r^2}= M_2,\;\text{say,}
\label{eq:DK}
\end{align}
where $\mu_0^{C}=\infty$ and $M_2$ is a finite positive real independent of $n$ and $T$, and we used also the fact that $\mu_{r+1}^{C} =0$.
This proves part (iv). This completes the proof.  $\Box$

\begin{lem}\label{lem:covarianzerighe}
Let $\bm\varepsilon_i$ be the $n$-dimensional vector with one in entry $i$ and zero elsewhere. Under Assumptions \ref{ass:common} through \ref{ass:eval} and Assumption~\ref{eq:ortho}, and either Assumption \ref{ass:Wold} or \ref{ass:Hannan} or \ref{ass:Wu} in Appendix \ref{sec:hannan}: 
\begin{compactenum}[(i)]
\item  for all $i=1,\ldots, n$ and all $T\in\mathbb N$, as $n\to\infty$,
$\frac {\min(\sqrt n,\sqrt T)}{\sqrt n} \Vert\bm\varepsilon^\prime_i (\wh{\bm\Gamma}^x-\bm\Gamma^{C})\Vert = O_{\mathrm {m.s.}}(1)$;
\item  for all $i=1,\ldots, n$, as $n\to\infty$, $\sqrt n\Vert\mbf v_i^{C}\Vert = O(1)$;
\item for all $i=1,\ldots, n$ and all $T\in\mathbb N$, as $n\to\infty$, ${\min(\sqrt n,\sqrt T)} \sqrt n \Vert\wh{\mbf v}^{x\prime}_i-\mbf v^{{C}\prime}\mbf J\Vert = O_{\mathrm {m.s.}}(1)$.
\end{compactenum}
\end{lem}

\noindent
\textbf{Proof.} First notice that
\begin{align}
\max_{i=1,\ldots, n}\E\l[\l\Vert \frac 1{\sqrt n}\bm\varepsilon^\prime_i\l( \sum_{t=1}^T\bm e_t\bm e_t^\prime -\bm\Gamma^ e\r) \r\Vert^2\r]
&\le\max_{i=1,\ldots, n}\frac 1{n}\sum_{j=1}^n\E\l[\l(\frac 1T\sum_{t=1}^T  e_{it} e_{jt}-[\bm\Gamma^ e]_{ij}\r)^2\r]\nn\\
&\le \max_{i,j=1,\ldots, n} \E\l[\l(\frac 1T\sum_{t=1}^T  e_{it} e_{jt}-[\bm\Gamma^ e]_{ij}\r)^2\r]
\le \frac {K_ e}{T}\label{eq:Gammaxiappriga},
\end{align}
since $K_ e$ is independent of $i$ and $j$. Therefore, from Lemmas \ref{lem:FTLN}(i) and \ref{lem:covarianzeF}(i), and using \eqref{eq:Gammaxiappriga}, and the $C_r$-inequality with $r=2$,
\begin{align}
\max_{i=1,\ldots, n} \E\l[\l\Vert\frac  1{\sqrt n}\bm\varepsilon^\prime_i\l( \wh{\bm\Gamma}^x-\bm\Gamma^x\r)\r\Vert^2\r]
=&\,
\max_{i=1,\ldots, n} \E\l[\l\Vert \frac 1{\sqrt n}\l\{ \bm\lambda_i^\prime \l( \frac 1T\sum_{t=1}^T \mbf F_t\mbf F_t^\prime-\bm\Gamma^F\r)\bm\Lambda^\prime 
 +\bm\varepsilon^\prime_i\l( \frac 1T\sum_{t=1}^T\bm e_t\bm e_t^\prime-\bm\Gamma^ e\r)\r\}\r\Vert^2
 \r]\nn\\
\le&\,2\max_{i=1,\ldots, n}\Vert \bm\lambda_i\Vert^2\, \l\Vert\frac{\bm\Lambda}{\sqrt n}\r\Vert^2  \, \E\l[\l\Vert \frac 1T\sum_{t=1}^T \mbf F_t\mbf F_t^\prime-\bm\Gamma^F\r\Vert^2\r]\nn\\
&+2\max_{i=1,\ldots, n}\E\l[\l\Vert \frac 1{\sqrt n}\bm\varepsilon^\prime_i\l( \frac 1T\sum_{t=1}^T\bm e_t\bm e_t^\prime-\bm\Gamma^ e\r)\r\Vert^2
 \r]\nn\\
\le&\, 2\frac{M_\Lambda^4 K_F+K_ e}{T},\label{eq:rigacov}
\end{align}
since $M_\Lambda$ and $K_F$ are independent of $i$. Then, following the same arguments as Lemma \ref{lem:covarianze}(ii), because of \eqref{eq:rigacov} and Lemma \ref{lem:Gxi}(v):
\begin{align}
\max_{i=1,\ldots, n}\E\l[\l\Vert\frac  1{\sqrt n}\bm\varepsilon^\prime_i\l( \wh{\bm\Gamma}^x-\bm\Gamma^{C}\r)\r\Vert^2\r]& \le\max_{i=1,\ldots, n} \E\l[\l\Vert\frac  1{\sqrt n}\bm\varepsilon^\prime_i\l( \wh{\bm\Gamma}^x-\bm\Gamma^x\r)\r\Vert^2\r] +\max_{i=1,\ldots, n}\l\Vert\bm\varepsilon^\prime_i\frac  {\bm\Gamma^ e}{\sqrt n}\r\Vert^2\nn\\
 &\le  \frac{M_\Lambda^4K_F+K_ e}{T}+\max_{i=1,\ldots, n}  \Vert \bm\varepsilon_i\Vert^2\, \l\Vert\frac{\bm\Gamma^ e}{\sqrt n}\r\Vert^2\nn\\
 &=\frac{M_\Lambda^4K_F+K_ e}{T}+ \frac{\mu_1^ e}{n} \le \frac{M_\Lambda^4K_F+K_ e}{T}+ \frac{M_{2 e}}{n}\le M_1\max\l(\frac 1T,\frac 1n\r),\;\text{say,}\nn
 \end{align}
 since $\Vert \bm\varepsilon_i\Vert=1$ and where $M_1$ is a finite positive real independent of $n$ and $T$ defined in Lemma \ref{lem:covarianze}(ii). This proves part (i).

For part (ii), notice that for all $t\in\mathbb Z$, by Assumptions \ref{ass:common}(a) and \ref{ass:common}(b), we must have:
\beq\label{eqfinitevarchi}
\max_{i=1,\ldots, n}\Var({C}_{it}) =\max_{i=1,\ldots, n}\bm\lambda_i^\prime\bm\Gamma^F\bm\lambda_i\le \Vert\bm\lambda_i\Vert^2\,\l\Vert\bm\Gamma^F\r\Vert
\le M_\Lambda^2 M_F.
\eeq
Now, by \citet[Theorem 7]{MK04}, 
\begin{align}\label{eqfinitevarchi2}
\max_{i=1,\ldots, n}\Var({C}_{it})&= \max_{i=1,\ldots, n}\sum_{j=1}^r \mu_j^{C} [\mbf V^{C}]_{ij}^2\ge  \mu_r^{C} \max_{i=1,\ldots, n}\sum_{j=1}^r  [\mbf V^{C}]_{ij}^2\nn\\
& \ge \frac{\mu_r\l(\bm\Lambda^\prime\bm\Lambda\r)}n \mu_r(\bm\Gamma^F)\ 
n \max_{i=1,\ldots, n}\Vert\mbf v_i^{C}\Vert^2 \ge  \frac{\mu_r\l(\bm\Lambda^\prime\bm\Lambda\r)}n m_F
\ n \max_{i=1,\ldots, n}\Vert\mbf v_i^{C}\Vert^2,
\end{align}
for some finite positive real $m_F$ independent of $T$, as implied by Assumption \ref{ass:common}(b).

Moreover, from Assumption \ref{ass:common}(a), 
\beq\label{eq:urla}
\lim_{n\to\infty} \frac{\mu_r\l(\bm\Lambda^\prime\bm\Lambda\r)}n = \mu_r (\bm\Sigma_\Lambda) \ge c_\Lambda,
\eeq
for some finite positive real $c_\Lambda$ independent of $n$ (see also Lemma \ref{lem:ranghi}).
Therefore, from \eqref{eqfinitevarchi}, \eqref{eqfinitevarchi2}, and \eqref{eq:urla}, 
\[
c_\Lambda m_F \lim_{n\to\infty} n \max_{i=1,\ldots, n}\Vert\mbf v_i^{C}\Vert^2 \le \lim_{n\to\infty} \max_{i=1,\ldots, n}\Var({C}_{it})\le M_\Lambda^2 M_F,
\]
which implies
\[
\lim_{n\to\infty} n \max_{i=1,\ldots, n}\Vert\mbf v_i^{C}\Vert^2 \le \frac{M_\Lambda^2 M_F}{c_\Lambda m_F}.
\]
By noticing that  $M_\Lambda$,  $M_F$, $c_\Lambda$, and $m_F$ are all finite positive reals, we prove part (ii).

Finally, using the same arguments in \eqref{eq:DK} in the proof of Lemma \ref{lem:covarianze},  using  part (i) and Lemma \ref{lem:Gxi}(iv), and since $\mu_0^{C}=\infty$ and $\mu_{r+1}^{C} =0$, 
\begin{align}
\lim_{n\to\infty}\max_{i=1,\ldots, n}
\min(n,T)\E\l[\Vert\sqrt n(\wh{\mbf v}_i^{x\prime}-\mbf v_i^{{C}\prime}\mbf J)\Vert^2\r]
&=\lim_{n\to\infty} \max_{i=1,\ldots, n}\min(n,T)\E\l[\Vert\sqrt n\bm\varepsilon_i^\prime(\wh{\mbf V}^x-\mbf V^{C}\mbf J)\Vert^2\r]\nn\\
&\le\lim_{n\to\infty}\max_{i=1,\ldots, n}\ \frac
{\min(n,T) 2^3 \frac r{n^2}  n \E\l[\l\Vert\bm\varepsilon_i^\prime(\wh{\bm\Gamma}^x-\bm\Gamma^{C}) \r\Vert^2\r]}
{\frac 1{n^2}\l\{\min(\vert \mu_{0}^{C}-\mu_{1}^{C} \vert,\vert \mu_{r}^{C}-\mu_{r+1}^{C} \vert)\r\}^2}\nn\\
&=\lim_{n\to\infty}\max_{i=1,\ldots, n}\ \frac
{\min(n,T) 2^3 \frac r{n} \E\l[\l\Vert\bm\varepsilon_i^\prime(\wh{\bm\Gamma}^x-\bm\Gamma^{C}) \r\Vert^2\r]}
{\frac 1{n^2}\l\{\min(\vert \mu_{0}^{C}-\mu_{1}^{C} \vert,\vert \mu_{r}^{C}-\mu_{r+1}^{C} \vert)\r\}^2}\nn\\
&\le \frac{8r M_1}{\underline C_r^2}= M_2,\nn
\end{align}
where $M_2$ is a finite positive real independent of $n$ and $T$  defined in Lemma \ref{lem:covarianze}(iv).
This proves part (iii) and completes the proof.  $\Box$

\begin{lem}\label{lem:MO1} Under Assumptions \ref{ass:common} and \ref{ass:idio}:
\begin{compactenum}[(i)]
\item for all $n\in\mathbb N$, $\Vert\frac{{\mbf M}^{C}}{n}\Vert =O(1)$;
\item as $n\to\infty$, $\Vert(\frac{{\mbf M}^{C}}{n})^{-1}\Vert=O(1)$, and if also Assumption \ref{ass:sign_eval}(b) holds this holds for all $n\in\mathbb N$;
\item for all $n,T\in\mathbb N$, $\Vert\frac{\wh{\mbf M}^x}{n}\Vert=O_{\mathrm{P}}(1)$;
\item as $n,T\to\infty$, $\Vert(\frac{\wh{\mbf M}^x}{n})^{-1}\Vert=O_{\mathrm{P}}(1)$.
\end{compactenum}
\end{lem}

\noindent
\textbf{Proof.} For part (i) from \eqref{eq:KUMAR} in the proof of Lemma \ref{lem:Gxi}, by Lemma \ref{lem:LLN}(iv)
 and Assumption \ref{ass:common}(b),
 \[
\l\Vert\frac{{\mbf M}^{C}}{n}\r\Vert =
\frac{\mu_1^{C}}{n} 
\le \mu_1\l(\frac{\bm\Lambda^\prime\bm\Lambda}n\r)\mu_1(\bm\Gamma^F)= \l\Vert\frac{\bm\Lambda^\prime\bm\Lambda}n \r\Vert\, \Vert \bm\Gamma^F\Vert \le M_\Lambda^2 M_F.
\]

For part (ii) from \citet[Theorem 7]{MK04}, by Lemma \ref{lem:ranghi}
 and Assumption \ref{ass:common}(b),
\[
\lim_{n\to\infty}\l\Vert\l(\frac{{\mbf M}^{C}}{n}\r)^{-1}\r\Vert =\lim_{n\to\infty} \frac n{\mu_r^{C}}\le \frac 1{\mu_r\l(\frac{\bm\Lambda^\prime\bm\Lambda}n\r) \mu_r(\bm\Gamma^F) } = \lim_{n\to\infty}\l\Vert\l(\frac{\bm \Lambda^\prime\bm \Lambda}{n} \r)^{-1}\r\Vert \, \l\Vert(\bm\Gamma^F)^{-1}\r\Vert \le \frac {m_F}{c_\Lambda},
\]
for some finite positive reals $m_F$ and $c_\Lambda$ independent of $T$ and $n$, respectively. If Assumption \ref{ass:sign_eval}(b) holds too, then 
the above holds without taking the limit for $n\to\infty$.

For part (iii), because of part (i) and Lemma \ref{lem:covarianze}(iii),
\begin{align}
\l\Vert\frac{\wh{\mbf M}^x}{n}\r\Vert\le 
\l\Vert\frac{{\mbf M}^{C}}{n}\r\Vert+
\l\Vert\frac{\wh{\mbf M}^x}{n}-\frac{{\mbf M}^{C}}{n}\r\Vert\le \overline C_1+O_{\mathrm P}\l(\max\l(\frac1{ n},\frac 1{\sqrt T}\r)\r).\nn
\end{align}

For part (iv), letting $\wh{\mu}_r^x$ be the $r$th largest eigenvalue of $\wh{\bm\Gamma}^x$, by Lemma \ref{lem:covarianze}(iii) and part (ii), 
\begin{align}
\l\Vert\l(\frac{\wh{\mbf M}^x}{n}\r)^{-1}\r\Vert &= \frac n{\wh{\mu}_r^x}= \frac 1{1+\frac{\wh{\mu}_r^x-\mu_r^C}{\mu_r^C}}\frac n{\mu_r^C}\nn\\
&= \frac n{\mu_r^C}\l\{1- \frac{n(\wh{\mu}_r^x-\mu_r^C)}{n \mu_r^C} \r\} + O_{\mathrm P}\l(\max\l(\frac 1{n^2},\frac 1T\r)\r)\nn\\
&=\frac n{\mu_r^C}+ \frac{n^2(\mu_r^C-\wh{\mu}_r^x)}{n(\mu_r^C)^2}+ O_{\mathrm P}\l(\max\l(\frac 1{n^2},\frac 1T\r)\r)\nn\\
&\le \frac 1{\underline C_r}+\frac{\vert \mu_r^C-\wh{\mu}_r^x\vert}{n \overline C_r^2}+ O_{\mathrm P}\l(\max\l(\frac 1{n^2},\frac 1T\r)\r)\nn\\
&\le \frac 1{\underline C_r}+O_{\mathrm P}\l(\max\l(\frac 1{n},\frac 1{\sqrt T}\r)\r).\nn
\end{align}
This completes the proof. $\Box$

\begin{lem}\label{lem:Vzero} Under Assumptions \ref{ass:common} and \ref{ass:idio}, 
\begin{compactenum}[(i)]
\item as $n\to\infty$, $\Vert\frac{\mbf M^{C}}{n}-\bm V_0 \Vert=o(1)$ and $\Vert \bm V_0\Vert = O(1)$;

\item as $n,T\to\infty$, $\Vert\frac{\wh{\mbf M}^x}{n}-\bm V_0 \Vert=o_{\mathrm {m.s.}}(1)$;

\item as $n\to\infty$, $\Vert(\frac{\mbf M^{C}}{n})^{-1}-\bm V_0^{-1} \Vert=o(1)$ and $\Vert \bm V_0^{-1}\Vert = O(1)$;
\item as $n,T\to\infty$, $\Vert(\frac{\wh{\mbf M}^x}{n})^{-1}-\bm V_0^{-1} \Vert=o_{\mathrm {m.s.}}(1)$;
\end{compactenum}
where $\bm V_0$ is $r\times r$ diagonal with entries the eigenvalues of $\bm\Sigma_\Lambda\bm\Gamma^F$ sorted in descending order.
If Assumption \ref{ass:common}(a) holds with rate $\sqrt n$ and Assumption \ref{ass:common}(c-ii) holds with rate $\sqrt T$, then the rate in parts (i) and (iii) is $\sqrt n$ and the rate in parts (ii) and (iv) is $\min(\sqrt n,\sqrt T)$.
\end{lem}

\noindent
\textbf{Proof.} For part (i), first notice that the $r$ non-zero eigenvalues of $\frac{\bm\Gamma^{C}}{n}$ are also the $r$ eigenvalues of \linebreak $(\bm\Gamma^F)^{1/2}\frac{\bm\Lambda^\prime\bm\Lambda}{n}(\bm\Gamma^F)^{1/2}$ which in turn are also the entries of $\bm V_0$. 
Then, by Weyl's inequality \citep[Theorem 1]{MK04}, because of Assumptions \ref{ass:common}(a) and \ref{ass:common}(b),
\[
\lim_{n\to\infty}\l\Vert\frac{\mbf M^{C}}{n}-\bm V_0 \r\Vert
\le 
\lim_{n\to\infty}\l\Vert(\bm\Gamma^F)^{1/2}\frac{\bm\Lambda^\prime\bm\Lambda}{n}(\bm\Gamma^F)^{1/2}
-(\bm\Gamma^F)^{1/2}\bm\Sigma_\Lambda(\bm\Gamma^F)^{1/2}
\r\Vert\le \l\Vert(\bm\Gamma^F)^{1/2}\r\Vert^2 \lim_{n\to\infty}\l\Vert \frac{\bm\Lambda^\prime\bm\Lambda}{n}-\bm\Sigma_\Lambda\r\Vert
=0,\nn
\]
and the rate is $\sqrt n$ if Assumption \ref{ass:common}(a) holds with rate $\sqrt n$. Moreover, by definition, $\bm V_0 = \bm\Upsilon_0^\prime (\bm\Gamma^F)^{1/2}\bm\Sigma_\Lambda(\bm\Gamma^F)^{1/2}  \bm\Upsilon_0$ (see also \eqref{eq:U0V0U0}). Then, by Assumptions \ref{ass:common}(a) and \ref{ass:common}(b),
\[
\Vert \bm V_0\Vert \le \Vert \bm\Gamma^F\Vert \, \Vert \bm\Sigma_\Lambda\Vert \le M_F M_\Lambda.
\]

Part (ii) is a consequence of part (i) and Lemma \ref{lem:covarianze}(iii). And if Assumption \ref{ass:common}(a) holds with rate $\sqrt n$ and  Assumption \ref{ass:common}(c-ii) holds with rate $\sqrt T$, so that Lemma \ref{lem:covarianze}(iii) holds with rate $\min(n,\sqrt T)$, the rate in part (ii) is $\min(\sqrt n,\sqrt T)$. 

For part (iii), first note that, by \citet[Theorem 7]{MK04},
\beq\label{eq:paint}
\Vert \bm V_0^{-1}\Vert =\frac 1{\mu_r(\bm\Sigma_\Lambda\bm\Gamma^F)}\le \frac 1{\mu_r(\bm\Sigma_\Lambda)\mu_r(\bm\Gamma^F)}\le \frac 1{m_\Lambda m_F},
\eeq
since $\bm\Sigma_\Lambda$ and $\bm\Gamma^F$ are positive definite by Assumptions \ref{ass:common}(a) and \ref{ass:common}(b).
Then,
 \[
 \l\Vert\l(\frac{\mbf M^{C}}{n}\r)^{-1}-\bm V_0^{-1} \r\Vert \le\l \Vert\l(\frac{\mbf M^{C}}{n}\r)^{-1}\r\Vert\,
 \l\Vert\frac{\mbf M^{C}}{n}-\bm V_0 \r\Vert
\, \l\Vert
\bm V_0^{-1}
 \r\Vert,
 \] 
 and the proof follows from part (i), Lemma \ref{lem:MO1}(ii), and \eqref{eq:paint}. The rate is the same as in part (i).
 
 For part (iv), 
 \[
 \l\Vert\l(\frac{\wh{\mbf M}^{x}}{n}\r)^{-1}-\bm V_0^{-1} \r\Vert \le\l \Vert\l(\frac{\wh{\mbf M}^{x}}{n}\r)^{-1}\r\Vert\,
 \l\Vert\frac{\wh{\mbf M}^{x}}{n}-\bm V_0 \r\Vert
\, \l\Vert
\bm V_0^{-1}
 \r\Vert,
 \]  
  then the proof follows from part (ii), Lemma \ref{lem:MO1}(iv), and \eqref{eq:paint}.  The rate is the same as in part (ii).
  This completes the proof. $\Box$

\begin{lem}\label{lem:KO1} 
Under Assumption  \ref{ass:common}, for all $n\in\mathbb N$,
\begin{compactenum}[(i)]
\item  $\Vert{\mbf K}\Vert=O(1)$;
\item  $\Vert{\mbf K}^{-1}\Vert=O(1)$;
\item $\mbf K^\prime \mbf K=\mbf I_r$ and its columns are the eigenvectors of $(\bm\Gamma^F)^{1/2}\frac{\bm\Lambda^\prime \bm\Lambda}n(\bm\Gamma^F)^{1/2}$;
\end{compactenum}
where $\mbf K$ is defined in \eqref{eq:kappaproj} in the proof of Proposition \ref{prop:L}.
\end{lem}

\noindent
\textbf{Proof.} From \eqref{eq:kappaproj}  in the proof of Proposition \ref{prop:L}, 
\begin{align}
&\mbf K=(\bm\Gamma^F)^{-1/2} (\bm\Lambda^\prime\bm\Lambda)^{-1}\bm\Lambda^\prime\mbf V^{C}(\mbf M^{C})^{1/2},\label{eq:kappa}\\
&\mbf K^{-1}= (\mbf M^{C})^{-1/2}{\mbf V^{{C}\prime}\bm\Lambda}(\bm\Gamma^F)^{1/2}.\label{eq:kappainv}
\end{align}
From \eqref{eq:kappa}
\beq
\mbf K=(\bm\Gamma^F)^{-1/2} \l(\frac{\bm\Lambda^\prime\bm\Lambda}n\r)^{-1}\frac{\bm\Lambda^\prime}{\sqrt n}\mbf V^{C}\l(\frac{\mbf M^{C}}n\r)^{1/2}.\nn
\eeq
Then, part (i) follows from Assumption \ref{ass:common}(b), Lemmas  \ref{lem:FTLN}(i), \ref{lem:ranghi}, and \ref{lem:MO1}(i), and since eigenvectors are normalized.


Let us then consider first to part (iii). From \eqref{eq:kappaproj} in the proof of Proposition \ref{prop:L}, 
$\mbf V^{{C}} = \bm\Lambda (\bm\Gamma^F)^{1/2}\mbf K ({\mbf M^{C}})^{-1/2}$, which, jointly with \eqref{eq:kappainv}, implies
\begin{align}
 \frac{\mbf M^{C}}n  \mbf K^{-1}&=
 \frac{\mbf M^{C}}n (\mbf M^{C})^{-1/2}{\mbf V^{{C}\prime}\bm\Lambda}(\bm\Gamma^F)^{1/2}\nn\\
&=
\frac{\mbf M^{C}}n ({\mbf M^{C}})^{-1}\mbf K^\prime(\bm\Gamma^F)^{1/2}\bm\Lambda^\prime \bm\Lambda(\bm\Gamma^F)^{1/2}\nn\\
&= \mbf K^\prime(\bm\Gamma^F)^{1/2}\frac{\bm\Lambda^\prime \bm\Lambda}n(\bm\Gamma^F)^{1/2}.
\label{eq:a34}
\end{align}
And, from \eqref{eq:kappa},
\begin{align}
\mbf K\mbf K^\prime &= (\bm\Gamma^F)^{-1/2} (\bm\Lambda^\prime\bm\Lambda)^{-1}\bm\Lambda^\prime\mbf V^{C}(\mbf M^{C})^{1/2}
(\mbf M^{C})^{1/2}\mbf V^{{C}\prime}\bm\Lambda(\bm\Lambda^\prime\bm\Lambda)^{-1}(\bm\Gamma^F)^{-1/2} 
\nn\\
&= (\bm\Gamma^F)^{-1/2} (\bm\Lambda^\prime\bm\Lambda)^{-1}\bm\Lambda^\prime\mbf V^{C}
\mbf M^{C}\mbf V^{{C}\prime}\bm\Lambda(\bm\Lambda^\prime\bm\Lambda)^{-1}(\bm\Gamma^F)^{-1/2} 
\nn\\
&= (\bm\Gamma^F)^{-1/2} (\bm\Lambda^\prime\bm\Lambda)^{-1}\bm\Lambda^\prime\bm\Gamma^{C}\bm\Lambda(\bm\Lambda^\prime\bm\Lambda)^{-1}(\bm\Gamma^F)^{-1/2} 
\nn\\
&= (\bm\Gamma^F)^{-1/2} (\bm\Lambda^\prime\bm\Lambda)^{-1}\bm\Lambda^\prime\bm\Lambda\bm\Gamma^F\bm\Lambda^\prime\bm\Lambda(\bm\Lambda^\prime\bm\Lambda)^{-1}(\bm\Gamma^F)^{-1/2} 
\nn\\
&=  (\bm\Gamma^F)^{-1/2} \bm\Gamma^F(\bm\Gamma^F)^{-1/2} 
=\mbf I_r,\nn
\end{align}
thus $\mbf K$ is an orthogonal matrix for all $n\in\mathbb N$.
Therefore, 
$\mbf K^\prime = \mbf K^{-1}$ and from \eqref{eq:a34}  we see that the columns of $\mbf K$ are the normalized eigenvectors of $(\bm\Gamma^F)^{1/2}\frac{\bm\Lambda^\prime \bm\Lambda}n(\bm\Gamma^F)^{1/2}$ with eigenvalues $\frac{\mbf M^{C}}n$, i.e.,
\beq\nn
 \frac{\mbf M^{C}}n  \mbf K^{\prime} = \mbf K^\prime(\bm\Gamma^F)^{1/2}\frac{\bm\Lambda^\prime \bm\Lambda}n(\bm\Gamma^F)^{1/2}.
\eeq
Finally, part (ii) follows from the fact that $\mbf K^{-1}=\mbf K^\prime$ by part (iii) and from part (i). This completes the proof. $\Box$

\begin{lem}\label{lem:HO1bis} Under Assumptions  \ref{ass:common} and \ref{ass:idio}, for all $n,T\in\mathbb N$,
\begin{compactenum}[(i)]
\item $\Vert{\bm{\mathcal H}}\Vert=O_{\mathrm P}(1)$;
\item $\Vert{\bm{\mathcal H}}^{-1}\Vert=O_{\mathrm P}(1)$;
\end{compactenum}
where $\bm{\mathcal H}$ is defined in Proposition \ref{prop:L}.
\end{lem}

\noindent
\textbf{Proof.} From \eqref{eq:mcHbis} in the proof of Proposition \ref{prop:L} $\bm{\mathcal H}=(\bm\Gamma^F )^{1/2}\mbf K\mbf J$. Then, part (i) and (ii) follow immediately from Assumption \ref{ass:common}(b), Lemma \ref{lem:KO1}(i), \ref{lem:KO1}(ii), and since $\mbf J$ is obviously finite and invertible for all $T\in\mathbb N$. This completes the proof. $\Box$

\begin{lem}\label{lem:HO1} Under Assumptions  \ref{ass:common} through \ref{ass:ind}, as $n,T\to\infty$,
\begin{compactenum}[(i)]
\item $\Vert\wh{\mbf H}\Vert=O_{\mathrm {P}}(1)$;
\item $\Vert\wh{\mbf H}^{-1}\Vert=O_{\mathrm {P}}(1)$;
\end{compactenum}
where $\wh{\mbf H}$ is defined in \eqref{eq:acca}.
\end{lem}

\noindent
\textbf{Proof.} From \eqref{eq:acca}, by Proposition \ref{prop:L}(a), Lemma \ref{lem:LLN}(v), \ref{lem:MO1}(iv) \ref{lem:HO1bis}(i), 
\begin{align}
\l\Vert\wh{\mbf H}\r\Vert &\le \l\Vert\frac{\bm F^\prime\bm F}{T}\r\Vert\,
\l\Vert\frac{\bm\Lambda^\prime\wh{\bm\Lambda}}{n}\r\Vert\,
\l\Vert\l(\frac{\wh{\mbf M}^x}{n}\r)^{-1}\r\Vert\le \l\Vert\frac{\bm F}{\sqrt T}\r\Vert^2\, \l\Vert\frac{\bm\Lambda}{\sqrt n}\r\Vert\, \l\Vert\frac{\wh{\bm\Lambda}}{\sqrt n}\r\Vert\,\l\Vert\l(\frac{\wh{\mbf M}^x}{n}\r)^{-1}\r\Vert\nn\\
&\le\l\Vert\frac{\bm F}{\sqrt T}\r\Vert^2\, \l\Vert\frac{\bm\Lambda}{\sqrt n}\r\Vert^2\, \l\Vert\bm{\mathcal H}\r\Vert\, \l\Vert\l(\frac{\wh{\mbf M}^x}{n}\r)^{-1}\r\Vert+ \l\Vert\frac{\bm F}{\sqrt T}\r\Vert^2\, \l\Vert\frac{\bm\Lambda}{\sqrt n}\r\Vert\, \l\Vert\frac{\wh{\bm\Lambda}-\bm\Lambda\bm{\mathcal H}}{\sqrt n}\r\Vert\,\l\Vert\l(\frac{\wh{\mbf M}^x}{n}\r)^{-1}\r\Vert\nn\\
&= 
O_{\mathrm P}(1)+ O_{\mathrm P}\l(\max\l(\frac 1{\sqrt n},\frac 1{\sqrt T}\r)\r),\nn
\end{align}
which proves part (i). 

For part (ii), 
\begin{align}
\l\Vert\wh{\mbf H}^{-1}\r\Vert &\le 
\l\Vert\frac{\wh{\mbf M}^x}{n}\r\Vert
\l\Vert\l(\frac{\bm\Lambda^\prime\wh{\bm\Lambda}}{n}\r)^{-1}\r\Vert\,
\l\Vert\l(\frac{\bm F^\prime\bm F}{T}\r)^{-1}\r\Vert.\label{eq:HINVapp}
\end{align}
Now, because of Proposition \ref{prop:L}(a),  Lemma \ref{lem:FTLN}(i), \ref{lem:HO1bis}(i), and Assumption \ref{ass:common}(a),
\[
\l\Vert \frac{\bm\Lambda^\prime\wh{\bm\Lambda}}{n}-\bm\Sigma_\Lambda\bm{\mathcal H}\r\Vert  \le
\l\Vert \frac{\bm\Lambda^\prime\wh{\bm\Lambda}}{n}-\frac{\bm\Lambda^\prime{\bm\Lambda}\bm{\mathcal H}}{n}
\r\Vert+
\l\Vert\frac{\bm\Lambda^\prime{\bm\Lambda}\bm{\mathcal H}}{n}-
\bm\Sigma_\Lambda\bm{\mathcal H}\r\Vert  = o_{\mathrm P}(1).
\]
And since by Assumption \ref{ass:common}(a) and Lemma \ref{lem:HO1bis}(ii), $\bm\Sigma_\Lambda\bm{\mathcal H}$ is invertible as $n,T\to\infty$, then  $\frac{\bm\Lambda^\prime\wh{\bm\Lambda}}{n}$ is invertible with probability tending to one, as $n,T\to\infty$, i.e.,
\beq\label{eq:LLHATINV}
\l\Vert\l(\frac{\bm\Lambda^\prime\wh{\bm\Lambda}}{n}\r)^{-1}\r\Vert = O_{\mathrm P}(1).
\eeq
Then, because of Lemmas \ref{lem:FFinv} and \ref{lem:MO1}(iii),  and \eqref{eq:LLHATINV}, from \eqref{eq:HINVapp}, we prove part (ii). 

Alternatively, notice that the limiting quantities of all three terms in $\wh{\mbf H}$ are finite and invertible because of Assumption \ref{ass:common}(b) and  \ref{ass:common}(c-ii), Proposition \ref{prop:KKK}, and Lemma \ref{lem:covarianze}(iii) (jointly with Lemmas \ref{lem:Vzero}(i) and \ref{lem:Vzero}(iii)). This completes the proof. $\Box$

\end{document}